\documentclass[symmetry,review,accept,moreauthors,pdftex,12pt,a4paper]{mdpi}
\firstpage{1}
\articlenumber{70}
\doinum{10.3390/sym8080070}
\pubvolume{8}
\pubyear{2016}
\copyrightyear{2016}
\externaleditor{Academic Editor: Sergei D. Odintsov}
\history{Received: 13 May 2016 / Accepted: 11 July 2016 / Published: 26 July 2016}

\newcommand{\lb}{\left[}
\newcommand{\rb}{\right]}

\newcommand{\ba}{\begin{eqnarray}}
\newcommand{\ea}{\end{eqnarray}}
\newcommand{\be}{\begin{equation}}
\newcommand{\ee}{\end{equation}}

\newcommand{\al}{\alpha}
\newcommand{\bt}{\beta}

\newcommand{\ka}{\kappa}

\newcommand{\R}{\mathcal{R}}

\newcommand{\vv}{\bar{v}}
\newcommand{\cc}{\tilde{c}}
\usepackage{amssymb}
\usepackage{soul}
\usepackage{color}

\def\widebar{\overline}

\Title{Unveiling the Dynamics of the Universe}

\Author{Pedro Avelino $^{1,2}$, Tiago Barreiro $^{3,4}$, C. Sofia Carvalho $^{5,6}$,
Antonio da Silva $^{3,7}$, \mbox{Francisco S.N. Lobo $^{3,7,}$*,} Prado Mart\'{\i}n-Moruno $^{8}$, Jos\'{e} Pedro Mimoso $^{3,7}$, Nelson J. Nunes $^{3,7}$, Diego Rubiera-Garc\'{i}a $^{3}$, Diego S\'{a}ez-G\'{o}mez $^{3}$, Lara Sousa $^{1}$, Ismael Tereno $^{5,7}$ and \mbox{Arlindo Trindade $^{1}$}
}

\address{%
\textsuperscript{1} \quad Instituto de Astrof\'{\i}sica e Ci\^encias do Espa{\c c}o, Universidade do Porto, CAUP, Rua das Estrelas, PT4150-762 Porto, Portugal; pedro.avelino@astro.up.pt (P.A.); Lara.Sousa@astro.up.pt (L.S.); arlindo.trindade@gmail.com (A.T.)\\

\textsuperscript{2} \quad Departamento de F\'{\i}sica e Astronomia, Faculdade de Ci\^encias, Universidade do Porto, Rua do Campo Alegre 687, PT4169-007 Porto, Portugal\\
\textsuperscript{3} \quad Instituto de Astrof\'isica e Ci\^encias do Espa\c{c}o, Departamento de F\'isica
da Faculdade de Ci\^encias da Universidade de Lisboa, Edif\'icio C8, Campo Grande, P-1749-016
Lisbon, Portugal; tmbarreiro@gmail.com (T.B.); ajosilva@fc.ul.pt (A.S.); prado.moruno@gmail.com (P.M.-M.); jpmimoso@fc.ul.pt (J.P.M.); njnunes@fc.ul.pt (N.J.N.); drgarcia@fc.ul.pt (D.R.-G.); dsgomez@fc.ul.pt (D.S.-G.)\\
\textsuperscript{4} \quad Departamento de Matem\'{a}tica, ECEO, Universidade Lus\'{o}fona de Humanidades e Tecnologias, Campo Grande, 376, 1749-024 Lisboa, Portugal\\
\textsuperscript{5} \quad Instituto de Astrof\'isica e Ci\^encias do Espa\c{c}o, Universidade de Lisboa,
Tapada da Ajuda, 1349--018 Lisboa, Portugal; cscarvalho@oal.ul.pt (C.S.C.); iatereno@fc.ul.pt (I.T.)\\
\textsuperscript{6} \quad Research Center for Astronomy and Applied Mathematics, Academy of Athens, Soranou Efessiou 4, 11--527 Athens, Greece \\
\textsuperscript{7} \quad Departamento de F\'{\i}sica, Faculdade de
Ci\^encias da Universidade de Lisboa, Edif\'{\i}cio C8, Campo Grande,
P-1749-016 Lisboa, Portugal\\
\textsuperscript{8} \quad Departamento de F\'{\i}sica Te\'{o}rica I, Universidad Complutense de Madrid, E-28040 Madrid,
Spain

}

\corres{Correspondence: fslobo@fc.ul.pt; ; Tel.: +351-217-500-484; Fax: 351-217-500-977.}

\abstract{We explore the dynamics and evolution of the Universe at early and late times, focusing on both dark energy and extended gravity models and their astrophysical and cosmological consequences. Modified theories of gravity not only provide an alternative explanation for the recent expansion history of the universe, but they also offer a paradigm fundamentally distinct from the simplest dark energy models of cosmic acceleration. In this review, we perform a detailed theoretical and phenomenological analysis of different modified gravity models and investigate their consistency. We also consider the cosmological implications of well motivated physical models of the early universe with a particular emphasis on inflation and topological defects. Astrophysical and cosmological tests over a wide range of scales, from the solar system to the observable horizon, severely restrict the allowed models of the Universe. Here, we review several observational probes---including gravitational lensing, galaxy clusters, cosmic microwave background temperature and polarization, supernova and baryon acoustic oscillations measurements---and their relevance in constraining our cosmological description of the Universe.
}


\begin{document}

\newpage

\section{Introduction}

During the last few decades Cosmology has evolved from being mainly a theoretical area of physics to become a field supported by high precision observational data. Recent experiments call upon state of the art technology in Astronomy and Astrophysics to provide detailed information on the contents and history of the Universe, which has led to the measurement of parameters that describe our Universe with increasing precision. The standard model of Cosmology is remarkably successful in accounting for the observed features of the Universe. However, a number of fundamental open questions remains at the foundation of the standard model. In particular, we lack a fundamental understanding of the recent acceleration of the Universe \cite{Perlmutter:1998np,Riess:1998cb}. What is the so-called ``dark energy'' that is driving the cosmic acceleration? Is it vacuum energy or a dynamical field? Or is the acceleration due to infra-red modifications of Einstein's theory of General Relativity (GR)? How is structure formation affected in these alternative scenarios? What are the implications of this acceleration for the future of the Universe?

The resolution of these fundamental questions is extremely important for theoretical cosmology. Dark energy models are usually assumed to be responsible for the acceleration of the cosmic expansion in most cosmological studies. However, it is clear that these questions involve not only gravity, but also particle physics. String theory provides a synthesis of these two branches of physics and is widely believed to be moving towards a viable quantum gravity theory. One of the key predictions of string theory is the existence of extra spatial dimensions. In the brane-world scenario, motivated by recent developments in string theory, the observed 3-dimensional universe is embedded in a higher-dimensional spacetime \cite{Maartens:2003tw}. The new degrees of freedom belong to the gravitational sector, and can be responsible for the late-time cosmic acceleration \cite{Dvali:2000hr,deRham:2007rw}. On the other hand, generalizations of the Einstein-Hilbert Lagrangian, including quadratic Lagrangians which involve second order curvature invariants have also been extensively explored \cite{Sotiriou:2008rp,DeFelice:2010aj,Capozziello:2011et,Nojiri:2010wj,Lobo:2008sg}. These modified theories of gravity not only provide an alternative explanation for the expansion history of the Universe \cite{Capozziello:2002rd,Nojiri:2003ft,Carroll:2003wy}, but they also offer a paradigm fundamentally distinct from the simplest dark energy models of cosmic acceleration \cite{Copeland:2006wr}, even from those that perfectly mimic the same expansion history. Nevertheless, it has been realized that a large number of modified gravity theories are amenable to a scalar-tensor formulation by means of appropriate metric re-scalings and field redefinitions.

It is, therefore, not surprising that we can think of scalar-tensor gravity theories as a first stepping stone to explore modifications of GR. They have the advantage of apparent simplicity and of a long history of examination. First proposed in its present form by Brans and Dicke for a single scalar field \cite{BD 61}, they have been extensively generalised and have maintained the interest of researchers until the present date. For instance, an extensive field of work has been developed in the cosmological dynamics of scalar-tensor theories. This can be elegantly summarised by carrying out a unified qualitative analysis of the dynamical system for a single scalar field. We also have an established understanding of the observational bounds in these models, where we can use the  Parametrized Post-Newtonian formalism to constrain the model parameters. Finally, with a conformal transformation, these theories can be recast  as matter interacting scalar fields in General Relativity. In this format, they can still play an important role in dark energy modelling, such as in coupled quintessence models. The consideration of a multi-scalar fields scenario, which can be perceived as the possible reflection of a multi-scalar tensor gravity theory allows for cooperative effects between the fields yielding assisted quintessence.

Indeed, scalar fields are popular building blocks used to construct models of present-day cosmological acceleration. They are appealing because such fields are ubiquitous in theories of high energy physics beyond the standard model and, in particular, are present in theories which include extra spatial dimensions, such as those derived from string theories. Recently, relative to scalar-tensor theory, much work has been invested in the Galileon models and their generalizations \cite{deRham:2014zqa}. The latter models allow nonlinear derivative interactions of the scalar field in the Lagrangian and lead to second order field equations, thus removing any ghost-like instabilities. The Lagrangian was first written down by Horndeski in 1974 \cite{Horndeski:1974wa}, which contains four arbitrary functions of the scalar field and its kinetic energy. The form of the Lagrangian is significantly simplified by requiring specific self-tuning properties (though it still has four arbitrary functions), however, the screening is too effective, and will screen curvature from other matter sources as well as from the vacuum energy \cite{Charmousis:2011ea}. An alternative approach consists of searching for a de Sitter critical point for any kind of material content \cite{stmodels}. These models might alleviate the cosmological constant problem and can deliver a background dynamics compatible with the latest observational data.

A promising alternative to explain the late-time cosmic acceleration is to assume that at large scales Einstein's theory of GR breaks down, and a more general action describes the gravitational field. Thus, one may generalize the Einstein-Hilbert action by including second order curvature invariants such as $R^2$, $R^{\mu\nu}R_{\mu\nu}$, $R^{\mu\nu\alpha\beta}R_{\mu\nu\alpha\beta}$, $C^{\mu\nu\alpha\beta}C_{\mu\nu\alpha\beta}$, etc. Some of the physical motivations for these modifications of gravity were inspired on effective models raised in string theory, which indeed may lead to the possibility of a more realistic representation of the gravitational fields near curvature singularities \cite{Nojiri:2003rz}. Moreover, the quantization of fields in curved space-times tell us that the high-energy completion of the Einstein-Hilbert Lagrangian of GR involves higher-order terms on the curvature invariants above \cite{QFT}. This is in agreement with the results provided from the point of view of treating GR as an effective field theory \cite{Cembranos-effective}. Among these extensions of GR the so-called $f(R)$ gravity has drawn much attention over the last years, since it can reproduce late-time acceleration and in spite of containing higher order derivatives, it is free of the Ostrogradsky instability, as can be shown by its equivalence with scalar-tensor theories (for a review on $f(R)$ gravity see Refs.\cite{DeFelice:2010aj,Sotiriou:2008rp,Capozziello:2011et,Nojiri:2010wj}). 
Moreover, $f(R)$ gravities have been also proposed as solutions for the inflationary paradigm \cite{Bamba:2015uma}, where the so-called Starobinsky model is a successful proposal, since it satisfies the latest constraints released by Planck \cite{Planck:2013jfk}. In addition, the equivalence of $f(R)$ gravities to some class of scalar-tensor theories has provided an extension of the so-called chameleon mechanism to $f(R)$ gravity, leading to some viable extensions of GR that pass the solar system constraints \cite{Hu:2007nk,Nojiri:2007as}.
Other alternative formulations for these extensions of GR have been considered in the literature, namely, the Palatini formalism, where metric and affine connection are regarded as independent degrees of freedom, which yields an interesting phenomenology for Cosmology \cite{Olmo}; and the metric-affine formalism, where the matter part of the action now depends and is varied with respect to the connection  \cite{Sotiriou:2006qn}. Recently, a novel approach to modified theories of gravity was proposed that consists of adding to the Einstein-Hilbert Lagrangian an $f({\cal R})$ term constructed {\it a la} Palatini \cite{Harko:2011nh}. It was shown that the theory can pass the Solar System observational constraints even if the scalar field is very light. This implies the existence of a long-range scalar field, which is able to modify the cosmological and galactic dynamics, but leaves the Solar System unaffected. 

Note that these modified theories of gravity are focussed on extensions of the curvature-based Einstein-Hilbert action. Nevertheless, one could equally well modify gravity starting from its torsion-based formulation and, in particular, from the Teleparallel Equivalent of General Relativity (TEGR) \cite{Linder:2010py}. The interesting point is that although GR is completely equivalent with TEGR at the level of equations, their modifications (for instance $f(R)$ and $f(T)$ gravities, where $T$ is the torsion) are not equivalent and they correspond to different classes of gravitational modifications. Hence, $f(T)$ gravity has novel and interesting cosmological implications, capable of describing inflation, the late-time acceleration, large scale structure, bouncing solutions, non-minimal couplings to matter, etc \cite{Cai:2015emx,Harko:2014sja,Harko:2014aja}.

Another gravitational modification that has recently attracted much interest is the massive gravity paradigm, where instead of introducing new scalar degrees of freedom, such as in $f(R)$ gravity, it modifies the graviton itself. Massive gravity is a well-defined theoretical problem on its own and has important cosmological motivations, namely, if gravity is massive it will be weaker at large scales and thus one can obtain the late-time cosmic acceleration. Fierz and Pauli presented the first linear massive gravity. However, it was shown to suffer from the van Dam-Veltman-Zakharov (vDVZ) discontinuity \cite{vanDam:1970vg,Zakharov:1970cc}, namely the massless limit of the results do not yield the massless theory, namely, GR. The incorporation of nonlinear terms cured the problem but introduced the Boulware-Deser (BD) ghost. This fundamental problem puzzled physicists until recently, where a specific nonlinear extension of massive gravity was proposed by de Rham, Gabadadze and Tolley (dRGT), in which the BD ghost is eliminated by a Hamiltonian constraint \cite{deRham:2014zqa}. This new nonlinear massive gravity has interesting cosmological implications, for instance, it can give rise to inflation, late-time acceleration \cite{deRham:2014zqa}. However, the basic versions of this theory exhibit instabilities at the perturbative level, and thus suitable extensions are necessary. These could be anisotropic versions, $f(R)$ extensions, bigravity generalizations, partially-massive constructions. The crucial issue is whether one can construct a massive gravity and cosmology that can be consistent as an alternative to dark energy or other models of modified gravity, and whether this theory is in agreement with high-precision cosmological data, such as the growth-index or the tensor-to-scalar ratio, remains to be explored in detail.

Quantum field theory predicts that the universe underwent, in its early stages, a series of symmetry breaking phase transitions, some of which may have led to the formation of topological defects. Different types of defects may be formed depending on the (non-trivial) topology of the vacuum manifold or the type of symmetry being broken. For instance, {\it Domain Walls}---which are surfaces that separate domains with different vacuum expectation values---may arise due to the breaking of a discrete symmetry, whenever the vacuum manifold is disconnected.  Line-like defects, or \textit{Cosmic strings}, are formed if the vacuum is not simply connected or, equivalently, if it contains unshrinkable loops. This type of vacuum manifold results, in general, from the breaking of an axial symmetry. Moreover, if the vacuum manifold contains unshrinkable surfaces, the field might develop non-trivial configurations corresponding to point-like defects, known as \textit{Monopoles}. The spontaneous symmetry breaking of more complex symmetry groups may lead to the formation of textures, delocalized topological defects which are unstable to collapse. The production of topological defects networks as remnants of symmetry breaking phase transitions is thus predicted in several grand unified scenarios and in several models of inflation. Moreover, recent developments in the braneworld realization of string theory suggest that its fundamental objects---p-dimensional D-branes and fundamental strings---may play the cosmological role of topological defects.

Topological defects networks, although formed in the early universe, may in most instances survive throughout the cosmological history and leave a variety of imprints on different observational probes. The observational consequences of topological defect networks can be very diverse, depending both on the type of defects formed and on the evolution of the universe after they are generated. Although the possibility of a significant contribution to the dark energy budget has been ruled out both dynamically \cite{PinaAvelino:2006ia} and observationally \cite{Ade:2015xua}, light domain walls may leave behind interesting astrophysical and cosmological signatures. For instance, they may be associated to spatial variations of the fundamental couplings of nature (see, e.g., \cite{Avelino:2014xsa}). On the other hand, cosmic strings may contribute significantly to small-scale cosmological perturbations and have consequently been suggested to have significant impact on the formation of ultracompact minihalos \cite{Anthonisen:2015tda}, globular clusters \cite{Barton:2015zra}, super-massive black holes \citep{Bramberger:2015kua} and to provide a significant contribution to the reionization history of the Universe \cite{Avelino:2003nn}. Both cosmic strings and domain walls may be responsible for significant contributions to two of the most significant observational probes: the temperature and polarization anisotropies of the cosmic microwave background and the stochastic gravitational wave background. This fact---alongside the possibility of testing string theory through the study of topological defects---greatly motivates the interest on the astrophysical and cosmological signatures of topological defects.

An extremely important aspect of modern cosmology is the synergy between theory and observations. Dark energy models and modified gravity affect the geometry of the universe and cosmological structure formation, impacting the background expansion and leaving an imprint on the statistical properties of the large-scale structure. There are a number of well-established probes of cosmic evolution, such as type Ia supernovae, baryon acoustic oscillations (BAO), weak gravitational lensing, galaxy clustering and galaxy clusters properties \cite{weinberg:probes}. Different methods measure different observables, probing expansion and structure formation in different and often complementary ways and have different systematic effects. In particular, joint analyses with Cosmic Microwave Background (CMB) data are helpful in breaking degeneracies by constraining the standard cosmological parameters.  Indeed, CMB has revolutionized the way we perceive the Universe. The information encoded in its temperature and polarization maps provides one of the strongest evidences in favour of the hot Big-Bang theory and has enabled ways to constrain cosmological models with unprecedented accuracy \citep{2015arXiv150201589P}.

The CMB also encodes additional information about the growth of cosmological structure and the dynamics of the Universe through the secondary CMB anisotropies. These are originated by physical effects acting on the CMB after decoupling \citep{2008RPPh...71f6902A}, such as the integrated Sachs-Wolfe effect and the Sunyaev-Zel'dovich (SZ) effect, manifest respectively on the largest and arc-minute scales of the CMB.
In this review, we will discuss in some detail both a well-established acceleration probe (weak lensing) and a few promising ones related to galaxy cluster properties and the SZ effect. Galaxy clusters are the largest gravitationally bound objects in the Universe and are among the latest bound structures forming in the Universe. For this reason, their number density is highly sensitive to the details of structure formation as well as to cosmological background parameters and cluster abundance is a well established cosmological probe. The Sunyaev-Zel'dovich effect \cite{1972CoASP...4..173S, 1999PhR...310...97B} is the scattering of CMB photons by electrons in hot reservoirs of ionized gas in the Universe, such as galaxy clusters. In particular, SZ galaxy cluster counts, profiles, scaling relations, angular power spectra and induced spectral distortions are promising probes to confront model predictions with observations. Weak gravitational lensing \cite{bands} describes the deflection of light by gravity in the weak regime. Its angular power spectrum is a direct measure of the statistical properties of gravity and matter on cosmological scales. Weak lensing, together with galaxy clustering, is the core method of the forthcoming Euclid mission to map the dark universe. Euclid \cite{redbook} will provide us with weak lensing measurements of unprecedented precision. To obtain high-precision and high-accuracy constraints on dark energy or modified gravity properties, with both weak lensing and SZ clusters, non-linearities on structure formation must be taken into account. While linear covariant perturbations equations may be evolved with Boltzmann codes, non-linearities require dedicated N-body cosmological simulations \cite{nbodymodgrav}. There currently exist a number of simulations for various modified gravity and dark energy models, together with a set of formulas that fit a non-linear power spectrum from a linear one. Hydrodynamic simulations, commonly used in cluster studies, are increasingly needed in weak lensing applications to model various baryonic effects on lensing observables, such as supernova and AGN feedback, star formation or radiative cooling \cite{semboloni}.

The $\Lambda$CDM framework provides a very good fit to various datasets, but it contains some open issues \cite{bulloslo}. As an example, there are inconsistencies between probes, such as the tension between CMB primary signal (Planck) and weak lensing (CFHTLenS) \cite{joudaki2016}, as well as problems with the interpretation of large-scale CMB measurements (the so-called CMB anomalies) \cite{anomalies}. Alternatives to $\Lambda$CDM or deviations to General Relativity are usually confronted with data using one of two approaches: model selection or parameterizations. In model selection a specific model is analyzed and its parameters constrained. Such analyses have a narrower scope but may be better physically motivated. Parameterizations are good working tools and are helpful in highlighting a particular feature and in ruling out larger classes of models, however they must be carefully defined in a consistent way. Parameterizations are commonly applied to the dark energy equation of state and to deviations from General Relativity. An example of the latter is the gravitational slip, which provides an unambiguous signature of modified gravity, and can be estimated combining weak lensing measurements of the lensing potential with galaxy clustering measurements of the Newtonian potential. Model parameters in both of the approaches are usually estimated with Monte Carlo techniques, while the viability of different models may be compared using various information criteria.

Besides model testing, cosmological data is also useful to test foundational assumptions, such as the (statistical) cosmological principle and the inflationary paradigm. The common understanding is that cosmological structures are the result of primordial density fluctuations that grew under gravitational instability collapse. These primordial density perturbations would have originated during the inflationary phase in the early universe. Most single field slow-roll inflationary models produce nearly Gaussian distributed perturbations, with very weak possible deviations from Gaussianity at a level beyond detection \cite{2003NuPhB.667..119A,2003JHEP...05..013M,2004PhR...402..103B}. However, non-Gaussianities may arise in inflationary models in which any of the conditions leading to the slow-roll dynamics fail \cite{2009astro2010S.158K}, such as the curvaton scenario
\cite{2004PhRvD..69d3503B,2006PhRvD..74j3003S,2002PhLB..524....5L}, the ekpyrotic inflacionary scenario \cite{2010AdAst2010E..67L,2008PhRvD..77f3533L}, vector field populated inflation \cite{2008JCAP...08..005Y,2009PhRvD..80b3509K, 2010AdAst2010E..65D} and multi-field inflation \cite{2010AdAst2010E..76B,1998PhLB..422...52M,1994PhRvD..50.6123P,2006PhRvD..73h3522R}. Tests of non-Gaussianity are thus a way to discriminate between inflation models and to test the different proposed mechanisms for the generation of primordial density perturbations. Likewise, the assumption of statistical homogeneity may be tested. Locally, matter is distributed according to a pattern of alternate overdense regions and underdense regions. Since averaging inhomogeneities in the matter density distribution yields a homogeneous description of the Universe, then the apparent homogeneity of the cosmological parameters could also result from the averaging of inhomogeneities in the cosmological parameters, which would reflect the inhomogeneities in the density distribution. The theoretical setup closest to this reasoning is that of backreaction models, where the angular variations in the parameters could also source a repulsive force and potentially emulate cosmic acceleration. Hence the reasoning is to look for these inhomogeneities, not in depth, but rather across the sky and then to use an adequate toy model to compute the magnitude of the acceleration derived from angular variations of the parameters compared to the acceleration driven by a cosmological constant \cite{carvalho_2015}.

This work will focus on all of the above-mentioned topics. More specifically, this article is outlined in the following manner: In Section \ref{sec2}, we present scalar-tensor theories, and in Section \ref{sec3}, we consider Horndeski theories and the self-tuning properties. In Section \ref{sec4}, $f(R)$ modified theories of gravity and extensions are reviewed. In Section \ref{sec5}, an extensive review on topological defects is carried out. The following sections are dedicated to observational cosmology. In particular, in Section \ref{sec6}, cosmological tests with galaxy clusters at CMB frequencies are presented and in Section \ref{sec7}, gravitational lensing will be explored. In Section \ref{sec8}, the angular distribution of cosmological parameters as a measurement of spacetime inhomogeneities will be presented. Finally, in Section \ref{sec:concl} we conclude.

\section{Scalar-tensor Theories}\label{sec2}

{\textit {Scalar-tensor theories of gravity}}  stem from the original proposal of Brans-Dicke theory (BD)~\cite{BD 61}, although theories involving a scalar field with a gravitational role in addition to the tensor metric fields can be traced back to the earlier Kaluza-Klein type theories upon dimensional reduction \cite{Billyard:1997yb}. One notorious example was Jordan's  proposal for a field theory realization of Dirac's Large Number Hypothesis~\cite{Barrow:1988yia}. In the course of time, it has been realized  that a large number of modified gravity theories are amenable to a scalar-tensor formulation~\cite{Wands:1993uu} by means of appropriate metric re-scalings and field redefinitions.
Interestingly, this possibility of a scalar-tensor description occurs regardless of the field equations being derived through the metric or, alternatively, the Palatini variational prescription \cite{Harko:2011nh}.
For instance, it is well known that the $f(R)$ higher-order gravity theories can be cast as a scalar-tensor theory \cite{Wands:1993uu,Conf_equiv-fR,Conf_equiv-2}. The same happens for the theories based on Lagrangeans with non-linear kinetic terms of a scalar field dubbed $k$-essence~\cite{Malquarti:2003nn}.

\subsection{General Formalism}

The general {\it scalar-tensor} class of gravity theories encompassing BD's and a plethora of modified theories that we just alluded to, were first examined by Bergmann \cite{Bergm 68}, Wagoner \cite{Wagoner 70}, and Nordvedt \cite{Nordt 70} (BWN). Their fundamental action can be cast in the following form \cite{Will:2014xja}
\begin{equation}
S= \int \sqrt{-g}\, d^4x\, \left[\phi R -
\frac{\omega(\phi)}{\phi}\,
\phi_{,\mu} \phi^{,\mu} - 2 \phi \lambda(\phi)\right] + S_M \,,
\label{eSTTaction}
\end{equation}
where $R\,$ is the usual Ricci curvature scalar of the spacetime,
$\phi\,$ is a scalar field, $\omega(\phi)\,$ is a dimensionless function
of $\phi\,$ which calibrates the coupling between the scalar field and
gravity (in what follows we shall refer to it as the \textit{coupling
parameter}), $\lambda(\phi)\,$ is another  function of the scalar field
which can be interpreted both as a potential for $\phi\,$ and as a
cosmological parameter and, finally, $S_M\,$ represents the action
for the matter fields. It is worth mentioning at this point that the action
should  include an additional term involving the extrinsic curvature,
as in GR \cite{Wald:1984rg}, to account for a boundary term associated
with the variation of $R_{\alpha \beta}\,$ with respect to $g_{\alpha \beta}$. A
discussion of this term in the scalar-tensor theories is provided in 
\cite{Wands:1993uu}. It becomes immediately
clear that the scalar field $\phi$ plays  the  role which in Einstein's GR is
confined to the gravitational constant, and has dimensions of a mass
squared, $\phi\sim 1/G$. Since $\phi\,$ is now a
dynamical variable,
we thus see that the trademark of these  theories is the fact that they exhibit a varying gravitational
``constant''. In the archetypal Brans-Dicke theory the coupling $\omega(\phi)$ is constrained to be a constant.

The consideration of couplings between a scalar field and the spacetime geometry envisaged by this class of theories allows a generalization to multiple scalar fields so that we may consider {\it multi-scalar-tensor theories}. The Lagrangian can then be cast as
\begin{equation}
L_g = f(\phi^\mu)\, R -
\frac{\omega_{\mu\nu}(\phi^\lambda)}{f(\phi^\mu)}\, g^{\alpha\beta}
\nabla_\alpha\phi^\mu \nabla_\beta\phi^\nu +  V(\phi^\mu) + 16\pi \, L_m \,,
\label{e:MSTaction}
\end{equation}
where $R$ is the usual Ricci curvature scalar of a space-time
endowed with the metric $g_{\alpha\beta}$,  $\phi^\mu$ ($\mu=1,\ldots,n$)
represents $n$ scalar fields, $f(\phi^\mu)$ is a function of
those scalar fields, $\omega_{\mu\nu}(\phi^\beta)$ is the metric of the
internal space of the $\phi^\mu$ fields, $V(\phi^\mu)$ is the scalar
potential. The Lagrangian of the matter fields $L_m$ does
not carry any explicit dependence on $\phi^\mu$ which guarantees
that free-falling test particles follow the geodesics of the
spacetime (this amounts to keeping valid the conservation of the energy-momentum tensor 
$\nabla_\beta T^{\alpha\beta} =0$). We leave aside  a detailed consideration of this extended class of theories, which have been addressed in \cite{Damour+Nordtvedt 93,Dam+E-Farese 92}. However, in subsection \ref{assist_infl} we shall analyse a model of quintessence stemming from the cooperative behaviour of several scalar fields coupled to several matter components which as it will be argued can be interpreted as a particular realisation of the models characterised by Equation (\ref{e:MSTaction}). 

Taking the variational derivatives of the action (\ref{eSTTaction}) with respect to the two dynamical  variables $g_{\alpha \beta}\,$ and $\phi\,$, yields the field equations
\begin{eqnarray}
R_{\alpha \beta} - \frac{1}{2}\,g_{\alpha \beta}\, R+\lambda(\phi)\,g_{\alpha \beta} &=&\frac{\omega(\phi)}{\phi^2}
\left[ \nabla_{\alpha}\phi  \nabla_{\beta}\phi  -\frac{1}{2}\, g_{\alpha \beta} \nabla_{\gamma}\phi  \nabla^{\gamma}\phi  \right]
\nonumber\\
&& + \frac{1}{\phi}\,\left[ \nabla_{\beta} \nabla_{\alpha}\phi -g_{\alpha \beta} \nabla_{\gamma}\phi  \nabla^{\gamma}\phi  \right]+
8\pi G_N\,{T_{\alpha \beta}\over \phi} \label{eSTTFE1} ,
\end{eqnarray}
\begin{eqnarray}
\Box{\phi} - \frac{2\phi^2 \lambda'(\phi)-2\phi \lambda(\phi)} {2\omega(\phi)+3}
 = \frac{1}{2\omega(\phi)+3}\left[ 8\pi G_N\, T-\omega'(\phi)  \nabla_{\gamma}\phi  \nabla^{\gamma}\phi \right]\,, \label{eSTTFE2}
\end{eqnarray}
where $T\equiv {T^\gamma}_\gamma \,$ is the trace of the energy-momentum tensor,
${T_\alpha}^\beta$, of the matter content of spacetime, and $G_N\,$ is
the gravitational constant normalized to its present value (in what follows in this section, we set $G_N=1\,$).

From these equations the role of $\lambda(\phi)$ becomes more apparent. As pointed out in \cite{Bergm 68}, this term enters equations (\ref{eSTTFE1}) as a cosmological function, but intervenes in the second equation (\ref{eSTTFE2}) as a mass term. A consequence of this latter feature is that the scalar waves propagate with a speed different from the speed of the electromagnetic waves. Another interesting aspect to point out in connection with this term is that the inverse square law for the static gravitational field is modified in the presence of a non-vanishing $\lambda(\phi)\,$. This manifests itself in that a Yukawa-type term  arises 
\cite{Wagoner 70,Will:2014xja}.

It is important to notice two additional points. First, we assume that the usual
relation $\nabla_\beta T^{\alpha\beta} =0$ establishing the conservation laws
satisfied by the matter fields holds. This is ensured by requiring that the
matter fields Lagrangian
$L_m$  be independent of $\phi\,$ \cite{Bergm 68} 
The role of the scalar field is then that of helping to generate
the spacetime curvature associated with the metric. Matter may create
this field, but the latter cannot act back directly on the matter, which
thus responds only to the metric \cite{Thorne:1970wv}.

The second point follows from Equation~(\ref{eSTTFE2}). By allowing the coupling parameter $\omega(\phi)\,$ to tend to infinity, it can be seen that
the right-hand side of the equation may vanish. This will be the case when
$\omega\to \infty\,$ dominates over $\omega'(\phi)\,$, $\lambda(\phi)\,$ and $\lambda'(\phi)\,$ \cite{Mimoso:1998dn}. Indeed, in  the absence of $\lambda(\phi)$, one requires $\omega'/w^3\to 0$ \cite{Nordt 70}. A solution  with constant $\phi$ is then asymptotically admitted by the Klein-Gordon equation, and thus GR is recovered  in the $\omega \rightarrow \infty\,$  limit of scalar-tensor theories (see also \cite{Damour+Nordtvedt 93}). However, it is important, on the one hand, to separate the cases where the trace of the energy-momentum tensor is trivially vanishing, i.e., vacuum  and radiation. In the latter cases, and again when there is no cosmological potential $\lambda(\phi)$, it has been argued that there are  asymptotic behaviors that differ from GR~\cite{Damour+Nordtvedt 93,Billyard:1998kg,Mimoso:1998dn,Omeg_limit2}. On the other hand, in the presence of the potential $\lambda(\phi)$, the asymptotic behaviour is very much dependent on it. In Ref.~\cite{Mimoso:1998dn}, it was shown that when the potential is of a power-law nature, in particular, when $\phi\,\lambda(\phi)\propto\phi^2$, it governs the approach to GR, superseding the mechanism due to the divergence of $\omega(\phi)$.  Morevover, as shown below,  BD theory can also be an attractor~\cite{2003Ap&SS.283.661M}.

\subsection{Conformal Picture}

A convenient device to deal with modified gravity theories is to make use
of an appropriate conformal transformation \cite{Wald:1984rg,Wands:1993uu,Conf_equiv-fR} to bring the theory into the form of GR plus a minimally coupled field.
Consider the transformation
\begin{equation}
\tilde{g}_{\alpha \beta} = \frac{\phi}{\phi_0} g_{\alpha \beta} ,
\end{equation}
where $\phi_0\,$ is a constant with the dimensions of mass squared
which we introduce to make the scaling factor dimensionless. In order
to keep the next results as simple as possible, we shall assume
that $\phi_0=1$ and that $\phi$ in the following expressions is
dimensionless (i.e., whenever we write $\phi$ it corresponds to $\phi/\phi_0$ where $\phi_0$ is some value of reference, for instance $\phi_0=1/G$).
Using the transformed metric $\tilde{g}_{\alpha \beta}$ and redefining the scalar field as
\begin{equation}
\varphi \equiv \int\, \sqrt{\frac{2\omega(\phi)+3}{2}}\;\frac{{\rm d}\phi}{\phi},
\label{ePhi}
\end{equation}

we find that the action adopts the form
\begin{equation}
\tilde{S}= \int \ d^4x\, \sqrt{-\tilde{g}}\, \left[\tilde{R} -
\tilde{\nabla}_\alpha{\varphi}\tilde{\nabla}^\alpha{\varphi}
- 2{V}(\varphi)\right] +16\pi \int \ d^4x\, \sqrt{-\tilde{g}}\, \tilde{L}_m\; ,
\label{eCTSTTactionb}
\end{equation}
where $V(\varphi)=U\left( \phi(\varphi)\right)/\phi^2(\varphi)$, and also
$ \tilde{L}_m=  \phi^{-2} L_m$.
Thus, we realize that the action of the STT becomes the action of
GR with matter and a massive scalar field.
Writing the Einstein field equations in the transformed frame (dubbed the {\em Einstein frame}),
we obtain
\begin{eqnarray}
\tilde{R}_{\alpha \beta} - \frac{1}{2}\, \tilde{g}_{\alpha \beta}\tilde{R} = 8\pi
\tilde{T}_{\alpha \beta}  + \left[ \tilde{\nabla}_\alpha{\varphi}\tilde{\nabla}_\beta{\varphi}
-\frac{1}{2}\,\tilde{g}_{\alpha \beta}\, \tilde{\nabla}_\gamma{\varphi}\tilde{\nabla}^\gamma{\varphi}
-{V}(\varphi)\,\tilde{g}_{\alpha \beta}
\right],
\label{eEFEcframe}
\end{eqnarray}
\begin{equation} \label{Efield}
\tilde{\Box}\varphi - \frac{\rm d}{{\rm d}\varphi}{V}(\varphi) =
\sqrt{\frac{2}{2\omega +3}}
\;  8\pi \, \tilde{T},
\end{equation}
\begin{equation}
\tilde{\nabla}_\beta{{\tilde{T}}^\beta}{}_{\alpha} = - \frac{1}{2}\,
\left(\frac{\phi_{;\alpha}}{\phi}\right)\;
\tilde{T}{\sqrt{\frac{2}{2\omega +3}}} . \label{edivTcframe}
\end{equation}
It becomes immediately apparent that, when the trace of the energy-momentum tensor of the matter fields in the original frame is vanishing, there is an equivalence between the two pictures related via the conformal transformation. This happens trivially for a vacuum model, and also in the \mbox{radiation case.}

In all other cases, the GR-like frame exhibits a coupling between matter and the redefined scalar field. One is then drawn to conclude that the equivalence principle is violated. However, since this variation is induced by the presence of the scalar field which also acts as a source of the Einstein equations (\ref{eEFEcframe}) the latter conclusion seems arguable and has been much debated~\cite{Faraoni:1999hp,Flanagan:2004bz,Olmo:2006zu}. If one considers the scalar field as a component of the total energy-momentum tensor on the right-hand side of the field equations, then the realization that the original matter does not exhibit the geodesic behaviour is a mere consequence of the fact that it couples to the other source component.

At this stage we should mention that the conformal transformation mixes the geometric and matter degrees of freedom, which results in many interpretational ambiguities \cite{Capozziello:1996xg}. One may also mention that if one only restricts attention to the Einstein frame, one may also lose sight of the original motivations and modifications of gravity in the geometrical sector \cite{CLM1,Capozziello:2014bqa}. However, the Einstein and Jordan frames are physically equivalent, as can be traced back to Dicke's original paper \cite{Dicke:1961gz}, where the conformal transformation technique was introduced. In fact, both conformal frames are equivalent, in the spirit of Dicke's paper, provided that in the Einstein frame the units of mass, time and space scale as appropriate powers of the scalar field, and are thus varying. We will discuss the issues of discriminating the geometric and matter sectors in Section \ref{DE_vs_MG}.

\subsection{Cosmological Dynamics}

Here we consider the dynamics of cosmological models in general scalar-tensor gravity theories. A number of important, exact solutions of Brans-Dicke models~\cite{Nariai 68,O'Hanlon+Tupper 72}, and of more general class of theories~\cite{Barrow+Mimoso 94,GeneralSTsols,GeneralSTsols2} have been derived in the literature. In what follows we resort to a unified qualitative analysis of the major dynamical features of these scalar field cosmological models~\cite{Nunes:2000yc,Mimoso:1998dn,Charters:2001hi,2003Ap&SS.283.661M,Mimoso:1999ai,Nunes:2000ka} which has the advantage of relying on a closed, autonomous  system of equations where the functional dependence on $\phi$ of both the coupling $\omega(\phi)$ and cosmological potential $\lambda(\phi)$ is left unspecified. The analysis of the relevant features of the dynamics, namely, the set of possible asymptotic behaviors, leads to a classification of the classes of couplings and potentials vis-\`a-vis to their cosmological impact. Alternative dynamical system analyses that are found in the literature usually envisage specific classes of scalar-tensor theories~\cite{DynSis_BD1,DynSis_BD2,DynSis_BD3} or, when more general~\cite{DynSis_GenSTT,DynSis_GenSTT2,DynSis_GenSTT3,DynSis_GenSTT4}, are not so transparent as the treatment that follows.

In order to investigate the cosmological dynamics, it is most convenient to introduce dimensionless expansion normalized variables akin to the square roots of density parameters. This associates to the equilibrium points of the dynamical system (also dubbed {\em singular}, {\em critical}, or {\em fixed} points in the jargon of the theory of dynamical systems) non-static solutions of the cosmological evolution. An alternative qualitative approach based on the bare dimensional variables of the cosmological model, namely, on the scale factor of the universe, Hubble factor and/or on the energy densities of the various source components is also possible, but is of more limited scope in the scrutiny of non-static asymptotic regimes. It is however required  whenever the coupling function $\omega(\phi)$ or the cosmological potential $\lambda(\phi)$ have vanishing minima.

In what follows, we shall restrict to the homogeneous and isotropic universes given by the Friedmann-Robertson-Walker (FRW) metric
\begin{equation}
{\rm d}s^2 = - {\rm d}t^2 + a^2(t)\,\left[\frac{{\rm
d}r^2}{1-k\,r^2}+r^2({\rm d}\theta^2+\sin^2\theta\,{\rm d}\phi^2)
\right] \,, \label{Fried_met}
\end{equation}
where $k=0,\pm 1$ distinguishes the curvature of the spatial hypersurfaces.

In the Einstein frame the field equations with a matter content given by a perfect fluid with $p=(\gamma-1)\rho$ can be cast as~\cite{Mimoso:1998dn}
\begin{eqnarray}
3\frac{\dot{\tilde{a}}^2}{\tilde{a}^2} +  3\frac{k}{\tilde{a}^2} &=&
\frac{8\pi}{\Phi_\ast}\,  \left[ \frac{\dot{\varphi}^2}{2} + {V}(\varphi)+ \tilde{a}^{-3\gamma} \tilde{m}(\varphi) \right]
\label{eLPamV2} \\
\ddot{\varphi}+\frac{3\dot{\tilde{a}}}{\tilde{a}}\,\dot{\varphi}&=&
-\tilde{a}^{-3\gamma} \frac{{\rm d}\tilde{m}(\varphi)}{{\rm d}\varphi}-  \frac{{\rm d}{V}(\varphi)}{{\rm d}\varphi}  \; ,
\label{eLPbmV2}
\end{eqnarray}
where $\tilde a=\sqrt{\Phi/\Phi_\ast}\,a$, the overdots stand for the derivatives with respect to the conformally transformed time $\tilde {\rm d}\tilde t  =\sqrt{\Phi/\Phi_\ast} {\rm d}t$, $V(\varphi)=U(\Phi(\varphi))/(\Phi(\varphi)/\Phi_\ast)^2$, $\tilde{m}(\varphi)=\mu_0\,(\Phi(\varphi)/\Phi_\ast)^{(3\gamma-4)/2}$, where $\mu_0=8\pi\rho_0$ sets the arbitrary initial condition for the energy density of the perfect fluid, and, more importantly, $\partial_\varphi\ln m(\varphi) \propto (4-3\gamma)\alpha(\varphi)$ where $\alpha=(\sqrt{2\Omega+3})^{-1}$ is the PPN function that translates the coupling between the scalar field and matter in the Einstein frame as used in Damour and Nordtvedt~\cite{Damour+Nordtvedt 93}.

These equations are those of GR with a scalar field subject both to a time-independent potential, $V(\varphi)$, and to a time-dependent potential $\tilde{m}(\varphi)\,a^{-3\gamma}$ illustrating the fact that in the Einstein frame the masses of particles vary. (Hereafter we shall set $8\pi/\phi_\ast=1$).

Introducing the new time variable $N=\ln{a}$ and the dimensionless density parameters
\begin{eqnarray}
\displaystyle x^2 = \frac{\dot{\varphi}^2}{6H^2}\,, \qquad
\displaystyle y^2 = \frac{V(\varphi)}{3H^2},
\end{eqnarray}
as well as the expansion normalized curvature term $K=k/(aH)^2$, the Einstein field equations become a fourth order, autonomous dynamical system
\begin{eqnarray}
x' &=& -(3+K)\, x-\sqrt{\frac{3}{2}}\,
\left(\frac{\partial_{\varphi}V}{V}\right)\, y^2 +
\frac{3}{2}x\,\left[2x^2+\gamma\,
(1-x^2-y^2+K)\right] \nonumber \\
& & \hspace{1.5cm}- \sqrt{\frac{3}{2}}\,
 \left(\frac{\partial_{\varphi}m}{m}\right)\, (1-x^2-y^2 + K)  \label{sd_x}
\\
y' &=& \sqrt{\frac{3}{2}}\, \left(\frac{\partial_\varphi
V}{V}\right)\, x y + \frac{3}{2}y\,\left[2
x^2+\gamma\,(1-x^2-y^2+K)\right]-yK \; , \label{sd_y}\\
K' &=& -2K\,\left(1-\frac{3}{2}\,\left[2x^2+\gamma\,
(1-x^2-y^2+K)\right]\right) - 2 K^2 \,, \label{sd_K}  \\
\varphi' &=& \sqrt{6} \, x \; , \label{sd_varphi}~
\end{eqnarray}
where we have used $\rho/3H^2=\Omega_m= 1-x^2-y^2+K$.

GR models correspond to the case where $\partial_\varphi \ln m(\varphi)=0$, and Brans-Dicke models are characterised by an exponential coupling $m$, i.e., by $\partial_\varphi \ln m(\varphi)=\alpha_0$ (as well as by $V(\varphi)=0$ in BD original version, but we do not require it here). The crucial point regarding the qualitative study of general models with scalar fields lies in the $\varphi'$-equation, since it allows the consideration of arbitrary choices of $V(\varphi)$ and of ${m}(\varphi)$~\cite{Nunes:2000yc}. We compactify the phase space of (\ref{sd_x}--\ref{sd_K}) by considering $\varphi \in \Re \cup \{\infty \}$, and distinguish the fixed points arising at finite values of $\varphi$, which require $x=0$, from those at $\varphi=\infty$ (which we shall denote $\varphi_\infty$) that require either $x=0$ or $\psi=\varphi^{-1}=0$. The $K=0$ and the $y=0$ subspaces are invariant manifolds.

For $K=0$ the system reduces to the Equations (\ref{sd_x}), (\ref{sd_y}) and \ref{sd_varphi}) and we find the following fixed points, namely, at finite $\varphi$:
\begin{itemize}
\item $x=0, \; y=1,\;  \varphi_0 \quad \rm{where}\quad \partial _{\varphi }V(\varphi_0)=0$. This case corresponds to de Sitter solutions, dominated by the scalar field, and arise at non-vanishing maxima or minima of the potential $V$. Notice that $ \partial_{\varphi }m$ may be different from $0$. These solutions are attractors at minima of $V$  and repellors at maxima of $V$.
\item $x=y= 0,\; \varphi_0  \quad \rm{where}  \partial _{\varphi }m=0$. The second case corresponds to matter dominated solutions with $V=0$ at maxima or minima of $m$. Note however that the system (\ref{sd_x})--(\ref{sd_varphi}) is singular when $V=0$, translating the fact that the variables in use are not regular. In the original variables, this second class of solutions exists provided that $V$ has also an extremum, or is identically zero.
\end{itemize}

\begin{figure} [H]
\centering
\includegraphics[scale=.43]{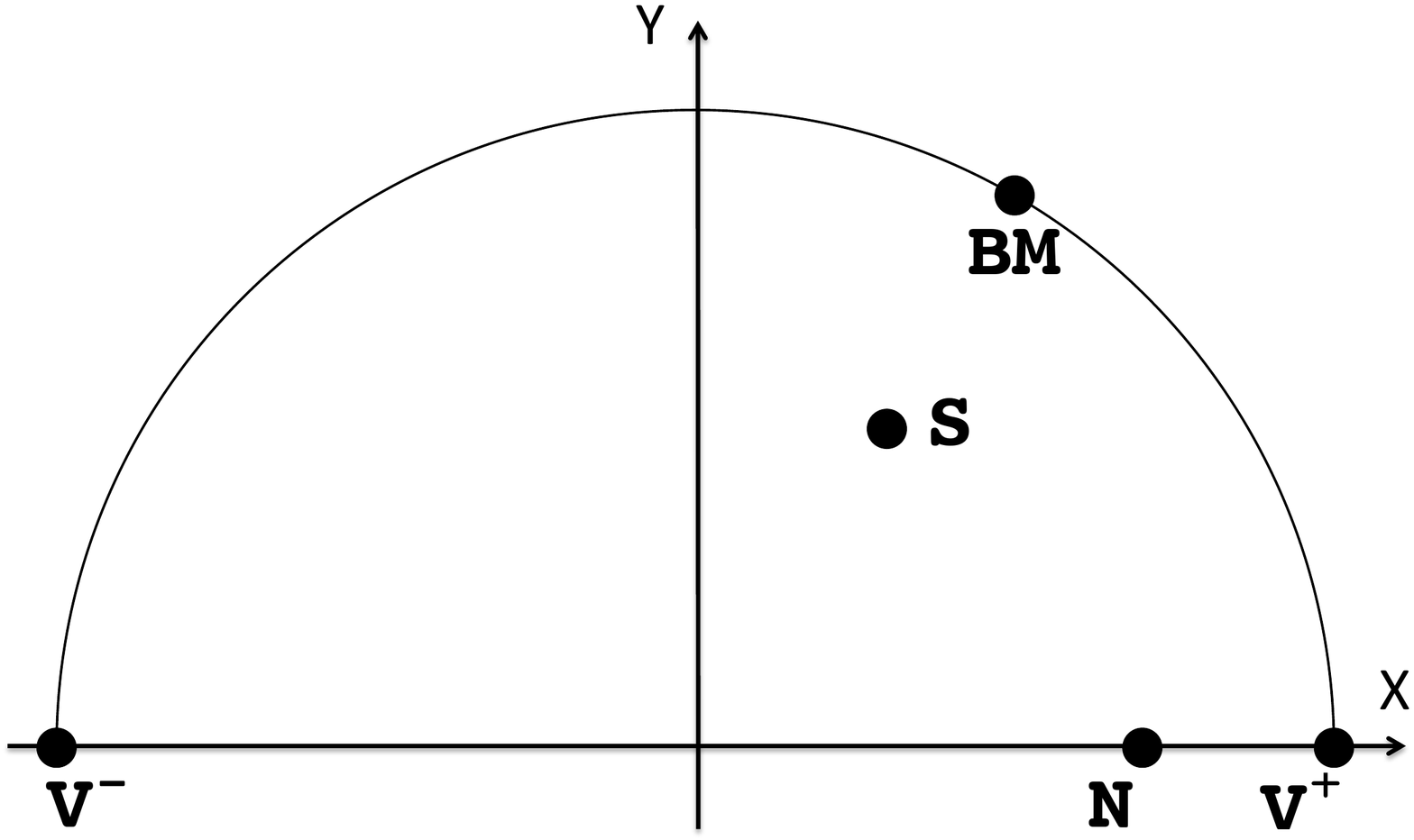}

\caption[]{The figure represents the fixed points at the $\varphi_\infty$ phase-plane for $V$ and $m$ asymptotically exponential.}
\label{Figure1}
\end{figure}

\begin{figure}[H] 
\centering
\includegraphics[scale=.43]{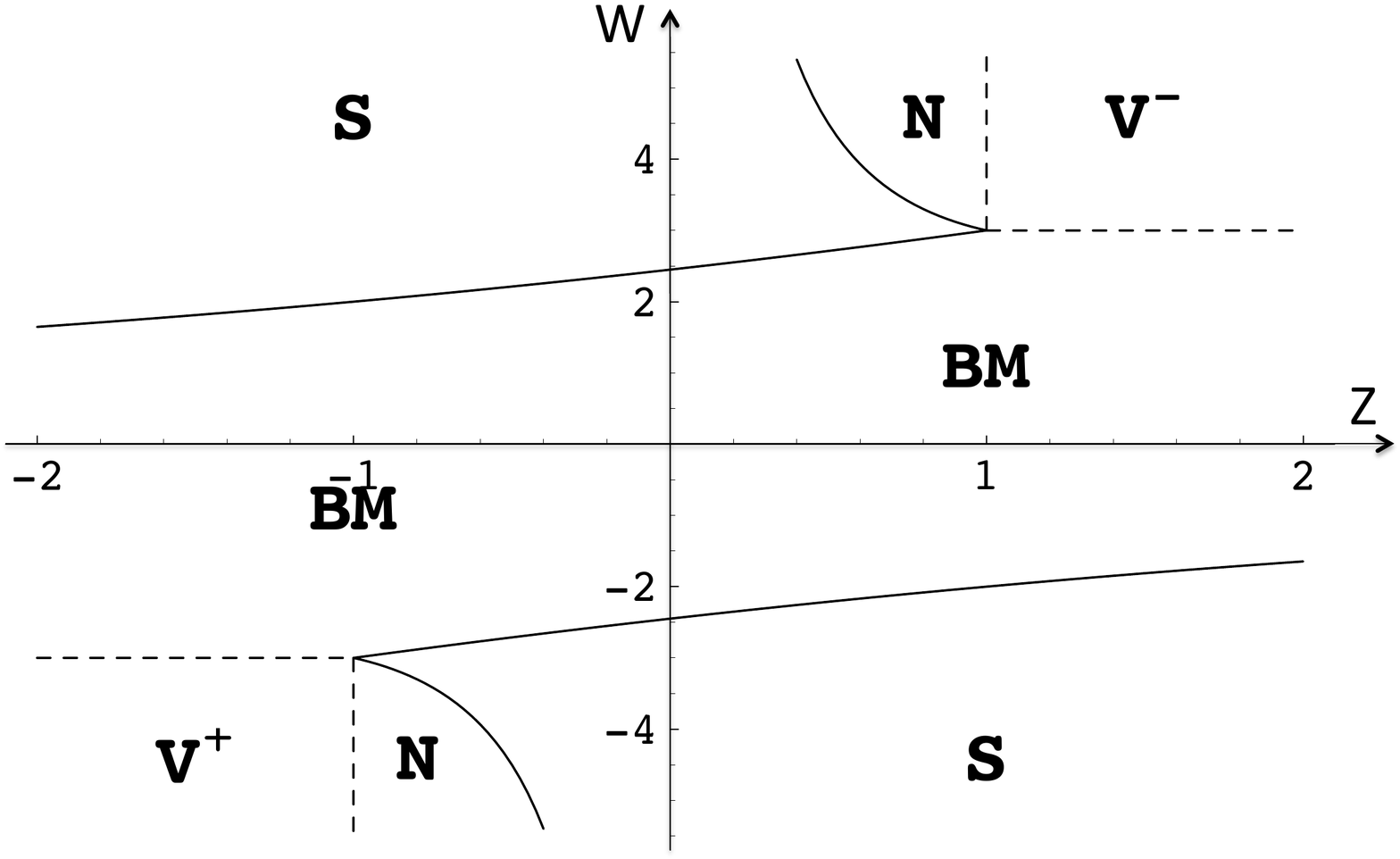}

\caption[]{The plot exhibits the regions in parameter space that correspond to the asymptotic Brans-Dicke behaviour. The horizontal axis is the $Z$-axis and the vertical axis is the $W$-axis. The regions S and BM are separated by the line $9\gamma/2-W^2+WZ=0$, and the separations between the S and N regions is given by the line $2Z^2-2WZ+9\gamma(2-\gamma)/2=0$.}
\label{Figure1b}
\end{figure}

At $\varphi_\infty$, the finite $\varphi$ solutions may exist also at $\varphi = \infty $ provided that $V$ or $m$ are asymptotically flat. However, other solutions may show up at $\varphi = \infty $, if both $V$ and $m$ are asymptotically exponential. In fact, in this case the equilibrium in $\psi = 1/\varphi $ is trivially satisfied at $\psi = 0$ \cite{Nunes:2000yc}. Defining $W = \sqrt{\frac{3}{2}} \lim_{\varphi \rightarrow \infty}\frac{\partial _{\varphi }V}{V}$ and $Z = \sqrt{\frac{3}{2}} \lim_{\varphi \rightarrow \infty}\frac{\partial _{\varphi }m}{m}$. The exponential asymptotic behaviour of $m(\varphi)$ amounts to having a late-time Brans-Dicke behaviour. One finds the following fixed points (see Figures~\ref{Figure1} and \ref{Figure1b})
~\cite{Amendola 99}:
\begin{itemize}
\item $x_{\infty}^{V^\pm}= \pm 1$, $y^{V^\pm}=0$, that lie on the intersection of  the invariant lines $x^2 + y^2 = 1$ and $y=0$. These points correspond to the vacuum solutions of Brans-Dicke theory found by \cite{O'Hanlon+Tupper 72,Barrow+Mimoso 94}.
\item $x_{\infty}^{BM}= -W/3$, which lies on the invariant line $x^2 + y^2 = 1$, provided $|W|<3$. These solutions (BM) correspond to those found in \cite{Barrow+Maeda 90}.
\item $x_{\infty}^{N}= -2Z/(3(2-\gamma))$, which lies on the invariant line $y=0$ and exists if $|Z| <3(2-\gamma)/2$. These solutions (N) correspond to the matter dominated solutions found by \cite{Nariai 68,Barrow+Mimoso 94}.
\item $x_{\infty}^{S}= \frac{3\gamma /2}{Z-W}, (x_{\infty}^{S})^2 + (y_{\infty}^{S})^2 = \frac{3x_{\infty}^{S}+Z}{Z-W}$, that lies in the interior of the phase space domain, provided that
\begin{equation}
0 < (x_{\infty}^{S})^2 + (y_{\infty}^{S})^2 = \frac{9\gamma /2 + Z(Z-W)}
{(Z-W)^2}<1.
\nonumber
\end{equation}
These are scaling solutions (S) (see \cite{2003Ap&SS.283.661M,Amendola 99,Uzan 99} and references therein)
\end{itemize}

For the $K\neq 0$ case, as $K'=0$ for $K=0$, we see from the linearization of the system that
\begin{equation}
\frac{\partial K'}{\partial K} = -2 + 2\, Q_{K=0} \; , \quad Q_{K=0} \equiv \frac{3}{2}\,\left[2x^2+\gamma\,
(1-x^2-y^2)\right] \; ,
\end{equation}
which shows that the $K=0$ subspace is stable whenever $Q<1$ which is precisely the condition for inflationary behaviour. Thus inflation is required to guarantee both the stability of the $K=0$ asymptotic solutions and the attraction to GR, whenever the latter applies.

As previously referred, when the potential $V(\phi)$ yields vanishing minima in the Einstein frame the expansion normalized variables adopted here are not suitable, as they yield a dynamical system that is not defined at those minima. In such a case it is preferable to use variables such as the scale factor, the kinetic energy of the scalar field and its potential, as done in~\cite{Mimoso:1998dn}, or in some of the references in \cite{DynSis_BD1,DynSis_GenSTT}.

Summing up the qualitative analysis of the dynamical system (\ref{sd_x})--(\ref{sd_K})  associates exact solutions to a classification of the fixed points, and also allows the consideration of how extended gravity theories dynamically relate to GR~\cite{2003Ap&SS.283.661M}. It reveals the interplay of the two main mechanisms that yield general relativity as a cosmological attractor of scalar-tensor gravity theories: (i) the vanishing of the coupling function $\alpha$, and (ii) the existence of a minimum of the scalar field potential $V$. The latter mechanism is shown to supersede the former one. Moreover the approach to GR is then characterized by a de Sitter inflationary behavior. This guarantees the stability of the $k=0$ GR solutions. We have also shown that at $\varphi \to \infty$ there is a variety of power-law solutions that correspond to Brans-Dicke behavior and that can be attractors of the cosmological dynamics.

In what follows we will address a multifield scenario, which can be understood as a particular case of the multi-scalar tensor gravity models of Equation (\ref{e:MSTaction}).

\subsection{Assisted Quintessence} \label{assist_infl}

Assisted inflation is a model of early universe inflation where multiple scalar fields cooperate to fuel the accelerated inflationary expansion \cite{Liddle:1998jc,Malik:1998gy,Copeland:1999cs}. Scalar potentials that are not viable to provide inflation as a single field can be seen to work when several such fields are present in parallel. A similar idea can be implemented for our present observed accelerated expansion \cite{Kim:2005ne,Tsujikawa:2006mw,Ohashi:2009xw,Karwan:2010xw}.

In the usual quintessence scenario, a single scalar field, uncoupled from dark matter, is responsible for the dark energy component in our universe \cite{Zlatev:1998tr}. In assisted quintessence, we have a scenario where the dark energy sector is composed of multiple scalar fields, with different scalar potentials. Since so little is known about the structure of the dark sector, there is the possibility of a  coupling between the scalar fields responsible for inflation and dark matter. Coupled quintessence, where a scalar field interacts with dark matter was introduced in Refs. \cite{Amendola:1999dr,Holden:1999hm,Amendola:1999er}. In \cite{Brookfield:2007au,Baldi:2012kt} it was further suggested that the scalar field can couple in a different way to several separate dark matter species. To keep the model as general as possible, we  assume  the dark matter sector to consist of several matter components with different couplings to the dark energy fields \cite{Amendola:2014kwa}.

\subsubsection{General Equations}

We consider an ensemble of $n$ scalar fields $\phi_i$ cross-coupled to an ensemble of $m$ dark matter components $\rho_\alpha$. The cross-couplings are described by the matrix $C_{i\alpha}$ where indexes $i,j$ identify scalar field indexes and indexes $\alpha,\beta$ identify the dark matter components. The equation of motion for the fields and the various dark matter components, in a spatially flat Friedmann-Robertson-Walker metric with scale factor $a(t)$, are then written as
\begin{eqnarray}
\ddot{\phi}_i + 3 H \dot{\phi}_i + V_{,\phi_i} &=& \kappa \sum_\alpha  C_{i\alpha} \rho_\alpha,
\\
\dot{\rho}_\alpha +3 H \rho_\alpha &=& - \kappa \sum_i C_{i\alpha} \dot{\phi}_i \rho_\alpha.
\end{eqnarray}

Notice that the ensemble of $n$ scalar fields $\phi_i$ cross-coupled to $m$ dark matter components $\rho_\alpha$ can be perceived as a particular case of the multi-scalar-tensor models of Equation (\ref{e:MSTaction}). Indeed, if we set the metric of the internal space of the scalar fields to be flat, i.e., $\omega_{\mu\nu}=\delta_{\mu\nu}$ and allow for  a combination of dark matter components in the matter sector,  the $C_{i\alpha}$ coupling coefficients are related to the couplings $\alpha_i =\partial \ln f/\partial \phi^i$ of the multi-scalar-tensor theory in the Einstein frame, and we obtain the latter equations. 

The solution for the dark matter component evolution can be given immediately in
terms of the values of the fields as
\begin{equation}
\rho_\alpha = {\rho_\alpha}_0  \exp\left(-3N - \kappa \sum_i C_{i\alpha} (\phi_i -
{\phi_i}_0)\right).
\end{equation}

The rate of change of the Hubble function is
\begin{equation}
\dot{H} = - \frac{\kappa^2}{2} \left( \sum_\alpha \rho_\alpha + \sum_i \dot\phi_i^2
\right) ,
\end{equation}
which is subject to the Friedmann constraint
\begin{equation}
H^2 = \frac{\kappa^2}{3} \left( \sum_\alpha \rho_\alpha + \sum_i \rho_{\phi} \right)\,,
\end{equation}
where $\rho_{\phi} =  \sum_i \dot{\phi}_i^2/2 + V(\phi_1,...,\phi_n)$.

It is possible to  define an effective equation of state parameter for the dark sector,
\begin{equation}
w_{\rm eff} = \frac{p_{\phi}}{\rho_m + \rho_\phi}\,,
\end{equation}
where $p_\phi = \sum_i \dot{\phi}_i^2/2 - V(\phi_1,...,\phi_n)$ is the total dark energy pressure and
$\rho_m = \sum_\alpha \rho_\alpha$ the total dark matter energy density.

We are interested in looking at the critical points of the evolution to look for viable attractors of late time evolution. For this, we will consider two possible forms for the scalar potential, both
leading to scaling solutions: a sum of exponential terms with
$V_{1}(\phi_1,...,\phi_n) = M^4 \sum_i e^{-\kappa \lambda_i \phi_i}$
and an exponential of a sum of terms with
$V_{2}(\phi_1,...,\phi_n) = M^4 e^{-\sum_i \kappa \lambda_i \phi_i}$.

\subsubsection{Scalar Field Dominated Solution}

We start our investigation with the case of a negligible matter contribution ($\rho_\alpha = 0$). We can show that the effective equation of state for the attractor is
\begin{equation}
w_{\rm eff} = -1 + \frac{1}{3} \lambda_{\rm eff}\,,
\end{equation}
where $\lambda_{\rm eff}$ can be interpreted as an effective logarithmic slope for an dynamically equivalent single field exponential potential.

For the first potential, $V_1$, we get that $\lambda_{\rm eff}$ has to obey
\begin{equation}\label{acq-leff}
\frac{1}{\lambda_{\rm eff}^2} = \sum_i \frac{1}{\lambda_i^2}\,.
\end{equation}
In this case  incre
asing the number of fields has the effect of decreasing the value of $\lambda_{\rm eff}$, so that an accelerated expansion can be achieved even in the case of each single field having a slope too large to fuel acceleration.

For the second potential, $V_2$, $\lambda_{\rm eff}$ is given by
\begin{equation}
\lambda_{\rm eff}^2 = \sum_i \lambda_i^2 \,.
\end{equation}

Here we see that increasing the number of fields actually increases the value of $\lambda_{\rm eff}$, making an accelerated expansion more difficult.

Both of these results are similar to what is obtained for assisted inflation \cite{Liddle:1998jc,Malik:1998gy,Copeland:1999cs}. This was to be expected since the matter sector is being neglected.

\subsubsection{Scaling Solution}

Here we will look for fixed points where all the sectors, kinetic  and potential energy densities for the scalar energy and the matter energy density, have non-vanishing contributions.
We can again define an effective scalar potential logarithmic slope, $\lambda_{\rm eff}$. But now we also have to take into account the coupling of the scalar fields to the matter sector. This can also be represented by an effective coupling $C_{\rm eff}$. In terms of these parameters, we can obtain the effective equation of state
\begin{equation}
w_{\rm eff} = \frac{C_{\rm eff}}{\lambda_{\rm eff} - C_{\rm eff}} \,,
\end{equation}
and for the dark energy sector
\begin{equation}
\Omega_{\phi} = \frac{3 - \lambda_{\rm eff} C_{\rm eff} + C^2_{\rm eff}}{(\lambda_{\rm eff}-C_{\rm eff})^2}  \,.
\end{equation}

For the first potential, the value of $\lambda_{\rm eff}$ is again obtained from Equation~(\ref{acq-leff}), whereas the value of $C_{\rm eff}$ is
\begin{equation}
C_{\rm eff} = \lambda_{\rm eff} \sum_{j=1}^n \frac{C_{j\alpha}}{\lambda_j} \,,
\end{equation}
where $\alpha$ can be taken to be any of the matter species. This is because the sum is constrained to yield the same result for all the matter species, otherwise there is no exact scaling solution. It is instructive to consider these quantities for the simple case of all the potentials being replicas, that is, having the same slopes and couplings. Then $\lambda_i = \lambda_1$ and the couplings $C$ are a diagonal matrix with $C_{ii} = C_{11}$. We then get
\begin{align}
w_{\rm eff} = \frac{C_{11}}{\lambda_1-C_{11}}\,, \qquad
\Omega_{\phi} = \frac{3 n - \lambda_1 C_{11} + C_{11}^2}{(\lambda_1 - C_{11}^2)} \,,
\end{align}
where $n$ is the number of scalar fields. The equation of state obtained is the usual result  for a single coupled field, but in the scalar field energy we can see a mild dependence in the number of fields.

For the second potential, the effective slope $\lambda_{\rm eff}$ obtained in scalar dominance is no longer valid. Instead it is possible to show that we can combine the fields through a rotation so that the kinetic energy of all fields is zero except for the first field. In terms of this rotation matrix $Q$ we get the effective equation of state as
\begin{equation}
\lambda_{\rm eff} = \sum_{j} Q_{1j} \lambda_j  \,,
\end{equation}
and the effective coupling to be
\begin{equation}
C_{\rm eff} = \sum_{j} Q_{1j} C_{j1} \,.
\end{equation}

The precise form of $Q$ is hard to get in the general case. But the simple case of a diagonal coupling matrix $C$ can give us an idea of what is to be expected. In this instance, we get
\begin{equation}
Q_{1i} = \frac{1}{C_{ii} \sqrt{\sum_j 1/C_{ll}^2}} \,,
\end{equation}
and so
\begin{align}
\lambda_{\rm eff} = C_{\rm eff} \sum_{i} \frac{\lambda_i}{C_{ii}}\,, \qquad
\frac{1}{C_{\rm eff}} = \sum_i \frac{1}{C_{ii}^2}  \,.
\end{align}

Finally, if we consider again the case of $n$ fields all with the same slope and couplings, we get $\lambda_{\rm eff} = \sqrt{n} \, \lambda_1$ and $C_{\rm eff} = C_{11}/\sqrt{n}$. Here we have a strong dependence in the number of fields in the effective equation of state and the dark energy density,
\begin{align}
w_{\rm eff} = \frac{C_{11}}{n \lambda_1 - C_{11}} \,, \qquad
\Omega_{\phi} = \frac{3 n - C_{11} \lambda_1 n + C_{11}^2}{(n \lambda_1 - C_{11})^2} \,.
\end{align}

This is clearly in contrast to what we have for the first potential type.

\subsubsection{Scalar Potential Independent Solutions}

Finally we look for solutions where the scalar potential energy is negligible. These solutions, therefore, are independent of the choice we make for the scalar potential. Note that the scalar field still contributes to the overall energy density through its kinetic energy.
These solutions can be grouped into three different types:
\begin{itemize}
\item First, we have the kinetic dominated solutions where all the energy density for the dark sector is due to the kinetic energy of the scalar fields. In this instance, we have $w_{\rm eff}  = \Omega_{\phi} = 1$.
\item As a second possibility, we have a contribution from the kinetic energy and the matter energy densities for the dark sector. In this case we have $w_{\rm eff} = \Omega_{\phi} = \frac{2}{3} \sum_i C^2_{i\alpha}$. This will be true for any matter species $\alpha$ since the sum is constrained to yield the same for all matter species.
\item Finally, as a third possibility we have a matter dominated scenario. This has been investigated previously in the literature for the case of a single field with two matter components \cite{Baldi:2012kt,Piloyan:2013mla,Piloyan:2014gta,Baldi:2014tja}. If we extend this to more than one field, we will see that there is a consistency condition imposed on the values of the couplings to allow a flat direction in the field space, otherwise this solution will be non existent. This is by its nature a transient solution since the matter contribution will eventually decay away and the scaling solution or the scalar field dominated solution will take over. However, this can be an interesting scenario for model building since it allows the dark matter sector to have a significant contribution before the acceleration of the Universe.
\end{itemize}

If we extend our analysis to the linear perturbations in these models we can show that there is a term proportional to $\sum_i C_{i\alpha} C_{i \beta}$ that can act as a growth or damping term to the density contrast of the matter components. This yields a strong constraint on the types of models that can be compatible with observations \cite{Amendola:2014kwa}.

\section{Horndeski Theories and Self-tuning}\label{sec3}

In the previous sections we have considered the simplest scalar-tensor theories of gravity, which are given by the action (\ref{eSTTaction}). Those theories can be re-written as a scalar field, which is coupled to the geometry through a term $f(\phi)R$ in the action, with a canonical kinetic term and a potential. This is:
\begin{equation}
S= \int \sqrt{-g}\, d^4x\, \left[f(\phi) R -\frac{1}{2}X+V(\phi)\right] + S_M,
\label{eSTTaction2}
\end{equation}
where $X=\nabla_\mu\phi\nabla^\mu\phi$. Nevertheless, {\it there are more things in heaven and earth} for scalar-tensor theories of gravity. K-essence models \cite{ArmendarizPicon:2000ah} assume a scalar field minimally coupled to gravity ($f(\phi)=1$) with an action that is not necessarily the sum of a kinetic term and a potential but a more general function $K(\phi,\,X)$. One can go further and consider interaction terms in the Lagrangian containing second order derivatives of the scalar field, as in the kinetic braiding models \cite{Deffayet:2010qz} that have a scalar field Lagrangian given by $K(\phi,X)+G(\phi,X) \Box \phi$.
These models can be stable because their field equations are only second order.
However, theories with Lagrangians containing second order derivatives generically have field equations which contain higher than second order derivatives. Therefore, such theories usually propagate an extra (ghostly) degree of freedom which means that they are affected by the Ostrogradski instability \cite{Woodard:2006nt}.
In fact, in 1974, Horndeski found the most general scalar-tensor action leading to second order equations of motion \cite{Horndeski:1974wa}. Nevertheless,
Horndeski work passed mostly unnoticed until Deffayet et al.~\cite{Deffayet:2011gz} rediscovered his theory when generalizing the covariantized version \cite{Deffayet:2009wt} of the galileons models \cite{Nicolis:2008in}.

The Horndeski Lagrangian can be written as \cite{Horndeski:1974wa}
\begin{small}
\begin{eqnarray}\label{H}
 \mathcal{L}_H&=&\delta^{\alpha\beta\gamma}_{\mu\nu\sigma}\left[\kappa_1\left(\phi,\,X\right)\nabla^\mu\nabla_\alpha\phi \,R_{\beta\gamma}{}^{\nu\sigma} -\frac{4}{3}\kappa_{1,X}\left(\phi,\,X\right)\nabla^\mu\nabla_\alpha\phi\nabla^\nu\nabla_\beta\phi\nabla^\sigma\nabla_\gamma\phi\right.\nonumber\\
 &&+\left.\kappa_3\left(\phi,\,X\right)\nabla_\alpha\phi\nabla^\mu\phi\,R_{\beta\gamma}{}^{\nu\sigma}-4\kappa_{3,X}\left(\phi,\,X\right)\nabla_\alpha\phi\nabla^\mu\phi\nabla^\nu\nabla_\beta\phi\nabla^\sigma\nabla_\gamma\phi\right]\nonumber\\
 &&+\delta_{\mu\nu}^{\alpha\beta}\left[F\left(\phi,\,X\right)\,R_{\alpha\beta}{}^{\mu\nu}-4F_{,X}\left(\phi,\,X\right)\nabla^\mu\nabla_\alpha\phi \nabla^\nu\nabla_\beta\phi +2\kappa_8\left(\phi,\,X\right)\nabla_\alpha\phi\nabla^\mu\phi\nabla^\nu\nabla_\beta\phi\right]\nonumber\\
 &&-3\left[2F_{,\phi}\left(\phi,\,X\right)+X\,\kappa_8\left(\phi,\,X\right)\right]\nabla_\mu\nabla^\mu\phi+\kappa_9\left(\phi,\,X\right),
\end{eqnarray}
 \end{small}
where $\kappa_i\left(\phi,\,X\right)$ and $F(\phi,\,X)$ are arbitrary functions, satisfying
$F_{,X}=\kappa_{1,\phi}-\kappa_3-2X\kappa_{3,X}$.
Brans--Dicke theory, k-essence, kinetic braiding and many other models can be seen as particular cases of the most general Horndeski family.
Thus, although the Horndeski Lagrangian restricts the number of stable scalar-tensor theories. Actually it has been shown that there are still stable theories beyond the Horndeski Lagrangian \cite{Zumalacarregui:2013pma,Gleyzes:2014dya,Gleyzes:2014qga}. Those theories have second order equations of motion only when one particular gauge is fixed.

On the other hand, the vacuum energy also gravitates in modified theories of gravity. Thus, the cosmological constant problem \cite{Weinberg:1988cp,Carroll:2000fy,Kaloper:2014dqa} is still present if, as Weinberg suggested, the extra degree of freedom could screen only a given value of that constant \cite{Weinberg:1988cp}.
In this context, the {\it fab four} models are based in the observation by Charmousis {\it et al.}~\cite{Charmousis:2011bf,Charmousis:2011ea} that Weinberg's no-go theorem can be avoided by relaxing one of the assumptions, that is allowing the field to have a non-trivial temporal dependence once the cosmological constant has been screened. Their dynamical screening is based in requiring Minkowski to be a critical point of the dynamics and, when the critical point is an attractor, it may alleviate the cosmological constant problem.
Nevertheless, these models are forced by construction to decelerate its expansion while approaching the Minkowski final state. Thus, a late time accelerating cosmology does not naturally arise in this scenario. When considering the compatibility of dynamical screening the vacuum energy with a late time phase of accelerated expansion, the concept of self-adjustment was extended to non-Minkowskian final states \cite{stmodels}. As we will now show in detail, a scalar field self-tuning to de Sitter can lead to very promising scenarios from a phenomenological point of view \cite{attracted1,attracted2}. In addition, it may alleviate the cosmological constant problem if the field is able to screen the vacuum energy before the material content in a particular model.

\subsection{Dynamical Screening}\label{general}

{In a cosmological context, one can consider} a FLRW geometry to express the Horndeski Lagrangian in the minisuperspace. Once the dependence on higher derivatives is integrated by parts the point-like Lagrangian takes the simple form \cite{Charmousis:2011ea}
\begin{equation}\label{Lsimple}
 L\left(a,\,\dot a,\,\phi,\,\dot\phi\right)=a^3\sum_{i=0..3}Z_i\left(a,\,\phi,\,\dot\phi\right)\,H^i,\qquad {\rm where}\qquad
 L=V^{-1}\int {\rm d}^3 x\,\mathcal{L}_H,
\end{equation}

$V$ is the spatial integral of the volume element,
$H=\dot a/a$ is the Hubble expansion rate, and an over-dot represents a derivative with respect the cosmic time $t$. The functions $Z_i$ are given by
\begin{equation}\label{Z}
Z_i\left(a,\,\phi,\,\dot\phi\right)=X_i\left(\phi,\,\dot\phi\right)-\frac{k}{a^2} Y_i\left(\phi,\,\dot\phi\right),
\end{equation}
where $X_i$ and $Y_i$ are written in terms of the Horndeski free functions \cite{Charmousis:2011ea}.
Taking into account Equation~(\ref{Lsimple}), the Hamiltonian density can be written as
\begin{equation}\label{Hamiltonian}
 \mathcal{H}\left(a,\,\dot a,\,\phi,\,\dot\phi\right)=\frac{1}{a^3}\left[\frac{\partial L}{\partial \dot{a}}\dot{a}+\frac{\partial L}{\partial \dot{\phi}}\dot{\phi}-L\right]=\sum_{i=0..3}\left[(i-1)Z_i+Z_{i,\dot\phi}\dot\phi\right]H^i.
\end{equation}

Assuming that the matter content, described by $\rho_{\rm m}(a)$, is minimally coupled with the geometry and non-interacting with the field, the modified Friedmann equation is given by
\begin{equation}\label{MFE}
 \mathcal H\left(a,\,\dot a,\,\phi,\,\dot\phi\right)=-\rho_{\rm m}(a).
\end{equation}

Now, one can follow a similar argument as that presented in reference~\cite{Charmousis:2011ea} for screening to Minkwoski, the only difference being that we require self-tuning to a more general given on-shell solution with $ H^2=H_{\rm on}^2\neq0$ (where the subscript ``on'' means evaluated {\it on-shell-in-a}), as done in reference~\cite{stmodels}.
Thus, assuming that the field is continuous in any phase transition that changes the value of the vacuum energy and noting that the screening of any value of the vacuum energy would lead to the screening of any material content,
one has to require $H_{\rm on}$ to be approached dynamical in order to have acceptable cosmological solutions. That is,  we want $H_{\rm on}$  to be an attractor solution, although this particular adjustment mechanism only ensures that it is a critical point.
As it is subtly discussed in reference~\cite{Charmousis:2011ea}, this requirement leads to the following three conditions:
\begin{itemize}
\item The field equation evaluated at the critical point has to be trivially satisfied leading the value of the field free to screen. This implies that the minisuperspace Lagrangian density at the critical point takes the form
 \begin{equation}\label{c1}
  \sum_{i=0..3}Z_i\left(a_{\rm on},\,\phi,\,\dot\phi\right)H_{\rm on}^i=c(a_{\rm on})+\frac{1}{a_{\rm on}^3}\dot\zeta\left(a_{\rm on},\,\phi\right).
 \end{equation}
\item The modified Freedman equation evaluated once screening has taken place has to depend on $\dot\phi$ to absorb possible discontinuities of the cosmological constant appearing on the r.~h.~s.~of this equation. Thus, taking into account Equations~(\ref{MFE}) and (\ref{c1}) (and its $\dot\phi$-derivative), it leads to
 \begin{equation}\label{c2}
  \sum_{i=1..3}i\,Z_{i,\dot\phi}\left(a_{\rm on},\,\phi,\,\dot\phi\right)H_{\rm on}^i\neq0.
 \end{equation}
\item The full scalar equation of motion must depend on $\ddot a$ to allow for a non-trivial cosmological dynamics before screening. This leads to a condition equivalent to (\ref{c2}) if $H_{\rm on}\neq0$.
\end{itemize}

A particular Lagrangian which satisfies conditions can be written as
 \begin{equation}\label{Lbarra}
 \widebar L\left(a,\,\dot a,\,\phi,\,\dot\phi\right)=a^3\left[c(a)+\sum_{i=1..3}\widebar Z_{i}\left(a,\,\phi,\,\dot\phi\right)\left(H^i-H_{\rm on}^i\right)\right].
\end{equation}

This Lagrangian (which is a trivial generalization of that presented in reference \cite{Charmousis:2011ea}) has a critical point at the solution $H_{\rm on}$ by construction \cite{stmodels}. In order to obtian some information about the $Z_i$'s, we note that, as it has been explicitly proven in \cite{Charmousis:2011ea} (it must be noted that such a proof is independent of the particular $H_{\rm on}$),
two Horndeski theories which self-tune to $H_{\rm on}$ are related through a total derivative of a function $\mu\left(a,\,\phi\right)$. Therefore, one can consider
\begin{equation}\label{relacionL}
 L\left(a,\,\dot a,\,\phi,\,\dot\phi\right)=\widebar L \left(a,\,\dot a,\,\phi,\,\dot\phi\right)+\dot\mu\left(a,\,\phi\right),
\end{equation}
with $\widebar L\left(a,\,\dot a,\,\phi,\,\dot\phi\right)$ given by Equation~(\ref{Lbarra}).
As this equation has to be valid during the whole evolution, we can equate equal powers of $H$ to obtain
\begin{equation}\label{Zs}
 Z_0=c(a)-\sum_{i=1..3}\widebar Z_{i}H_{\rm on}^i+\frac{\dot\phi}{a^3}\mu_{,\phi}, \qquad
 Z_1=\widebar Z_1+\frac{1}{a^2}\mu_{,a},\qquad Z_2=\widebar Z_2,\qquad Z_3=\widebar Z_3,
\end{equation}
which can be combined to yield \cite{Charmousis:2011ea,stmodels}
\begin{equation}\label{Zgral}
Z_i\left(a,\,\phi,\,\dot\phi\right)H_{\rm on}^i=c(a)+\frac{H_{\rm on}}{a^2}\mu_{,a}\left(a,\,\phi\right)+\frac{\dot\phi}{a^3}\mu_{,\phi}\left(a,\,\phi\right).
\end{equation}

\subsection{Requiring the Existence of a de Sitter Critical Point: $H_{\rm on}^2=\Lambda$}\label{particular}

In the first place, we consider $k=0$. In this case the dependence on the scale factor $a$ of the $Z_i$'s vanishes, as can be easily noted from Equation~(\ref{Z}),
and the point-like Lagrangian (\ref{Lsimple}) is independent of $Y_i$'s. For $k=0$, therefore, Equation~(\ref{Zgral}) is equal to the $L_{\rm on}\left(a_{\rm on}\right)/a_{\rm on}^3$.
In the second place, we demand $H_{\rm on}^2=\Lambda$, this leads to \cite{stmodels}
\begin{equation}
 \sum_{i=0..3}Z_i\left(\phi,\,\dot\phi\right)\Lambda^{i/2}=c(a)+\frac{\sqrt{\Lambda}}{a^2}\mu_{,a}+\frac{\dot\phi}{a^3}\mu_{,\phi}.
\end{equation}

As the l.~h.~s.~of this equation is independent of $a$, the r.~h.~s.~should also be independent of $a$ for any value of $\dot\phi$, which allows us to obtain $ \mu\left(a,\,\phi\right)$. Thus, we have
\begin{equation}\label{Lonshell}
 \sum_{i=0..3}X_i\left(\phi,\,\dot\phi\right)\Lambda^{i/2}=\frac{1}{a_{\rm on}^3}L_{\rm on}\left(a_{\rm on}\right)=3\sqrt{\Lambda}\,h(\phi)+\dot\phi\, h_{\phi}(\phi).
\end{equation}

Therefore, there are three different kind of terms which can appear in the Lagrangian. These are: (i) $X_i$-terms linear on $\dot\phi$;
(ii) $X_i$-terms with non-linear dependence on $\dot\phi$, which contribution has to vanish in the on-shell point-like Lagrangian; and, (iii) terms that are not able to self-tune because they contribute through total derivatives or terms multiplied by $k$ in the Lagrangian. Thus, the point-like Lagrangian takes the form \cite{stmodels}
\begin{equation}
 L\left(a,\,\dot a,\,\phi,\,\dot\phi\right)=L_{\rm linear}\left(a,\,\dot a,\,\phi,\,\dot\phi\right)+L_{\rm nl}\left(a,\,\dot a,\,\phi,\,\dot\phi\right)+\frac{{\rm d}}{{\rm d}t}\mu(a,\,\phi)+k\,G(a,\,\dot a,\,\phi,\,\dot\phi),
\end{equation}
with $L_{\rm linear}$ and $L_{\rm nl}$ to be given in the following discussion.

\subsubsection{Linear Terms: ``The Magnificent Seven"}

In order to satisfy equation (\ref{Lonshell}) considering only terms linear on $\dot\phi$, it is enough to have
\begin{equation}\label{lineargral}
 X_i^{\rm ms}\left(\phi,\,\dot\phi\right)=3\sqrt{\Lambda} \,U_i(\phi)+\dot\phi\,W_i(\phi),
\end{equation}
with the potentials $U_i$ and $W_i$ fulfilling the constraint
\begin{equation}\label{condnl}
  \sum_{i=0..3}W_i(\phi)\Lambda^{i/2}=\sum_{i=0..3}U_{i,\phi}(\phi)\Lambda^{i/2}.
\end{equation}

The Lagrangian of the linear terms is
\begin{equation}\label{Llinear}
 L_{\rm linear}=a^3 \sum_{i=0..3}\left[3\sqrt{\Lambda} \,U_i(\phi)+\dot\phi\,W_i(\phi)\right]H^i,
\end{equation}
with the potentials satisfying condition (\ref{condnl}). As there are a total of eight functions and only one constraint, there
are effectively only seven free functions which we coined `the magnificent seven''. The field equation for these models is
\begin{equation}\label{fieldg}
   \varepsilon_{\rm linear}=\sum_{i=0..3} \left\{3\sqrt{\Lambda} \left[U_{i,\phi}(\phi)-\frac{H}{\sqrt{\Lambda}}W_i(\phi)\right]-i\frac{\dot H}{H} W_i(\phi)\right\} H^i,
\end{equation}
where we have divided all the equations by $a^3$ for simplicity and
we require that the term in the square bracket does not vanish.
On the other hand, the Hamiltonian density is
\begin{equation}\label{Hglinear}
 \mathcal{H}_{\rm linear}=\sum_{i=0..3} \left[3(i-1)\sqrt{\Lambda}\,U_i(\phi)+i\,\dot\phi\,W_i(\phi)\right]H^i.
\end{equation}

When only this kind of terms are present the modified Friedmann equation is $\mathcal{H}_{\rm linear}=-\rho_m(a)$.
These models are able to screen any material content dynamically as long as there is at least one
$W_i\neq0$ with $i\neq0$. The models with only $W_0$ and $U_i$ potentials and those with only $U_i$'s would not spoil the screening of other models
when combined with them, but they are not able to self-tune by themselves \cite{stmodels}. It is worthy to note that a non-self-tuning Einstein--Hilbert term is contained in these models and can be explicitly written redefining $U_2(\phi)$ as $U_2(\phi)-1/(8\pi G\sqrt{\Lambda})$.

\subsubsection{Non-linear Terms}

We now consider terms with an arbitrary dependence on $\phi$ and $\dot\phi$. These terms have a minisuperspace Lagrangian given by
\begin{equation}\label{Lnl}
 L_{\rm nl}=a^3 \sum_{i=0}^3X^{\rm nl}_i\left(\phi,\,\dot\phi\right)\,H^i.
\end{equation}

The functions $X^{\rm nl}_i$ are, however, not arbitrary. Taking into account Equation~(\ref{Lonshell}), the contribution of these terms has to vanish when $L_{\rm nl}$ is evaluated at the critical point. Therefore, they have to satisfy the condition
\begin{equation}\label{condnnl}
 \sum_{i=0}^{3} X^{\rm nl}_i\left(\phi,\,\dot\phi\right)\Lambda^{i/2}=0.
\end{equation}

The Hamiltonian density of this models is
\begin{equation}\label{Hnlg}
 \mathcal{H}_{\rm nl}=\sum_{i=0}^{3}\left[(i-1)X^{\rm nl}_i\left(\phi,\,\dot\phi\right)+\dot\phi \,X^{\rm nl}_{i,\dot\phi}\left(\phi,\,\dot\phi\right)\right]H^i.
\end{equation}

Taking into account condition (\ref{condnnl}) and its derivatives, it can be seen that $ \mathcal{H}_{\rm nl,\,os}$ depends on $\dot\phi$; thus, these models are always able to self-tune. If only these nonlinear terms are present, the modified Friedmann equation is $\mathcal{H}_{\rm nl}=-\rho_m(a)$, and $\mathcal{H}_{\rm linear}+\mathcal{H}_{\rm nl}=-\rho_m(a)$ otherwise.

On the other hand, taking into account Lagrangian (\ref{Lnl}) the field equation can be written as
\begin{eqnarray}\label{fieldnong}
 \varepsilon_{\rm nl}=\sum_{i=0}^{3}  \left[X^{\rm nl}_{i,\phi}-3X^{\rm nl}_{i,\dot\phi}H- iX^{\rm nl}_{i,\dot\phi}\frac{\dot H}{H}-
 X^{\rm nl}_{i,\dot\phi \phi}\dot\phi-X^{\rm nl}_{i,\dot\phi\dot\phi}\ddot\phi\right] H^i,
\end{eqnarray}
with $\varepsilon_{\rm linear}+\varepsilon_{\rm nl}=0$.

\subsection{Cosmology of the Models}

\subsubsection{Example of a Linear Model}
 
The field equation  and the Friedmann equation read \cite{attracted1}
\begin{eqnarray*}
\label{H'}
 H'=3\frac{\sum_iH^i\left(\sqrt{\Lambda}\,  U_{i,\phi}(\phi)-H\,  W_{i}(\phi)\right)}{\sum_ii\,H^{i}  W_i(\phi)}, \qquad
 \phi'=\sqrt{\Lambda}\frac{\left(1-\Omega\right)H^2-3\sum_i(i-1)\,H^i\,  U_i(\phi)}{\sum_ii\,H^{i+1}  W_i(\phi)},
\end{eqnarray*}
where a prime represents derivative with respect to $\ln a$.

As an example of a linear model let us consider the three potentials $U_2$, $U_3$ and $W_2$. The constraint equation imposes $U_{2,\phi} \Lambda + U_{3,\phi} \Lambda^{3/2} = W_2 \Lambda$, and then
\begin{equation}
\frac{H'}{H} = -3 \frac{U_{2,\phi}}{W_2} \left(1- \frac{\sqrt{\Lambda}}{H} \right).\nonumber
\end{equation}

\begin{figure}[H]
\centering
\includegraphics[width=0.6\textwidth]{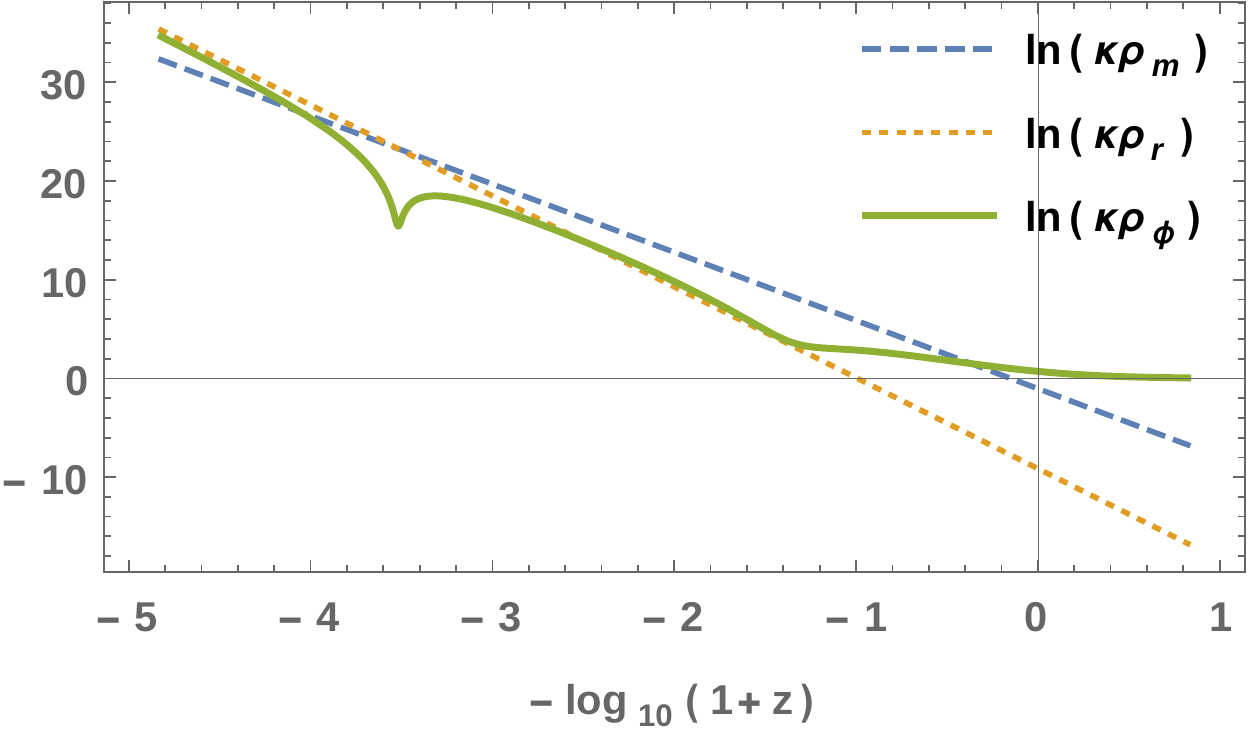}
\caption{\label{tripod1} Energy densities evolution for the models of Section 3.3.1.}

\end{figure}

For $H\gg \sqrt{\Lambda}$, \hspace{0.5cm }
\begin{equation}
\frac{H'}{H} = -\frac{3}{2}\frac{U_{2,\phi}}{W_2}, \nonumber
\end{equation}
therefore we need:
$U_{2,\phi}/W_2 = 1$, during a matter domination epoch, and
$U_{2,\phi}/W_2 = 4/3$, for a radiation domination epoch.
For example, the choice of the potentials, $U_2 = e^{\lambda \phi} + \frac{4}{3} e^{\beta \phi}$, and
$W_2 = \lambda e^{\lambda \phi} + \beta e^{\beta \phi}$, give us the desired behaviour, as shown in Figure~\ref{tripod1}.
The de Sitter evolution is attained when $H \rightarrow \sqrt{\Lambda}$.
Unfortunately, the contribution of the field at early times is too large to satisfy current constraints.

\subsubsection{Example of a Non-linear Model}

We will restrict the analysis to the  shift-symmetric cases, which means no dependence on $\phi$, and make use of the convenient redefinition $\psi = \dot\phi$. Under these assumptions we obtain the equations of motion \cite{attracted2},
\begin{eqnarray}
H'=\frac{3(1+w)Q_0P_1-Q_1P_0}{Q_1P_2-Q_2P_1}, \qquad
\psi'=\frac{3(1+w)Q_0P_2-Q_2P_0}{Q_2P_1-Q_1P_2},\nonumber
 \end{eqnarray}
where $Q_0$, $Q_1$, $Q_2$, $P_0$, $P_1$, $P_2$, are non-trivial functions of $X_i$ and $H$,
and the average equation of state parameter of matter fluids is
\begin{equation}
1+w=\frac{\sum_s\Omega_{s}(1+w_s)}{\sum_s\Omega_{s}}.\nonumber
 \end{equation}

\begin{figure}[H]
\centering
\includegraphics[width=0.6\textwidth]{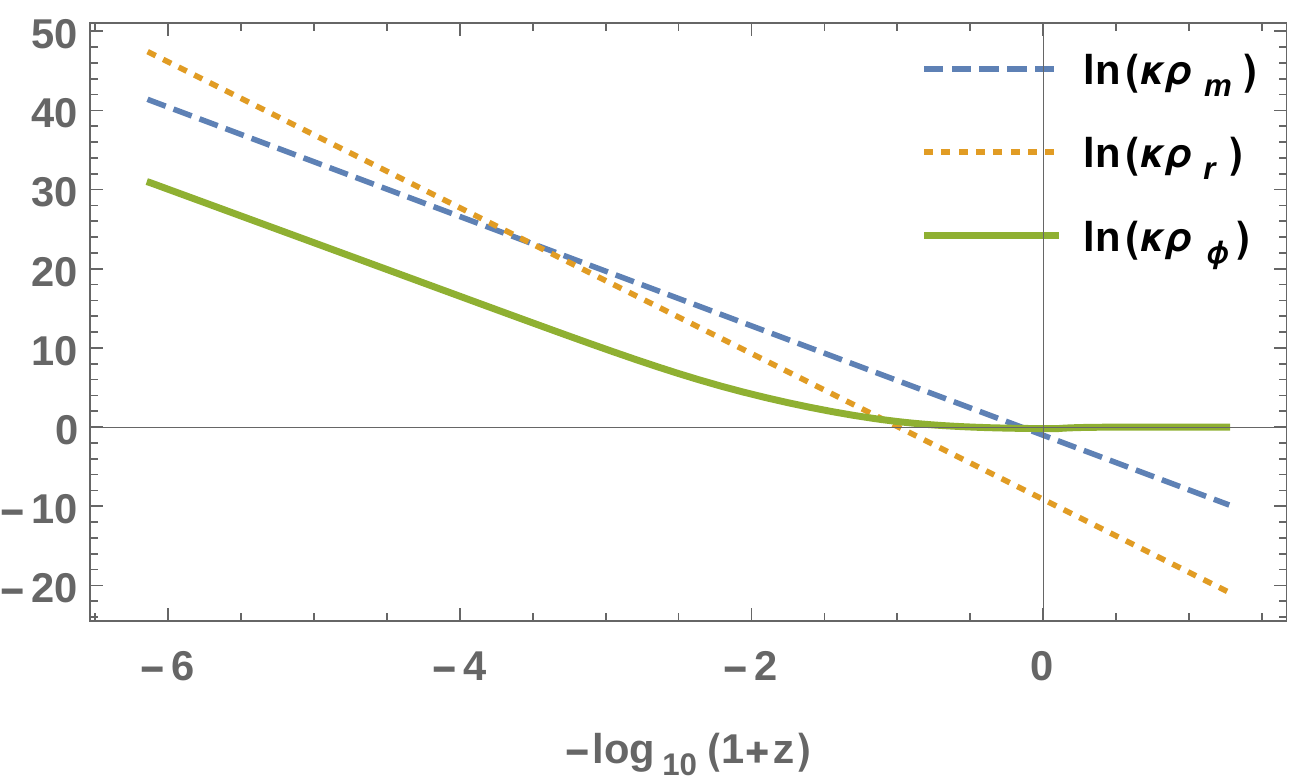}
\caption{\label{x0x1x2} Energy densities evolution for the model with non-vanishing $X_0$, $X_1$ and $X_2$. }

\end{figure}

Let us consider a case involving the three potentials $X_0$, $X_1$ and $X_2$ such that
\begin{equation}
X_2(\psi)=\alpha\psi^n,\qquad X_1(\psi)=-\alpha\psi^n+\frac{\beta}{\psi^m},\qquad X_0(\psi)=-\frac{\beta}{\psi^m}.\nonumber
\end{equation}
We can obtain a model with $w_\psi = w_0 + w_a (1-a)$, such that, $w_0 = -0.98$ and $w_a = 0.04$, which is compatible with current limits and moreover, has a negligible dark energy contribution at early times. The evolution of the energy densities is illustrated in Figure~\ref{x0x1x2}.

\subsection{Summary}

In this section we have considered a subclass of the Horndeski Lagrangian that leads to a late-time de Sitter evolution of the Universe and that may provide a mechanism to alleviate the cosmological constant problem. We have presented examples of a linear and a non-linear model. \mbox{In particular,} the shift-symmetric non-linear models, the de Sitter critical point is indeed an attractor. We can, thus,  understand the current accelerated expansion of our Universe as the result of the dynamical approach of the field to the critical point. The model presented satisfies current observational bounds on the background evolution and appear very promising.

\section{$f(R)$ Modified Theories of Gravity and Extensions}\label{sec4}

Nowadays, a great multidisciplinary effort has firmly established experimentally the standard $\Lambda$CDM model of cosmology, which is based on Einstein's General Theory of Relativity (GR), a cold dark matter source and a tiny cosmological constant. GR itself has been tested on scales ranging from the submilimeter to the solar system \cite{Will:2014xja}. Nonetheless, a number of conceptual issues still refuses to naturally merge into this paradigm. Since the establishment of the theorems by Hawking and Penrose \cite{Theorems,Theorems2,Theorems3}, space-time singularities both deep inside black holes and in the early universe have been a disturbing issue from a theoretical point of view since they represent regions where Physics looses predictability. On the other hand, introduction of a dark sector in the model allows to fit the data, but implies that up to 96\% of the total energy-matter content of the universe is of unknown nature. Though several candidates for the microscopic origin of the dark matter component have been considered, like supersymmetric partners \cite{SUSY}, this new physics has not been directly detected at particle accelerators so far. Moreover, the accumulated data in favour of a late-time accelerated expansion of the universe only hints at an exotic form of energy whose equation of state is nearly equal to that of a cosmological constant.

In view of these difficulties one may wonder whether this is a signal that the conceptual and mathematical framework represented by GR is reaching its limit of validity and that the increasing complexity of the additional ad hoc hypothesis needed to fit the data could be removed by finding a new framework for gravity where the difficulties (singularities and dark sources) are naturally encompassed. At the more fundamental level, great theoretical progress has been made within the unifying frameworks of both string theory \cite{ST} and loop quantum gravity \cite{LQG} but, given the high energy/curvature scales at which the new effects would take place, they are troubled with large difficulties to make contact with experimental results. In this sense, many authors have focused their attention on phenomenological models where specific issues of the $\Lambda$CDM model (singularities, inflation, dark sources, etc) are dealt with. In the high-energy regime, these \emph{modified gravity} models including higher-order powers of the curvature invariants (see e.g., \cite{Capozziello:2011et} for a review) are supported by a high-energy completion of the Einstein-Hilbert Lagrangian of GR within the theory of quantized fields in curved space-times \cite{QFT,QFT2}.

\subsection{$f(R)$ Modified Theories of Gravity}\label{sec:IIa}

\subsubsection{General Formalism}

Recently, $f(R)$ modified theories of gravity have been extensively explored in the literature, where the standard Einstein-Hilbert action is replaced by an arbitrary function of the Ricci scalar $R$~\cite{first,first2}. These theories can be formulated through different approaches: (i) one could use the \emph{metric} formalism, which consists in varying the action with respect to the metric $g^{\mu\nu}$ only; (ii) in the Palatini formalism \cite{Palatini,Palatini2,Palatini3,Palatini4,Palatini5,Palatini6,Palatini7,Sotiriou:2006qn}, where the metric and the affine connection are treated as independent variables (see Section (\ref{sec:IV.3}) below for more details on this formulation); (iii) and the metric-affine formalism, where the matter part of the action now depends on and is varied with respect to the connection~\cite{Sotiriou:2006qn}.

The action for the $f(R)$ modified theories of gravity is given by
\begin{equation}
S=\frac{1}{2\kappa}\int d^4x\sqrt{-g}\;f(R)+S_M(g^{\mu\nu},\psi)\,,
\label{FRaction}
\end{equation}
where $\kappa =8\pi G$. $S_M(g^{\mu\nu},\psi)$ is the matter
action, defined as $S_M=\int d^4x\sqrt{-g}\;{\cal
L}_m(g_{\mu\nu},\psi)$, where ${\cal L}_m$ is the matter
Lagrangian density, in which matter is minimally coupled to the
metric $g_{\mu\nu}$ and $\psi$ collectively denotes the matter
fields.

Using the metric approach, by varying the action with respect to
$g^{\mu\nu}$ yields the following field equation
\begin{equation}
FR_{\mu\nu}-\frac{1}{2}f\,g_{\mu\nu}-\nabla_\mu \nabla_\nu
F+g_{\mu\nu}
\nabla_{\alpha}\nabla^{\alpha}
 F=\kappa\,T^{(m)}_{\mu\nu} \,,
    \label{field:eq}
\end{equation}
where $F\equiv df/dR$. The matter energy-momentum tensor, $T_{\mu
\nu}^{(m)}$, is defined as
\begin{equation}
T_{\mu \nu
}^{(m)}=-\frac{2}{\sqrt{-g}}\frac{\delta(\sqrt{-g}\,{\cal
L}_m)}{\delta(g^{\mu\nu})} ~.
 \label{defSET}
\end{equation}

Now, considering the contraction of Equation (\ref{field:eq}), provides
the following relationship
\begin{equation}
FR-2f+3\,
\nabla_{\alpha}\nabla^{\alpha} F=\kappa\,T \,,
 \label{trace}
\end{equation}
which shows that the Ricci scalar is a fully dynamical degree of
freedom.


$f(R)$ gravity may be written as a scalar-tensor theory, by introducing a transformation $\left\{ R,f\right\}
\rightarrow \left\{ \phi ,V\right\} $ defined as
\begin{equation}
\phi \equiv F\left( R\right), \quad V\left( \phi \right) \equiv
R\left( \phi \right) F-f\left( R\left( \phi \right) \right) \,,
	\label{ScR}
\end{equation}
so that the action (\ref{FRaction}) can be written as 
\begin{equation}
S=\frac{1}{2\kappa}\int \left[ \phi R-V(\phi) +L_{m}\right]
\sqrt{-g}\; d^{4}x.
\label{BDactfR}
\end{equation}

Note that action is simply a Brans-Dicke type action (\ref{eSTTaction}) with the parameter $\omega
=0$.
The gravitational field equations of $f(R)$ gravity, in the scalar-tensor representation, simply reduce to (\ref{eSTTFE1}) and (\ref{eSTTFE2}), with a redefinition of the potential $V(\phi)=2 \phi \lambda(\phi)$.

\subsubsection{Discriminating between Dark Energy and Modified Gravity Models}
\label{DE_vs_MG}

Due to the fact that subsets of scalar-tensor theories can be obtained from modifications to the gravitational sector through specific mappings, it is fundamental to understand how one may differentiate these modified theories of gravity from dark energy models. This ambiguity requires a practical classification, which is not a trivial issue as one cannot discriminate between the dark energy models and modified gravity, solely from the expansion rate of the Universe. However, as structure formation behaves differently by these two classes of theories, information on the growth of structure, at different scales and redshifts, will break the degeneracy and will serve to discriminate between both models of dark energy and modified gravity \cite{Amendola:2012ys}. In this context, generic modifications of the dynamics of scalar perturbations, with respect to the $\Lambda$CDM background, can be represented by  two new degrees of freedom in the Einstein constraint equations, namely, through the functions $Q(a,k)$ and $\eta(a,k)$, where $a$ is the scale factor and $k$ the perturbation scale.
In modified theories of gravity, $Q(a,k)$ results from a mass-screening effect due to local modifications of gravity, and effectively modifies Newton's constant. In the context of dynamical dark energy models, the function $Q(a,k)$ incorporates the interaction with other fields, due to the perturbations. The function $\eta$, absent in $\Lambda$CDM, parameterizes the effective stresses due to the modification of gravity or specific dynamical dark energy models. Finally, the scale and time-dependence of both functions, $Q$ and $\eta$, can be derived in the considered model and traced on a $(Q,\eta)$ plane. 

Following \cite{Amendola:2012ys}, a practical manner is to denote the term ``modified gravity'' if additional contributions to the Poisson equation are presented, which induces $Q \neq 1$, and if extra effective stresses arise, implying $\eta \neq 1$. In this context, ``modified gravity'' is related to models in which modifications are present in the gravitational sector and in which dark energy clusters or interacts with other fields. Following this practical classification \cite{Amendola:2012ys}, in the context of first order perturbation theory, models with $Q=\eta =1$ are denoted standard dark energy models, such as, a minimally-coupled scalar field with standard kinetic energy \cite{Amendola:2012ys}. On the other hand, models for $Q \neq 1$ and $\eta \neq 1$ are denoted ``modified gravity'', such as scalar-tensor theories, $f(R)$ gravity, massive gravity and generalized galileons, Horndeski interactions, bi-(multi-) gravity, etc. Thus, in the context of the EUCLID mission \cite{Amendola:2012ys}, the definitions of the functions $Q$ and $\eta$ are extremely convenient, for instance, EUCLID can distinguish between standard dynamical dark energy and modified gravity by forecasting the errors on $Q$ and $\eta$, and several combinations of these functions, such as $Q/\eta$.

\subsubsection{Late-time Cosmic Acceleration}

In this subsection, we show that $f(R)$ gravity may lead to an effective dark energy, without the need to introduce a negative pressure ideal fluid. Consider the FLRW metric (\ref{Fried_met}). Taking into account the perfect fluid description for matter, we verify that the gravitational field equation, Equation (\ref{field:eq}), provides the generalised Friedmann equations in the following form
\cite{Capozziello:2003tk,Sotiriou:2007yd}:
\begin{eqnarray}
\left(\frac{\dot{a}}{a}\right)^2-\frac{1}{3F(R)}\left\{\frac{1}{2}
\left[f(R)-RF(R)\right]-3\left(\frac{\dot{a}}{a}\right)\dot{R}
F'(R)\right\}&=&\frac{\kappa}{3}\rho \,,  \\
\left(\frac{\ddot{a}}{a}\right)+\frac{1}{2F(R)}\left\{\frac{\dot{a}}{a}\dot{R}
F'(R)+\ddot{R}F'(R)+\dot{R}^2F''(R)-\frac{1}{3}\left[f(R)-RF(R)\right]\right\}
&=&-\frac{\kappa}{6}(\rho+3p) \,.
\end{eqnarray}

These modified Friedmann field equations may be rewritten in a more familiar form, as
\begin{eqnarray}
\left(\frac{\dot{a}}{a}\right)^2&=&\frac{\kappa}{3}\rho_{\rm tot} \,,  \\
\left(\frac{\ddot{a}}{a}\right)&=&-\frac{\kappa}{6}(\rho_{\rm
tot}+3p_{\rm tot}) \,,
    \label{rho+3p}
\end{eqnarray}
where $\rho_{\rm tot}=\rho+\rho_{(c)}$ and $p_{\rm
tot}=p+p_{(c)}$, and the curvature stress-energy components,
$\rho_{(c)}$ and $p_{(c)}$, are defined as
\begin{eqnarray}
\rho_{(c)}&=&\frac{1}{\kappa F(R)}\left\{\frac{1}{2}
\left[f(R)-RF(R)\right]-3\left(\frac{\dot{a}}{a}\right)\dot{R}
F'(R)\right\} \,,  \\
p_{(c)}&=&\frac{1}{\kappa
F(R)}\left\{2\left(\frac{\dot{a}}{a}\right)\dot{R}
F'(R)+\ddot{R}F'(R)+\dot{R}^2F''(R)-\frac{1}{2}\left[f(R)-RF(R)\right]\right\}
 \,,
\end{eqnarray}
respectively. The late-time cosmic acceleration is achieved if the
condition $\rho_{\rm tot}+3p_{\rm tot}<0$ is obeyed, which follows
from Equation (\ref{rho+3p}).

For simplicity, consider the absence of matter, $\rho=p=0$. Now,
taking into account the equation of state $\omega_{\rm
eff}=p_{(c)}/\rho_{(c)}$, with $f(R)\propto R^n$ and a generic
power law $a(t)=a_0(t/t_0)^\alpha$ \cite{Capozziello:2003tk}, the
parameters $\omega_{\rm eff}$ and $\alpha$ are given by
\begin{equation}
\omega_{\rm eff}=-\frac{6n^2-7n-1}{6n^2-9n+3} \,, \qquad
\alpha=\frac{-2n^2+3n-1}{n-2} \,,
\end{equation}
respectively, for $n\neq 1$. Note that a suitable choice of $n$
can lead to the desired value of $\omega_{\rm eff}<-1/3$,
achieving the late-time cosmic acceleration.

It is interesting to note that observations might slightly favour the fact that the effective dark energy equation-of-state parameter lies in the phantom regime, i.e., $\omega_{\rm eff}<-1$ \cite{Ade:2015xua}, although the data are still far from being conclusive.
In fact, recent fits to supernovae, CMB and weak gravitational lensing data indicate that an evolving equation of state crossing the phantom divide, is mildly favored, and several models have been proposed in the literature. In particular, models considering a redshift dependent equation of state, possibly provide better fits to the most recent and reliable SN Ia supernovae Gold dataset. In a cosmological setting, it has been shown that the transition into the phantom regime, for a single field is probably physically implausible \cite{Vikman:2004dc}, so that a mixture of various interacting non-ideal fluids is necessary. In this context, we refer the reader to \cite{Cai:2009zp} for a recent review on specific models in modified gravity. If confirmed in the future, this behavior has important implications for theoretical models of dark energy and modified gravity. For instance, this implies that dark energy is dynamical and excludes the cosmological constant and the models with a constant parameter, such as quintessence and phantom models, as possible candidates for dark energy. All of these models present an extremely fascinating aspect for future experiments focussing on supernovae, cosmic microwave background radiation and weak gravitational lensing and for future theoretical research.

\subsection{Curvature-matter Couplings in $f(R)$ Gravity}

Recently, an extension of $f(R)$ gravity was presented, by considering an explicit  curvature-matter coupling. The latter coupling induces a non-vanishing covariant derivative of the energy-momentum tensor, $\nabla_\mu T^{\mu\nu} \neq 0$, which potentially leads to a deviation from geodesic motion, and consequently the appearance of an extra force \cite{Bertolami:2007gv}. Implications, for instance, for stellar equilibrium have been studied in Ref. \cite{Bertolami:2007vu}. The equivalence with scalar-tensor theories with two scalar fields has been considered in Ref. \cite{Bertolami:2008im}, and a viability stability criterion was also analyzed in Ref. \cite{Faraoni:2007sn}. It is interesting to note that nonlinear couplings of matter with gravity were analyzed in the context of the accelerated expansion of the Universe \cite{Odintsov,Odintsov2,Odintsov3}, and in the study of the cosmological constant problem \cite{Lambda}.

The action for $f(R)$ curvature-matter coupling \cite{Bertolami:2007gv} takes the following form
\begin{equation}
S=\int \left\{\frac{1}{2}f_1(R)+\left[1+\lambda f_2(R)\right]{\cal
L}_{m}\right\} \sqrt{-g}\;d^{4}x~,
\end{equation}
where $f_i(R)$ (with $i=1,2$) are arbitrary functions of the
curvature scalar $R$. For notational simplicity we consider
$\kappa=1$ throughout this subsection.

Varying the action with respect to the metric $g^{\mu\nu}$ yields the field equations, given by
\begin{eqnarray}
F_1R_{\mu \nu }-\frac{1}{2}f_1g_{\mu \nu }-\nabla_\mu \nabla_\nu
\,F_1+g_{\mu\nu}\nabla_{\alpha}\nabla^{\alpha} F_1=-2\lambda F_2{\cal L}_m R_{\mu\nu}
   \nonumber \\
+2\lambda(\nabla_\mu \nabla_\nu-g_{\mu\nu}\nabla_{\alpha}\nabla^{\alpha}){\cal L}_m F_2
+(1+\lambda f_2)T_{\mu \nu }^{(m)}\,, \label{field}
\end{eqnarray}
where we have denoted $F_i(R)=f'_i(R)$, and the prime represents the derivative with respect to the scalar curvature.
We refer the reader to \cite{Harko:2010hw} for the curvature-matter coupling in the Palatini formalism.

As mentioned above, an interesting feature of this theory is the non-conservation of the energy-momentum tensor. More specifically, by taking into account the generalized Bianchi identities~\cite{Bertolami:2007gv,Koivisto,Bertolami:2008zh}, one deduces the
following conservation equation
\begin{equation}
\nabla^\mu T_{\mu \nu }^{(m)}=\frac{\lambda F_2}{1+\lambda
f_2}\left[g_{\mu\nu}{\cal L}_m- T_{\mu \nu
}^{(m)}\right]\nabla^\mu R ~, \label{cons1}
\end{equation}
where the coupling between the matter and the higher derivative curvature terms describes an exchange of energy and momentum between both.

Consider the equation of state for a perfect fluid $T_{\mu\nu}^{(m)}=\left(\rho +p\right)U_{\mu}U_{\nu}+pg_{\mu\nu}$, where $\rho$ is the energy density and $p$, the pressure, respectively. The four-velocity, $U_{\mu }$, satisfies the conditions $U_{\mu }U^{\mu }=-1$ and $U^{\mu }U_{\mu ;\nu }=0$. Introducing the projection operator $h_{\mu\nu}=g_{\mu\nu}+U_{\mu}U_{\nu}$, gives rise to non-geodesic motion governed by the following equation of motion for a fluid element: $dU^{\mu}/ds+\Gamma _{\alpha \beta}^{\mu}U^{\alpha}U^{\beta}=f^{\mu}$, where the extra force, $f^{\mu}$, is given by
\begin{equation}
f^{\mu}=\frac{1}{\rho +p}\left[\frac{\lambda F_2}{1+\lambda
f_2}\left({\cal L}_m-p\right)\nabla_\nu R+\nabla_\nu p \right]
h^{\mu \nu }\,.
     \label{force}
\end{equation}

Note that an intriguing feature is that a specific choice for the matter Lagrangian yields different dynamics. For instance, in \cite{Sotiriou:2008it}, it was argued that
a ``natural choice'' for the matter Lagrangian density for perfect
fluids is ${\cal L}_m=p$, based on Refs.
\cite{Schutz:1970my,Brown:1992kc}, where $p$ is the pressure. This
choice has a particularly interesting application in the analysis
of the $R-$matter coupling for perfect fluids, which implies in
the vanishing of the extra force \cite{Bertolami:2007gv}. However,
it is important to point out that despite the fact that ${\cal
L}_m=p$ does indeed reproduce the perfect fluid equation of state,
it is not unique \cite{BLP}. Other choices include, for instance,
${\cal L}_m=-\rho$ \cite{Brown:1992kc,HawkingEllis}, where $\rho$
is the energy density, or ${\cal L}_m=-na$, were $n$ is the
particle number density, and $a$ is the physical free energy
defined as $a=\rho/n-Ts$, with $T$ being the temperature and $s$
the entropy per particle (see Ref. \cite{BLP,Brown:1992kc} for
details).

Hence, it is clear that no immediate conclusion may be extracted
regarding the additional force imposed by the non-minimum coupling
of curvature to matter, given the different available choices for
the Lagrangian density. One may conjecture that there is a deeper
principle or symmetry that provides a unique Lagrangian density
for a perfect fluid \cite{BLP}. This has not been given due
attention in the literature, as arbitrary gravitational field
equations depending on the matter Lagrangian have not always been
the object of close analysis. See Ref. \cite{BLP} for more
details.

\subsection{The Palatini Approach} \label{sec:IV.3}

The discussion outlined in the beginning of this section hints at reconsidering the foundational aspects of gravity as a geometric phenomenon. The key point of such an approach is to make one step back and reconsider the role that basic geometric objects play in gravitational physics. The first such step is to recognize the different role that metric and affine connection have as geometrical structures, and take them as independent objects (Palatini approach \cite{Olmo}), which implies the presence of non-metricity (this is actually the simplest extension towards more general geometric theories of gravitation), i.e., the failure of the independent connection to be metric, $Q_{\lambda\mu\nu}=\nabla_{\lambda} g_{\mu\nu} \neq 0$. Note that whether the underlying structure of spacetime is Riemannian or not is a fundamental question of gravity as a geometric phenomenon which has received little attention so far, and its associated phenomenology has been scarcely studied. Indeed, a glance at the modern description of ordered structures with defects in condensed matter physics (like Bravais crystals) teaches us that the presence of point defects induces a continuous description in terms of a geometry with independent metric and affine structures \cite{Kroner,Kroner2,Kroner3}, which could provide important tools for gravitational \mbox{physics \cite{lor15,lor152}.} In the case of GR, one finds that the Palatini approach gives exactly the same result as the standard approach of fixing the Levi-Civita connection \emph{a priori} and making variations with respect to the metric $g_{\mu\nu}$ (\emph{metric} approach); this result being attached to the particular functional form of the Einstein-Hilbert Lagrangian of GR. Nonetheless, for modified gravity it turns out that it avoids many of the problems of the metric formulation, and generically yields a set of second-order field equations and absence of ghost-like instabilities.

To fix ideas let us consider an $f({\cal R})$ extension of GR, defined by the standard action
\begin{equation} \label{eq:actionfR}
S=\frac{1}{2\kappa^2}\int d^4x\sqrt{-g}\;f({\cal R})+ S_m
\,,
\end{equation}
where $\kappa^2 =8\pi G$, $g$ is the determinant of the space-time metric $g_{\mu\nu}$, which is independent of the connection $\Gamma_{\mu\nu}^{\lambda}$. The Ricci tensor, ${\cal R}_{\mu\nu}\equiv {{\cal R}^\alpha}_{\mu\alpha\nu}= -\partial_{\mu}\Gamma^{\lambda}_{\lambda \nu} + \partial_{\lambda} \Gamma^{\lambda}_{\mu\nu}+\Gamma^{\lambda}_{\mu\nu}\Gamma^{\rho}_{\rho\lambda}-\Gamma^{\lambda}_{\nu\rho}\Gamma^{\rho}_{\mu\lambda}$, is an object entirely constructed out of the independent connection, while the matter sector, $S_m =\int d^4x \sqrt{-g} L_m(\psi_m,g_{\mu\nu})$ (where $\psi_m$ denotes the matter fields) only couples to the metric $g_{\mu\nu}$. We point out that the independence between metric and connection in the Palatini approach also allows for the presence of torsion (the antisymmetric part of the connection). However, for simplicity, in this section, we will assume that torsion vanishes, $\Gamma_{[\mu\nu]}=0$, and, in addition, that the Ricci tensor is purely symmetric, ${\cal R}_{[\mu\nu]}=0$ (for a derivation of the field equations in full generality under the presence of torsion see \cite{Torsion}).

Under these conditions, variation of the action (\ref{eq:actionfR}) yields
\begin{equation} \label{eq:var-total}
\delta S= \frac{1}{2\kappa^2} \int d^4 x  \sqrt{-g} \Big[ \Big(f_{\cal R} {\cal R}_{\mu\nu}-\frac{1}{2} f g_{\mu\nu}\Big) \delta g^{\mu\nu} + f_{\cal R} g^{\mu\nu} \delta {\cal R}_{\mu\nu}(\Gamma)\Big] + \delta S_m,
\end{equation}

Since we are working on the Palatini approach, the terms on the variations with respect to $ \delta g^{\mu\nu}$ and $ \delta {\cal R}_{\mu\nu}(\Gamma)$ must vanish independently. This brings two systems of equations of the form
\begin{eqnarray}
f_{\cal R} {\cal R}_{\mu\nu} - \frac{f}{2}g_{\mu\nu}&=&\kappa^2 T_{\mu\nu}, \label{eq:metric}\\
\nabla_{\lambda}^{\Gamma} (\sqrt{-g} f_{\cal R} g^{\mu\nu})&=&0, \label{eq:connection}
\end{eqnarray}
where $f_{\cal R} \equiv df/d{\cal R}$ and the matter energy-momentum tensor $T_{\mu\nu}$ is defined by Equation (\ref{defSET}). Note that in the GR case, $f({\cal R})={\cal R}$, Equation (\ref{eq:connection}) simply expresses the standard metric-connection compatibility condition, while Equation (\ref{eq:metric}) become the Einstein equations. In the $f({\cal R})$ case, tracing with $g_{\mu\nu}$ in (\ref{eq:metric}) yields ${\cal R} f_{\cal R}-2f=\kappa^2 T$, which is an algebraic relation whose solution, ${\cal R}={\cal R}(T)$, generalizes the GR relation ${\cal R}=-\kappa^2 T$. The field equations (\ref{eq:metric}) can be solved by introducing a new rank-two tensor $h_{\mu\nu}=f_{\cal R} g_{\mu\nu}$ in such a way that the connection equation (\ref{eq:connection}) is rewritten as $\nabla_{\lambda}^{\Gamma} (\sqrt{-h} h^{\mu\nu})=0$, i.e., the independent connection $\Gamma_{\mu\nu}^{\lambda}$ becomes the Levi-Civita connection of $h_{\mu\nu}$. 
This tensor is connected to the physical metric $g_{\mu\nu}$ via the conformal transformations given above. 
With this result, the remaining equations (\ref{eq:metric}) can be written as \cite{or11}
\begin{equation} \label{eq:Rmunu}
{{\cal R}_\mu}^{\nu}(h)=\frac{\kappa^2}{f_{\cal R}^2} \left(\frac{f({\cal R})}{2\kappa^2} {\delta_\mu}^{\nu} + {T_\mu}^{\nu} \right).
\end{equation}
where ${{\cal R}_\mu}^{\nu}(h)$ is the Ricci tensor computed with the \emph{auxiliary} metric $h_{\mu\nu}$. This is a set of second-order Einstein-like equations with all the objects in the right-hand-side depending only on the matter sources (because $f({\cal R}) \equiv f({\cal R}(T))$ is a function of the trace of the energy-momentum tensor), which can thus be read as a modified matter source, namely, one can write $G_{\mu\nu}=\tau_{\mu\nu}$, where $\tau_{\mu\nu}$ accounts for an \emph{effective} energy-momentum tensor. This property, which results from the special way gravity and matter couple in the Palatini approach, is in sharp contrast with the generic fourth order equations and ghosts that one finds in the (more conventional) metric formulation of these theories [see e.g., \cite{Borunda,Borunda2} for a comparison between both approaches]. Note that in the absence of matter, ${T_\mu}^{\nu}=0$, the field equations boil down to those of GR with a cosmological constant term and thus these theories contain no extra propagating degrees of freedom (in particular, the theory is free of ghosts). As a consequence, the dynamics of these theories can be only excited in regions of large curvatures/short distances or for strong fields. Solving the equations (\ref{eq:Rmunu}) just requires to specify the matter source, ${T_\mu}^\nu$, and the object $f({\cal R}(T))$, and to transform the solution $h_{\mu\nu}$ back to the physical metric $g_{\mu\nu}$ using the conformal transformations above.

\subsubsection{Quadratic Cosmology from Gravity-matter Coupling}

This framework can be extended to include a gravity-matter coupling term $g({\cal R}) L_m(\psi_m,g_{\mu\nu})$ replacing the matter sector in the action (\ref{eq:actionfR}), which yields a new set of equations \cite{orbw}
\begin{equation} \label{eq:Rmunub}
{{\cal R}_\mu}^{\nu}(h)=\frac{\kappa^2}{\Phi^2} \left(\frac{f({\cal R})}{2\kappa^2} {\delta_\mu}^{\nu} +g({\cal R}) {T_\mu}^{\nu} \right).
\end{equation}
where now $\Phi \equiv f({\cal R})+2\kappa^2 g_{\cal R} L_m$, $h_{\mu\nu}=\Phi g_{\mu\nu}$ and $h^{\mu\nu}=\Phi^{-1} g^{\mu\nu}$. Again, these equations are manifestly second-order due to the same considerations of the previous case.

In the following, we consider a FRW cosmology, in the presence of a perfect fluid. Assuming an equation of state $P=\omega \rho$ (with $\omega=$constant), and inserting these inputs into the field equations (\ref{eq:Rmunub}) one gets two sets of equations \cite{orbw}
\begin{eqnarray}
&&H^2= \frac{f(R)+ \kappa^2 g(R) (1+3\omega)\rho}{6\Phi(1-\Delta)^2}\,, 
	\\
&&\dot{H}(1-\Delta)+\frac{H^2}{2} [1-\Delta^2 +6(\rho+p) \Delta_{\rho} ] 
=\frac{1}{12 \Phi}[f(R)-\kappa^2 g(R)(5\rho+3p)]\,,
\end{eqnarray}
where $\Delta \equiv \frac{3}{2} (1+\omega)\rho \frac{\Phi_{\rho}}{\Phi}$ and a subindex means a derivative. Both the $f({\cal R})$ and the matter-gravity coupling frameworks can successfully reproduce the background evolution of loop quantum cosmology, whose chief accomplishment is the resolution of the Big Bang singularity applying non-perturbative canonical quantization techniques \cite{LQC,LQC2}. The singularity is replaced by a non-singular cosmic bounce at an energy of Planck's density order $\rho_c \equiv c^5/(\hbar G^2)$ via a quadratic cosmology model of the form
\begin{equation}
H^2=\frac{\kappa^2}{6} \rho \left( 1+\epsilon \frac{\rho}{\rho_c} \right)\,,
\end{equation}
with $\epsilon \pm 1$, where $\epsilon=+1$ corresponds to the solution arising in braneworld scenarios \cite{bws} and $\epsilon=-1$ in loop quantum cosmology \cite{lqcs}. In $f({\cal R})$ gravity, quadratic cosmology can be numerically fit with a specific model \cite{bos} (which is consistent with the fact that space-time singularities can also be avoided in spherically symmetric charged black hole scenarios of Palatini $f({\cal R})$ gravity \cite{orf,orf2}), while in $f(R)$ theories with gravity-matter coupling different choices of $\Phi \equiv \Phi(\rho)$ yields different models able to analytically reproduce this cosmic evolution \cite{orbw}. This result can further be extended to isotropic and anisotropic cosmologies in Palatini gravity including quadratic ${\cal R}_{\mu\nu} {\cal R}^{\mu\nu}$ corrections \cite{bo}.

\subsubsection{Bouncing Cosmologies in Born-Infeld-like Gravity Models}

A somewhat different proposal for modified gravity models was introduced by Deser and Gibbons, coined Born-Infeld gravity \cite{DG,DG2}, where the action is given by
\begin{equation} \label{eq:BIg}
S_{BI} = \lambda^4 \int d^4x \left( \sqrt{g_{\mu\nu} + \lambda^{-2} R_{\mu\nu}} - \sqrt{-g} \right)\,,
\end{equation}
where $\lambda$ is a small parameter, which must be suitably tuned to recover GR at low energies/curvatures. This means that, from an experimental point of view, the action (\ref{eq:BIg}) represents a viable theory of gravity, while encompassing new effects at high energies. Born-Infeld gravity has raised great interest due to its many astrophysical and cosmological applications, including the existence of singularity-free cosmologies.

To investigate the latter point, let us rewrite, by convenience, the first term in Born-Infeld gravity as $\sqrt{-g} \det \hat{M}$, where a hat denotes a matrix and $\hat{M}=\sqrt{\hat{g}^{-1} \hat{q}}$, with $q_{\mu\nu} \equiv g_{\mu\nu}+ \lambda^{-2} R_{\mu\nu}$. Therefore we can rewrite (\ref{eq:BIg}) as
\begin{equation} \label{eq:BIg2}
S_{BI}=\lambda^4 \int d^4x \sqrt{-g} \left( \det \hat{M}- 1 \right) \,.
\end{equation}
Writing in this way, it is clear that Born-Infeld gravity (\ref{eq:BIg2}) contains two of the invariant polynomials of the matrix $\hat{M}$ and, therefore, a natural extension of this model would be \cite{jho}
\begin{equation}
S=\lambda^4 \int d^4x \sqrt{-g} \sum_{n=0}^4 \beta_n (e_n(\hat{M}))\,,
\end{equation}
where $e_n$ are the five invariant polynomials of the matrix $\hat{M}$ (see \cite{jho} for details) and $\beta_n$ is a set of free dimensionless parameters. In this formulation, the original Born-Infeld gravity comes from the contributions $e_0(\hat{M})=1$ and $e_4(\hat{M})=\det \hat{M}$.  When testing the non-singular character of solutions in the Palatini approach it is important to establish its \emph{robustness}, namely, that the singularity resolution is not attached to a particular choice of the functions defining the model, or to some ad hoc procedure to obtain a desired result, like a fine tuning of the parameters. This program has been very successful when applied to spherically symmetric black holes replacing the point-like singularity by a wormhole structure giving continuity to all null and time-like geodesics in a number of gravitational and matter scenarios \cite{or-sing,or-sing2,or-sing3}. Similarly, to study bouncing solutions, let us focus on the simple extension of Born-Infeld gravity given by
\begin{equation} \label{eq:BIc}
S=\frac{1}{\kappa^2 \epsilon} \int d^4x \sqrt{-g} \left[ f( (\det \hat{\Omega})^{1/2})- \tilde{\lambda} \sqrt{-g}  \right] + S_m(\psi_m,g_{\mu\nu})\,,
\end{equation}
where we have redefined the constants to match with GR results at low energies, $\epsilon \ll 1$, introduced a new matrix $\hat{\Omega} \equiv \hat{M}^{1/2}$, for notational simplicity, and added the matter contribution. As usual, in the Palatini formalism, we treat metric and connection as independent entities and obtain the field equations by variation with respect to both of them. After a bit of (lengthy) algebra one gets the field equations for the auxiliary metric $t_{\mu\nu}$ as \cite{oor}
\begin{equation}\label{eq:Rmn-t}
{R^\mu}_\nu(t)=\frac{\kappa^2}{2 \phi^2 (\det \hat{\Omega})^{3/2}}\left(L_G{\delta^{\mu}}_\nu+{T^{\mu}}_\nu\right) \,,
\end{equation}
with the following definitions: $\phi \equiv df/dA$ and $V(\phi)=A(\phi)f_A -f(A)$ are functions of the auxiliary scalar (Brans-Dicke-like) field $A$, the auxiliary metric $t_{\mu\nu}$ (associated with a Levi-Civita connection, $\nabla_{\lambda}[\sqrt{-t} t^{\mu\nu}]=0$) is related to the physical metric $g_{\mu\nu}$ as $t_{\mu\nu}=\phi (\det \hat{\Omega})^{1/2} g_{\mu\alpha} {\hat{\Omega}^\alpha}_{\nu}$, and $L_G\equiv (\phi (\det \hat{\Omega})^{1/2} -V(\phi)-\tilde{\lambda})/(\epsilon\kappa^2)$ is the Lagrangian density. Note that the equation resulting from variation of Equation (\ref{eq:BIc}), namely, $\det\hat{\Omega}=dV/d\phi$, establishes an algebraic relation $\phi=\phi(\det\hat{\Omega})$ for every choice of $f(A)$. Moreover, due to the fact that the equations for the metric $g_{\mu\nu}$ can be written as
\begin{equation}\label{eq:Omega-T}
\phi (\det\hat\Omega) {[{\hat\Omega}^{-1}]^\mu}_\nu=\frac{\left(\phi (\det\hat\Omega)+V(\phi)+\tilde{\lambda}\right)}{2}{\delta^{\mu}}_\nu-\frac{\kappa^2\epsilon}{2}{T^{\mu}}_\nu \,,
\end{equation}
this implies that $\hat{\Omega}=\hat{\Omega}({T_\mu}^{\nu})$ and, consequently, $\phi=\phi({T_\mu}^{\nu})$. Therefore, all terms appearing on the right-hand-side of the field equations (\ref{eq:Rmn-t}) are just functions of the matter, like in the case of $f({\cal R})$ theories. Thus, finding a solution of (\ref{eq:Rmn-t}) for $t_{\mu\nu}$ immediately yields a solution for $g_{\mu\nu}$.

Let us assume a perfect fluid energy-momentum tensor with $N$ species of non-interacting fluids, namely, $\rho=\sum_{i=1}^N \rho_i$ and $p=\sum_{i=1}^N p_i$, with equations of state $p_i=\omega_i(\rho_i) \rho_i$, in a flat FRW spacetime. The Hubble function then reads \cite{oor}
\begin{equation}\label{eq:H2final}
\epsilon H^2=\frac{1}{6 \left(1+\frac{\dot\Delta}{2\Delta H}\right)^2}\left[\frac{\epsilon \kappa^2(\rho+3p)+2(\phi (\det\hat\Omega)-V-\tilde{\lambda} )}{\phi (\det\hat\Omega)+V+\tilde{\lambda}+\epsilon \kappa^2 \rho}\right] \ ,
\end{equation}
where $\Delta=\frac{2\phi^2 (\det\hat\Omega)^{3/2}}{\phi (\det \hat\Omega)+V+\tilde{\lambda}-\epsilon \kappa^2 P}$ and
\begin{equation}\label{eq:DenomH2}
\frac{\dot\Delta}{2\Delta H}=-\frac{3}{2\Delta}\sum_{i=1}^N\left[\frac{\partial \Delta}{\partial  (\det\hat\Omega)} \frac{\partial  (\det\hat\Omega)}{\partial \rho_i}+\frac{\partial \Delta}{\partial \rho_i} \right] (1+\omega_i)\rho_i \ .
\end{equation}

To obtain explicit solutions one needs to further specify the family of Born-Infeld gravity extensions. Let us consider the power-law family $f(\det\hat{\Omega})=(\det\hat{\Omega})^n$ where the case $n=1/2$ corresponds to the original Born-Infeld gravity. Expanding in series of $\epsilon \ll 1$, the corresponding Lagrangian yields the result
\begin{equation} \label{eq:BIlow}
\lim_{\epsilon \ll 1 }S=\int d^4 x\sqrt{-g}\left[\frac{(1-\tilde{\lambda})}{\epsilon\kappa^2}+\frac{n}{\kappa^2}R+O(\epsilon)\right]+S_m \,,
\end{equation}
which, via a redefinition of constants, is just GR plus a cosmological constant term. Therefore, this kind of theories is in agreement with observational data in the low energy/curvature regime, but provides new physics regarding the early universe. In this sense, at high energies two kinds of solutions are found (see \cite{oor} for details):
(i) those with $\epsilon \rho >0$ and $\omega>0$  have a vanishing Hubble parameter for some finite energy density $\rho_B$ characterized by $H^2 \propto (\rho_B-\rho) \propto (a-a_B)^2$, and represent unstable solutions with a minimum value at past infinity, where $a(-\infty)=a_B$; (ii) those with $\epsilon \rho <0$ represent genuine non-singular bouncing solutions (as long as $\omega>-1$) with $H^2 \propto (\rho_B-\rho) \propto (a-a_B) \propto (t-t_B)^2$. This is in agreement with the fact that the avoidance of spacetime singularities in black holes within Born-Infeld gravity is achieved when $\epsilon<0$.

As the short discussion above shows, high-energy modifications of GR within the Palatini approach ($f({\cal R})$, quadratic, Born-Infeld-type, and other generalizations) are able to avoid the big bang singularity via a cosmic bounce followed by a period of de Sitter expansion in a generic and robust manner. At the same time, Palatini theories generically produce an effective cosmological constant at low energies [see Equation (\ref{eq:BIlow})]. The specific sign and value of this constant depends both on the gravitational action and on the specific choices of parameters. This suggests to consider the combination of these two properties to achieve a model in which a sufficient small cosmological constant compatible with cosmological observations arises at late times as a result of high-energy or low-energy modifications of the theory. This could provide a unified and consistent scenario accommodating the two known de Sitter phases of the cosmic expansion without invoking dark sources of energy. On the other hand, as first suggested by Hawking \cite{Hawking}, the existence of electrically charged microscopic black holes originated by large density fluctuations during the early universe (primordial black holes) could have a substantial impact in accounting for the matter density of the hypothesized dark matter particles. The prediction of some Palatini models of the existence of singularity-free, horizonless massive objects below a certain charge/mass scale \cite{lor13}, could offer new avenues for the consideration of black hole remnants \cite{Chen} as a reasonable source of dark matter. Obtaining precise estimates on the abundance of such objects, as resulting from their production in the early universe, could allow to make detailed cosmological predictions according to this viewpoint.

\subsection{Hybrid Metric-Palatini Gravity}

The hybrid metric-Palatini gravitational theory is a novel approach to modified theories of gravity that consists of adding to the Einstein-Hilbert Lagrangian an $f({\cal R})$ term constructed {\it a la} Palatini \cite{Harko:2011nh}. Using the respective dynamically equivalent scalar-tensor representation, it was shown that the theory can pass the Solar System observational constraints even if the scalar field is very light. This implies the existence of a long-range scalar field, which is able to modify the cosmological and galactic dynamics, but leaves the Solar System unaffected. The absence of instabilities in perturbations was also verified and the theory provides explicit models which are consistent with local tests and lead to the late-time cosmic acceleration.

\subsubsection{General Formalism}

The hybrid metric-Palatini gravity theory proposed in \cite{Harko:2011nh}, is given by the following action
\be
\label{action} S= \int d^4 x \sqrt{-g} \lb R + f(\R) + 2\ka^2
{L}_m \rb\,. \ee
can be recast into a scalar-tensor representation
\cite{Harko:2011nh,Capozziello:2012ny,Capozziello:2012hr} given by
the action
\begin{equation} \label{scalar2}
S= \frac{1}{2\kappa^2}\int d^4 x \sqrt{-g} \left[ (1+\phi)R
+\frac{3}{2\phi}\nabla _\mu \phi \nabla ^\mu \phi -V(\phi)
\right] +S_m \,,
\end{equation}
where $S_m$ is the matter action, $\kappa ^2=8\pi G/c^3$, and $V(\phi )$ is the scalar field potential. Note that the gravitational theory given by Equation~(\ref{scalar2}) is similar to a Brans-Dicke scalar-tensor action with parameter $w=-3/2$, but differs in the coupling to curvature.

The variation of this action with respect to the metric tensor
provides the field equations
\begin{eqnarray}\label{einstein_phi}
(1+\phi)G_{\mu\nu}=\kappa^2T_{\mu\nu} + \nabla_\mu\nabla_\nu\phi -
\nabla_{\alpha}\nabla^{\alpha}
\phi\/g_{\mu\nu}
 -\frac{3}{2\phi}\nabla_\mu\phi \nabla_\nu\phi +
\frac{3}{4\phi}\nabla_\lambda\phi\nabla^\lambda\phi g_{\mu \nu}-
\frac{1}{2}Vg_{\mu\nu},
\end{eqnarray}
where $T_{\mu \nu}$ is the matter energy-momentum tensor. The scalar
field $\phi$ is governed by the second-order evolution equation
\begin{equation}\label{eq:evol-phi}
-
\nabla_{\alpha}\nabla^{\alpha}
\phi+\frac{1}{2\phi}\nabla _\mu \phi \nabla ^\mu
\phi+\frac{1}{3}\phi\left[2V-(1+\phi)\frac{dV}{d\phi}\right]=
\frac{\phi\kappa^2}{3}T\,,
\end{equation}
which is an effective Klein-Gordon equation \cite{Harko:2011nh,Capozziello:2012ny}.

\subsubsection{Unifying Local Tests and the Late-time Cosmic Acceleration}

In the weak field limit and far from the sources, the scalar field
behaves as $\phi(r) \approx \phi_0 + ( 2G\phi_0 M /3r)
e^{-m_\varphi r}$; the effective mass is defined as $m_\varphi^2
\equiv \left. (2V-V_{\phi}-\phi(1+\phi)V_{\phi\phi})/3\right|
_{\phi=\phi_0}$, where $\phi_0$ is the amplitude of the background
value. The metric perturbations yield
\begin{eqnarray}
h_{00}^{(2)}(r)= \frac{2G_{\rm eff} M}{r} +\frac{V_0}{1+\phi_0}
\frac{r^2}{6}\,, \qquad h_{ij}^{(2)}(r)= \left(\frac{2\gamma
G_{\rm eff} M}{r}-\frac{V_0}{1+\phi_0}\frac{r^2}
{6}\right)\delta_{ij} \label{cor3}\ ,
\end{eqnarray}
where the effective Newton constant $G_{\rm eff}$ and the
post-Newtonian parameter $\gamma$ are defined as
\begin{eqnarray}
G_{\rm eff}\equiv
\frac{G}{1+\phi_0}\left[1-\left(\phi_0/3\right)e^{-m_\varphi
r}\right]\,, \qquad \gamma \equiv
\frac{1+\left(\phi_0/3\right)e^{-m_\varphi
r}}{1-\left(\phi_0/3\right)e^{- m_\varphi r}} \,.
\end{eqnarray}

As it is clear from the above expressions, the coupling of the scalar
field to the local system depends on $\phi_0$. If $\phi_0 \ll 1$,
then $G_{\rm eff}\approx G$ and $\gamma\approx 1$ regardless of
the value of $m_\varphi^2$. This contrasts with the result
obtained in the metric version of $f(R)$ theories, and,  as long as $\phi_0$ is sufficiently small, allows to
pass the Solar System tests, even if the scalar field is very light.

In the modified cosmological dynamics, consider the spatially flat
Friedman-Robertson-Walker (FRW) metric $ds^2=-dt^2+a^2(t) d{\bf
x}^2$, where $a(t)$ is the scale factor. Thus, the modified
Friedmann equations take the form
\begin{eqnarray}
3H^2&=& \frac{1}{1+\phi }\left[\kappa^2\rho
+\frac{V}{2}-3\dot{\phi}\left(H+\frac{\dot{\phi}}
{4\phi}\right)\right] \,,\label{field1} \\
2\dot{H}&=&\frac{1}{1+\phi }\left[
-\kappa^2(\rho+P)+H\dot{\phi}+\frac{3}
{2}\frac{\dot{\phi}^2}{\phi}-\ddot{\phi}\right] \,\label{field2}
\end{eqnarray}
respectively.
The scalar field equation (\ref{eq:evol-phi}) becomes
\begin{equation}
\ddot{\phi}+3H\dot{\phi}-\frac{\dot{\phi}^2}{2\phi}+\frac{\phi}{3}
[2V-(1+\phi)V_\phi]=-\frac{\phi\kappa^2}{3}(\rho-3P) .  \label{3}
\end{equation}

As a first approach, consider a model that arises by demanding
that matter and curvature satisfy the same relation as in GR.
Taking
\begin{equation} \label{pot1}
V(\phi)=V_0+V_1\phi^2\,,
\end{equation}
the trace equation automatically implies $R=-\kappa^2T+2V_0$
\cite{Harko:2011nh,Capozziello:2012ny}. As $T\to 0$ with the
cosmic expansion, this model naturally evolves into a de Sitter
phase,  which requires $V_0\sim \Lambda$ for consistency with
observations. If $V_1$ is positive, the de Sitter regime
represents the minimum of the potential.  The effective mass for
local experiments, $m_\varphi^2=2(V_0-2 V_1 \phi)/3$, is then
positive and small as long as $\phi<V_0/V_1$. For sufficiently
large $V_1$ one can make the field amplitude small enough to be in
agreement with Solar System tests. It is interesting that the
exact de Sitter solution is compatible with dynamics of the scalar
field in this model.

Relative to the galactic dynamics, a generalized virial theorem,
in the hybrid metric-Palatini gravity, was extensively analyzed
\cite{Capozziello:2012qt}. More specifically, taking into account
the relativistic collisionless Boltzmann equation, it was shown
that the supplementary geometric terms in the gravitational field
equations provide an effective contribution to the gravitational
potential energy. The total virial mass is
proportional to the effective mass associated with the new terms
generated by the effective scalar field, and the baryonic mass.
This shows that the geometric origin in the generalized virial
theorem may account for the well-known virial mass
discrepancy in clusters of galaxies. In addition to this,
astrophysical applications of the model where explored, and it was
shown that the model predicts that the effective  mass associated to the
scalar field, and its effects, extend beyond the virial radius of
the clusters of galaxies. In the context of the galaxy cluster
velocity dispersion profiles predicted by the hybrid
metric-Palatini model, the generalized virial theorem can be an
efficient tool in observationally testing the viability of this
class of generalized gravity models. Thus, hybrid metric-Palatini
gravity provides an effective alternative to the dark matter
paradigm of present day cosmology and astrophysics.

In a monistic view of physics, one would expect Nature to make somehow a choice between the two distinct possibilities offered by metric and Palatini formalisms. We have shown, however, that a theory consistent with observations and combining elements of these two standards is possible. Hence gravity admits a diffuse formulation where mixed features of both formalisms allow one to successfully address large classes of phenomena.

\subsubsection{Future Outlook: More General Hybrid Metric-Palatini Theories}

The ``hybrid'' theory space is a priori large. In addition to the metric and its Levi-Civita connection, one has also an additional independent connection as a building block to construct curvature invariants from \cite{Capozziello:2015lza}. Thus one can consider various new terms such as
\begin{align}
{\R}^{\mu\nu}{\R}_{\mu\nu}\,, \quad R^{\mu\nu}{\R}_{\mu\nu}\,, \quad {\R}^{\mu\nu\al\bt}{\R}_{\mu\nu\al\bt}\,, \quad \quad R^{\mu\nu\al\bt}{\R}_{\mu\nu\al\bt}\,, \quad \R R\,, \quad \text{etc}\,.
\end{align}

Though an exhaustive analysis of such hybrid theories has not been performed, there is some evidence that the so called hybrid class of theories we are exclusively focusing our attention upon here is a unique class of viable higher order hybrid gravity theories. In the more restricted framework of purely metric theories, it is well known that the $f(\R)$ class of theories is exceptional by avoiding the otherwise generic Ostrogradski instabilities by allowing for a separation of the additional degrees of freedom in a harmless scalar degree of freedom: as we have already seen, such a separation is possible also for the hybrid theories. Furthermore, it turns out that this feature is a similar exception in the larger space of metric-affine theories, since a generic theory there is inhabited by ghosts, superluminalities or other unphysical degrees of freedom.

\subsection{Slow-roll Inflation in $f(R)$ Gravity}

Cosmological inflation has been a great success over the last decades as it solves in an elegant way the problems raised in the Big Bang model: the so-called horizon and flatness  problems, basically initial conditions problems. Inflation consists of initial super accelerated phase, during which the above-mentioned problems are solved by requiring a sufficient duration of such a phase (see Ref.~\cite{Mukhanov,Mukhanov2,Mukhanov3} and references therein). After the end of inflation, the universe reheats and the original hot state of the Big Bang model is recovered. The original idea was proposed by Alan Guth in 1981, which was later improved by Andrei Linde in 1982. Over the years inflation has become a predictable theory, since it provides the seeds, coming from quantum fluctuations, that form the basis of the formation of large scale structures as well as the anisotropies observed in the CMB. Here, we review the essentials of the inflationary paradigm and we focus on the inflationary models within the context of extended theories of gravity, particularly the so-called $f(R)$ gravity. Furthermore, the usual observables predicted by inflation are obtained and compared with the latest data released by the Plank mission \cite{Ade:2015lrj}.

\subsubsection{Inflationary Parameters}

Slow-roll inflation has been widely studied in the literature, particularly within the framework of scalar fields, since by the appropriate scalar potential, inflation can be easily reproduced and its exit achieved at the corresponding number of e-folds (see Ref.~\cite{Lidsey:1995np} and references therein). This is probably the simplest model for slow-roll inflation, whose general Lagrangian is given by a single minimally coupled scalar field,
\be
\label{2.1}
S_{\phi} = \int d^4 x \sqrt{-g} \left(- \frac{1}{2}\partial_\mu \phi \partial^\mu \phi - V(\phi) \right)\, ,
\ee

Then, by assuming a flat FLRW metric, the FLRW equations are obtained
\ba
\frac{3}{\kappa^2} H^2 = \frac{1}{2}{\dot \phi}^2 + V(\phi)\, ,   \qquad - \frac{1}{\kappa^2} \left( 3 H^2 + 2\dot H \right)
= \frac{1}{2}{\dot \phi}^2 - V(\phi)\, .
\label{2.2}
\ea

Whereas the scalar field equation is given by
\be
\ddot{\phi}+3H\dot{\phi}+\frac{\partial V(\phi)}{\partial\phi}=0 \,.
\label{2.3}
\ee

Inflation occurs basically when the scalar field behaves approximately as an effective cosmological constant, and the friction term in (\ref{2.3}) dominates $H\dot{\phi}\gg \ddot{\phi}$ while $V \gg \dot{\phi}^2$. Inflation ends when the kinetic term of the scalar field increases and dominates over the potential, oscillating about the minimum of the potential and reheating the universe. A useful way of describing slow-roll inflation is to define the so-called slow-roll parameters:
\be
\epsilon=
\frac{1}{2\kappa^2} \left( \frac{V'(\phi)}{V(\phi)} \right)^2\, ,\qquad
\eta= \frac{1}{\kappa^2} \frac{V''(\phi)}{V(\phi)}\, , \qquad
\lambda^2 = \frac{1}{\kappa^4} \frac{V'(\phi) V'''(\phi)}{\left(V(\phi)\right)^2}\,,
\label{2.4}
\ee
where the primes denote derivatives with respect to the scalar field $\phi$. The magnitude of the quantities in (\ref{2.4}) have to be small enough during inflation in order to attain the sufficient number of e-folds, or in other words $\epsilon \ll 1$ and $\eta<1$. When inflation is ending, $\epsilon\geq 1$ is required. Thus, for the above model the  spectral index $n_\mathrm{s}$ of the curvature perturbations, the tensor-to-scalar ratio $r$ of the density perturbations, and the running of the spectral index $\alpha_\mathrm{s}$ can be written in terms of the slow-roll parameters as follows
\be
n_\mathrm{s} - 1= - 6 \epsilon + 2 \eta\, ,\qquad r = 16 \epsilon \, , \qquad \alpha_\mathrm{s} = \frac{d n_\mathrm{s}}{d \ln k} \sim 16\epsilon \eta - 24 \epsilon^2 - 2 \xi^2\, .
\label{2.5}
\ee

The values of the above parameters can be compared then with the values provided by the latest Planck results (see \cite{Ade:2015lrj}). Note that the aim here is to provide the above expressions (\ref{2.5}) in terms of a particular underlying model, capable of reproducing slow-roll inflation, in order to compare with the observational data. For the simple model (\ref{2.1}), the slow-roll parameters are already expressed in terms of the scalar potential in (\ref{2.4}), while the reconstruction of the scalar field Lagrangian (\ref{2.1}) given a particular inflationary model $H(N)$, is provided by the FLRW equations (\ref{2.2}), which yield \cite{Elizalde:2008yf}
\ba
\label{2.6bis}
\omega(\phi) &=& -\frac{2 H'(\phi)}{\kappa^2 H(\phi)}\, , \\ \nonumber
V(\phi) &=&  \frac{1}{\kappa^2}\left[ 3 \left(H(\phi)\right)^2 + H(\phi) H'(\phi) \right]\, ,
\ea
where $\omega(\phi)$ is the kinetic term and we have redefined the scalar field as $\phi=N$, where $N$ is the number of e-folds $N=\log(a/a_i)$. Hence,  by assuming a particular Hubble parameter, the corresponding Lagrangian is easily obtained. Note also that the slow-roll parameters (\ref{2.4}) can be expressed in terms of the Hubble parameter by using (\ref{2.6bis}), leading to~\cite{Bamba:2014daa,Bamba:2014daab,Bamba:2014daac}
\ba
\epsilon &=&- \frac{H(N)}{4 H'(N)} \left[ 6\frac{H'(N)}{H(N)}+ \frac{H''(N)}{H(N)} + \left( \frac{H'(N)}{H(N)} \right)^2 \right]^2\left( 3 + \frac{H'(N)}{H(N)} \right)^{-2} \, , \\ 
\eta &=&-\frac{1}{2} \left( 3 + \frac{H'(N)}{H(N)} \right)^{-1} \left[9 \frac{H'(N)}{H(N)} + 3 \frac{H''(N)}{H(N)} + \frac{1}{2} \left( \frac{H'(N)}{H(N)} \right)^2 \right.\\ \nonumber
&&\left.-\frac{1}{2} \left( \frac{H''(N)}{H'(N)} \right)^2+ 3 \frac{H''(N)}{H'(N)} + \frac{H'''(N)}{H'(N)} \right] \, , \\ 
\xi^2 &=&\frac{1}{4} \left( 3 + \frac{H'(N)}{H(N)} \right)^{-2}\left[6 \frac{H'(N)}{H(N)} + \frac{H''(N)}{H(N)}+ \left( \frac{H'(N)}{H(N)} \right)^2 \right]\left[ 3 \frac{H(N) H'''(N)}{H'(N)^2}\right .\nonumber \\
&&+ 9 \frac{H'(N)}{H(N)} - 2 \frac{H(N) H''(N) H'''(N)}{\left(H'(N)\right)^3} + 4 \frac{H''(N)}{H(N)}+ \frac{H(N) \left(H''(N)\right)^3}{\left(H'(N)\right)^4}\\ \nonumber
&& \left.+ 5 \frac{H'''(N)}{H'(N)} - 3 \frac{H(N) \left(H''(N)\right)^2}{\left(H'(N)\right)^3} - \left( \frac{H''(N)}{H'(N)} \right)^2+ 15 \frac{H''(N)}{H'(N)}+ \frac{H(N) H''''(N)}{\left(H'(N)\right)^2} \right]\,,
\label{2.7}
\ea
respectively.
By assuming a particular Hubble parameter, the slow-roll conditions are straightforward calculated, while the corresponding Lagrangian for the scalar field is also obtained through the expressions (\ref{2.6bis}). In the next section, we review previous works where instead of considering a scalar field, inflation is driven by modifications of the Hilbert-Einstein action.

\subsubsection{Inflation in $f(R)$ Gravity}

Over the last years, $f(R)$ gravity has been widely analyzed, since it can reproduce late-time acceleration as well as inflation with no need of an extra field and/or cosmological constant (for a review see \cite{Nojiri:2010wj}). Particularly, the reconstruction of the corresponding $f(R)$ action that leads to the appropriate cosmological evolution has drawn much attention (see Ref.~\cite{SaezGomez:2008uj,Nojiri:2009kx}). Here we review some aspects of $f(R)$ gravities and specifically its roll within the inflationary paradigm.

Let us start with the general action for $f(R)$ gravity, given in the action (\ref{FRaction}). The field equations are obtained by varying the above action with respect to the metric tensor, yielding Equation (\ref{field:eq}). Then, by assuming a spatially flat FLRW metric, the modified FLRW equations are obtained for $f(R)$ gravity:
\ba
\label{JGRG15}
0 &=& -\frac{f(R)}{2} + 3\left(H^2  + \dot H\right) f'(R)
 - 18 \left( 4H^2 \dot H + H \ddot H\right) f''(R) ,\\
\label{Cr4b}
0 &=& \frac{f(R)}{2} - \left(\dot H + 3H^2\right)f'(R)
+ 6 \left( 8H^2 \dot H + 4 {\dot H}^2 + 6 H \ddot H + \dddot H \right) f''(R)
   \nonumber \\
&&+ 36\left( 4H\dot H + \ddot H\right)^2 f'''(R)\,.
\ea
As shown in Ref.~\cite{Bamba:2014wda,DeFelice:2011jm}, the corresponding slow-roll parameters (\ref{2.7}) can be expressed in terms of the $f(R)$ action as well. Nevertheless, here we are interested in reviewing slow-roll inflation in $f(R)$ gravity by using the scalar-tensor equivalence of $f(R)$ gravity and expressing the action (\ref{FRaction}) in the Einstein frame.

Firstly, note that the action (\ref{FRaction}) in the Jordan frame, can be rewritten by action (\ref{BDactfR}), so that by applying the conformal transformation
\be
\tilde{g}_{\mu\nu}=\Omega^2 g_{\mu\nu}\ \quad \text{where} \quad \Omega^2=\phi \,,
\label{ConfT}
\ee
the action (\ref{BDactfR}) is transformed as follows
\be
\tilde{S}=\int d^4x\sqrt{-\tilde{g}}\left[\frac{\tilde{R}}{2\kappa^2}-\frac{1}{2}\partial_{\mu}\tilde{\phi}\partial^{\mu}\tilde{\phi}-\tilde{V}(\tilde{\phi})\right]\ ,
\label{action-Einstein-frame}
\ee
where
\be
\phi=e^{\kappa\sqrt{\frac{2}{3}}\tilde{\phi}}\,,
\qquad 
\tilde{V}=\frac{e^{-2\kappa\sqrt{\frac{2}{3}}\tilde{\phi}}}{2\kappa^2}V\ .
\label{Potential-Einstein-frame}
\ee

Hence, the Lagrangian is now written in the form of (\ref{2.1}), such that slow-roll inflation occurs when the conditions described above are satisfied. In addition, the slow-roll parameters (\ref{2.4}) can be expressed in terms of the original $f(R)$ action by using the correspondence expressions (\ref{ScR}) and (\ref{Potential-Einstein-frame}), leading to
\ba
\epsilon&=&\frac{1}{3}\frac{(-2 f+f_R R)^2}{f_R^4\left(f-Rf_R\right)^2}\ , \\ \nonumber
\eta&=&-\frac{2}{3}\frac{f_R^2-4f_{RR}+Rff_Rf_{RR}}{f_{RR}f_R^2\left(f-Rf_R\right)}\ .
\label{SRPFR}
\ea

Then, given a particular form of the $f(R)$ action, the corresponding slow-roll parameters can be calculated in terms of the Hubble parameter, which is obtained eventually by solving the FLRW equations. In addition, the corresponding spectral index and scalar-to-tensor ratio are obtained and for a particular number of e-folds, (most of inflationary models require $N_e\sim55-65$), both magnitudes can be compared with the constraints from the Planck results \cite{Ade:2015lrj}. 

Let us now consider the so-called Starobinsky inflation \cite{staro}, which has been a very successful inflationary model, taking into account the observational constraints of the Planck satellite. The model has also been analysed from the quantum point of view, where it was shown to provide a correct behaviour and a successful inflationary period \cite{Copeland:2013vva}. 
The model corresponds to a particular form of the $f(R)$ function consisting of the Einstein--Hilbert term and a quadratic correction, given by
\begin{align}
f(R) = R + \frac{R^2}{6 m^2}.
\end{align}

The parameter $m^2$ in the Starobinsky model is a free parameter which is constrained by the observations, and it is in fact related to the mass of the new, dynamical scalar degree of freedom present in the model, usually denoted as the ``scalaron''. Although the inflationary dynamics of the model can be analysed either in the Jordan or in the Einstein frame, here we will focus our discussion on the Einstein-frame action, applying the procedure described earlier. Notice that according to the previous discussion, the requirement of the positivity of the second derivative of $f$, $f_{RR}  >0$, translates into the requirement that the squared mass $m^2$ remains positive, i.e., $m^2 >0$.

In order to re-express the model in the Einstein frame, we note that for the Starobinsky model the Einstein-frame scalar is given by
\begin{align}
\tilde \phi = \sqrt{\frac{3}{2}} \kappa^{-1} \log \left(1 + \frac{\phi}{3m^2} \right).
\end{align}

Given the last relation, and using Equation (\ref{Potential-Einstein-frame}), it is straightforward to derive the Einstein frame potential as
\begin{align}
\tilde{V}( \tilde{\phi}) = \frac{3}{4 \kappa^2} m^2 \left( 1 - e^{- \sqrt{2/3} \kappa \tilde{\phi}} \right)^2 . \label{V-Star-Einstein}
\end{align}

Therefore, the Einstein-frame action will be described by the action (\ref{action-Einstein-frame}), with the potential defined through Equation (\ref{V-Star-Einstein}).

During inflation, the background spacetime is described by a FRW spacetime, which after the conformal transformation (\ref{ConfT}) is expressed as follows
\begin{align}
ds^{2} = -dt^2 + \tilde{a}(t)^2 d{\bf x}^2,
\end{align}
where $\tilde{a}(t)$ denotes the scale factor in the Einstein frame. Similarly to Equation (\ref{2.2}), the Friedmann equation in this spacetime is described by
\begin{align}
\tilde{H}^2 \equiv \left( \frac{\dot{\tilde{a}}}{\tilde{a}} \right)^2 = \frac{\kappa^2}{3}\left(\frac{ \dot{\tilde{\phi}}^2 }{2 } +  \tilde{V}(\tilde{\phi})\right), \label{Friedman-Starobinsky}
\end{align}
while the equation of motion for the scalar $\phi$ takes the form
\begin{align}
\ddot{\tilde{\phi}} + 3\tilde{H} \dot{\tilde{\phi}} +  \tilde{V}'(\tilde{\phi}) = 0, \label{Phi-eq-Starobinsky}
\end{align}
where $\tilde{V}'(\tilde{\phi}) \equiv d \tilde{V}(\tilde{\phi})/d \tilde{\phi}$.
The first and second term on the right hand side of the Friedmann equation (\ref{Friedman-Starobinsky}) describe the kinetic and potential energy of the inflaton. Recall that slow--roll inflation occurs in the regime where $\kappa \phi \gg 1$, i.e., as long as the field $\tilde{\phi}$ rolls down the flat part of the potential (\ref{V-Star-Einstein}) sufficiently slowly, and the kinetic energy of the inflaton is much smaller than its potential energy, $\dot{\tilde{\phi}}^2 \ll \tilde{V}(\tilde{\phi})$. Then the Friedmann and the scalar field equation, Equations (\ref{Friedman-Starobinsky}) and (\ref{Phi-eq-Starobinsky}), respectively, become approximately
$\tilde{H}^2  \simeq  \frac{\kappa^2}{6}  \tilde{V}(\tilde{\phi})$, and $3\tilde{H} \dot{\tilde{\phi}} \simeq -  \tilde{V}'(\tilde{\phi})$.

The Hubble function $\tilde{H}$ is approximately constant during inflation, and its time variation is described by the slow--roll parameter $\epsilon$ defined as
\begin{align}
\epsilon & \equiv - \frac{\dot{\tilde{H}}}{\tilde{H}^2}.
\end{align}

In the slow-roll approximation it is required that $\epsilon$ is small, $\epsilon \ll1$. We can approximate it as
\ba
\epsilon &\simeq& \frac{1}{2 \kappa^2} \left( \frac{\tilde{V}'(\tilde{\phi})}{\tilde{V}(\tilde{\phi})} \right)^2\; \simeq \frac{4}{3} \left(e^{\sqrt{\frac{2}{3}}\kappa \tilde{\phi}} -1 \right)^{-2}  \simeq \frac{4}{3} e^{-2\sqrt{\frac{2}{3}}\kappa \tilde{\phi}}\ , \\ \nonumber
\eta &\simeq& \frac{1}{\kappa^2}\frac{\tilde{V}''(\tilde{\phi})}{\tilde{V}(\tilde{\phi})}\; \simeq \frac{4}{3} \frac{2-e^{\sqrt{\frac{2}{3}}\kappa \tilde{\phi}}}{\left(-1+e^{\sqrt{\frac{2}{3}}\kappa \tilde{\phi}}\right)^{2}}  \simeq -\frac{4}{3} e^{-\sqrt{\frac{2}{3}}\kappa \tilde{\phi}}.
\ea

The last step in the above relation assumed that $\kappa \tilde{\phi} \gg 1$. Inflation ends as long as $\epsilon \simeq 1$, which corresponds to the inflaton's kinetic energy becoming important.

The amount of inflation is measured by the number of e-foldings $N$ which is defined as
\begin{align}
N  \equiv  \int_{t_{start}}^{t_{end}} \tilde{H} dt\,,
\end{align}
and for agreement with observations one usually requires that $N \simeq 55 - 65$. Now, under the slow-roll approximation, the above relation takes the following approximate form
\begin{align}
N &\simeq -\kappa^2 \int_{\tilde{\phi}_{start}}^{\tilde{\phi}_{end}} \frac{\tilde{V}(\tilde{\phi})}{\tilde{V}'(\tilde{\phi})} d\phi
\simeq \frac{3}{4}e^{\sqrt{2/3}\kappa \; \tilde{\phi}_{start}}.
\end{align}

In the last step, we used the fact that in the slow-roll approximation, $\kappa \tilde{\phi} \gg 1$, and we performed the integration with respect to $\tilde{\phi}$. Notice that the number of e-foldings is related to the slow-roll parameters as
\be
\epsilon \simeq \frac{3}{4}\frac{1}{N^2}\ , \qquad \eta \simeq -\frac{1}{N}\ .
\label{SRparamStaro}
\ee

The scalar and tensor fluctuations produced during inflation have a characteristic amplitude and scale dependence, which in the Einstein frame, under the slow-roll approximation, are given by the standard relations for GR minimally coupled to a scalar field \cite{Mukhanov,Mukhanov2,Mukhanov3},
\begin{align}
& P_{\mathcal{R}} \simeq  \frac{1}{4\pi^2}\frac{\tilde{H}^4}{\dot{\tilde{\phi}}^2 } \simeq \frac{N^2}{3 \pi}  \frac{m^2}{m_{p}^2}, \\
& P_{GW} \simeq \frac{16}{\pi} \frac{\tilde{H}^2}{m_{p}^2} \simeq \frac{4}{ \pi}  \frac{m^2}{m_{p}^2},
\end{align}
while the tensor to scalar ratio follows from the above relations as $r \equiv P_{GW} /P_{\mathcal{R}}\simeq  (48/\pi)N^{-2} \simeq 16 \epsilon$. Notice that in the final approximate equalities of the above relations we restricted ourselves to the Starobinsky potential for $\kappa \tilde{\phi} \gg 1$, under the slow-roll approximation.
The spectral indices corresponding to the scalar and tensor fluctuations, respectively, under the slow-roll regime are given by the relations (\ref{2.5}).

Note that Planck provides a value for the spectral index $n_s=0.968\pm 0.006$ and an upper bound for the scalar-to-tensor ratio $r< 0.07$ coming from a joint analysis of the BICEP2/Keck Array and Planck data \cite{Ade:2015lrj}. In order to illustrate the power of Starobinsky inflation, let us assume a number of e-folds $N_e=65$, which leads to the following values of the inflationary observables:
\be
n_s=0.968\,, \qquad r=0.00284\,.
\label{SpRStaro}
\ee

Hence, Starobinsky inflation satisfies quite well the constraints coming from Planck data. \\

Let us consider now a slight modification of Starobinsky inflation to show the uniqueness of the above model,
\begin{align}
f(R) = R + \left(\frac{R}{6 m}\right)^{2+\lambda},
\end{align}
where $\lambda$ is a positive parameter that accounts for the deviation from Starobinsky inflation. In this case, and following the same steps as above, the scalar potential in the Einstein frame yields,
\be
\tilde{V}(\tilde{\phi})=A\left[B e^{\kappa\sqrt{\frac{2}{3}}\tilde{\phi}}-1\right]^{\frac{2+\lambda}{1+\lambda}}\; e^{-2\kappa\sqrt{\frac{2}{3}}\tilde{\phi}}\ ,
\label{PotentialBS}
\ee

Here $A=\frac{\lambda}{2\kappa^2(6m)^{2+\lambda}}$ and $B=\frac{(6m)^{2+\lambda}}{2+\lambda}$. This actually does not lead to an effective cosmological constant when $\kappa\tilde{\phi}\gg 1$ since at that limit, the scalar potential becomes
\be
\tilde{V}(\tilde{\phi})\simeq AB^{\frac{2+\lambda}{1+\lambda}} e^{-\kappa\sqrt{\frac{2}{3}}\left(\frac{\lambda}{1+\lambda}\right)\tilde{\phi}}\ .
\label{PotentialBSappr}
\ee

Since $\lambda$ is positive, the potential is an exponential potential at large values of the scalar field which only mimics an effective cosmological constant in the case of $\lambda=0$ when Starobinsky inflation is recovered. Otherwise, for $\kappa\tilde{\phi}\gg 1$, the scalar potential is exponentially suppressed, and inflation does not occur. In the case of $\lambda<0$, the scalar field will roll fast, invalidating the slow-roll conditions. Hence, deviations from $\lambda=0$ will not reproduce slow-roll inflation, what may lead to other kinds of inflationary models as chaotic inflation which present deviations from the values of the spectral index and the scalar-to-tensor ratio as pointed by Planck Collaboration \cite{Ade:2015lrj}).

Other extensions of $f(R)$ gravity can also generate slow-roll inflation. In this sense, the inflationary paradigm has been recently explored within the so-called $f(R,\mathcal{L}_{\phi})$ gravities, where $\mathcal{L}_{\phi}=-\frac{1}{2}\partial_{\mu}\phi\partial^{\mu}\phi-V(\phi)$ (see \cite{delaCruz-Dombriz:2016bjj}). Such kind of theories extends $f(R)$ gravities by including a nonminimally coupling scalar field. As shown in Ref.~\cite{delaCruz-Dombriz:2016bjj}, this model is equivalent to multifield inflationary scenarios, where isocurvature modes may become important, while adiabatic perturbations can be analytically calculated (see \cite{Sasaki:1995aw}). However, following \cite{delaCruz-Dombriz:2016bjj} the adiabatic perturbations can be obtained and the isocurvature modes may be ignored under certain conditions on the action.

\section{Topological Defects}\label{sec5}

The Standard Model of Particle Physics describes three of the fundamental interactions between particles --- the electromagnetic, weak and strong forces --- in a unified quantum field theory. According to this model, the universe underwent, in its early stages, a series of symmetry breaking phase transitions, some of which may have led to the formation of topological defects.

\subsection{Spontaneous Symmetry Breaking and Topological Defects}

To illustrate the process of topological defect formation as a result of spontaneous symmetry breaking, let us consider the simplest model that admits topological defect solutions: the Goldstone model with a single scalar field, $\phi$. In this model, the Lagrangian density is given by
\begin{equation}
\mathcal{L}=\frac{1}{2}\phi_{,\mu}\phi^{,\mu}-\frac{\lambda}{4}\left(\phi^2-\eta^2\right)^2\,,
\end{equation}
where $\lambda$ is a coupling constant. Although this Lagrangian is invariant under transformations of the form $\phi\rightarrow-\phi$ --- thus having a $Z_2$ symmetry --- the Goldstone potential has two degenerate minima at $\phi=\pm\eta$. Hence, once $\phi$ acquires a vacuum expectation value (VEV),
\begin{equation}
\phi_{\pm}=\left\langle 0\right|\phi\left|0\right\rangle=\pm\eta\,,
\end{equation} the $Z_2$ symmetry is spontaneously broken, leading to the formation of a kink separating regions where the scalar field has different VEVs (here, $\left|0\right\rangle$ is the ground state of the model).

In a $1+1$ dimensional Minkowski spacetime, the equation of motion of the scalar field,
\begin{equation}
\frac{\partial^2\phi}{\partial t^2}-\frac{\partial^2\phi}{\partial x^2}=-\frac{dV}{d\phi}\,,
\end{equation}
admits the following solution
\begin{equation}
\phi(x)=\eta\tanh\left[\eta{(\lambda /2)^{1/2}} \gamma(x-x_0-vt)\right]\,,
\label{staticsol}
\end{equation}
with $\gamma=(1-v^2)^{-1/2}$, that corresponds to a kink moving with velocity $v$ in the positive $x$-direction. Since $\phi$ interpolates between $\phi=-\eta$ at $x=-\infty$ and $\phi=\eta$ at $x=\infty$, there is a region (the domain wall) centred at $x=x_0+vt$ which has non-vanishing energy.  This configuration has a conserved topological charge, $Q=\int dx j^0=\phi(+\infty)-\phi(-\infty)=2\eta$ (where $j^{\mu}=\epsilon^{\mu\nu}\phi_{,\nu}$ is the topological current), and, for this reason, it is stable. The constant vacuum state  has \mbox{$Q=0$,} and, as a result, this configuration is unable to relax into the vacuum while conserving its topological charge. \mbox{In $N+1$ dimensions,} the solution in Equation (\ref{staticsol}) remains valid and represents a \mbox{$(N-1)$-dimensional} planar surface. These surfaces also separate two domains with different VEVs and, for this reason, they are commonly denominated \textit{Domain Walls}.

Domain walls arise due to the breaking of a discrete symmetry, whenever the vacuum manifold is disconnected. However, different types of defects may be formed depending on the (non-trivial) topology of the vacuum manifold (or the type of symmetry being broken). Line-like defects, or \textit{Cosmic strings}, are formed if the vacuum is not simply connected or, equivalently, if it contains unshrinkable loops. This type of vacuum manifold results, in general, from the breaking of an axial symmetry. Moreover, if the vacuum manifold contains unshrinkable surfaces, the field might develop non-trivial configurations corresponding to point-like defects, known as \textit{Monopoles}. The spontaneous symmetry breaking of more complex symmetry groups may lead to the formation of textures, delocalized topological defects which are unstable to collapse.

\subsection{A VOS Model for Topological Defect Networks of Arbitrary Dimensionality}

The production of topological defects is predicted in several grand unified scenarios. \mbox{In most} instances, these topological defects may survive throughout cosmological history until late cosmological times, and leave a variety of imprints on different observational probes. A unified phenomenological framework for the description of the cosmological evolution of topological defects of arbitrary dimensionality was developed in \cite{Avelino:2008mv,Sousa:2011ew,Sousa:2011iu}. This model---which is a generalization of the Velocity-dependent One-Scale (VOS) model for cosmic strings \cite{Martins:1996jp}---provides a statistical description of the large-scale dynamics of these networks. This generalized VOS model characterizes quantitatively the macroscopic dynamics of the $p$-dimensional topological defect network in $(N+1)$-dimensional Friedmann-Robertson-Walker backgrounds by describing the evolution of the characteristic length of the network, $L$, and of its root-mean-squared velocity (RMS), $\vv$.

The  characteristic length of a statistically homogeneous defect network may be defined as
\be
\bar{\rho}=\frac{\sigma_p}{L^{N-p}}\,,
\ee
where $\bar{\rho}$ is the average defect energy density, and $\sigma_p$ is the defect mass per unit $p$-dimensional area. For infinitely thin and featureless topological defects, the following equations describing the evolution of the characteristic length and velocity of the network ($L$ and $\vv$, respectively) may be obtained directly from the generalized Nambu-Goto action \cite{Sousa:2011iu}
\begin{eqnarray}
\frac{d\vv}{dt} & = & \left(1-\vv ^2\right)\left[\frac{k}{L}-\frac{\vv}{\ell_d}\right]\,,\\
\frac{dL}{dt} & = & HL+\frac{L}{D\ell_d}\vv^2+\frac{\cc}{D}\vv\,,\label{VosL}
\end{eqnarray}
where $D=N-p$, and $\ell_d^{-1}=(p+1)H+\ell_f^{-1}$ is a damping length scale that includes, not only the effects of cosmological expansion, but also the effect of the frictional forces caused by the scattering of particles off the defects (which is encoded in the friction length scale, $\ell_f$). Here, $k$ and $\cc$ are phenomenological parameters that account, respectively, for the effect of defect curvature and the energy loss associated to defect reconnection on the macroscopic dynamics of the network. \mbox{The main} advantage of this phenomenological model lies precisely on these two parameters. They may either be calibrated using numerical simulations---as was done for cosmic strings \cite{Martins:2000cs} and domain walls \cite{Leite:2012vn}---to accurately describe the large scale dynamics of standard networks, or be used in a phenomenological description of a large variety of alternative defect scenarios.

This simplest version of the VOS model has been extended in order to allow for the description of non-standard defect scenarios, with additional energy-momentum decay channels (e.g., direct energy loss through gravitational radiation \cite{Martins:2000cs}) or additional degrees of freedom (e.g., the existence of conserved currents \cite{Oliveira:2012nj}). Other extensions allow for the description of more complex defect scenarios---such as multi-tension cosmic (super-)string networks with junctions \cite{Avgoustidis:2007aa}, semi-local cosmic string networks \cite{Nunes:2011sf}, or hybrid defect networks \cite{Martins:2009hj}---by including terms to account for the energy-momentum transfer between different network components.

\subsection{Observational Signatures of Topological Defects\label{alpha}}

The observational consequences of topological defect networks can be very diverse, depending both on the type of defects formed and on the evolution of the universe after they are generated. For instance, the average domain wall density grows more rapidly than the background density in the radiation and matter eras and, consequently, domain walls must necessarily be very light in order not to completely dominate the energy density of the universe at late times. Although the possibility of a significant contribution to the dark energy budget has been ruled out both dynamically \cite{PinaAvelino:2006ia} and observationally \cite{Ade:2015xua}, it is possible for light domain walls to leave behind interesting astrophysical and cosmological signatures.

Other topological defects, such as cosmic strings in a linear scaling regime, have an average density that is roughly proportional to the background density. Hence, standard cosmic string networks do not tend to dominate the energy density of the universe and they naturally generate a roughly scale-invariant spectrum of density perturbations on cosmological scales. Although recent cosmological data, in particular the cosmic microwave background anisotropies, rules out cosmic strings as the main source of large-scale cosmological perturbations (see, however, \cite{Avelino:2000iy}), it remains plausible that they may play an important role on small scales. In fact, it has been shown that the power spectrum of a string-induced density perturbation component is expected to dominate over a primordial inflationary component on small enough scales \cite{Wu:1998mr}. In this context, cosmic strings have been suggested to have significant impact on the formation of ultracompact minihalos \cite{Anthonisen:2015tda}, globular clusters \cite{Barton:2015zra}, super-massive black holes \citep{Bramberger:2015kua} and to provide a significant contribution to the reionization history of the Universe \cite{Avelino:2003nn}. The dynamics of cosmic defects, in particular of domain walls, has also been associated to possible variations of the fundamental couplings of nature (see, e.g., \cite{Avelino:2014xsa}).

In this section, we shall focus on the contribution of both cosmic strings and domain walls to two of the most significant observational probes: the temperature and polarization anisotropies of the cosmic microwave background and the stochastic gravitational wave background.

\subsubsection{CMB Anisotropies}

The presence of a topological defect network at late cosmological times would necessarily leave imprints on the Cosmic Microwave Background (CMB). Although current CMB observations seem to be consistent with the inflationary paradigm (in which the fluctuations are seeded in the very early universe) they also allow for a subdominant topological defect contribution \cite{Ade:2013xla}. Topological defects source metric perturbations actively throughout the cosmological history. For this reason, their CMB signatures are expected to be fundamentally different from those due to primordial fluctuations. In particular, cosmic defects are expected to generate a significant vector component that would not be present in inflation-seeded scenarios, because vector modes decay rapidly in the absence of a source \cite{Bardeen:1980kt}. The B-mode polarization signal originated by topological defects has then contributions from both tensor and vector modes, and may, for this reason, produce an observationally relevant signal in this channel, despite providing only subdominant contributions to the temperature and E-mode power spectra. The B-mode polarization channel thus offers a relevant observational window for the detection of topological defects. With several CMB experiments planned for the near future---e.g., LITEBIRD, PRISM and CoRE+ (see \cite{Martin:2014rqa} for a recent overview)---there is the prospect either for the detection of topological defects or for the tightening of current observational constraints in this scenario.

\begin{figure}[H]
\centering
\includegraphics[width=6in]{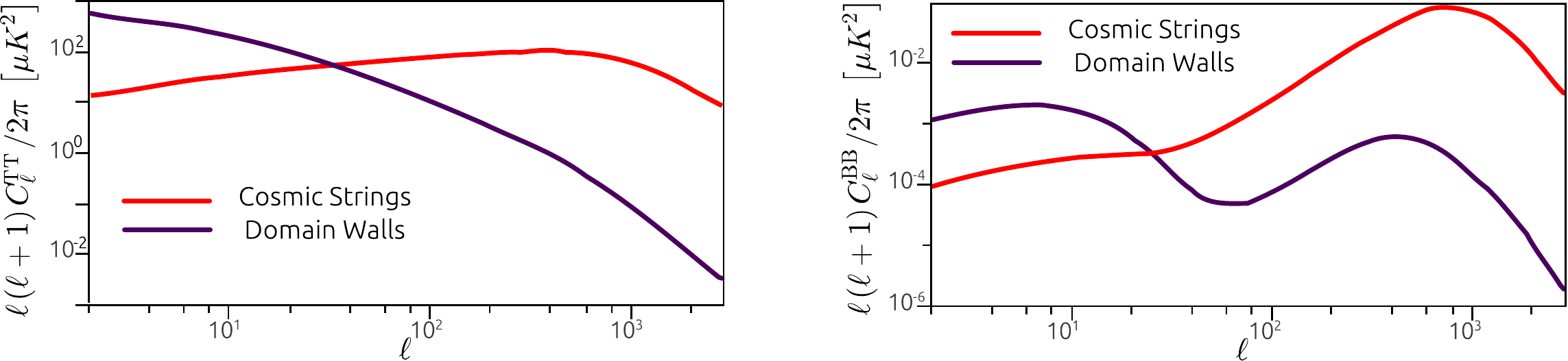}
\caption{The total temperature (left panel) and B-mode (right panel) angular power spectra generated by cosmic string and domain networks with the maximum tension allowed by current observational data: $G\sigma L_0=5.6 \times 10^{-6}$ and $G\mu=2.4 \times 10^{-7}$ for domain walls and cosmic string, respectively. The spectra were obtained using, respectively, the CMBACT code \cite{Pogosian:1999np,Pogosian:2006hg}, and the extension introduced in \cite{Sousa:2015cqa}. These plots have been adapted from \cite{Sousa:2015cqa}.}
\label{CMB}
\end{figure}

Cosmic string and domain wall networks produce temperature anisotropies characterized by significantly different angular power spectra. Domain walls only become cosmologically relevant at late times, when the characteristic length scale of the network is large. For this reason, these networks contribute mostly to the TT angular power spectrum at large scales (or small multipole modes $\ell$). Since low-$\ell$ observational data is highly affected by cosmic variance and has thus large associated uncertainties, current data allows for a significant contribution of standard domain wall networks to the temperature power spectrum. In Ref. \cite{Sousa:2015cqa}, the authors have derived a conservative constraint on the domain wall mass per unit area of $\sigma<3.52\times 10^{-5}\,{\rm kg\ m^{-2}}$ allowed by current observational data (which corresponds to a constraint on the wall-forming symmetry breaking scale of $\eta<0.92\,{\rm MeV}$). Cosmic strings, on the other hand, generate a TT angular power spectrum that has a broad peak at $\ell\sim 300$, and therefore contribute dominantly to the angular power spectrum at higher multipoles (which have small observational uncertainties). 
For this reason, observational data allow for a significantly larger contribution of domain walls to the temperature power spectrum on large angular scales than it does of cosmic strings.
Current CMB observational data constrain the cosmic string tension to be  $G\mu<1.3\,(2.4)\times 10^{-7}$ for Nambu-Goto (Abelian-Higgs) strings \cite{Ade:2015xua} (which is significantly smaller than the corresponding constraint on domain wall mass per unit area $G\sigma L_0=5.6\times 10^{-6}$). Theses differences in the shape and amplitude of the temperature power spectra generated by domain wall and cosmic string networks are clearly illustrated in the left panel of Figure \ref{CMB}, where the TT power spectra generated by cosmic string and domain wall networks with the maximum tensions allowed by current observational data are plotted. We chose $G\sigma L_0 = 5.6 \times 10^{-6}$ for the domain wall networks, and the weakest constraint on cosmic string tension obtained using Planck data \cite{Ade:2015xua}, $G\mu = 2.4 \times 10^{-7}$ (which corresponds to the constraint on Abelian-Higgs strings).

Although cosmic defect contributions to the temperature power spectrum are necessarily subdominant when compared to the primordial inflationary contribution, they may produce a significant B-mode polarization signal. As for primary temperature anisotropies, cosmic strings and domain walls contribute to the BB angular power spectrum at different scales: domain walls contribute dominantly at low-$\ell$, while strings have a significant contribution at small scales. Moreover, the B-mode signal generated by domain wall networks has, in general, smaller amplitude than that of cosmic strings.  In the right panel of Figure \ref{CMB}, we plot the total BB power spectrum generated by domain wall networks and cosmic string networks that have the maximum fractional contribution to the TT power spectrum allowed by current observational data. This figure shows that, since the observational constraints on $G\mu$ are more stringent than those on $G\sigma L_0$, the domain wall contribution may dominate over that of cosmic strings for low multipole modes (particularly if their energy density is close to the maximum allowed by current data). Moreover, the fact that domain walls and cosmic strings have spectra with different shapes and that these defects contribute mostly at different scales should, in principle, allow one to distinguish between their contributions if a signal is detected.

\subsubsection{The Stochastic Gravitational Wave Background}

Cosmic string interactions often result in the formation of loops that detach from the long string network (for instance, in self-intersections or when kinky strings collide). After formation, these loops oscillate relativistically under the effect of their tension, and lose their energy in the form of gravitational waves. These loops, thus, have a finite life span and, as a consequence, the cosmic string network loses energy through this channel. The dynamical effects caused by this energy loss are taken into account in the VOS model through the inclusion of the last (phenomenological) term in Equation (\ref{VosL}).

The VOS model may be used to estimate the rate of loop production (per comoving volume) as a result of cosmic string interactions
\begin{equation}
\frac{dn_c}{dt}=\frac{\tilde c}{\alpha}\frac{\bar{v}}{L^4}\,,
\end{equation}
where $\alpha$ is a parameter that characterizes the size of loops, $l_c=\alpha L(t_c)$, at the moment of creation, $t_c$. This rate of loop production is strongly dependent on the large-scale properties of the network (through its dependence on $\vv$ and $L$) and, therefore, its accurate characterisation requires an accurate description of cosmic string network dynamics.

The production of cosmic string loops is expected to occur copiously throughout the evolution of cosmic string networks. At any given time, there are several loops emitting gravitational waves in different directions. The superimposition of these emissions generates a stochastic gravitational wave background (SGWB) that spans a wide range of frequencies \cite{Vilenkin:1981bx,Brandenberger:1986xn,Sanidas:2012ee,Sousa:2013aaa}. The SGWB is often quantified using the spectral density of gravitational waves per logarithmic frequency ($f$) interval, in units of critical density,
\begin{equation}
\Omega_{\rm GW}=\frac{1}{\rho_{\rm c}}\frac{d\rho_{\rm GW}}{d\log f}\,,
\end{equation}
where $\rho_{\rm GW}$ is the gravitational radiation energy density. This quantity is strongly dependent on the rate of loop production as a function of time, and, consequently, on the large-scale cosmic string parameters.

\begin{figure}[H]
\centering
\includegraphics[width=5.9in]{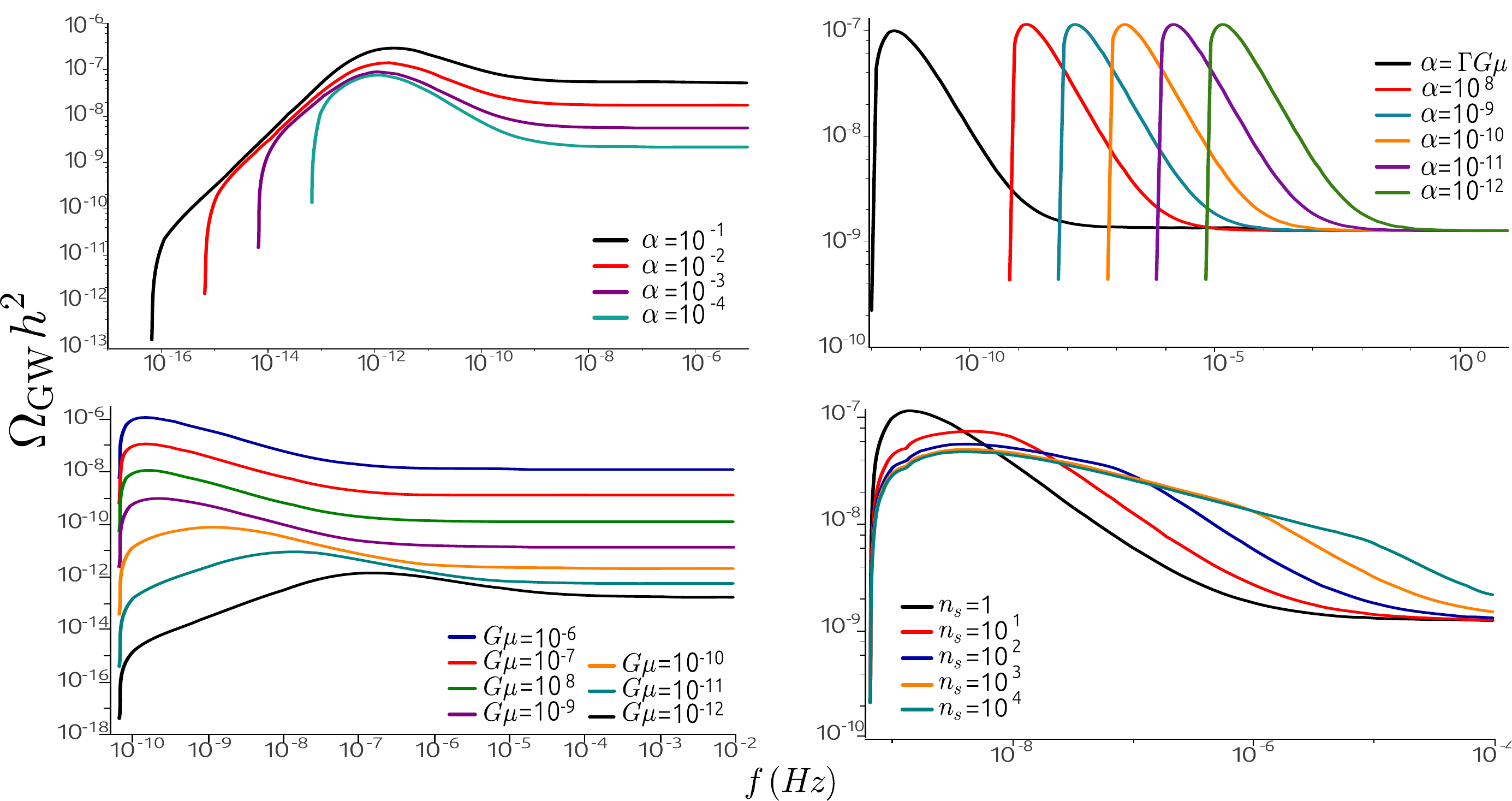}
\caption{\textit{Clockwise, from the top-left Panel:} The stochastic gravitational wave background generated by cosmic string networks: 1) For different loop sizes, in the large loop regime (where $\alpha$ is larger than the gravitational back-reaction scale, $\Gamma G\mu$). Here, cosmic string tension is fixed to $G\mu=10^{-7}$, and we have only considered one harmonic mode of emission ($n_s$). 2) For different values of $\alpha$ in the small-loop regime ($\alpha<\Gamma G\mu$). Here, $G\mu=10^{-7}$ and $n_s=1$. 3) For different value of cosmic string tension, $G\mu$. Here, the loop size parameter is set to $\alpha=10^{-7}$, and $n_s=1$. 4) For an increasing number of harmonic modes of emission, $n_s$. For larger values of $n_s$, the shape of the spectrum saturates and is identical to that of the spectrum with $n_s=10^4$. Here, $\alpha=10^{-8}$, $G\mu=10^{-7}$, and we have only considered loops with cusps. These plots were adapted from \cite{Sousa:2013aaa}.}
\label{SGWB}
\end{figure}

This spectrum has a characteristic shape: it has a flat portion in the high frequency range, generated by the emissions of loops created during the radiation era, and a pronounced peak at low frequencies that is produced during the matter era. Note however that the amplitude of the spectrum and the characteristics of the peak (its slope, broadness, location and height) vary greatly: they are not only dependent on the cosmic string tension and the large scale properties of the network, but also on the small-scale properties of the loops that generate it. For instance, the size of loops and their emission spectrum dramatically affects the overall shape of the spectrum. The dependence of the SGWB spectrum on these parameters is illustrated in Figure \ref{SGWB}, where SGWB spectra generated by cosmic string networks are plotted, for a wide range of values of these macroscopic and microscopic parameters. There is the prospect of probing different parts of the SGWB spectra using current and upcoming gravitational wave detectors in the near future---most notably (Advanced) LIGO \cite{TheLIGOScientific:2014jea}, (Advanced) VIRGO \cite{Accadia:2015pda}, eLISA \cite{Dufaux:2012rs} and pulsar timing arrays \cite{Manchester:2007mx,Ferdman:2010xq}. Understanding the shape of the spectrum generated by cosmic strings and its dependence on macroscopic and microscopic parameters is, therefore, crucial.

As in the case of cosmic strings, as domain wall networks evolve and interact, a fraction of their energy is released in the form of gravitational waves. In fact, in \cite{Hiramatsu:2010yz,Hiramatsu:2013qaa}, it has been shown using field theory simulations that domain wall networks produce a spectrum of gravitational waves that peaks at the scale corresponding to the size of the Hubble radius at the time of their decay. Studies of the gravitational wave spectrum generated by domain wall networks are, however, preliminary and limited to the case of metastable networks.

\section{Cosmological Tests with Galaxy Clusters at CMB Frequencies}
\label{sec6}

The Cosmic Microwave Background (CMB) has revolutionized the way we perceive the Universe. The information encoded in the temperature and polarization of the CMB provides one of the strongest evidences in favour of the hot Big-Bang theory and has enabled ways to constrain cosmological models with unprecedented accuracy \citep{2015arXiv150201589P}. The CMB also provides a unique way to study the formation and evolution of cosmic structure. Physical effects acting on the CMB after decoupling originate {\it secondary CMB anisotropies} that encode additional information about the growth of cosmological structure and the dynamics of the Universe \citep{2008RPPh...71f6902A}. The Sunyaev-Zeldovich (SZ) effect \cite{1972CoASP...4..173S, 1999PhR...310...97B} is the scatter of CMB photons by plasma in galaxy clusters and filamentary structures. It is among the strongest and most studied sources of secondary anisotropies. Whereas the primordial CMB signal has been widely used to constrain cosmological parameters, the SZ effect is only now giving the first solid steps into adulthood as a probe of Cosmology and structure formation. 
Largely contributing to this state of affairs are ground-based and satellite observations such as those by SPT \citep{2015ApJS..216...27B} and Planck \cite{2015arXiv150201597P} and proposals for future experiments such as in \cite{2014JCAP...02..006A}.

In this section we review some of the key features of the SZ effect and its application to test cosmological models and the physics of galaxy clusters, in light of Planck cluster observations. We end with an example of the application of cluster data to the study alternative cosmological scenarios involving primordial non-Gaussianities.  

\subsection{Galaxy Clusters \label{sec:galaxy_clusters}}
%

\begin{figure}[H]
\centering
\includegraphics[scale=0.56]{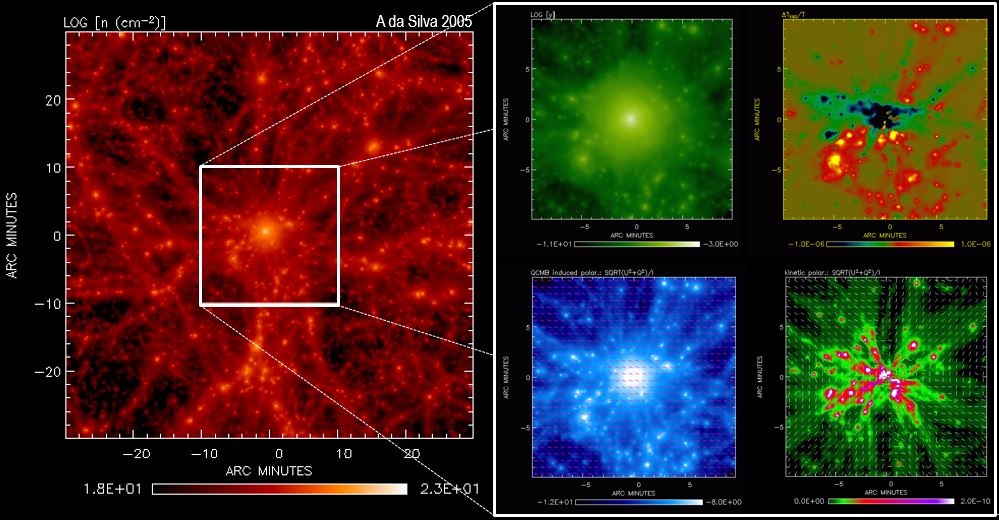}
\caption{Images of a Galaxy cluster in a hydrodynamic N-Body simulation.
The main panel shows a column density map, in logarithmic scale, of a region centered in the cluster. The smaller maps inside the zoomed box show the intensity and polarization of the CMB radiation that emerges from the cluster. The mapped quantities are the cluster thermal SZ effect (top left); kinetic SZ effect (top right); CMB quadrupole induced polarization (bottom left); and tangential velocity polarization  (bottom right), \cite{2012ApJ...757...44R}, due to the interaction of the CMB with the hot gas in the cluster.}
\label{fig:galaxy_cluster}

\end{figure}

Galaxy clusters are the largest gravitationally bound objects in the Universe. 
They are formed by gravitational accretion of infalling material at the intersection of filamentary structures. Their typical masses range from $10^{14}-10^{15}\,h^{-1}$~M$_{\odot}$ and contain hundreds to several thousand galaxies within radii of the order of $1-1.5\,h^{-1}$~Mpc. The intra-cluster medium (ICM) is filled with hot ionized gas, typically at temperatures 1--15 keV, which produces strong X-ray emission and microwave spectral distortions via the SZ effect, see Figure~\ref{fig:galaxy_cluster}.

Because clusters are among the latest bound structures forming in the Universe 
their number density is highly sensitive to the growth rate and power spectrum of 
density perturbations as well as to cosmological background parameters like the 
Universe's matter density and the Hubble parameter.  
The central quantity that goes into the computation of the expected 
abundance of clusters for a given cosmological scenario is the mass function,
$n(M,z)$. It gives the comoving number density of virialized halos 
forming at the redshift $z$ with mass $M$. As introduced by 
\cite{2001MNRAS.321..372J}, this can be expressed as: 
\begin{equation} 
\label{eq:massfna} 
f(\sigma, z) \equiv \frac{M}{\bar{\rho}}{{\rm d}n(M, z)\over{\rm d}\ln\sigma^{-1}}, 
\end{equation}
where  $f(\sigma, z)$ is the so called multiplicity function, $\bar{\rho}$ is the mean 
background density and $\sigma(M,z)$ is the variance of the linear density field at redshift $z$, smoothed on a scale $R$, enclosing the mass $M$,
\begin{equation}
\label{eq:sigma}
\sigma^2(M,z)  =  {D^2(z)\over2\pi^2}\int_0^\infty k^2P(k)W^2(k,M){\rm d}k.
\end{equation} 

Here, $D(z)$ is the growth factor of linear perturbations, $P(k)$ is the linear density power spectrum extrapolated to redshift zero, and $W(k,M)$ is the Fourier transform of the smoothing kernel in real space. The multiplicity function can be computed analytically or by assessing the number of halos forming in numerical N-body simulations. The following formulae give the multiplicity function for the popular  Press--Schecter, $f_{\rm PS}$, \cite{1974ApJ...187..425P} and Jenkins, $f_{\rm J}$, \cite{2001MNRAS.321..372J} mass functions:
\begin{equation}
\label{eq:massfn} 
f_{\rm PS} = \sqrt{2\over\pi}
{\delta_c\over\sigma}\exp\bigg[-{\delta_c^2\over2\sigma^2}\bigg]\,,
\qquad
f_{\rm J}=0.315\exp(-|\ln
\sigma^{-1}+0.61|^{3.8})\,.
\end{equation}

Here $\delta_c$ is the extrapolated linear overdensity of a spherical perturbation at the time of collapse. The Press--Schecter mass function was the first being derived using analytical arguments, but it overpredicts/underpredicts  the abundance of low/high mass halos in simulations. The Jenkins mass function was established using simulations. Several other simulation based mass functions have been proposed to improve the functional form of the multiplicity function, see e.g.,~\cite{2012ARA&A..50..353K} for a summary with other multiplicity function expressions. Most of these studies have been focusing on improving systematic uncertainties related to the halo mass definition and quantifying deviations from the universality of $f$ for different cosmologies. This includes extensive mass function studies using primordial non-Gaussian N-body simulations, see e.g., \cite{2012JCAP...03..002W}, and $f(R)$ modified gravity simulations, see e.g., \cite{2015arXiv151101494A}. 

The redshift distribution of observed galaxy cluster abundances can be compared against model predictions using the expression:
\begin{equation}
 \label{eq:dndmdz} 
 {d^2N\over dz\,d\Omega}=
 {d^2V\over dzd\Omega}\, \int_0^\infty  {dn(M,z)\over dM}\, f_{\rm survey}(M,z)\, dM\,.
\end{equation}

Here $d^2N/dzd\Omega$ is the number of clusters per unit redshift per unit solid angle, $d^2V/dzd\Omega$ is the volume element of the underlying model, and 
$f_{\rm survey}(M,z)$ is the cluster survey selection function.
Setting $f_{\rm survey}=1$ and assuming that its possible to observe all cluster masses, Equation~(\ref{eq:dndmdz}) gives the full redshift distribution of halos in the Universe, see e.g.,~\cite{2006A&A...450..899N} for an application to dark energy cluster counts.
The $f_{\rm survey}$ function depends on survey parameters such as sky coverage 
and noise distribution of the observed quantity $S_{\nu}$, which is usually a cluster flux 
or band luminosity. In a best case scenario, where 
the sky noise is uniform and the survey is complete above a given flux limit, $S_{\rm lim}$, 
Equation~(\ref{eq:dndmdz}) can be computed as an integral over the observed flux,  
$\int (dn/dM)\, (dM/dS_\nu) \, dS_{\nu}$, with  $S_{\nu}\ge S_{\nu ,\rm lim}$. Here 
$dM/dS_\nu $ is the derivative of a scaling relation between $M$ and $S_{\nu}$.
Galaxy cluster number counts are therefore a powerful probe of cosmology 
provided the relation between total mass and the observed survey flux are known.
Poor knowledge about the later may imply large uncertainties in the 
determination of cosmological parameters, as well as leaving a number of unanswered questions regarding the underlying cluster formation mechanism.

If gravity is the dominating force driving the formation and evolution of collapsed 
structures, virialized clusters should have their thermodynamic properties related 
to the total mass (through the viral theorem).
The simplest model describing cluster scaling relations was presented by \cite{1986MNRAS.222..323K}. This assumes, among other things, that gravitational collapse is scale-free, i.e., clusters of different masses are self-similar replicas of each-other. If non-gravitational effects, such as radiative gas physics and energy feedback, influence the thermodynamic state of clusters one should expect modified scalings. These are usually parameterized as power laws of the form \cite{2012ARA&A..50..353K},
\begin{equation}
\label{eq:scalings}
Y=A\, E(z)^{\beta_{\rm ss}}\,(X/X_0)^{\alpha} \,, 
\end{equation}
where $Y$, $X$ are cluster properties,  $A$ is a normalization parameter 
giving the amplitude of $Y$ at $X=X_0$, $E(z)=H(z)/H_0$, $\alpha$ is the 
power law index of the (independent) property $X$, and $\beta_{\rm ss}$ is 
a parameter giving the self-similar evolution of the normalization. 
In general, the amplitude parameter $A$ may itself be a function of redshift. 
In this case, $A$ evolves in a non self-similar way, usually 
parameterized by a power-law of redshift, $A(z)=A_0\,(1+z)^{\beta}$.  
The parameters $\alpha$ and $\beta_{\rm ss}$ can be computed 
analytically within an extended self-similar modelling that also assumes 
that the baryon gas component in clusters is in hydrostatic equilibrium 
\cite{2012ARA&A..50..353K}.
For example, under these 
conditions, the scaling between the SZ integrated flux and mass,  $Y-M$, is 
given by Equation~(\ref{eq:scalings}) with the indexes $\alpha=5/3$, and 
$\beta_{\rm ss}=2/3$.
Hydrodynamic N-body simulations are the appropriate method to study galaxy 
cluster scaling relations as they allow us to model the baryon physics throughout the 
full non-linear evolution of the ICM gas 
and determine the parameters $\alpha$, $\beta_{\rm ss}$,  $\beta$ and $A_0$. 
We refer the reader to references 
\cite{2007MNRAS.377..317K, 2009MNRAS.396..849D, 2009A&A...496..637A, 2016arXiv160309270T} 
for examples of simulations studies showing the impact of various gas physical 
models, dark energy and primordial non-Gaussianities on several galaxy cluster 
scalings.

Numerical N-body simulations have also been intensively used to study the
internal structure of galaxy clusters. These studies demonstrate that galaxy 
clusters profiles can be used as a sensitive probe of cosmology, cluster 
physics and structure formation mechanisms. One of the most well known 
results from simulations, in the context of hierarchical CDM models, is that 
the radial distribution of dark (collissionless) matter in clusters follows a simple 
expression known as the Navarro-Frenk-White (NFW) density profile \cite{1997ApJ...490..493N}
\begin{equation}
\label{eq:scalingsb}
\rho_{\rm NFW}(r)=\frac{4\rho_s}{x(1+x)^2}, 
\end{equation}
where $x\equiv r/r_s$,  $r_s$ is a scale radius, and $\rho_s$ is the characteristic 
density at $r=r_s$. Subsequent numerical simulations allowed one to improve this result and 
to study the impact on cluster profiles of alternative cosmological models such as modified 
gravity, see \cite{2012ARA&A..50..353K} for a review.
Gas profiles have also been extensively investigated with hydrodynamic N-body 
simulations. For example, \cite{2007ApJ...668....1N} found that the radial pressure 
profile, $P(r)$, in SZ observations of resolved clusters follows a universal profile, 
\begin{equation}
\frac{P(r)}{P_{500}} = \frac{P_0}{ x^{\gamma} ( 1 + x^{\alpha}
  )^{(\beta-\gamma) / \alpha}}\, ,
\label{eq:pmodel}
\end{equation}
where $x = r/r_s$, $r_s$ is the scale radius, and $P_0$, $\alpha, \beta, \gamma$ are 
fitting parameters to derive from observations. $P_{500}$ is the integrated pressure 
at $ r_{500}$ (the radius at which the cluster density exceeds 500 times
the universe mean background density). Equation (\ref{eq:pmodel}) is known as the 
generalized NFW (GNFW) profile of the SZ signal in clusters.

\subsection{The SZ Effect: Temperature and Polarization Signal}

\begin{figure}[H]
\centering
\includegraphics[scale=0.52]{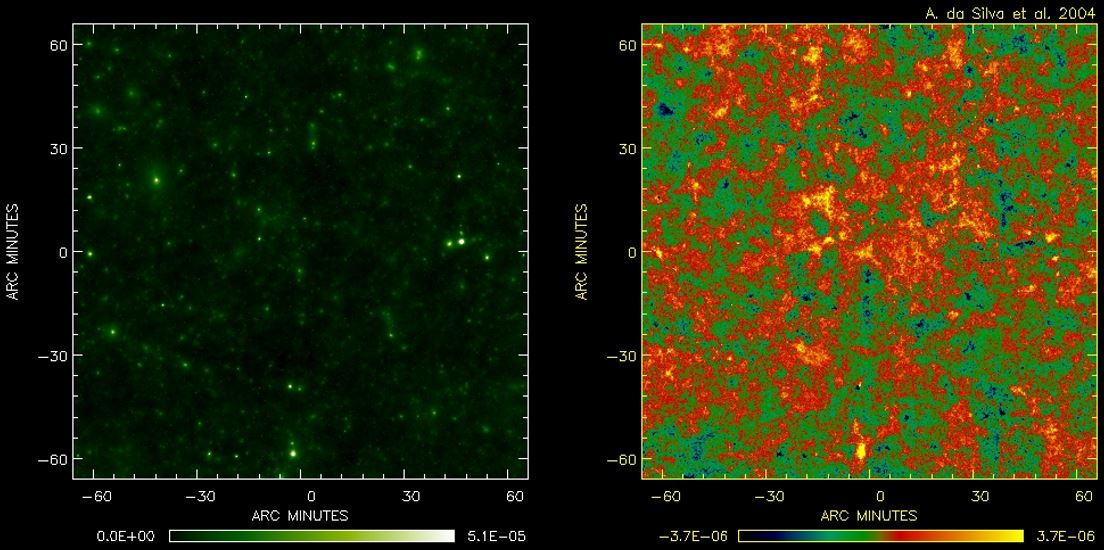}
\caption{Images of the thermal $y$-comptonization parameter (left panel)
and kinetic temperature fluctuations $(\Delta T/T)_k=-v_{\rm r}/c\, \tau$ (right panel) of
the SZ effect from hydrodynamic N-body simulations of large scale structure. 
Maps show the same sky patch, with an angular size of 5 degrees$^2$.
The bright objects in the $y$ map are galaxy clusters.
}
\label{fig:szmaps}

\end{figure}

The Sunyaev-Zel'dovich effect is the scattering of CMB photons by electrons in hot
reservoirs of ionized gas in the Universe. A particularly effective scattering medium is the ICM where the gas is at temperatures of the order of $10^8$ Kelvin. When the much colder and isotropic CMB radiation propagates through these plasma regions, a fraction of the original CMB photons scatter off the moving electrons and suffer frequency Doppler shifts towards higher energies (inverse Compton scattering) causing a spectral distortion in the CMB spectrum. In the process the total number of photons is conserved. 
In fact, the SZ effect can be regarded as a composition of two effects: the  {\it thermal} and {\it kinetic} SZ effects. The thermal effect is due to the thermal (random) motion of the electrons in the gas cloud. If the gas has also a bulk peculiar velocity in the CMB radiation frame, there is a kinematic Doppler effect in addition to the thermal effect. The spectral distortions caused by the 
thermal, $\Delta I_{\rm th}$, and kinetic, $\Delta I_{\rm k}$, SZ effects are given by 
simple formulae \cite{1999PhR...310...97B}:
\begin{eqnarray}
\Delta I_{\rm th} =I_0\, g(x) \,y \,,
\qquad
y={{k_{\rm B}\sigma _{\rm T}} \over {m_{\rm e}c^2}}\int {T_{\rm e}n_{\rm e}\,dl}\,, \\ 
\Delta I_{\rm k}=-I_0h(x){{v_{\rm r}} \over c}\tau \,,
\qquad 
\tau =\int {\sigma _T\, n_{\rm e}(l)\,dl}\,,
\label{cp3_20}
\end{eqnarray}
where $I_0=2 k_{{\rm B}}^3 T^3/h^2 c^2$, $T$ is the CMB temperature,
$y$ is the line-of-sight Compton SZ parameter, $\tau $ is the optical depth,
$v_{\rm r}$ is the line-of-sight velocity of the center of mass of the gas cloud 
(positive/negative when the cloud is receding/approaching), $T_{\rm e}$ is the 
electron density, $n_{\rm e}$ is the electron number density, and 
$g(x)$, $h(x)$ are functions giving the frequency dependence of the effects.
Figure~\ref{fig:szmaps} shows maps of the thermal and kinetic SZ distortions 
imprinted on  the CMB by cosmic structures in  
hydrodynamic N-body simulations. The mapped quantities are the $y$ parameter 
and the temperature fluctuations of the kinetic SZ effect, 
$(\Delta T/T)_k=\Delta I_{\rm k}/(I_0h(x))=-v_{\rm r}/c\, \tau$, see 
e.g.,~\cite{2001MNRAS.326..155D}.
For typical cluster velocities and optical depths, the thermal SZ effect is the dominant effect. Therefore the total SZ signal integrated over the sky angular size of a cluster is well approximate by   
%
\begin{equation}
Y\equiv \int {y\, d\Omega}=d_A^{-2}\int {y\,dA}=
{ {k_B\sigma _{\rm T}} \over {mc^2}}\,d_A^{-2}\,
\int _{V}\, T_{{\rm e}} n_{{\rm e}} \, dV\,,
\label{cp6_4}
\end{equation}
where $d_A$ is the angular diameter distance from the observer to the cluster.
This is known as the volume integrated SZ $Y$-flux or $Y$-luminosity and is a 
measure of the total thermal energy density of the cluster.

The scattering of CMB photons by hot gas in clusters also generates new (secondary) polarization signatures in the CMB that carry information about the intra-cluster medium. The dominant CMB secondary polarization effect in clusters is due to the existence of an intrinsic quadrupole component from the CMB temperature anisotropy around the scattering media. This effect is known as the CMB quadrupole induced polarization. In addition, the bulk motion of the gas cloud may also induce CMB polarization proportional to the square of the sky transverse velocity of the cloud. This effect is often referred to as induced kinetic polarization. Finally, double scattering events inside the cluster gas cloud also originate polarization, known as double scattering induced polarization. The bottom panels in Figure~\ref{fig:galaxy_cluster} are detailed simulations of the linear polarization degree and polarization angle of the CMB quadrupole and kinetic induced polarizations in a cluster. These simulations are an essential tool to assess the degree of contamination of the primordial CMB signal by secondary polarization effects and allow to prepare observational strategies to use CMB cluster polarization as an additional probe of cosmology and structure formation, see \cite{2012ApJ...757...44R} for a detailed description of cluster polarization simulations and their use as a new observing tool.

\subsection{Cosmological Tests Using SZ Cluster Surveys}

SZ galaxy cluster surveys are recognized as having a great potential to test a wide range of cosmological and astrophysical scenarios. SZ galaxy cluster counts, profiles, scaling relations and SZ angular power spectra are most promising probes to confront model predictions with observations. In this section we will briefly review the current state of cosmological constraints from observations with the Planck satellite.

\subsubsection{SZ Cluster Counts}

The Planck SZ cluster survey is the largest and deepest all-sky cluster survey yet produced at CMB frequencies. Its latest galaxy cluster catalogue (PSZ2, \cite{2015arXiv150201598P}) contains 1653 detections, of which 1203 are confirmed clusters with identified counterparts in other observing bands (X-rays, optical, near infra-red and other SZ experiments). More than a thousand of these clusters possess redshift information. 
From this catalogue, the authors in \cite{2015arXiv150201597P} selected a sample of 439 clusters with high signal to noise ($q=S/N>6$) to infer constraints on cosmological parameters using SZ galaxy cluster number counts. 

For a given cosmological model, the redshift distribution of clusters abundances detected by Planck above a given noise level, $q>q_{\rm cat}$ can be computed from Equation~(\ref{eq:dndmdz}) as:
\begin{equation}
\label{eq:dndz}
\frac{dN}{dz}(q>q_{\rm cat}) = \int {d^2V\over dzd\Omega}\, d\Omega 
\int {dn(M,z)\over dM}\, \hat{\chi}(M,z,l,b,q_{\rm cat}) \, dM \, , 
\end{equation}
see \cite{2015arXiv150201597P}. Here $M$ is the mass contained within a 
cluster radius, $R_{500}$, where the cluster density 
exceeds by 500 times the mean background density of the Universe,  
$\hat{\chi} (M,z,l,b,q_{\rm cat})  =  \int_{q_{\rm cat}}^\infty  P[q | \bar{q}_{\rm m}(M,z,l,b)]\, dq$ is the Planck selection function, (i.e., $\hat{\chi}=f_{survey}$ in Equation~(\ref{eq:dndmdz})) and $P[q | \bar{q}_{\rm m}(M,z,l,b)]$ is the distribution of $q$ given the mean signal-to-noise value, $\bar{q}_{\rm m}(M,z,l,b)$, predicted by the model for a cluster of mass $M$ and redshift $z$ located at Galactic coordinates $(l,b)$. This expression reflects the fact that the survey noise level is not uniform on the sky and that clusters are detected above a given noise level, $q_{\rm cat}$.
The distribution $P[q | \bar{q}_{\rm m}]$ incorporates noise fluctuations and intrinsic scatter in the actual cluster SZ $Y$ signal, (\ref{cp6_4}), computed within $R_{500}$, around the mean value, $\bar{Y}(M,z)$, predicted from a $Y-M$ scaling relation calibrated using Planck cluster data. 

This method was used in \cite{2015arXiv150201597P} to impose constraints on
cosmological and cluster parameters, such as the variance of the density fluctuations 
smoothed on a scale of $R=8\, h^{-1}Mpc$, $\sigma _8$, see Equation~(\ref{eq:sigma}), 
the matter density parameter, $\Omega_{m0}$, the Hubble constant, $H_0$, and the 
slope of the $Y-M$ scaling relation, $\alpha$. Assuming a   
$\Lambda CDM$ concordance (base) model their analysis show that Planck SZ cluster 
constraints on the $\sigma _8 - \Omega_{m0}$ plane are in tension with those obtained 
from Planck primary CMB anisotropies, unless priors on the 
hydrostatic mass bias parameter, $1-b$, (which relates the observed baryon mass with the total mass of the cluster) and the effect of gas physics on the growth of structure are changed to values that are in conflict with constraints from other cosmological probes and the predictions from numerical simulations, 
\cite{2015arXiv150201597P,2014A&A...571A..20P}. A way to try to alleviate the tension is to combine cluster and primary CMB data to explore constraints on extensions to the base flat CDM model. This was carried out in \cite{2015arXiv150201597P} by allowing an extra varying parameter (to the base model), namely, the neutrino mass scale, the Thomson optical depth to reionization, the dark energy equation of state, and the Universe curvature parameter. The analysis seems to favor non-minimal neutrino masses (see Figures 11 and 14 in \cite{2015arXiv150201597P}), but this ``solution'' also implies a decrease on the Hubble constant to values that further deviate from most constraints by astrophysical determinations of $H_0$.

The present Planck analysis also puts in evidence the  importance of improving the precision of the cluster mass bias parameter, which in fact may depend on cluster mass and redshift. In the future, large lensing surveys, such as Euclid, WFIRST and LSST, are expected to determine mass biases with one percent level accuracy. With this level of precision on $1-b$, the present cluster combined analysis of Planck data could by itself provide a stringent test of the $\Lambda CDM$ concordance model, see Figure 13 in \cite{2015arXiv150201597P}.

\subsubsection{SZ Power Spectra}

An important statistic to derive from sky maps of the SZ effect 
is the angular power spectrum of the thermal and kinetic SZ temperature fluctuations. 
The theoretical computation of these spectra is straightforward from 
simulations, such as those in Figure~\ref{fig:szmaps}, 
but the computational resources involved to run simulations limit their use  
if one is required 
to compute the SZ power spectra for many model realizations.
Analytical methods are much faster and allow to explore many model realizations,
but they are much less capable of modelling the cluster gas 
physics and the contribution from the diffuse gas in filamentary structures.

Analytically, the angular power spectrum of the thermal SZ signal  
can be computed as a sum of two terms 
$C_\ell^{\rm total} = C_\ell^{\rm Poisson} \,+\,C_\ell^{\rm clustering}$,
where $C_\ell^{\rm Poisson}$ is the power spectrum resulting from a 
Poisson distribution of objects (clusters) and $C_\ell^{\rm clustering}$ is a 
``2-halo'' or clustering term giving the power arising from object 
(cluster two-point) correlations.
These terms are given by (see \cite{2008RPPh...71f6902A} and references therein):
\begin{eqnarray} 
& &C_\ell^{\rm Poisson} =
\int_0^{z_{\rm max}} {{dV}\over{dzd\Omega}} \int_{M_{\rm min}}^{M_{\rm max}} 
{{dn(M,z)}\over{dM}} {|y_\ell(M,z)|}^2\, dM\, dz , \label{eq:cl1}\\
& &C_\ell^{\rm clustering} = \int_0^{z_{\rm max}} {{dV}\over{dzd\Omega}} P(k) \times 
 {\left[\int_{M_{\rm min}}^{M_{\rm max}} {{dn(M,z)}\over{dM}} 
b(M,z) y_\ell(M,z)\, dM\right]} ^2 \, dz \, , \label{eq:cl2}
\end{eqnarray} 
where $y_\ell$ is the 2D Fourier transform of the projected 3D Compton $y$-parameter radial profile of individual cluster, $P(k, z)$ is the 3D matter power spectrum at redshift z, and $b(M,z)$ is the linear bias factor that relates the matter power spectrum, $P(k, z)$, to the cluster correlation power spectrum $P_{\rm cluster}(k,M_1,M_2,z)=b(M_1,z)b(M_2,z)D^2(z)P(k,z=0)$. The mass and redshift range of the integrations are chosen so as to cover all relevant SZ source contributions up to high enough redshift. 

Observationally, the SZ all sky anisotropies can be separated from the overall 
CMB signal by applying component separation techniques that make use of the spectral 
signature and scale of the SZ effect. 
The first all-sky maps of the thermal SZ effect were produced from Planck observations \cite{2014A&A...571A..21P,2015arXiv150201596P}. These studies allowed 
to investigate the distribution of the $y$ signal and characterize its higher order 
statistics, including the power spectrum and bi-spectrum.
The observed thermal SZ power spectrum confirmed to be very sensitive to 
cosmological parameters, especially $\sigma_8$ and $\Omega_{\rm m0}$, but
these are strongly degenerate with the mass bias parameter $b$ (they found that 
$\sigma_8(\Omega_{\rm m0}/0.28)^{3/8} = 0.80^{+0.01}_{-0.03}$
if $b=0.2$ and $\sigma_8(\Omega_{\rm m0}/0.28)^{3/8} = 0.90^{+0.01}_{-0.03}$
if $b=0.4$). 
These results are consistent with the constraints on $\sigma_8(\Omega_{\rm m0}/0.28)^{3/8}$ from cluster number counts but are again in ``soft tension'' with the constraints obtained from primary CMB data assuming a $\Lambda$CDM concordance base model, see Figure 17 in \cite{2015arXiv150201596P}.

\subsubsection{Probing New Physics with the SZ Clusters}

The present best CMB spectral observations by COBE/FIRAS are
unable to determine if the CMB spectrum preserved its planckian
shape at all redshifts and if the CMB temperature-redshift scaling is
the one predicted by the standard Big-Bang model, $T=T_0(1+z)$, where $T_0$ is 
the CMB temperature observed at $z=0$.
However it is possible to consider alternative models where 
the CMB has a planckian spectrum at $z=0$, but a different 
temperature-redshift scaling.
Observational tests of this scaling require measurements of the CMB
temperature at high z, i.e., throughout the Universe's history. 
The SZ effect has long been recognized as a most promising method 
to measure the CMB temperature at the redshift and location of clusters, 
allowing to test the CMB temperature-redshift relation at high redshifts, 
see e.g.,~\cite{2012ApJ...757..144D} and references therein. 
Recent observations by Planck allowed to impose constraints on a simple deviation
of the standard scaling, $T=T_0(1+z)^{1-\beta}$, using CMB primary 
data \cite{2015arXiv150201589P} and SZ effect cluster data, see e.g., \cite{2015ApJ...808..128D}. 
The emerging view is that Planck data impose tight constraints on the $T(z)$ that are
consistent with the standard CMB temperature scaling predicted by the 
standard  Big-Bang model (i.e., $\beta = 0$). 

\subsection{Probing Primordial Non-Gaussianities with Galaxy Clusters}

The formation of the structures observed at large scales in the Universes is one of the most challenging and complex problems still open in modern cosmology. The widespread understanding is that these structures are the result of tiny primordial density fluctuations that might have existed at early times and that eventually grew under gravitational instability collapse. The origin of these tiny primordial density density perturbations is a cause of great debate, yet the inflationary paradigm remains the consensus framework for their generation.

The statistical properties of the primordial density perturbations predicted by most standard single-field-slow roll inflationary models are nearly Gaussian in nature, and thus the level of non-Gaussianity is not strong enough to be observationally detected \cite{2003NuPhB.667..119A,2003JHEP...05..013M,2004PhR...402..103B}. 
Although recent CMB anisotropies and LSS data seem to support such prediction, more complex models may lead to the generation of larger a non-Gaussianity amplitude, which substantially increases the chances of detection with future high precision observational experiments. For example, under certain conditions, some multi-field inflation models where slow-roll holds, can lead to larger departure from Gaussianity. However, non-Gaussianities may arise in inflationary models in which any of the conditions leading to the slow-roll dynamics fail \cite{2009astro2010S.158K}. The spectrum of models capable of generating larger non-Gaussianities amplitudes include the curvaton scenario \cite{2002PhLB..524....5L,2004PhRvD..69d3503B,2006PhRvD..74j3003S}, the ekpyrotic inflacionary scenario \cite{2008PhRvD..77f3533L,2010AdAst2010E..67L}, vector field populated inflation \cite{2008JCAP...08..005Y,2009PhRvD..80b3509K,2010AdAst2010E..65D} and multi-field inflation \cite{1994PhRvD..50.6123P,1998PhLB..422...52M,2006PhRvD..73h3522R,2010AdAst2010E..76B}. 
The current efforts that are being done to detect and constrain primordial non-Gaussianities (hereafter PNG) using a variety of cosmological observables is of the tremendous importance. A positive detection of non-Gaussianity signal would allow to asses the different proposed mechanisms for the generation of primordial density perturbations and in particular, it would make possible to discriminate between different inflation models and rule out some of them. Moreover, our understanding of the key physical processes at the early Universe would benefit greatly from such outcome.

There is a wide range of observables to which we can resort to, in order to search for primordial non-Gaussianities. The traditional way to proceed is to measure three-point statistics of the CMB temperature anisotropies maps. Using this observable, there have been in the past reports of a positive detection of non-Gaussianity at a significant level \cite{2009ApJ...701..369R}. 
However, although some contradicting analysis exist (see \citep{2009ApJS..180..330K,2009ApJ...706..399C} and more recently \cite{2014A&A...571A..24P,2015arXiv150201592P}), the interest in testing deviations from Gaussian initial conditions has gained momentum ever since. 
Nevertheless, the possible existence of primordial non-Gaussianities could also be tested using other cosmological probes such as the characterization of the statistical properties of the structures at large scales. These probes include the the bispectrum and/trispectrum of distributions of galaxies and galaxy clusters (e.g., \cite{2007PhRvD..76h3004S,2008ApJ...677L..77M,2012MNRAS.422.2854G}), weak-lensing observations,  (e.g., \cite{2012MNRAS.421..797S,2012MNRAS.426.2870H}) and CMB-LSS \cite{2013MNRAS.431.2017T} as well as  CMB-21cm line \cite{2012PhRvD..85d3518T} cross-correlations. 
Another set of cosmological of probes very discussed in the literature to contain PNG, is the time evolution of the numerical abundances both of massive collapsed objects such as  galaxy clusters, and its counterparts,  large voids (see e.g \cite{2000MNRAS.311..781R,2000ApJ...541...10M} and \cite{2009JCAP...01..010K,2009MNRAS.399.1482L,2011PhRvD..83b3521D,2012arXiv1204.2726S}).

\subsubsection{Parametrizing Primordial Non-Gaussianity}

The cosmological information contained in the power spectrum (or equivalently the two-point correlation function) of the primordial density perturbations is not enough if one wants to look for deviations from Gaussian initial conditions. Therefore we must resort to higher order correlation functions in order to gain further insight on nature of non-Gaussianities. Thus, the bispectrum, $B_{\Phi}$ is the lowest order statistics sensible to primordial non-Gaussianities and it is defined as
\begin{equation}
\label{eq:bispectrum}
\langle \Phi\left({\bf k}_{1}\right)\Phi\left({\bf k}_{2}\right)\Phi\left({\bf k}_{2}\right)\rangle = \left(2 \pi\right)^{3}\delta_{D}^{\left(3\right)}\left({\bf k}_{1}+{\bf k}_{2}+{\bf k}_{3}\right)B_{\Phi}\left({\bf k}_{1},{\bf k}_{2},{\bf k}_{3} \right),
\end{equation}
where $\Phi$ is the Bardeen's potential and $\delta_{D}^{\left(3\right)}$ is the three dimensional Dirac delta. The bispectrum can be be written in the alternative form \cite{2009PhRvD..80d3510F},
\begin{equation}
\label{eq:bispectrum_alternative}
B_{\Phi}\left({\bf k}_{1},{\bf k}_{2},{\bf k}_{3} \right) = f_{NL}\left(k_{1}k_{2}k_{3}\right)^{-2}\mathcal{A}\left({\bf k}_{1},{\bf k}_{2},{\bf k}_{3} \right),
\end{equation}
where the dimensionless parameter $f_{\rm NL}$  accounts for the amplitude and the shape function $\mathcal{A}$ stores the information on the shape of the bispectrum.

Due to the large number of inflationary models capable of generating primordial non-Gaussianities, it is common practice to group them according to their resulting bispectrum shape. There are generically four classes of bispectrum shapes frequently studied in the literature: Local, Equilateral, Folded and Orthogonal. However, from these shapes, the most studied in the literature is the Local one, which can be written in terms of Bardeen's potential and performing a Taylor expansion around an auxiliary Gaussian random field $\Phi$, (see eg. \cite{2004PhR...402..103B}) 
\begin{equation}
    \Phi\left(\bf x \right) = \phi\left(\bf x \right) + f^{local}_{\rm NL} \left( \phi^{2}\left(\bf x \right) - \langle \phi^{2}\left(\bf x \right)\rangle \right).
\label{eq:PNG local taylor}
\end{equation}

The  corresponding bispectrum of the Local shapes is given by,
\begin{align}
\label{eq:local_bispectrum}
B^{loc}_{\phi} =& 2 f_{\rm NL}^{local}\left[P_{\phi}\left(k_{1}\right)P_{\phi}\left(k_{2}\right) + P_{\phi}\left(k_{1}\right) P_{\phi}\left(k_{3}\right)+ P_{\phi}\left(k_{2}\right) P_{\phi}\left(k_{3}\right) \right]
\end{align}

To date, the strongest constraints on primordial non-Gaussianities of the Local type were obtained by the Planck collaboration \cite{2015arXiv150201592P} (see also \cite{2014A&A...571A..24P} for previous constraint PNG by Planck collaboration), $f_{NL}^{local}=2.5\pm5.7$ from temperature alone and $f_{NL}^{local}=0.8\pm5$ combining temperature and polarization data. Note however, that although these constraints are consistent with Gaussian initial conditions, they were obtained assuming that the non-linear parameter $f_{NL}$ is scale-independent. Nevertheless, some inflationary models may lead to a scale dependent $f_{NL}$ (see \cite{2008JCAP...04..014L} and references therein).

\subsubsection{The Non-Gaussian Mass Function}

\begin{figure}[H]
\centering
 \includegraphics[scale=0.45]{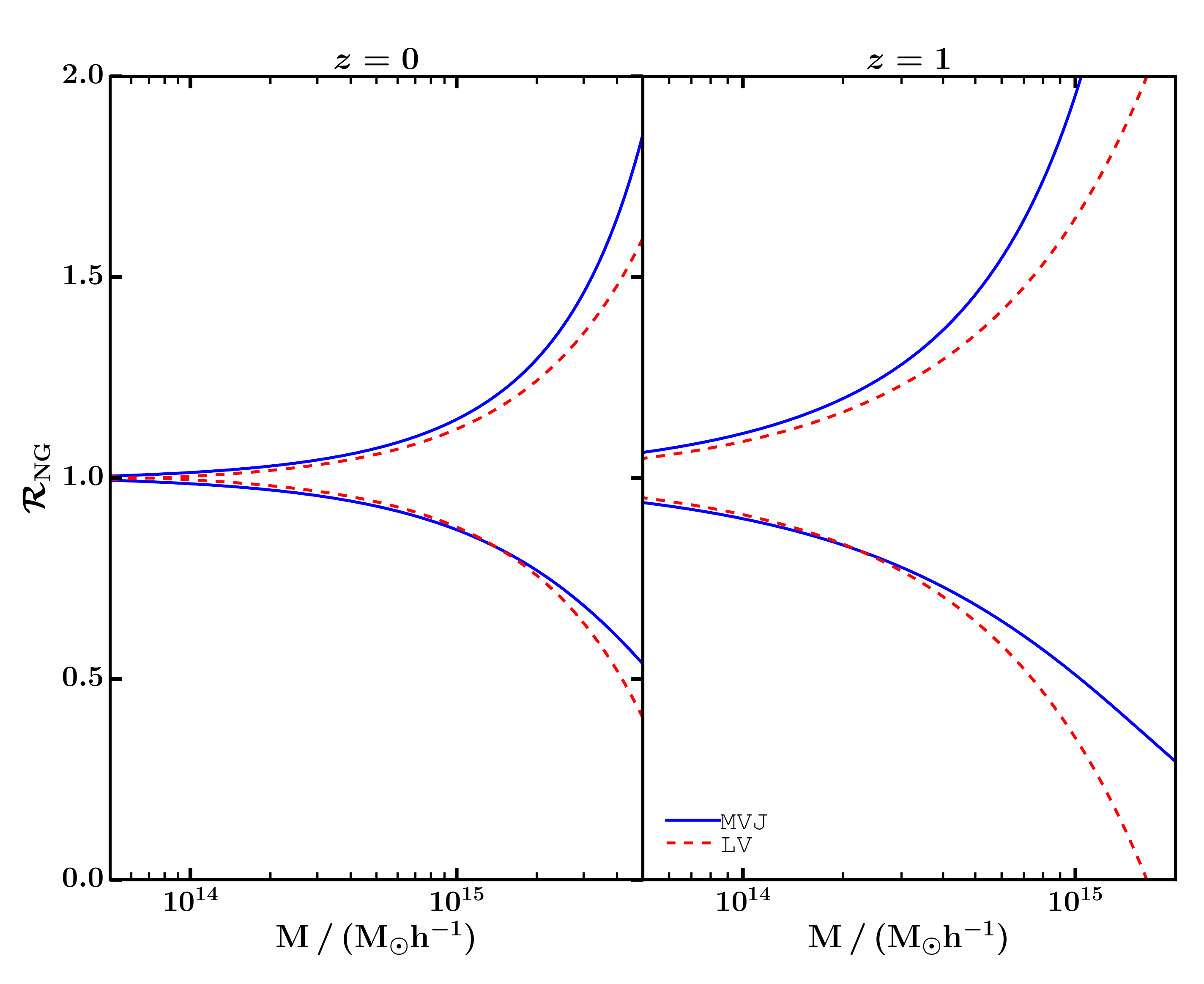}

\caption{The non-Gaussian correction to the mass function due to primordial non-Gaussianities as a function of mass and for two refshifts: on the left panel $z=0$ and on the right panel $z=1$. Blue line corresponds to the prescription of MVJ (Equation \eqref{eq:ratioMVJellips}), while  red dashed line corresponds to LV formula (Equation \eqref{eq:ratioLoVellips}). The level of non-Gaussianity of the Local type used was $f_{NL}\pm 100$.  \label{fig:R_NG}}  
\end{figure}

There is no unique prescription to incorporate the effects induces by primordial non-Gaussianities in the mass function (see \cite{2000ApJ...541...10M,2010PhRvD..81f3530G,2010ApJ...717..526M,2011JCAP...02..001D,2011JCAP...08..003L,2012JCAP...02..002A,2013MNRAS.428.2765D}). The majority of the of the non-Gaussian mass functions are obtained resorting to some expansion of the probability density distribution (PDF) of the primordial density field and either using an extension of the Press-Schecter formalism \cite{1974ApJ...187..425P} or the excursion set theory \cite{2007IJMPD..16..763Z,2010ApJ...711..907M}. 
Generically, the modified mass function can be written in the form
\begin{equation}
\label{eq:ng mass function}
\frac{dn_{\rm NG }\left(z,M\right)}{dM}  =  \frac{dn_{\rm G}\left(z,M\right)}{dM} \times \mathcal{R}_{\rm NG},
\end{equation}
where $\mathcal{R}_{\rm NG}$ encodes the correction to the Gaussian mass function due to PNG. In \eqref{eq:ng mass function}, the Gaussian mass function can be simply the one of the PS formalism or any other motivated by N-body numerical simulation with gaussian initial conditions (references of fitting formulas), such as Equation \eqref{eq:massfn}. Two of the most discussed and used expression for the $\mathcal{R}_{\rm NG}$ correction in the literature are the MVJ \cite{2000ApJ...541...10M} and the LV \cite{2008JCAP...04..014L}. Both rely on the extension of the PS formalism to non-Gaussian initial conditions. The respective functional forms are, 
\begin{equation}
\label{eq:ratioMVJellips}
\mathcal{R}_{\rm NG}=\exp \left[\delta_{c}^3 \frac{S_{3,M}}{6 \sigma_M^2}\right] \times \left[\frac{1}{6}
\frac{\delta_{c}^{2}}{\delta_{\ast}} \frac{dS_{3,M}}{d\ln \sigma_{M}} + \frac{\delta_{\ast}}{\delta_{c}}\right], 
\end{equation}
\begin{equation}
\label{eq:ratioLoVellips}
{\cal R}_{\rm NG}= 1+\frac{1}{6}\frac{\sigma_M^2}{\delta_{ec}} \left[S_{3,M}\left(\frac{\delta_{c}^4}{\sigma_M^4} 
-2\frac{\delta_{c}^2}{\sigma_M^2}-1\right)+\frac{dS_{3,M}}{d \ln \sigma_M}
\left(\frac{\delta_{c}^2}{\sigma_M^2}-1\right)\right] 
\end{equation}
where $\delta_{c}\left(z\right)$ is the collapse threshold at the redshift $z$, $S_{3}$ is the reduced skewness and $\delta_{\ast}\left(z\right) = \delta_{c}\left(z\right)\sqrt{1-S_{3}\delta_{c}\left(z\right)/3}$.
Figure \ref{fig:R_NG} show the dependence of the non-Gaussian corrective term to the mass function, $\mathcal{R}_{NG}$, the mass and  as a function of redshift, for the Equations \eqref{eq:ratioMVJellips} (blue line) and  \eqref{eq:ratioLoVellips} (red dashed line). In both cases $f_{NL}=\pm 100$ was used. It is clear that a higher positive/negative $f_{NL}$ value increases the number of predicted dark matter halos with higher mass. The effects are stronger at high redshift. The MJV prescription seems to predict a higher number of halos than LV for positive values of $f_{NL}$. The opposite occurs when negative $f_{NL}$ values are considered.

\subsubsection{Biased Cosmological Parameter Estimation with Cluster Counts}

The statistical properties of the primodial density perturbations has a great deal of influence on the redshift distribution of the number of galaxy clusters. Thus, these objects can be used as cosmological probes to search for deviations from Gaussian initial conditions. However, neglecting erroneously the existence of primordial non-Gaussianities, can lead to biases in the estimation of the cosmological parameters using clusters number counts. In \cite{2012MNRAS.424.1442T}, the authors investigated how primordial non-Gaussianities influence the estimations of the effective dark energy equation of state parameter, when the information contained in the abundance of galaxy clusters is used to explore different cosmological scenarios. In their work, they computed the effective dark energy equation of state per redshift bin, assuming Gaussian initial conditions, that would allow them to recover the galaxy clusters counts in different non-Gaussian models. From their findings, there resulted a new diagnostic feature for the presence of PNG, in the form of an apparent evolution with time of the effective dark energy equation of state, characterized by the appearance of a  discontinuity (more details see \cite{2012MNRAS.424.1442T}). The same authors also estimated the magnitude of the biases that may arise in the determination  of a larger set of cosmological parameters with galaxy clusters number counts, again ignoring that PNG may exist (see \cite{2013MNRAS.435..782T}). Their results indicate that, although the estimation of the present-day dark energy density and respective equation of state parameter do not seem to be sensible to non-Gaussian initial conditions in the primordial matter density field, the biases can be much larger in the case of the amplitude of the primordial density perturbations. Furthermore, the authors also argue that the conflicting constrains on the amplitude of the primordial perturbations obtained by the Planck collaboration using galaxy cluster number counts from the Planck Sunyaev-Zel'dovich Catalog  and  the primary Cosmic Microwave Background temperature anisotropies, could be alleviated, if a significant non-Gaussianity level at clusters scales exists (see \cite{2013MNRAS.435..782T} for more details).  

\subsubsection{The Impact of pRimordial Non-Gaussianities on Galaxy Clusters Scaling Relations}

\begin{figure}[H]
\centering
\includegraphics[scale=0.35]{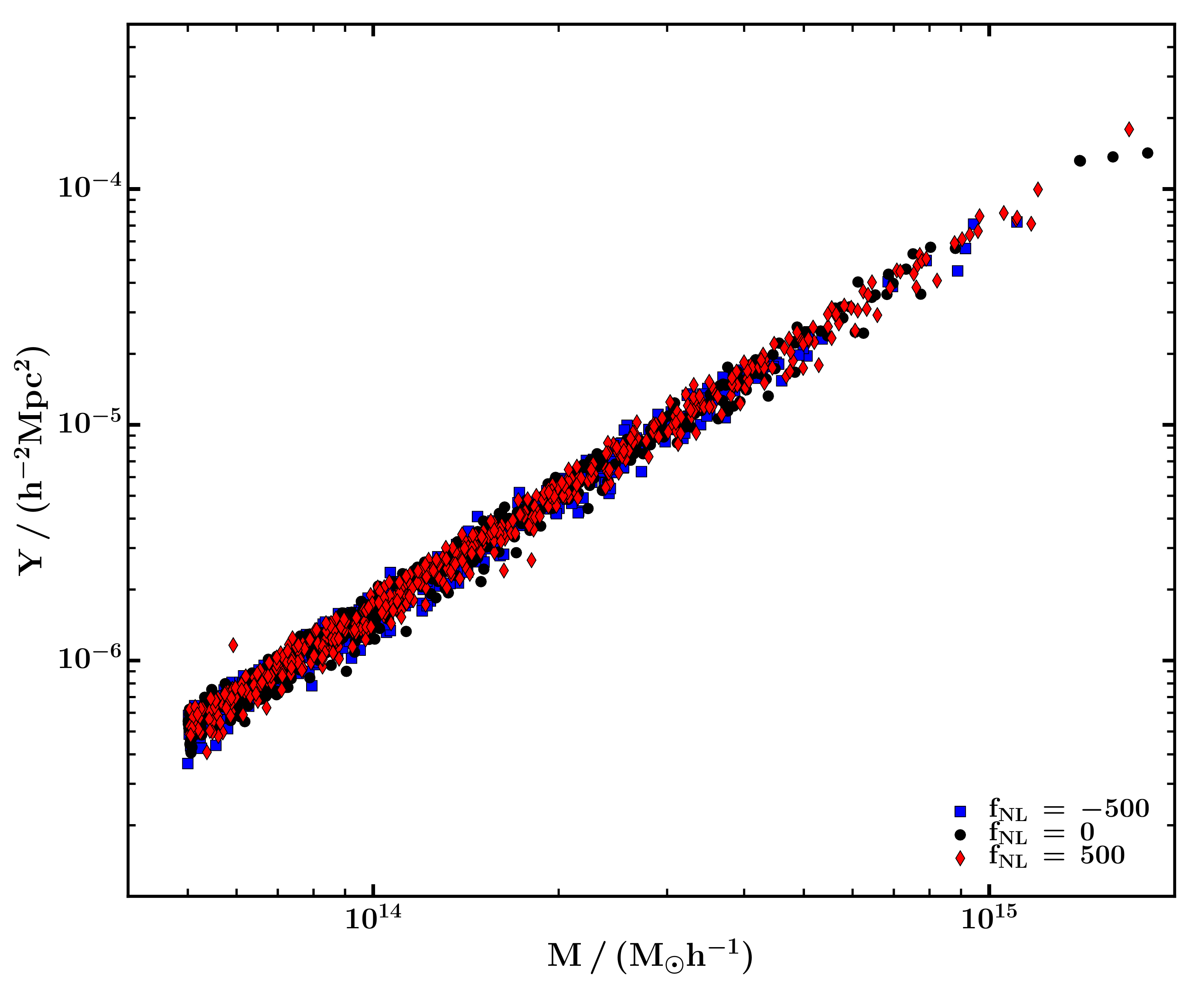}
\caption{The $Y-M$ cluster scaling relation at redshift zero for values of $f_{NL}$ of $-500$ (blue squares), $0$ (black circles) and $500$ (red pentagons).}\label{fig-1}

\end{figure}

\begin{figure}[H]
\centering
\includegraphics[scale=0.4]{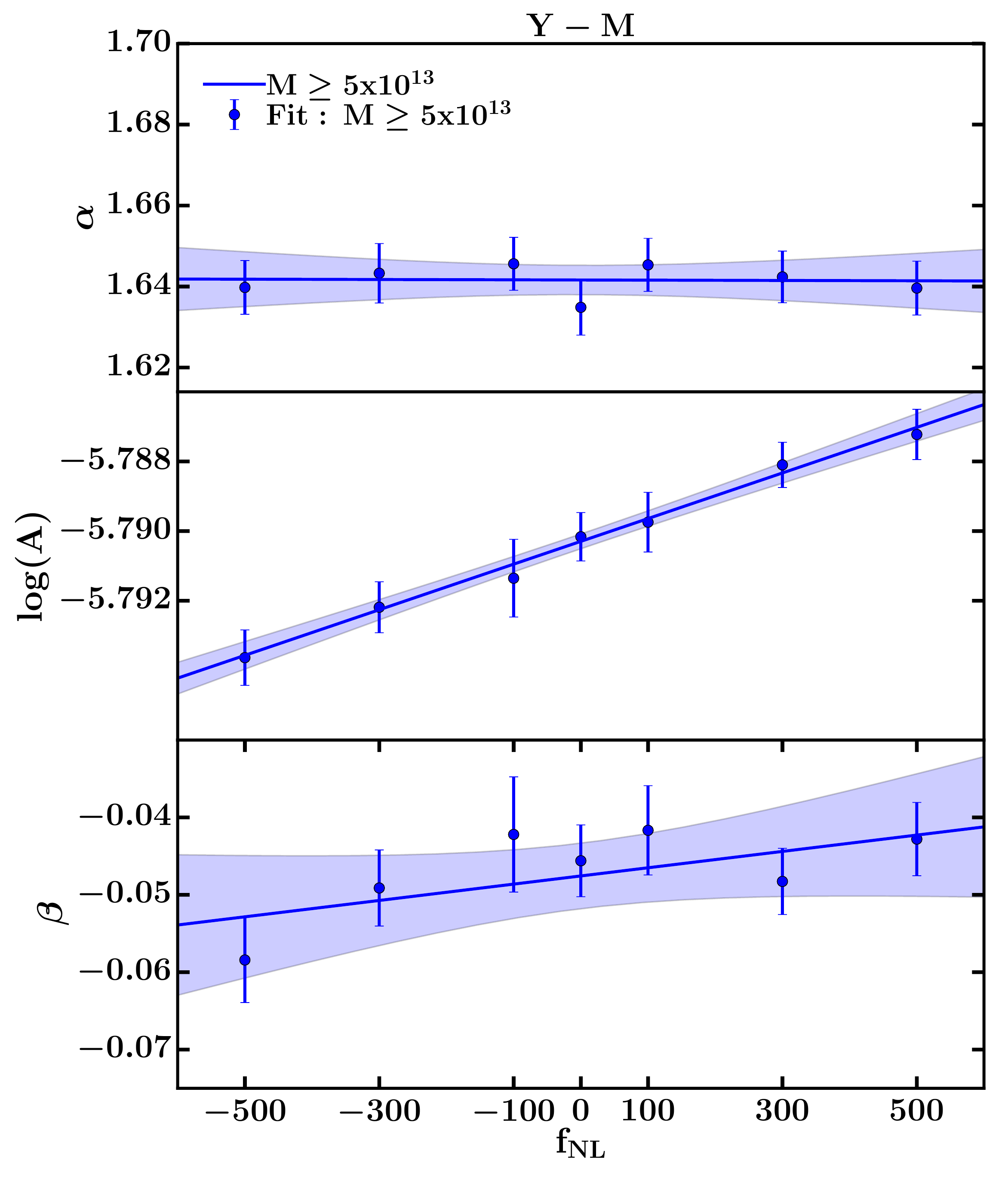}
\caption{The power-law index $\alpha$ of mass, normalization parameter $\log_{10}\left(A \right)$ (note that here $A$ is the $A_{0}$ defined in section \ref{sec:galaxy_clusters}), and power-law index of redshift $\beta$ for the $Y-M$ scaling, as a function of $f_{NL}$, for a mass cut of $5\times10^{13}\,M_{\odot}h^{-1}$, and respective $1 \sigma$ error bars. The blue solid line and shaded area are respectively, the linear fit to the data points and estimated $95\%$ C.L. confidence interval.}\label{fig-2}

\end{figure}

In order to use galaxy clusters as cosmological probes, it is essential to have an accurate estimate of their mass. However, this task is not easy, since it is not possible to directly estimate the mass of a galaxy cluster from observational data, and thus one must rely on indirect methods, such as the scaling properties of clusters, to obtain a reliable estimate. The effect of PNG on galaxy clusters scaling relations was investigated for the first time in \cite{2016arXiv160309270T}.  The authors resorted to a series of hydrodynamical N-body numerical simulations featuring adiabatic gas physics and allowing for different levels of non-Gaussian initial conditions within the fiducial $\Lambda$CDM model. In particular they studied the $T-M$, $S-M$, $Y-M$ and $Y_{X}-M$ scalings relating the total cluster mass with temperature, entropy and SZ cluster integrated pressure that reflect the thermodynamical state of the intra-cluster medium. Figure \ref{fig-1} shows as an example, the distribution of galaxy clusters for the $Y-M$, for three different levels of PNG, i.e., $f_{NG}\pm500,0$ obtained by \cite{2016arXiv160309270T}, while Figure \ref{fig-2} shows the result of fitting Equation \eqref{eq:scalings} for the $Y-M$ scaling to the corresponding  distribution of clusters, using the procedure suggested by \cite{2004MNRAS.348.1401D,2009A&A...496..637A}. The main findings of \cite{2016arXiv160309270T}, indicate that PNG do have an impact on clusters scaling laws, in particular on the amplitude and redshift evolution of the scalings normalization. In the case of $Y-M$, the redshift evolution of the its normalization can change as much as 22\% when $f_{NL}$ varies  from $-500$ to $500$. If $f_{NL}$ is a scale independent parameter and has a small value, as suggested by the current Planck constraints \cite{2014A&A...571A..24P,2015arXiv150201592P}, then the impact of PNG on clusters scaling laws is negligible when compared with the effect of additional gas physics and other cosmological effects, such as dark energy. On the other hand, if $f_{NL}$ is instead a scale dependent parameter, as suggested by some inflationary models, then PNG could have larger positive/negative amplitude at clusters scales. In this case, the effects of PNG must be taken into account when galaxy cluster data are used to assess the constraining power of future cluster surveys or to estimate the cosmological parameters from observational cluster counts data. 

\section{Testing Gravity with Weak Lensing}\label{sec7}

Gravitational lensing is the name commonly given to the phenomenon of deflection of light by gravitational fields. A lensing system consists of three parts: the source of light, the lens (a inhomogeneous distribution of mass, or a spacetime geometry) and the observer. Gravitational lensing of a point source by intervening matter produces image displacement, or even multiple images of the source. On extended sources, tidal gravitational fields lead to differential deflection, producing shape distortion and size magnification. Gravitational lensing on cosmological scales is a direct probe of Poisson-like constraint equations and probes structure formation independently of mass-to-light properties. It is thus a promising powerful way of testing gravity on cosmological scales and it is a core probe in the Euclid space mission (http://www.euclid-ec.org).

\subsection{Gravitational Lensing}

Gravitational lensing \cite {glbook,saasfee,narayan,bartelmann} involve different types of astrophysical objects that spread in a broad range of mass, mass density and length scale. Various types of systems have been studied, both theoretically and observationally, with the aim of investigating the source, the lens, or the geometrical properties of the spacetime, from the observed images. Sources may range from stars, galaxies, to the last scattering surface, while lenses may be stars, galaxies, galaxy clusters or the large scale structure. The lensing properties are the same in all cases: the effect depends only on the projected two-dimensional mass density distribution of the lens and is independent of the luminosity or composition of the lens; gravitational lenses do not have a well-defined focal point; the effect is achromatic; there is no emission or absorption of photons; the surface brightness is conserved. The conservation of brightness implies that a change in size changes the flux, as if the source was closer to the observer, creating thus a zooming effect without loss of resolution, the so-called natural telescope.

The lens equation is a mapping between source $(\beta)$ and image $(\theta)$ planes,
\begin{equation}
\vec\beta=\vec\theta-\vec\alpha.
\end{equation}

The deflection field $\vec\alpha(\theta)$ is a gradient of the gravitational potential and it is a central quantity of gravitational lensing. It is determined by the lensing potential, i.e., the gravitational potential integrated along the line-of-sight, and it depends thus on the geometry of the system (or the spacetime metric in cosmological-size systems). It is useful to Taylor-expand the lens equation, a procedure valid for small sources in angular size and small deflections (implying small angular size images). To linear order, the lens equation will thus be described by a Jacobian mapping, known as the amplification matrix:
\begin{equation}
A_{ij}(\theta)=\frac{\partial\beta_i}{\partial\,\theta_j}=\left(\delta_{ij}-\frac{\partial\,\alpha_i}{\partial\,\theta_j}\right)\,.
\label{ampmatrix}
\end{equation}

When applied to a light ray bundle emitted from an extended source, the symmetrical traceless part of the matrix produces a sheared image, the trace is responsible for an isotropically convergence (or divergence) and the antisymmetrical component introduces a rotation. This decomposition defines the optical components fields in the image plane: the scalar convergence field $\kappa(\vec\theta)$ and the shear field $\gamma(\theta)$. The convergence is given by the Laplacian of the lensing potential,
\begin{equation}
\kappa=\frac{1}{2}(\psi,_{11}+\psi,_{22})\,,
\label{kappa}
\end{equation}
and it corresponds thus, through the Poisson equation,  to the surface mass density in the effective two-dimensional plane where $\psi$ is defined. The shear is a spin-2 quantity and is given by,
\begin{equation}
\gamma_1=\frac{1}{2}(\psi,_{11}-\psi,_{22})\;, \qquad \gamma_2=\psi,_{12}.
\label{sheardef}
\end{equation}

The rotational field is not present, since the deflection field is a gradient and the amplification matrix is thus curl-free.

\begin{figure}[H]
\centering
\includegraphics[scale=0.75]{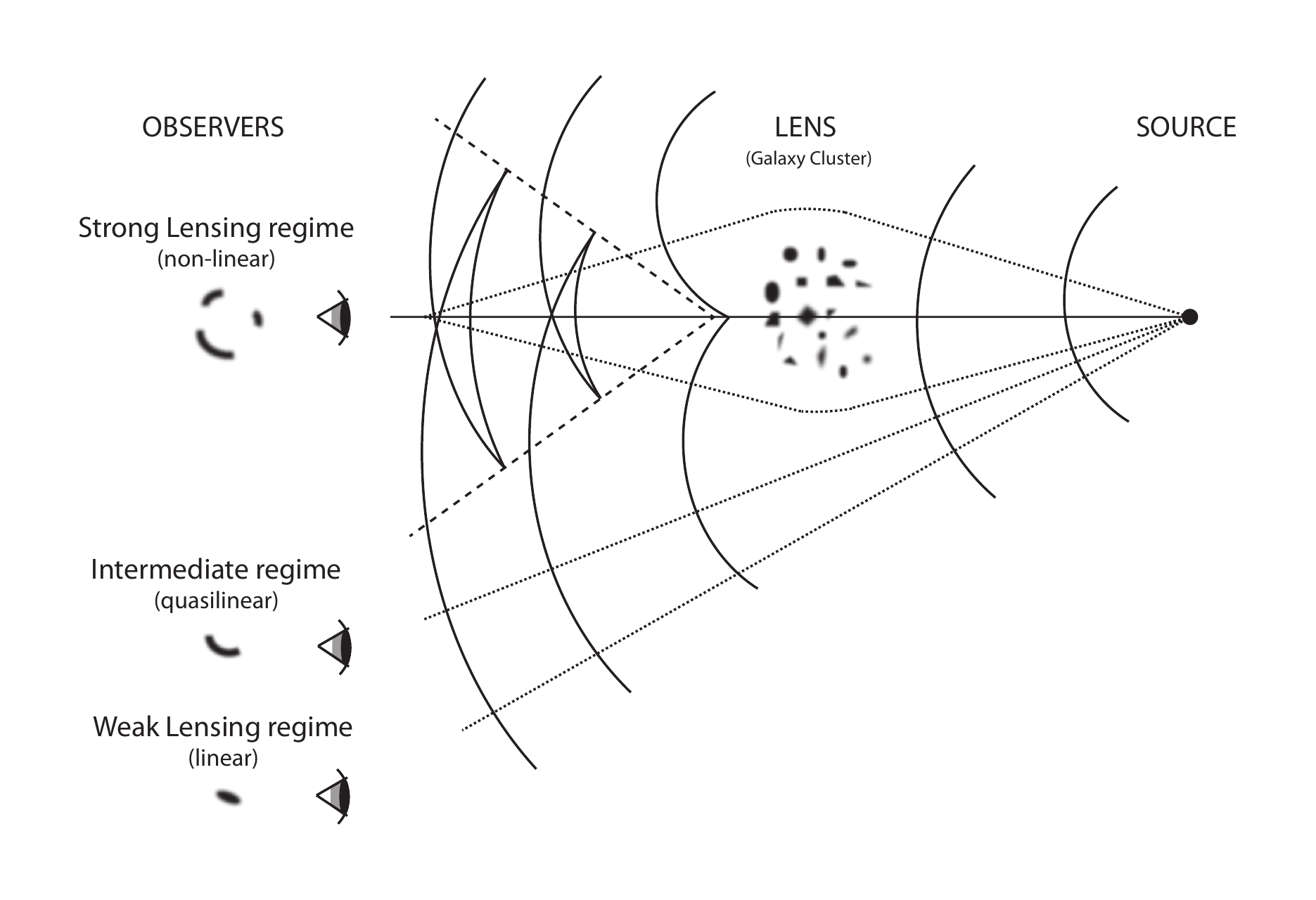}
\caption{\small Illustration of the 3 regimes of lensing.}
\label{typelenswave}
\end{figure}

Lensing effects are classified as strong or weak. Strong lensing effects occur at positions of the image plane where the convergence and shear fields take large values $(>1)$. In these cases multiple images may be formed, as well as strongly deformed multiple images that may merge forming giant arcs or even a full ring, the Einstein ring. The magnification of the images $\mu(\theta)$ is given by the inverse of the determinant of the amplification matrix. Images are magnified or demagnified depending on $\mu$ being greater or smaller than 1. The sign of the determinant defines the parity of the images, images in regions of $\det A < 0$ are inverted with respect to the source. The points in the image plane where $\det A = 0$ form closed curves named critical curves. The observed giant arcs form near critical curves. Strong lensing occurs at various scales: a quasar (source) - galaxy (lens) system produces multiple images of the quasar with a typical separation of arcsec, while a background star - foreground star system produces multiple images with a typical separation of miliarcsec, which is known as microlensing. If $\kappa \lesssim 1$ everywhere, multiple images do not form. An example in this intermediate regime are the arclets formed in the images of galaxy clusters, by lensing of background galaxies. The separation of the lensing regimes in an image depends mainly on the position with respect to the source-lens-observer alignment, as illustrated in Figure~\ref{typelenswave} (see also \cite{figtypesgl}).

In the linear regime (weak lensing), when $\kappa \ll 1$, the distortion on one source is not noticeable. However, the lensing effect can still be detected statistically when there is a large enough sample of sources suffering a coherent lensing effect. This is the case with images of dense populations of faint galaxies present in deep optical and near-infrared observations. The images of these background galaxies are coherently sheared by a large-scale inhomogeneous matter distribution: the large scale structure of the universe. The shear field of such images is known as cosmic shear and shows angular correlations up to degree scales for sources at $z\sim1$. Cosmic shear correlations are a direct measure of the statistical properties of the lensing potential, which depend on the statistical properties of the cosmological density field and the geometrical properties of spacetime. For this reason cosmic shear is a probe of both geometry and structure formation and it is the lensing effect we deal with in this contribution.

\subsection{Cosmic Shear}

Observationally, a pioneering attempt to measure a cosmological weak lensing signal was made in \cite{cosmicshear80}, having produced an upper limit for the signal. A decade later, \cite{cosmicshear90} measured shear correlations with error bars of order $100\%$ and verified the existence of instrumental and atmospheric anisotropical contaminations. A first significant detection, cosmic variance dominated, was made in \cite{schneiderobs} on a very small field (4 sqarcmin). Finally in 2000, the first definitive detection of the cosmological shear field signal was reported almost simultaneously by four independent teams \cite{2000vw,2000bacon,2000wittman,2000kaiser}. A large number of surveys from various ground-based observatories was soon undertaken, confirming the detections and using two-point cosmic shear correlations to constrain $\Omega_m$ and $\sigma_8$ parameters (see \cite{cscompilation} for a compilation). Those surveys culminated on the Canada-France-Hawaii Telescope Legacy Survey (CFHTLS), a dedicated cosmic shear survey covering 170 ${\rm deg^2}$ that successfully detected a signal on degree scales \cite{fucfhtls}. In the past 5 years the focus of cosmic shear observations moved from the detection itself to the robust estimate of residual systematics. Currently, the state-of-the-art cosmic shear survey is the Canada-France-Hawaii Telescope Lensing Survey (CFHTLenS) \cite{cfhtlenserben}, a reanalysis of CFHTLS where new methods have been applied to measure the galaxy distortions \cite{cfhtlenslensfit}, verify the source redshift distributions \cite{cfhtlensphotoz} and remove intrinsic alignment correlations \cite{cfhtlensia}. These robust data were used in various cosmological applications, including testing GR on cosmological scales, both alone and in combination with other probes, both for specific gravity theories and parametrized deviations to GR \cite{cfhtlensmodgravsimpson,cfhtlensdossett,cfhtlensmodgravvw}.

Cosmic shear studies deal with the light from background galaxies that propagates in the perturbed FRW spacetime \cite{bands,munshi,kilbinger}. In the Newtonian gauge, considering scalar perturbations and a flat universe, this spacetime is described by the metric
\begin{equation}
ds^2=-\left(1+2\Psi(x,t)\right)dt^2+a^2(t)\left(1-2\Phi(x,t)\right)\delta_{ij}dx^i dx^j\,,
\end{equation}
where $\Psi$ and $\Phi$ are the Newtonian gauge potentials.
The lens equation is the solution of the equation of movement for the comoving transverse separation between the perturbed null geodesic and a reference unperturbed one.  The deflection at a given point in the image may be seen as the integral of multiple local deflections over cosmological distances. The local deflection of a null geodesic at a given redshift depends on the traveling time of a light ray, and thus on the sum of the metric perturbations, $(\Psi+\Phi)\,.$ The resulting lens equation has a simple form in the Born approximation, i.e., when the line-of-sight integral is performed along the unperturbed geodesic, which is a good approximation for small deflections. In that case the lens equation is
\begin{equation}
\beta_i(\vec \theta,\chi)=\theta_i -\int_0^{\chi} d\chi'\,\frac{(\chi-\chi')\,\chi'}{\chi}(\Psi+\Phi),_i(\chi\theta,\chi')\,,
\end{equation}
where $\chi$ is the comoving radial coordinate. From here we can read off the local deflections as $\vec\alpha=\psi,_i$, defining the lensing potential as
\begin{equation}
\psi(\vec \theta,\chi)=\int_0^\chi d\chi'\,\frac{(\chi-\chi')\,\chi'}{\chi}
\left(\Psi(\chi\theta,\chi')+\Phi(\chi\theta,\chi')\right)\,.
\label{psi}
\end{equation}

Note that the dependence on the sum of the two potentials, which is a relativistic feature, is responsible for the well-known factor of 2 discrepancy between General Relativity and Newtonian physics predictions for the deflection of light. The convergence and shear fields may then be computed from the second-order derivatives of the lensing potential, according to Equations~(\ref{kappa}) and (\ref{sheardef}), with the following caveat. The convergence and shear fields at an image location $\theta$ have contributions from all light rays that may reach that point from a population of galaxies and not necessarily from only a single source at radial coordinate $w$. The converge then writes
\begin{equation}
\kappa(\vec \theta)=\frac{1}{2}
\int_0^{\chi_H} d\chi\,p(\chi)\,\left(\psi,_{11}(\vec \theta,\chi)+\psi,_{22}(\vec \theta,\chi)\right),
\end{equation}
where the integration is made over the sources, which are distributed over redshift as $p(\chi)\,d\chi=p(z)\,dz$.

Cosmic shear can only be detected statistically and the most used observables are two-point correlation functions and power spectra. Considering the angular correlation function of the convergence field in the Limber approximation (i.e., neglecting contributions from pairs at large radial separation) and assuming no source clustering (i.e., assuming two independent source distributions), the convergence power spectrum may be written as
\begin{equation}
P_{\kappa}(\ell)=\frac{1}{4}\int d\chi\,q_1(\chi)\,q_2(\chi)\,\left(\frac{l}{\chi}
\right)^4\,P_{\Psi+\Phi}\left(\frac{|\vec \ell|}{\chi},\chi\right).
\label{pkappa}
\end{equation}

The lensing power spectrum on an angular multipole scale $\ell$ has contributions from the power spectra of the Laplacian of the lensing potential at various redshifts $z(\chi)$ on a redshift-dependent scale $k=l/\chi$. The functions $q$ include the dependences on the sources distances and distributions. Notice also that the statistical properties of the convergence and shear fields are related and in particular their power spectra are identical: $P_\kappa(\ell)=P_\gamma(\ell)$.

\subsection{Testing Deviations from General Relativity}

In order to test a cosmological model with cosmic shear data, we need to be able to compute the evolution of the lensing potential. In General Relativity, Einstein equations at perturbation level provide a set of evolution equations for the metric and a set of constraint equations that relate the metric perturbations (the potentials $\Psi$, $\Phi$) to the perturbations of the cosmological fluid (matter density contrast $\delta$, pressure perturbation $\delta p$, fluid velocity $v$, and anisotropic pressure $\sigma$).

A combination of the $00$ and $0i$ Einstein equations relates $\Phi$ with the comoving matter density perturbation: it is a generalized Poisson equation \cite{mabertschinger}. In Fourier space, with $k^2$ being the spatial Laplacian, it writes
\begin{equation}
k^2\Phi=-4\pi\,G\,a^2\rho\Delta\,,
\label{poisson}
\end{equation}
where $\Delta$ is the comoving density perturbation $\Delta=\delta+3H(1+w)v/k^2$ and $w$ is the equation-of-state parameter $w=p/\rho$. The ij Einstein equation gives a second constraint, relating the difference $\Phi-\Psi$ with the stress $\sigma$: it is the anisotropy equation \cite{mabertschinger},
\begin{equation}
k^2(\Phi-\Psi)=12\pi\,G\,a^2\rho\,(1+w)\sigma.
\label{gslip}
\end{equation}

The lensing convergence is determined by the Laplacian of the lensing potential, and thus by,
\begin{equation}
k^2(\Phi+\Psi)=-8\pi\,G\,a^2\rho\Delta.
\label{modgrav3}
\end{equation}

Inserting this in Equation~(\ref{pkappa}), the dependence of the convergence power spectrum on the matter power spectrum becomes explicit. The latter needs then to be worked out from the evolution equations, possibly with the help of Boltzmann codes and N-body simulations.

Now, if we consider gravity theories that are modifications of General Relativity, the above equations are modified. Similar to the PPN formalism used in solar systems tests, it is useful to encapsulate the modifications in a set of parameters that can be computed in the various theories and can be easily constrained by experiments \cite{consistentPPF}. Various simplified versions of this so-called Parameterised Post-Friedmannian formalism have been proposed \cite{bertschingerPPF,caldwellPPF,husawicki,amendolaPPF}, where the modifications are described with two independent parameters. These are the gravitational screening $Q=G_{\rm eff}/G$, which parametrizes a change in the local gravitational force on cosmological scales, and the gravitational slip $\eta=\Psi/\Phi-1$, which parametrizes an anisotropy responsible for a difference between the two potentials. The 2 parameters may be functions of scale and redshift and their values in GR are $Q=1$ and $\eta=0$. This simple relations are valid for a large number of theories of gravity. Inserting these parameters in the constraint equations Equations~(\ref{poisson}) and (\ref{gslip}), we can combine them to get
\begin{equation}
k^2(\Phi+\Psi)=-8\pi\,G\,Q\left(1+\frac{\eta}{2}\right)\,a^2\rho\Delta.
\label{modgrav4}
\end{equation}

Inserting this in Equation~(\ref{pkappa}), the convergence power spectrum may be written as,
\begin{equation}
P_\gamma^{ij}(\ell)=\frac{9}{4}\,\Omega_m^2 H_0^4\int_0^{w_{\rm{h}}}dw\frac{g_i(w)g_j(w)}{a^2(w)}\,\Sigma^2(k,a)P_\delta\left(\frac{\ell}{w}\,,w\right).
\label{modpkappa}
\end{equation}

Here it was assumed that $\Delta \sim \delta$ and the $\Omega_m$ factor comes from the matter density in Equation~(\ref{modgrav3}). The parameter $\Sigma(a) = Q(a)(1+\eta(a)/2)$ is the combination of PPF parameters cosmic shear is sensitive to \cite{amendolaPPF} . The amplitude of the power spectrum depends thus on a degenerate combination of: modified gravity via constraint equations $(\Sigma)$; distances to the sources (functions $g$, which depend both on the knowledge of the source redshifts and on the cosmological background); matter density $(\Omega_mh^2)$; amplitude of the matter power spectrum ($\sigma_8$ and structure formation in modified gravity); and external systematic effects.

We applied cosmic shear data from the COSMOS survey to constrain the $\Sigma$ parameter \cite{thiswork}. The COSMOS cosmic shear survey covers an area of 1.64 ${\rm deg^2}$ where cosmic shear angular correlations were measured on a narrow scale range going up to 20 arcmin \cite{schrabbackCOSMOS}. It is however deep, with a limiting magnitude $i = 26.7$ and a high density of 76 galaxies/${\rm arcmin}^2$, allowing one to cross-correlate populations at different redshifts and perform a tomographic analysis \cite{hutomo}. We assume $\Sigma$ is scale-independent in the narrow range of scales probed. Following \cite{caldwellPPF}, we also assume that the departure from GR originates an effective smooth dark energy $\Lambda$ with constant $w=-1$, which implies the background evolution is identical to $\Lambda$CDM and the PPF parameters deviate from GR proportionally to $\rho_{\rm DE}(a)/\rho_{\rm m}(a)$. This implies that
\begin{equation}
Q(a)=1+Q_a\,a^3\,, \;\; \eta(a)=\eta_a\,a^3\,.
\end{equation}

Assuming the PPF parameters remain small, the evolution of $\Sigma$ also scales as
$\Sigma(a)\approx 1+\Sigma_a\,a^3\,.$

The last item needed for the evaluation of Equation~(\ref{modpkappa}) is the evolution of the matter overdensities, i.e., the growth function. We consider the system of the evolution equations and constraint equations for the case of sub-horizon scales, assuming time derivatives of the perturbations are negligible compared to the spatial derivatives, and $\Delta\approx \delta$. The evolution equation of the cold dark matter perturbation greatly simplifies, containing only a Hubble drag term and a term on $k^2\Psi$ \cite{daniel2010}. In this regime, the growth parameter approximation \cite{lindergrowth} applies, where
\begin{equation}
\frac{{d\rm } \ln(\delta/a)}{{\rm d} \ln(a)}=\Omega_m(a)^\gamma-1\,.
\label{eq:growth}
\end{equation}

The growth parameter $\gamma$ has the value $\gamma=6/11$ in GR and
\begin{equation}
\gamma=\frac{6}{11}\left[1-\frac{Q_{\rm a}+\eta_{\rm a}}{2}\frac{\Omega_m}{1-\Omega_m}\right]\,,
\label{eq:gamma}
\end{equation}
in PPF-parameterized modified gravity models \cite{lindercahn}.

With these assumptions, we attempt to constrain the parameters using only the cosmic shear data in a Bayesian analysis. The cosmic shear data consists of 15 two-point cross-correlation functions, between 6 galaxy redshift bins. All 15 functions have similar shapes but different amplitudes depending on their redshifts. Notice that our parametrization only affects the amplitude of the cosmic shear functions, since we work in the ansatz that $\Sigma$ and $\gamma$ are scale-independent. This implies that our constraint will necessarily be a degenerate combination of the 2 modified gravity parameters, the redshift and the standard cosmological parameters affecting the amplitude of the power spectrum. For this reason, we need to use either additional data to constrain the standard cosmology or to introduce strong priors. We introduce a prior in the form of importance sampling \cite{camb}. In particular, we take WMAP7 $\Lambda$CDM Monte Carlo Markov Chains \cite{wmap7} and evaluate a cube of models in a $(Q_a, \eta_a, f_z)$ space for each point of the MCMC chain. In this way, the $\Lambda$CDM MCMC weights will be changed as a function of the cosmic shear likelihoods in the cubes of models. The $(Q_a, \eta_a)$ likelihoods are marginalized over the redshift-calibration nuisance parameter $f_z$ that accounts for the source redshift uncertainty. The shear correlation functions are computed using the transfer function of Eisenstein $\&$ Hu \cite{transfer} with the modified linear growth $\gamma$ and the non-linear halofit prescription \cite{halofit}. Besides addressing the degeneracies with the standard cosmological parameters and the redshift uncertainties, we also minimise contamination from intrinsic alignments by not using cosmic shear auto-correlation functions in the analysis.

The result obtained in this analysis is compatible with GR. We find the following 1-$\sigma$ constraint for the following combination of the 2 parameters:
\begin{equation}
\Sigma+2.3\,(\gamma-0.55)=0.99\pm 0.31\,.
\end{equation}

\subsection{Testing Gravity with Future Cosmic Shear Data}

To take full advantage of future high-precision cosmic shear data, accurate modelling is needed at several levels, and the analyses become much more complex than the one described above.
When studying specific modified gravity theories, structure formation needs to be worked out for those models, including non-linear scales, which calls for model-specific perturbative analytical approaches \cite{valageas} or N-body simulations \cite{nbodymodgrav}. Numerical simulations are also required for other purposes. Lensing simulations are needed to estimate the covariance of lensing observables, including cosmic variance. Hydrodynamical simulations are increasingly needed to model various baryonic effects on lensing observables, such as supernova and AGN feedback, star formation or radiative cooling \cite{semboloni}.

Working in a model-independent approach requires detailed modelling of the deviations in Einstein equations. Besides applying the PPF parameters, it is also interesting to parametrize the deviations at the observables level. \cite{tessacosmicshear} parametrizes deviations of the lensing power spectrum from the non-linear GR matter power spectrum. The latter can be computed from the linear one with a fitting formula \cite{halofitnew} or from templates \cite{coyote}. Scale and time dependencies of the parametrized deviations may also be probed in bins by performing a principal components analysis \cite{zhaopca}.

Cosmic shear probes deviations from GR through the integrated effect of $\Psi$ and $\Phi$ on null geodesics. It has a high signal-to-noise on the scales where the deviations are maximal and is thus more promising than other probes of $\Phi+\Psi$, such as the integrated Sachs-Wolfe effect \cite{schmidt}. However, to be able to break the degeneracy between the two potentials in the modified gravitational effects, cosmic shear needs to be combined with a local probe of matter overdensities, which are sensitive to $\Psi$ alone. One possible combination of cosmic shear and redshift-space distortions is the so-called ${\rm E_G}$ test \cite {egproposal,egnew}, which has the additional advantage of being independent of galaxy bias. A first detection of the ${\rm E_G}$ signal was made in \cite{egdetection}. The combination of weak lensing and galaxy clustering observables is an important asset of the forthcoming Euclid space mission to be launched in 2020, which combines a weak lensing and galaxy clustering wide survey and aims at measuring the rate of cosmic structure growth to a $2\%$ precision \cite{redbook}.

\section{Angular Distribution of Cosmological Parameters as a Measurement of Spacetime Inhomogeneities}\label{sec8}

On large scales, the matter distribution is statistically homogeneous.
Locally however, matter is distributed according to a pattern of alternate overdensed regions and underdensed regions. Since averaging inhomogeneities in the matter density distribution yields a homogeneous description of the Universe, then the apparent homogeneity of the cosmological parameters could also result from the averaging of inhomogeneities in the cosmological parameters, which would reflect the inhomogeneities in the density distribution.
In the context of backreaction models, angular variations in the parameters could also source a repulsive force and potentially emulate cosmic acceleration. Hence the reasoning was to look for these inhomogeneities 
across the sky and then to use an adequate toy model to compute the magnitude of the acceleration derived from angular variations of the parameters compared to the acceleration driven by a cosmological constant.

Recently, a method was proposed to search for inhomogeneities based on a local estimation of cosmological parameters \cite{carvalho_2015}. We demonstrated the method with supernovae (SNe), since they are a sensitive probe of the late--time expansion history of the Universe by means of the dependence of the luminosity distance on both the geometry and the growth of structure. Hence the parameters to be estimated are those that affect the luminosity distance, namely, $\{\Omega_{M}, \Omega_{\Lambda}, H_{0}\}.$

Supernovae have previously been used to constrain inhomogeneities in the cosmic expansion, namely, by measuring the hemispherical anisotropy of $H_{0}$ \cite{kalus_2013}, by assessing the dependence of $H_{0}$ with the position on the cosmic web \cite{wojtak_2014} and by mapping $\{H_{0}, q_{0}\}$ in the sky \cite{bengaly_2015}. Supernovae have also been used to constrain inhomogeneities in the dark energy, namely by constraining fluctuations of a dynamical dark energy \cite{cooray_2010} and by constraining radial inhomogeneity for a Lema\^itre--Tolman--Bondi metric with a cosmological constant \cite{marra_2014}.

\begin{figure}[H]
\centerline{
\includegraphics[width=7.8cm]{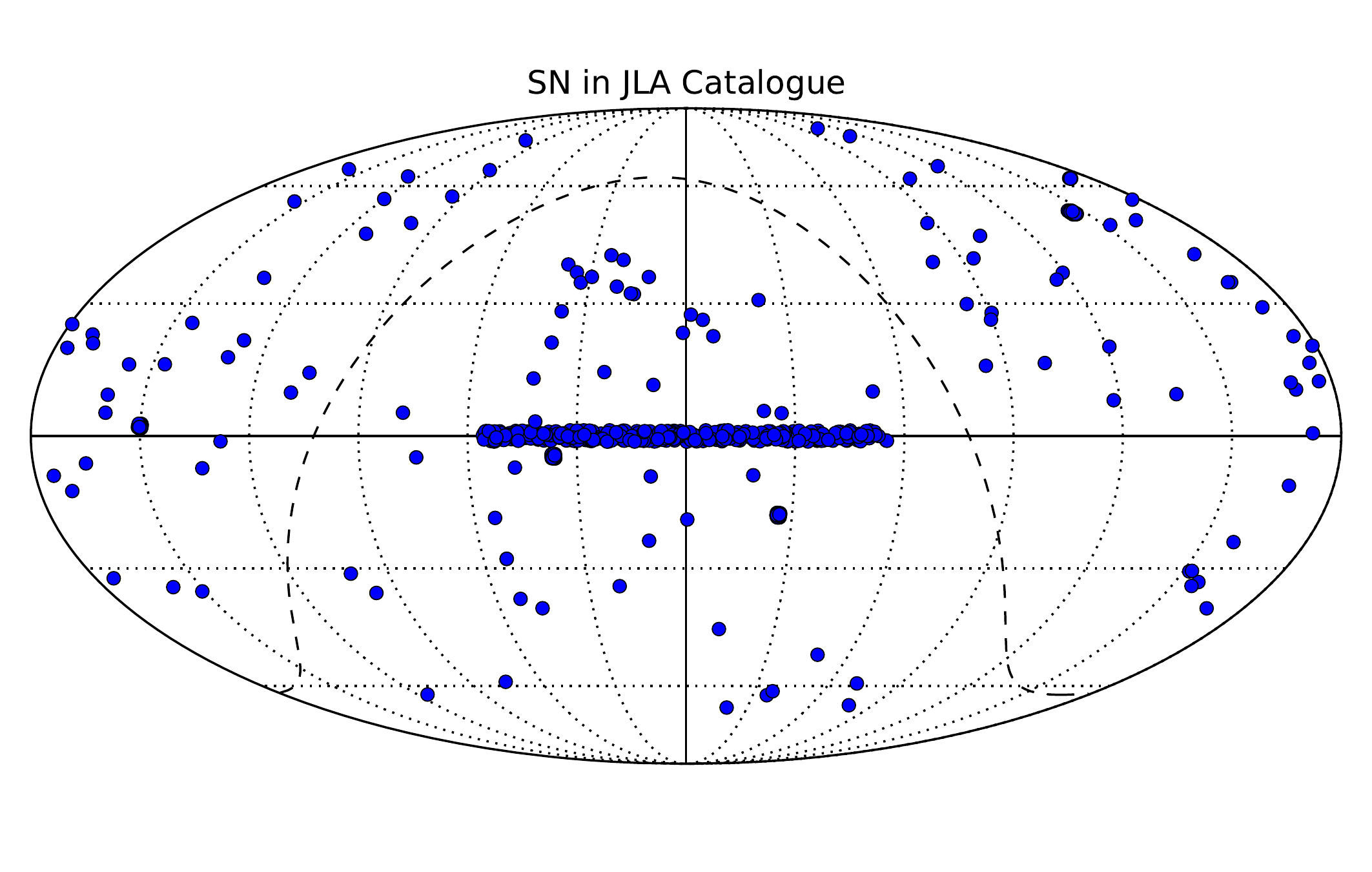}
\includegraphics[width=7.8cm]{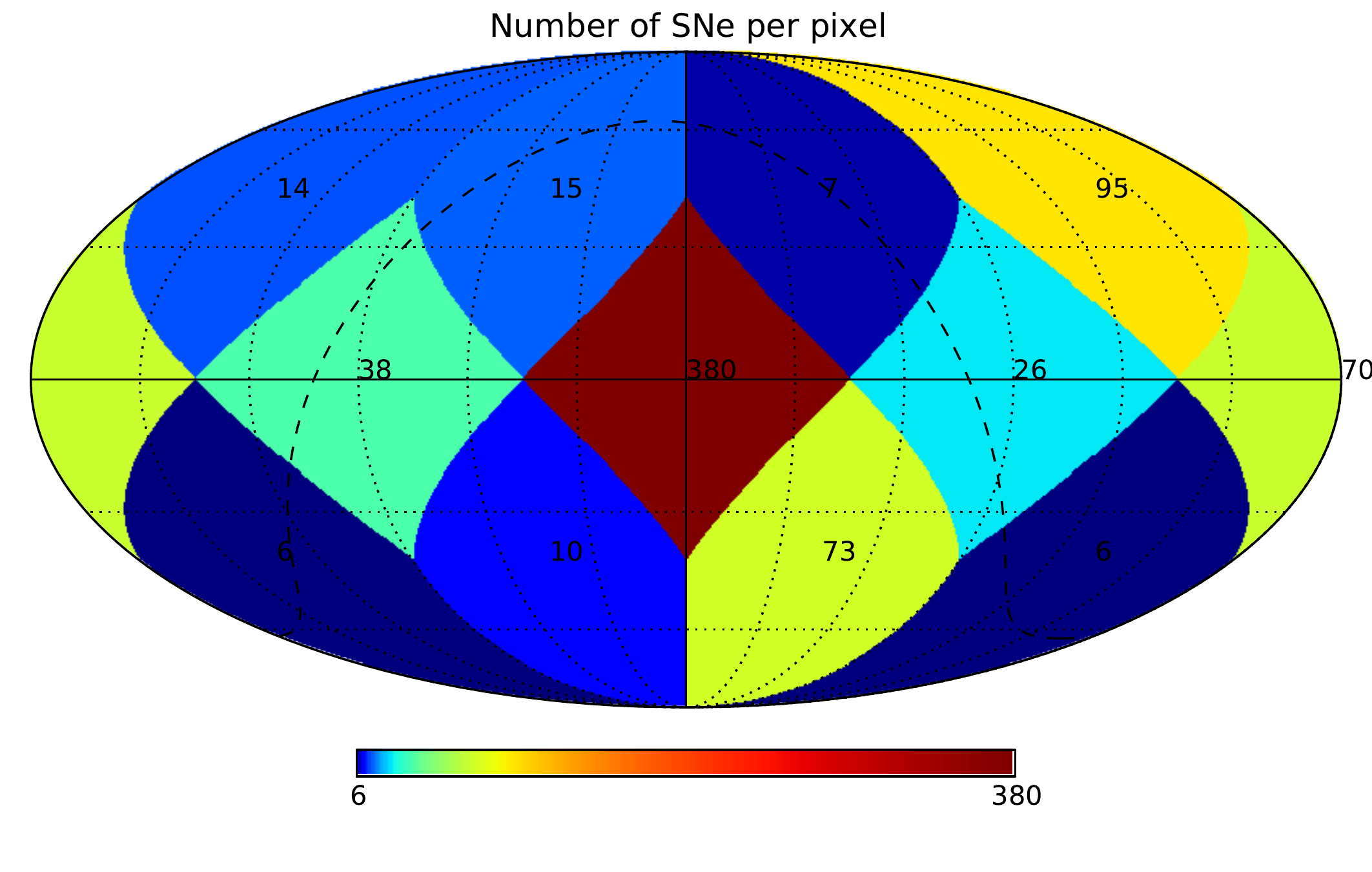}
}
\caption{\baselineskip=0.5cm{
{\bf Left panel}: Angular distribution of the JLA type Ia SN sample. The sample consists of type Ia SNe compiled from different surveys as described in Ref.~\cite{betoule_2014} and plotted in Celestial coordinates. 
{\bf Right panel}: Angular distribution of the number of type Ia SNe per pixel. The total SN sample is divided into subsamples grouped by pixels in the HEALPix pixelation scheme. The number of SNe in each pixel is printed in the centre of the pixel. The pixel indices $\{1,2,3,4,5,6,7,8,9,10,11,12\}$ correspond to the SN counts $\{15, 14, 95, 7, 380, 38, 70, 26, 10, 6, 6, 73\},$ as indicated by the colour bar in logarithmic scale. In both panels, the dashed line represents the Galactic equator.}
}
\label{fig:jla_sn}
\end{figure}

We used the type Ia supernova sample compiled by the SNLS--SDSS collaborative effort Joint Light--curve Analysis (JLA) \cite{betoule_2014} , totalling $N_{\text{SNe}}=740$ SNe with redshift $0.01 \le z \le 1.30$ and distributed on the sky according to Figure~\ref{fig:jla_sn}. The sample was obtained from {\tt http://supernovae.in2p3.fr/sdss\_snls\_jla/ReadMe.html}. 

\subsection{Method of Local Parameter Estimation}

Our method consists in dividing the SNe over a pixelated map of the sky, with pixels of equal surface area according to the HEALPix pixelation scheme \cite{gorski_2005}. The SNe subsample that falls in each pixel $k$ is used to estimate the cosmological parameters in that pixel, thus the designation of local estimation. For each cosmological parameter being estimated, there results a map with the same pixelation as that of the SNe subsamples. Each pixel is assumed to be described by a Friedmann--Lema\^itre--Roberston--Walker metric so that the full sky is an inhomogeneous ensemble of disjoint, locally homogeneous regions. In each pixel, we estimated the cosmological parameters that minimise the chi--square of the fit of the theoretical distance modulus
\ba
\mu_{\rm theo}(z;\Omega_{M},\Omega_{\Lambda},H_{0})
=5\log[d_{L}(z;\Omega_{M},\Omega_{\Lambda},H_{0})]+25 \,,
\ea

($d_L$ in units of Mpc) computed for the trial values of $\{\Omega_{M}, \Omega_{\Lambda}, H_{0}\},$ 
to the measured distance modulus
\ba
\mu_{\rm data}(z)=m_{B}(z)-(M_{B}-\alpha x_{1}+\beta c) \,,
\label{eqn:mu_data}
\ea
computed for the estimated light--curve parameters $\{m_{B},x_{1}, c\}$ for each SN.
From the local estimation there resulted the maps $x_{ik}=\{\Omega_{M},\Omega_{\Lambda}, H_{0}\}_{k}.$  From the estimation using the complete SN sample, designated as the global estimation, we obtained the values $x_{i}^{\text{fid}}=\{\Omega_{M}^{\text{fid}},\Omega_{\Lambda}^{\text{fid}}, H_{0}^{\text{fid}}\},$ which we adopt as fiducial values for the parameters. The fiducial values for the estimated parameters are: $\{\Omega_{M}^{\text{fid}},\Omega_{\Lambda}^{\text{fid}}, H_{0}^{\text{fid}}\}=\{0.256,   0.715,   71.17\}.$ For the derived parameters $\Omega_{\kappa}=1-\Omega_{M}-\Omega_{\Lambda}$ and $q_{0}=\Omega_{M}/2-\Omega_{\Lambda},$ the fiducial values are $\Omega_{\kappa}^{\text{fid}}=0.029$ and 
$q_{0}^{\text{fid}}=-0.586.$ For better visualization, we compare the local estimation with the fiducial values by defining the difference maps $\Delta x_{ik}=x_{ik}-x_{i}^{\text{fid}}.$

Since the SN sample used is a collection from different SN surveys, the SN subsampling in pixels is highly inhomogeneous. As a result, well sampled pixels return robust estimations, whereas poorly sampled pixels return weak constraints. This generates a noise bias.
We hypothesised that the noise bias due to the inhomogeneity in the SN subsampling is modelled by the local estimation obtained from randomizing the dependence of redshift with position and then averaging over the different randomizations. For each randomization in $z,$ the SN positions in the sky remain the same as in the original sample.
Subtracting the noise bias off the maps previously estimated we obtained unbiased maps $x_{ik}^{\text{unbias}}\equiv x_{ik}-\bar x_{ik}^{\text{bias}}+x_i^{\text{fid}},$ where $\bar x_{ik}^{\text{bias}}\equiv\big< x_{ij^{\prime}k}^{\text{bias}}\big>_{j^{\prime}}$ is the average over $j^{\prime}$ randomizations in $z.$ For better visualization, we define the unbiased difference maps as $\Delta x_{ik}^{\text{unbias}}\equiv x_{ik}^{\text{unbias}}-x_{i}^{\text{fid}}=x_{ik}-\bar x_{ik}^{\text{bias}}.$

\begin{figure}[H]
\centering{
\includegraphics[width=10cm]
{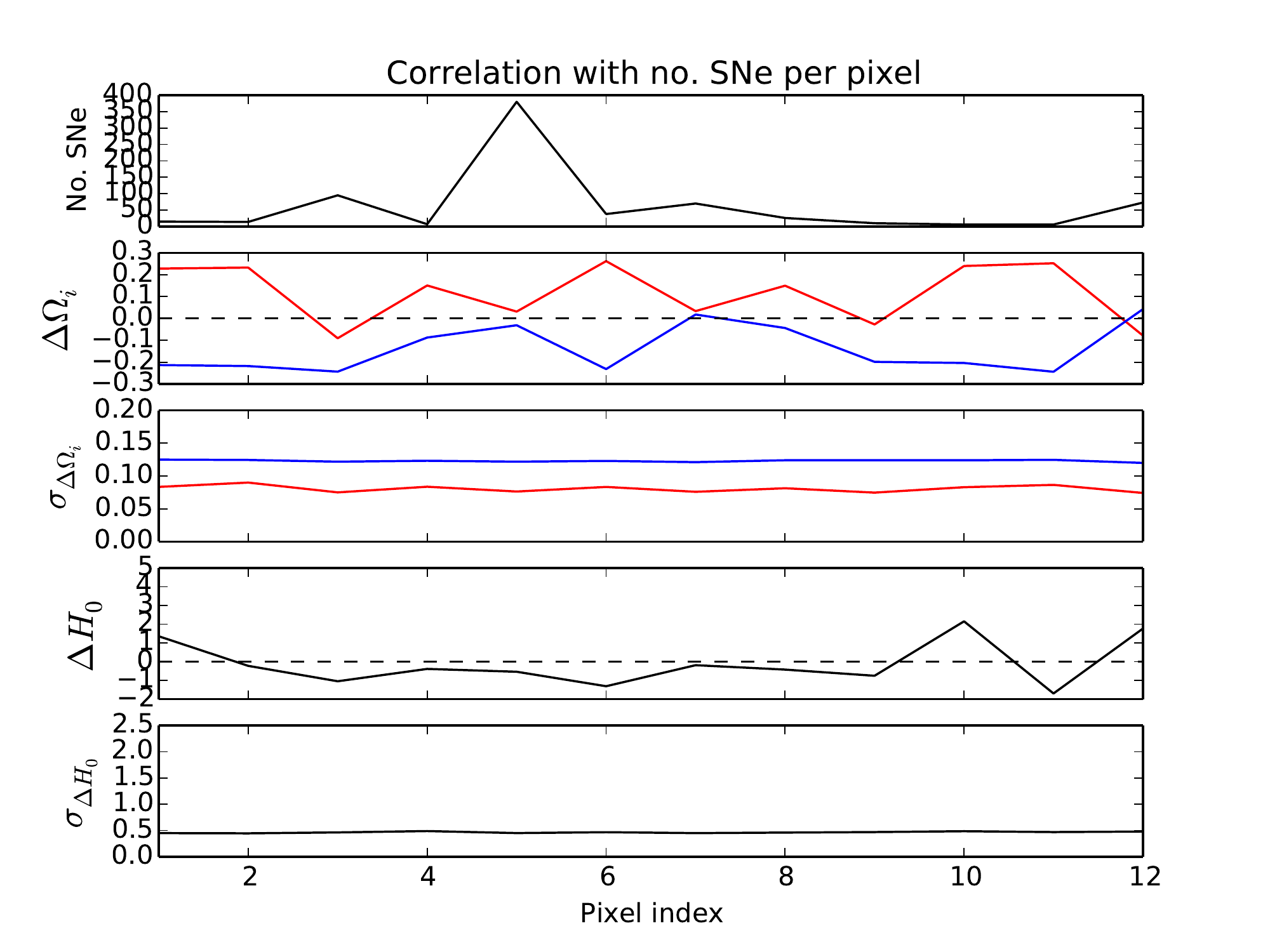}
\hspace{-1cm}
\includegraphics[width=10cm]{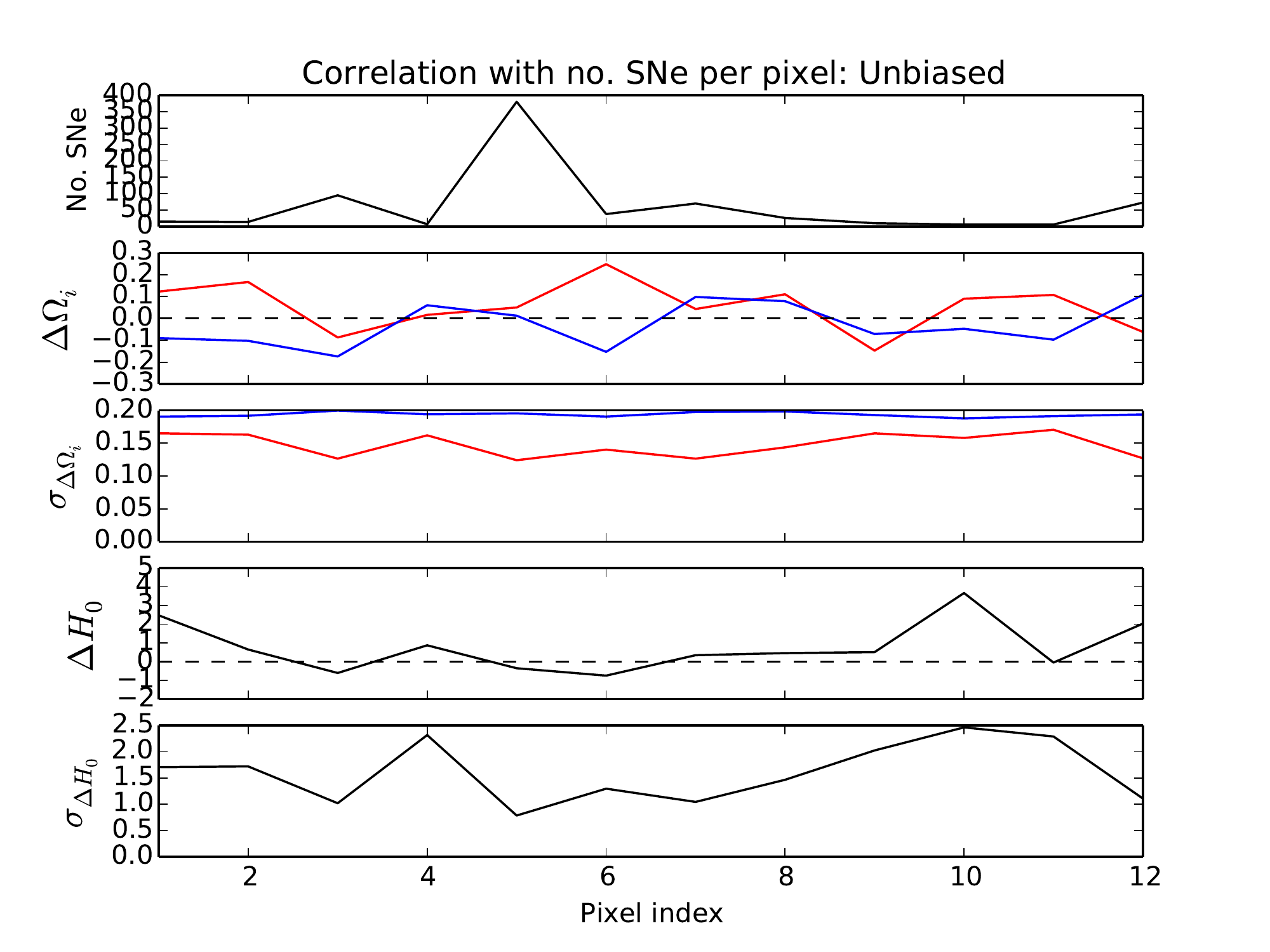}
}
\caption{\baselineskip=0.5cm{ Correlation of the number of SNe per pixel with the estimated parameters' fluctuations about the fiducial values, before (left panel) and after (right panel) the noise bias removal. At each pixel we plot: a) in the first panel, the number of SNe; b) in the second panel, the value of the parameters difference $\{\Delta\Omega_{M},\Delta\Omega_{\Lambda}\}$ respectively as the solid red and solid blue lines, with the dashed black line marking the zero difference; c) in the third panel, the standard deviation $\{\sigma_{\Delta\Omega_{M}},\sigma_{\Delta\Omega_{\Lambda}}\}$ respectively as solid red and solid blue lines; d) in the fourth plot, the value of $\Delta H_{0}$ as the solid black line, with the dashed black line marking the zero difference; e) in the fifth panel, the standard deviation $\sigma_{\Delta H_{0}}$ as the solid black line.}}
\label{fig:corr_per_pix}
\end{figure}

In Figure~\ref{fig:corr_per_pix} we show the values of the difference maps at each pixel, before (left panel) and after (right panel) the noise bias subtraction. We also show the corresponding errors.
We observe a negative correlation of the number of SNe in each pixel both with the fluctuations about the fiducial values, and with the corresponding errors.
These results reflect the poor sampling derived from an inhomogeneous coverage of the sky, both in angular and redshift space, by the SN surveys.
A poor sampling into pixels implies a poor sampling in redshift, which limits the constraint of the degeneracies among the parameters and hence compromises the parameters estimation.

We also observe that, after the bias subtraction, the fluctuations about the fiducial values are reduced, albeit the errors increase. Consequently, for the correlation of the number of SNe per pixel, the unbiased maps in comparison with the original maps show a smaller correlation with the difference values and a larger correlation with the errors, as expected. These results indicate the validity of $\bar x_{ik}^{\text{bias}}$ as a measure of the noise bias due to the inhomogeneity of the SN sampling. 
Comparing the unbiased difference maps with the fiducial values, we measure fluctuations of order 5\%--95\% for $\Omega_{M},$ 1\%--25\% for $\Omega_{\Lambda}$ and up to 5\% for $H_{0}.$

\subsection{Average over the Local Parameter Estimation}

When averaging over the homogeneous regions, angular fluctuations in the expansion factor (and consequently in $H_{0}$) induce a backreaction term in the form of an extra positive acceleration \cite{rasanen_2006}. We derived the analytical extra positive acceleration for a toy model of an arbitrary number of disjoint, homogeneous regions and computed the overall deceleration parameter assuming a) no backreaction and b) backreaction for the measured angular distribution of $H_{0}.$ 

In the absence of backreaction, the averaging consists in taking the mean weighted by the variance's inverse $w_k=1/\text{Var}[\bar x_{ik}]$ of parameter $x_{i}$ in each pixel $k,$ 
\ba
\bar x_i=\big< \bar x_{ik}\big>_{k}={\sum_{k}^{N_{\text{pixel}}}w_k~\bar x_{ik} \over {\sum_{k}^{N_{\text{pixel}}}w_k}}.
\ea

For the estimated parameters, we find
$\{\overline\Omega_{M},\overline \Omega_{\Lambda},\overline H_{0}\}
=\{0.242,0.607,71.06\}\pm \{0.099,0.115,0.87\}.$ 
These values are consistent with, but systematically smaller than, the fiducial values. 
After the noise bias removal, we find 
$\{\overline\Omega_{M},\overline \Omega_{\Lambda},\overline H_{0}\}_{\text{unbias}}
=\{0.294,0.679,71.37\}\pm \{0.115,0.120,0.94\}.$ 
These values are consistent with the fiducial values. For the derived parameters, we find 
$\overline\Omega_{\kappa}=0.078\pm0.131$ and $\bar q_{0}=-0.451\pm 0.159,$  
and after the noise bias removal we find
$\overline\Omega_{\kappa,\text{unbias}}=0.002\pm0.135$ and $\bar q_{0,\text{unbias}}=-0.526\pm 0.172.$ After the noise bias removal, the pixel average values of both $\Omega_{\kappa}$ and $q_{0}$ become closer to the corresponding values estimated using the complete sample.

In the presence of backreaction, the averaging consists in taking the mean weighted by the 3--volume $V_k$ of each pixel $k,$ 
\ba
\bar x_i=\big< \bar x_{ik}\big>_{V_k}={\sum_{k}^{N_{\text{pixel}}}V_k~\bar x_{ik} \over {\sum_{k}^{N_{\text{pixel}}}V_k}}.
\ea

Since all pixels have the same surface area and hence the angular directions expand the same way today, we can assume that the radial direction also expands the same way today, which implies that the 3--volume is the same today for all pixels. This amounts to identifying a volume as a pixel. 
Defining $v_{k}=V_{k}/\sum_{k}V_{k},$ then for $N_{\text{pixel}}$ disjoint regions, the average of the Hubble parameter is given by
\ba
\left< H_0\right>_{V_k}=\sum_{k}^{N_{\text{pixel}}}v_{k}H_{0,k}.
\ea

Taking the derivative, it follows that 
\ba
\left<{\ddot a \over a}\right>_{V_k}=\sum_{k}^{N_{\text{pixel}}}v_{k} {\ddot a \over a} 
+2\sum_{k}^{N_{\text{pixel}}}\sum_{l>k}^{N_{\text{pixel}}}v_{k}v_{l}\left(H_{0,k}-H_{0,l}\right)^2,
\label{eqn:average_Vk}
\ea
which decomposes into a linear term in the pixel average of $(\ddot a/a)$ and a quadratic term in differences of $H_{0}$ between pairs of pixels. The quadratic (backreaction) term generates an acceleration that is due not to regions speeding up locally, but instead to the slower regions becoming less represented in the average. Note that, in the absence of the quadratic term, the volume average reduces to the pixel average above. Then the volume average of $q_{0}$ becomes
\ba
\left< q_{0}\right>_{V_k}&=&\sum_{k}^{N_{\text{pixel}}}v_{k}q_{0,k}
-{2\over \left(\sum_{k}^{N_{\text{pixel}}} v_{k}H_{0,k}\right)^2}
\sum_{k}^{N_{\text{pixel}}}\sum_{l>k}^{N_{\text{pixel}}}v_{k}v_{l}\left(H_{0,k}-H_{0,l}\right)^2.
\ea

Using the fluctuations in $H_{0}$ measured in the local parameter estimation, we find 
$\bar q_{0}=-0.452\pm0.159$ and after the noise bias removal we find 
$\bar q_{0,\text{unbias}}=-0.527\pm0.173.$ The volume average of the other parameters is the same as the pixel average.
After the noise bias removal, the volume average values become closer to the value for the deceleration estimated from the complete sample. The results from both averaging methods strengthen the validity of $\bar x_{ik}^{\text{bias}}$ as a measure of the noise bias due to the inhomogeneity of the SN sampling.

The quadratic term is of order $10^{-3}$ (or lower) times the linear term, which means that the error of the difference is below the standard deviation. 
Hence for the angular fluctuations in $H_{0}$ measured with this SN sample, the contribution of the quadratic term in Equation~(\ref{eqn:average_Vk}) is insignificant, which renders the volume averaging equivalent to the pixel averaging. Hence in the context of this toy model of an inhomogeneous spacetime, backreaction is not a viable dynamical mechanism to emulate cosmic acceleration. A subsequent study using a kinematic parametrisation of the luminosity distance expressed in terms of time derivatives of the scale factor yielded concordant results \cite{carvalho_2016}.

\section{Summary and Conclusion}\label{sec:concl}

In this work, we have explored the dynamics and evolution of the Universe at early and late times, focusing on both dark energy and extended gravity models and their astrophysical and cosmological consequences.
 More specifically, we presented scalar-tensor theories, focussing on a brief review of the general formalism, the conformal picture, discussed the experimental and observational status of the theory, presented a detailed dynamical system analysis of the cosmological dynamics, 
and concluded with a quintessence scenario where a multi-scalar-tensor theory is responsible for the present accelerated expansion of the Universe.
 Furthermore, we considered Horndeski theories and the self-tuning properties, and explored realistic cosmologies that can be constructed, by studying the properties of this intriguing theory. We also reviewed $f(R)$ modified theories of gravity and extensions, where the foundational questions and astrophysical/cosmological issues of the Palatini approach were extensively explored; we also focussed on recently proposed theories such as on curvature-matter couplings and the hybrid metric-Palatini theory. 
The inflationary epoch was also presented in $f(R)$ gravity, where the so-called Starobinsky inflation is known as one of the more reliable candidates for explaining the inflationary paradigm as provided by the recent Planck/Bicep2 constraints. 
Moreover, we have reviewed the evolution and cosmological consequences of topological defect networks that may be formed in the early universe. In particular, we focused on two of the most significant signatures of these networks: the cosmic microwave background anisotropies and the stochastic gravitational wave background.

As mentioned in the Introduction, due to the extremely important aspect of the synergy between theory and observations, the second part of this review was dedicated to observational cosmology. In particular, cosmological tests with galaxy clusters at CMB frequencies were presented. More specifically, galaxy clusters and the thermal and kinetic Sunyaev-Zel'dovich (SZ) effects were reviewed, and the cosmological tests using SZ cluster surveys were presented, focussing on SZ cluster counts, SZ power spectra and the possibility of probing new physics with the SZ clusters; a review on probing primordial non-gausianities with galaxy clusters was also presented, with an emphasis on the parametrization of primordial non-Gaussianities, the non-Gaussian mass function, biased cosmological parameter estimation with cluster counts, the impact of primordial non-Gaussianities on galaxy clusters scaling relations. Furthermore, gravitational lensing was also explored, with a focus on cosmological scales, with the possibility of testing deviations from General Relativity, in particular, with future cosmic shear data. In conclusion, the angular distribution of cosmological parameters as a measurement of spacetime inhomogeneities was presented, using a method of averaging over the local parameter estimation.
Thus, cosmological tests were discussed, and their relevance in constraining our cosmological description of the Universe.

\vspace{6pt}

\acknowledgments
{We thank Ippocratis Saltas for helpful comments and suggestions. 
This work was supported by Funda\c{c}\~ao para a Ci\^encia e a Tecnologia (FCT) through the research grant UID/FIS/04434/2013.
PPA acknowledges financial  support from FCT through the Investigador FCT Contract No.~IF/00863/2012, funded by FCT/MCTES (Portugal).
CSC is funded by FCT, Grant No.~SFRH/BPD/65993/2009.
FSNL acknowledges financial  support from FCT through the Investigador FCT Contract No.~IF/00859/2012, funded by FCT/MCTES (Portugal).
PMM acknowledges financial support from the Spanish Ministry of Economy and Competitiveness through the postdoctoral training contract FPDI- 2013-16161, and the project FIS2014-52837-P.
DRG acknowledges support from FCT, Grant No. SFRH/BPD/102958/2014.
DSG acknowledges support from FCT, Grant No.~SFRH/BPD/95939/2013.
LS is supported by FCT through the grant SFRH/BPD/76324/2011.
IT acknowledges support from FCT through the Investigador FCT Contract No.~IF/01518/2014 and POPH/FSE (EC) by FEDER funding through the program Programa Operacional de Factores de Competitividade -- COMPETE.
The authors also acknowledge the COST Action CA15117, supported by COST (European Cooperation in Science and Technology).}



\authorcontributions{All the authors have substantially contributed to the present work.}



\conflictofinterests{The authors declare no conflict of interest.}

\bibliographystyle{mdpi}
\makeatletter
\renewcommand\@biblabel[1]{#1. }
\makeatother

\bibliographystyle{mdpi}
\renewcommand\bibname{References}


\begin{thebibliography}{999}

\bibitem{Perlmutter:1998np} 
  Perlmutter, S.; Aldering, G.; Goldhaber, G.; Knop, R.A.; Nugent, P.; Castro, P.G.; Deustua, S.; Fabbro, S.; Goobar, A.; Groom, D.E.; et al. 
  Measurements of Omega and Lambda from 42 high redshift supernovae.
  {\it Astrophys.  J.} {\bf 1999}, \textit{517}, 565--586.

\bibitem{Riess:1998cb} 
  Riess, A.G.; Filippenko, A.V.; Challis, P.; Clocchiattia, A.; Diercks, A.; Garnavich, P.M.; Craig, G.; Hogan, C.J.; Jha, S.; Kishner, R.P.; et al. 
  Observational evidence from supernovae for an accelerating universe and a cosmological constant.
  {\it Astron.\ J.} {\bf 1998}, {\it 116}, 1009--1038.

\bibitem{Maartens:2003tw} 
  Maartens, R.
  Brane world gravity.
  {\it Living Rev.\ Rel.} {\bf 2004}, {\it 7}, 7.

\bibitem{Dvali:2000hr} 
  Dvali, G.R.; Gabadadze, G.; Porrati, M.
  4-D gravity on a brane in 5-D Minkowski space.
  {\it Phys.\ Lett.\ B} {\bf 2000}, \textit{485}, 208.

\bibitem{deRham:2007rw} 
  de Rham, C.; Hofmann, S.; Khoury, J.; Tolley, A.J.
  Cascading Gravity and Degravitation.
  {\it J. Cosmol. Astropart. Phys.} {\bf 2008} {\it 0802}, 011.
 
\bibitem{Sotiriou:2008rp} 
  Sotiriou, T.P.; Faraoni, V.
  f(R) Theories of Gravity.
  {\it Rev.\ Mod.\ Phys.}  {\bf 2010}, {\it 82}, 451.

\bibitem{DeFelice:2010aj} 
  De Felice, A.; Tsujikawa, S.
  f(R) theories.
  {\it Living Rev.\ Rel.} {\bf 2010}, {\it 13}, 3.

\bibitem{Capozziello:2011et} 
  Capozziello, S.; De Laurentis, M.
  Extended Theories of Gravity.
  {\it Phys.\ Rept.} {\bf 2011}, {\it 509}, 167.

\bibitem{Nojiri:2010wj} 
  Nojiri, S.; Odintsov, S.D.
  Unified cosmic history in modified gravity: from F(R) theory to Lorentz non-invariant models.
  {\it Phys.\ Rept.} {\bf 2011}, {\it 505}, 59.

\bibitem{Lobo:2008sg} 
  Lobo, F.S.N.
  The Dark side of gravity: Modified theories of gravity. In
  {\it Dark Energy-Current Advances and Ideas}; Research Signpost: Kerala, India, 2009; pp. 173--204.

\bibitem{Capozziello:2002rd} 
  Capozziello, S.
  Curvature quintessence.
  {\it Int.\ J.\ Mod.\ Phys.\ D} {\bf 2002}, {\it 11}, 483.

\bibitem{Nojiri:2003ft} 
  Nojiri,~S.; Odintsov, S.D.
  Modified gravity with negative and positive powers of the curvature: Unification of the inflation and of the cosmic acceleration.
  {\it Phys.\ Rev.\ D} {\bf 2003}, {\it 68}, 123512.

\bibitem{Carroll:2003wy} 
  Carroll, S.M.; Duvvuri, V.; Trodden, M.; Turner, M.S.
  Is cosmic speed-up due to new gravitational physics?
  {\it Phys.\ Rev.\ D} {\bf 2004}, {\it 70}, 043528.

\bibitem{Copeland:2006wr} 
  Copeland, E.J.; Sami, M.; Tsujikawa, S.
  Dynamics of dark energy.
  {\it Int.\ J.\ Mod.\ Phys.\ D} {\bf 2006}, {\it 15}, 1753--1936.

\bibitem{BD 61}  
Brans,~C.; Dicke,~R.H.
 Mach's principle and a relativistic theory of gravitation.
 {\em Phys. Rev.} {\bf 1961}, {\it 124}, 925--935.

\bibitem{deRham:2014zqa} 
  de Rham, C.
  Massive Gravity.
  {\it Living Rev.\ Rel.} {\bf 2014}, {\it 17}, 7.

\bibitem{Horndeski:1974wa}
  Horndeski, G.W.
  Second-order scalar-tensor field equations in a four-dimensional space.
  {\it Int.\ J.\ Theor.\ Phys.} {\bf 1974}, {\it 10}, 363.

\bibitem{Charmousis:2011ea}
  Charmousis, C.; Copeland, E.J.; Padilla, A.; Saffin, P.M.
  Self-tuning and the derivation of a class of scalar-tensor theories.
  {\it Phys.\ Rev.\ D} {\bf 2012}, {\it 85}, 104040.

\bibitem{stmodels}
  Mart\'{\i}n-Moruno, P.; Nunes, N.J.; Lobo, F.S.N.
  Horndeski theories self-tuning to a de Sitter vacuum.
 {\it Phys.\ Rev.\ D} {\bf 2015}, {\it 91}, 8, 084029.

\bibitem{Nojiri:2003rz}
  Nojiri, S.; Odintsov, S.D. 
  Where new gravitational physics comes from: M
Theory? \textit{Phys.\ Lett.\ B} {\bf 2003}, \textit{576}, 5.

\bibitem{QFT}
Parker, L.; Toms, D.J.
\textit{Quantum Field Theory in Curved Spacetime: Quantized Fields and Gravity}; Cambridge University Press: Oxford, UK, 2009.

\bibitem{Cembranos-effective}
Cembranos, J.A.R. 
Dark matter from $R^2$ gravity. 
{\it Phys. Rev. Lett.} {\bf 2009}, {\it 102}, 141301.

\bibitem{Bamba:2015uma} 
  Bamba, K.; Odintsov, S.D. 
  Inflationary cosmology in modified gravity theories.
  {\it Symmetry} {\bf 2015}, {\it 7}, 220.


\bibitem{Planck:2013jfk}
Ade, P.A.R.; Aghanim, A.N.; Armitage-Caplan, C.; Ashdown, A.M.; Atrio-Barandela, F.; Aumont, J.; Baccigalupi, C.; Banday, A.J.; Barreiro, R.B.; Bartlett, J.G.; et al. Planck 2013 results. XXII. Constraints on inflation.
\textit{Astron. Astrophys.} {\bf 2014}, \textit{571}, A22.

\bibitem{Hu:2007nk}
Hu, W.; Sawicki, I. 
Models of f(R) Cosmic Acceleration that Evade Solar-System Tests.
\textit{Phys.\ Rev.\ D} {\bf 2007}, \textit{76}, 064004.

\bibitem{Nojiri:2007as}
Nojiri, S.; Odintsov, S.D. 
Unifying inflation with LambdaCDM epoch in modified f(R) gravity consistent with Solar System tests. \textit{Phys.\ Lett.\ B} {\bf 2007}, \textit{657}, 238.

\bibitem{Olmo}
Olmo, G.J. 
Palatini Approach to Modified Gravity: $f(R)$ Theories and Beyond.
  {\it Int.\ J.\ Mod.\ Phys.\ D} {\bf 2011}, {\it 20}, 413.

\bibitem{Sotiriou:2006qn} 
  Sotiriou, T.P.; Liberati, S.
  Metric-affine $f(R)$ theories of gravity.
  {\it Annals Phys.} {\bf 2007}, {\it 322}, 935.

\bibitem{Harko:2011nh}
  Harko, T.; Koivisto, T.S.; Lobo, F.S.N.; Olmo, G.J.
  Metric-Palatini gravity unifying local constraints and late-time cosmic acceleration.
  {\it Phys.\ Rev.\ D} {\bf 2012}, {\it 85}, 084016.

\bibitem{Linder:2010py} 
  Linder, E.V.
  Einstein's Other Gravity and the Acceleration of the Universe.
  {\it Phys.\ Rev.\ D} {\bf 2010}, {\it 81}, 127301.

\bibitem{Cai:2015emx} 
  Cai, Y.F.; Capozziello, S.; De Laurentis, M.; Saridakis, E.N.
  $f(T)$ Teleparallel Gravity and Cosmology. {\bf 2015},
  arXiv:1511.07586 [gr-qc].

\bibitem{Harko:2014sja} 
  Harko, T.; Lobo, F.S.N.;Otalora, G.; Saridakis, E.N.
  Nonminimal torsion-matter coupling extension of $f(T)$ gravity.
  {\it Phys.\ Rev.\ D} {\bf 2014}, {\it 89}, 124036.

\bibitem{Harko:2014aja} 
 Harko, T.; Lobo, F.S.N.;Otalora, G.; Saridakis, E.N.
  $f(T,\mathcal{T})$ gravity and cosmology.
  {\it J. Cosmol. Astropart. Phys.} {\bf 2014}, {\it 1412}, 021.


\bibitem{vanDam:1970vg} 
  van Dam, H.; Veltman, M.J.G.
  Massive and massless Yang-Mills and gravitational fields.
  {\it Nucl.\ Phys.\ B} {\bf 1970}, {\it 22}, 397.

\bibitem{Zakharov:1970cc} 
  Zakharov, V.~I.
  Linearized gravitation theory and the graviton mass.
  {\it JETP Lett.} {\bf 1970}, {\it 12}, 312.


\bibitem{PinaAvelino:2006ia}
  Avelino, P.P.; Martins, C.J.A.P.; Menezes, J.; Menezes, R.; Oliveira, J.C.R.E.
  Frustrated expectations: defect networks and dark energy.
  {\it Phys.\ Rev.\ D} {\bf 2006}, {\it 73}, 123519.


\bibitem{Ade:2015xua} 
  Ade, P.A.R.; Aghanim, N.; Arnaud, M.; Ashdown, M.; Aumont, J.; Baccigalupi, C.; Banday, A.J.; R.B. Barreiro, R.B.; J.G. Bartlett, J.G.; N. Bartolo, N.; {et al.} [Planck Collaboration].
  Planck 2015 results. XIII. Cosmological parameters. {\bf 2015},
 arXiv:1502.01589 [astro-ph.CO]. 
  
\bibitem{Avelino:2014xsa}
  Avelino, P.P.; Sousa, L.
  Observational Constraints on Varying-alpha Domain Walls.
  {\it Universe} {\bf 2015}, {\it 1}, 6--16.
    
  
\bibitem{Anthonisen:2015tda}
  Anthonisen, M.; Brandenberger, R.; Scott, P.
 Constraints on cosmic strings from ultracompact minihalos.
  {\it Phys.\ Rev.\ D} {\bf 2015}, {\it 92}, 023521.
  
\bibitem{Barton:2015zra}
  Barton, A.; Brandenberger, R.H.; Lin, L.
  Cosmic Strings and the Origin of Globular Clusters.
  {\it J. Cosmol. Astropart. Phys.} {\bf 2015}, {\it 1506}, 022.
  
\bibitem{Bramberger:2015kua}
  Bramberger, S.F.; Brandenberger, R.H.; Jreidini, P.; Quintin, J.
  Cosmic String Loops as the Seeds of Super-Massive Black Holes.
  {\it J. Cosmol. Astropart. Phys.} {\bf 2015}, {\it 1506}, 007.
  
\bibitem{Avelino:2003nn}
  Avelino, P.P.; Liddle, A.R.
  Cosmological perturbations and the reionization epoch.
  {\textit{Mon.\ Not.\ Roy.\ Astron.\ Soc.}} {\bf 2004}, {\it 348}, 105.
  
\bibitem{weinberg:probes} 
Weinberg, D.H.; Mortonson, M.J.; Eisenstein, D.J.; Hirata, C.; Riess, A.G.; Rozo, E. 
Observational probes of cosmic acceleration. {\em Phys. Rep.} {\bf 2013}, {\em 530}, 87--255.

\bibitem[Planck Collaboration et al.(2015)]{2015arXiv150201589P} 
Ade, P.A.R.; Aghanim, N.; Arnaud, M.; Ashdown, M.; Aumont, J.; Baccigalupi, C.; Barreiro, R.B.; Bartlett, J.G.; Bartolo, E.; Battaner, R.; et al.
Planck Collaboration, 
Planck 2015 results. XIII. Cosmological parameters.\ 2015, arXiv:1502.01589.

\bibitem[Aghanim et al.(2008)]{2008RPPh...71f6902A} 
Aghanim, N.; Majumdar, S.; Silk, J. 
Secondary anisotropies of the CMB. 
{\it Rep. Prog. Phys.} {\bf 2008}, {\it 71}, 066902.

\bibitem[Sunyaev \& Zeldovich(1972)]{1972CoASP...4..173S} 
Sunyaev, R.A.; Zeldovich, Y.B. The Sunyaev-Zel'dovich effect.
{\it Comments Astrophys. Space Phys.} {\bf 1972} {\it 4}, 173.

\bibitem[Birkinshaw(1999)]{1999PhR...310...97B} 
Birkinshaw, M. 
The Sunyaev-Zel'dovich effect. {\it Phys. Rep.} {\bf 1999}, {\it 310}, 97. 
   
  \bibitem{bands} 
  Bartelmann, M.; Schneider, P. 
Weak Gravitational Lensing. {\em Phys. Rep.} {\bf 2001}, {\em 340}, 297--472.

  \bibitem{redbook} 
  Laureijs, R.; Amiaux, J.; Arduini, S.; Brinchmann, J.; Cole, R.; Cropper, M.; Dabin, C.; Duvet, L.; Ealet, A.; Garilli, B.; et al. Euclid Definition Study Report. {\em arXiv preprint} {\bf 2011}, arXiv:1110.3193.

\bibitem{nbodymodgrav} 
Winther, H.A.; Schmidt, F.; Barreira, A.; Arnold, C.; Sownak Bose, S.; Claudio Llinares, C.; Marco Baldi, M.; Falck, B.; Hellwing, W. A.; Koyama, K.; {et al. }
Modified gravity N-body code comparison project. {\em Mon. Not. Roy. Astron. Soc.} {\bf 2015}, {\em 454}, 4208--4234.

\bibitem{semboloni} 
Semboloni, E.; Hoekstra, H.; Schaye, J.; van Daalen, M.~P.; McCarthy, I.~J.
Quantifying the effect of baryon physics on weak lensing tomography. {\em Mon. Not. Roy. Astron. Soc.} {\bf 2011}, {\em 417}, 2020--2035.

\bibitem{bulloslo} 
Bull, P.; Akrami, Y.; Adamek, J.; Baker, T.; Bellini, E.; Beltran-Jimenez, J.; Bentivegna, E.; Camera, S.; Clesse, S.; Davis, J. H.
Beyond {$\Lambda$} CDM: Problems, solutions, and the road ahead. {\em Phys. Dark Universe} {\bf 2016}, {\em 12}, 56--99.

\bibitem{joudaki2016} 
Joudaki, S.; Blake, C.; Heymans, C.; Choi, A.; Harnois-Deraps, J.; Hildebrandt, H.; Joachimi, B.; Johnson, A.; Mead, A.; Parkinson, D.; {et al.}
CFHTLenS revisited: Assessing concordance with Planck including astrophysical systematics. 
{\em arXiv preprint} {\bf 2016}, arXiv: 1601.05786.

\bibitem{anomalies} 
Schwarz, D.; Copi, C.; Huterer, D.; Starkman, G. 
CMB Anomalies after Planck. {\em arXiv preprint} {\bf 2015}, arXiv: 1510.07929.

\bibitem[Acquaviva et al.(2003)]{2003NuPhB.667..119A} 
Acquaviva, V.; Bartolo, N.; Matarrese, S.; Riotto, A. 
Gauge-invariant second-order perturbations and non-Gaussianity from inflation, {\em Nucl. Phys. B} {\bf 2003}, {\em 667}, 119.

\bibitem[Maldacena(2003)]{2003JHEP...05..013M} 
Maldacena, J. 
Non-gaussian features of primordial fluctuations in single field inflationary models, {\em J. High Energy Phys.} {\bf 2003}, {\em 5}, 013.

\bibitem[Bartolo et al.(2004)]{2004PhR...402..103B} 
Bartolo, N.; Komatsu, E.; Matarrese, S.; Riotto, A. 	
Non-Gaussianity from inflation: theory and observations. {\em Phys. Rep.} {\bf 2003}, {\em 402}, 103.

\bibitem[Komatsu et al.(2009)]{2009astro2010S.158K} 
Komatsu, E.; Afshordi, N.; Bartolo, N.; Baumann, D.; Bond, J.~R.; Buchbinder, E.~I.; Byrnes, C.~T.; Chen, X.; Chung, D.~J.~H.; Cooray, A.; {et al.} 
Non-Gaussianity as a Probe of the Physics of the Primordial Universe and the Astrophysics of the Low Redshift Universe. 
{\em Astron. Astrophys. Decadal Surv.} {\bf 2003}, arXiv:0902.4759 [astro-ph.CO].

\bibitem[Lyth \& Wands(2002)]{2002PhLB..524....5L} 
Lyth, D.H.; Wands, D. 
Generating the curvature perturbation without an inflaton. 
{\em Phys. Lett. B} {\bf 2002}, {\em 524}, 5.

\bibitem[Bartolo et al.(2004)]{2004PhRvD..69d3503B} 
Bartolo, N.; Matarrese, S.; Riotto, A. 
Non-Gaussianity in the curvaton scenario. 
{\em Phys. Rev. D} {\bf 2004}, {\em 69}, 043503. 

\bibitem[Sasaki et al.(2006)]{2006PhRvD..74j3003S} 
Sasaki, M.; V{\"a}liviita, J.; Wands, D. 
Non-Gaussianity of the primordial perturbation in the curvaton model. {\it Phys. Rev. D} {\bf 2006}, {\it 74}, 103003. 

\bibitem[Lehners \& Steinhardt(2008)]{2008PhRvD..77f3533L} 
Lehners, J.-L.; Steinhardt, P.J. 
Non-Gaussian density fluctuations from entropically generated curvature perturbations in ekpyrotic models. {\em Phys. Rev. D} {\bf 2008}, {\em 77}, 063533.

\bibitem[Lehners(2010)]{2010AdAst2010E..67L} 
Lehners, J.-L. 
Ekpyrotic Nongaussianity: A Review. {\em Adv. Astron.} {\bf 2010}, 903907.

\bibitem[Yokoyama \& Soda(2008)]{2008JCAP...08..005Y} 
Yokoyama, S.; Soda, J. 
Primordial statistical anisotropy generated at the end of inflation. {\em JCAP}, {\bf 2008}, {\em 8}, 005.

\bibitem[Kar{\v c}iauskas et al.(2009)]{2009PhRvD..80b3509K} 
Kar{\v c}iauskas, M.; Dimopoulos, K.; Lyth, D.H. 
Anisotropic non-Gaussianity from vector field perturbations. {\em Phys. Rev. D} {\bf 2009}, {\em 80}, 023509.

\bibitem[Dimastrogiovanni et al.(2010)]{2010AdAst2010E..65D} 
Dimastrogiovanni, E.; Bartolo, N.; Matarrese, S.; Riotto, A. 
Non-Gaussianity and Statistical Anisotropy from Vector Field Populated Inflationary Models. {\em Adv. Astron.} {\bf 2010}, 752670, doi:10.1155/2010/752670.

\bibitem[Polarski \& Starobinsky(1994)]{1994PhRvD..50.6123P} 
Polarski, D.; Starobinsky, A.A. 
Isocurvature perturbations in multiple inflationary models. {\em Phys. Rev. D} {\bf 1994}, {\em 50}, 6123.

\bibitem[Mukhanov \& Steinhardt(1998)]{1998PhLB..422...52M} 
Mukhanov, V.F.; Steinhardt, P.J. 	
Density perturbations in multifield inflationary models. {\em Phys. Lett. B} {\bf 1998}, {\em 422}, 52.

\bibitem[Rigopoulos et al.(2006)]{2006PhRvD..73h3522R} 
Rigopoulos, G.I.; Shellard, E.P.S.; van Tent, B.J.W.
Large non-Gaussianity in multiple-field inflation. {\em Phys. Rev. D} {\bf 2006}, {\em 73}, 083522.

\bibitem[Byrnes \& Choi(2010)]{2010AdAst2010E..76B} 
Byrnes, C.T.; Choi, K.-Y. 	
Review of Local Non-Gaussianity from Multifield Inflation. {\em Adv. Astron.} {\bf 2010}, 724525.


\bibitem{carvalho_2015}
Carvalho C.S.; Marques K. 
Angular distribution of cosmological parameters: measurement of inhomogeneities from type Ia supernovae. 
{\bf 2015}, arXiv: 1512.07869 [astro-ph/CO].


\bibitem{Billyard:1997yb}
  Billyard,~A.; Coley,~A. 
  On the correspondence between exact solutions in Kaluza-Klein theory and in scalar tensor theories. {\em Mod. Phys. Lett. A} {\bf 1997}, {\it 12}, 2121.

\bibitem{Barrow:1988yia}
  Barrow, ~J.D.; Tipler,~F.J. 
  \textit{The Anthropic Cosmological Principle}; 
  Cambridge U. Press: Cambridge, UK, 1988.

\bibitem{Wands:1993uu}
 Wands,~D.
  Extended gravity theories and the Einstein-Hilbert action.
  {\em Class.\ Quant.\ Grav.} {\bf 1994}, {\it 11}, 269.


 \bibitem{Conf_equiv-fR}
  Barrow,~J.D; Cotsakis,~S.
  Inflation and the Conformal Structure of Higher Order Gravity Theories.
  {\em Phys.\ Lett.\ B},  {\bf 1988} {214}, 515.
 
\bibitem{Conf_equiv-2}
  Kalara,~S.; Kaloper,~ N.; Olive,~K.A.
  Theories of Inflation and Conformal Transformations.
  {\em Nucl.\ Phys.\ B} {\bf 1990}, {\it 341}, 252.

\bibitem{Malquarti:2003nn}
  Malquarti,~M.; Copeland, E.J.; Liddle, A.R.; Trodden, M.
 A New view of k-essence.
  {\em Phys.\ Rev.\ D} {\bf 2003}, {\it 67}, 123503.


\bibitem{Bergm 68}
Bergmann,~P.G.
Comments on the scalar tensor theory.
{\it Int. J. Theor. Phys.} {\bf 1968}, {\it 1}, 25.

\bibitem{Wagoner 70}
Wagoner,~R.V. 
Scalar tensor theory and gravitational waves.
{\sl Phys. Rev. D} {\bf 1970}, 1, 3209.

\bibitem{Nordt 70}
Nordtvedt,~N.
PostNewtonian metric for a general class of scalar tensor gravitational theories and observational consequences.
{\sl Astrophys. J.} {\bf 1970}, {\it 161}, 1059.

\bibitem{Will:2014xja}
Will,~C.M.
The Confrontation between General Relativity and Experiment.
 {\sl Living Rev.\ Rel.} {\bf 2014}, {\it 17}, 4.

\bibitem{Wald:1984rg}
 Wald,  R.M.,
 \textit{General Relativity}; {University of Chicago Press: Chicago, IL, USA}, 1984; p. 491.

\bibitem{Damour+Nordtvedt 93}
Damour,~T.; Nordtvedt,~ K. 
Tensor-scalar cosmological models and their relaxation toward general relativity. {\it Phys. Rev. D} {\bf 1993}, {\it 48}, 3436.

\bibitem{Dam+E-Farese 92} Damour,~T.; Esposito-Far\`ese,~ G.
Tensor multiscalar theories of gravitation.
{\it Class. Quantum Grav.} {\bf 1992}, {\it 9}, 2093.


\bibitem{Thorne:1970wv}
  Thorne, K.S.; Will, C.M.~
 Theoretical Frameworks for Testing Relativistic Gravity. I. Foundations.
  {\sl Astrophys.\ J.} {\bf 1971}, {163}, 595.

\bibitem{Mimoso:1998dn}
  Mimoso,~J.P.; Nunes, A.M.
 General relativity as a cosmological attractor of scalar tensor gravity theories.
  {\sl Phys.\ Lett.\ A} {\bf 1998}, {\it{248}}, 325.

  \bibitem{Omeg_limit2}
  Banerjee,~N.; Sen,~S.
  Does Brans-Dicke theory always yield general relativity in the infinite omega limit? \textit{Phys.\ Rev.\  D} {\bf 1997}, {\it 56}, 1334.

\bibitem{Billyard:1998kg}
 Billyard,~A.; Coley,~A.; Ibanez,~J.
 On the asymptotic behavior of cosmological models in scalar tensor theories of gravity.
  {\sl Phys.\ Rev.\ D} {\bf 1999}, {\it 59}, 023507.

 \bibitem{2003Ap&SS.283.661M}
 Mimoso, J.P.; Nunes, A.
A Qualitative Analysis of the Attractor Mechanism of General relativity.
  {\sl Astrophys.\ Space Sci.} {\bf 2003}, {\it 283},  661.


\bibitem{Faraoni:1999hp}
Faraoni,~V.; Gunzig,~E. 
Einstein frame or Jordan frame? {\sl Int.\ J.\ Theor.\ Phys.} {\bf 1999}, {\it 38}, 217.

\bibitem{Flanagan:2004bz}
 Flanagan,~E.E. 
 The Conformal frame freedom in theories of gravitation. 
 {\sl Class.\ Quant.\ Grav.} {\bf 2004}, {\it 21}, 3817.

\bibitem{Olmo:2006zu}
Olmo,~G.J. 
Violation of the Equivalence Principle in Modified Theories of Gravity. {\sl Phys.\ Rev.\ Lett.} {\bf 2007}, {\it 98}, 061101.


\bibitem{Capozziello:1996xg} 
  Capozziello, S.; de Ritis, R.; Marino, A.A.
  Some aspects of the cosmological conformal equivalence between `Jordan frame' and `Einstein frame'.
  {\it Class.\ Quant.\ Grav.} {\bf 1997}, {\it 14}, 3243.


\bibitem{CLM1} 
  Capozziello, S.; Lobo, F.S.N.; Mimoso, J.P.
  Energy conditions in modified gravity.
{\it Phys.\ Lett.\ B} {\bf 2014}, {\it 730}, 280.

\bibitem{Capozziello:2014bqa} 
  Capozziello, S.; Lobo, F.S.N.; Mimoso, J.P.
  Generalized energy conditions in Extended Theories of Gravity.
  {\it Phys.\ Rev.\ D} {\bf 2015}, {\it 91}, 124019.

\bibitem{Dicke:1961gz} 
  Dicke,  R.H.
 Mach's principle and invariance under transformation of units.
  {\it Phys.\ Rev.} {\bf 1962} {\it 125}, 2163.







\bibitem[\protect\citeauthoryear{Nariai}{1968}]{Nariai 68} 
Nariai, H.
On the Green's function in an expanding universe and its role in the problem of Mach's principle.
 {\sl Progr. Theor. Phys.} {\bf 1968}, {\it 40}, 49.

\bibitem[\protect\citeauthoryear{O'Hanlon and Tupper}{1972}]
{O'Hanlon+Tupper 72} 
O'Hanlon, J.; Tupper, B.O.J. 
Vacuum-field solutions in the Brans-Dicke theory. 
{\sl Il Nuovo Cimento} {\bf 1972}, {\it 7}, 305;


\bibitem{Barrow+Mimoso 94}
Barrow,~J.D.; Mimoso,~J.P. 
Perfect fluid scalar-tensor cosmologies. 
{\sl Phys. Rev. D} {\bf 1994}, {\it 50}, 3746.

\bibitem{GeneralSTsols}
Barrow,~J.D.
Scalar-tensor cosmologies.
{\sl  Phys. Rev. D} {\bf 1993}, {\it 47}, 5329.

\bibitem{GeneralSTsols2}
  Mimoso,~J.P.; Wands,~D.
  Massless fields in scalar---Tensor cosmologies.
  {\sl Phys.\ Rev.\ D} {\bf 1995}, {\it 51}, 477.


\bibitem{Nunes:2000yc}
Nunes,~A.;  Mimoso, ~J.P.
 On the potentials yielding cosmological scaling solutions.
  {\sl Phys.\ Lett.\ B} {\bf 2000}, {\it 488}, 423.

\bibitem{Charters:2001hi}
 Charters, T.C.; Nunes,~A.; Mimoso,~J.P. 
 Stability analysis of cosmological models through Liapunov's method, 
 {\sl Class.\ Quant.\ Grav.} {\bf 2001}, {\it 18}, 1703.

\bibitem{Mimoso:1999ai}
Mimoso,~J.P.; Nunes,~A. 
General relativity as an attractor to scalar tensor gravity theories. 
{\sl Astrophys.\ Space Sci.} {\bf 1999}, {\it 261}, 327.
 

\bibitem{Nunes:2000ka}
Nunes,~A.; Mimoso,~J.P.; Charters,~T.C. 
Scaling solutions from interacting fluids. {\sl Phys.\ Rev.\ D} {\bf 2001} {\it 63}, 083506.

\bibitem{DynSis_BD1}
  Santos,~C.; Gregory,~R.
 Cosmology in Brans-Dicke theory with a scalar potential.
 {\sl Ann. Phys.} {\bf 1997}, {\it 258}, 111.

\bibitem{DynSis_BD2}
 Holden, ~D.J.; Wands,~D. 
 Phase-plane analysis of Friedmann-Robertson-Walker cosmologies in Brans-Dicke gravity. {\sl Class.\ Quant.\ Grav.} {\bf 1998}, {\it 15}, 3271.

\bibitem{DynSis_BD3}
 Carloni,~S.; Dunsby,~P.K.S.; Capozziello,~S.; Troisi,~A. 
 Cosmological dynamics of {$R^n$} gravity. {\sl Class.\ Quant.\ Grav.} {\bf 2005}, {\it 22}, 4839.
 
\bibitem{DynSis_GenSTT}
 Kuusk,~P.; Jarv,~L.; Saal,~M.
 Scalar-tensor cosmologies: General relativity as a fixed point of the Jordan frame scalar field. {\sl Int.\ J.\ Mod.\ Phys.\ A} {\bf 2009}, {\it 24}, 1631.
  
\bibitem{DynSis_GenSTT2}
 Jarv,~L.; Kuusk,~P.; Saal,~M. 
 Potential dominated scalar-tensor cosmologies in the general relativity limit: Phase space view. {\sl Phys.\ Rev.\ D} {\bf 2010}, {\it 81}, 104007.

\bibitem{DynSis_GenSTT3}
  Faraoni,~V.
  Phase space geometry in scalar-tensor cosmology. {\sl Ann. Phys.} {\bf 2005},  {\it 317}, 366.
 
 \bibitem{DynSis_GenSTT4}
  Tsujikawa,~S. 
  Modified gravity models of dark energy. {\sl Lect.\ Notes Phys.},  {\bf 2010}, {\it 800}, 99.

\bibitem{Amendola 99} Amendola,~L.
Scaling solutions in general nonminimal coupling theories. {\sl Phys. Rev. D} {\bf 1999}, {\it 60}, 043501.

\bibitem[\protect\citeauthoryear{Barrow and Maeda}{1990}]
{Barrow+Maeda 90} 
Barrow, J.D.; Maeda, K. 
Extended inflationary universes. {\sl Nucl. Phys.} {\bf 1990}, {\it B341}, 294.


\bibitem{Uzan 99} Uzan,~J.P. 
Cosmological scaling solutions of nonminimally coupled scalar fields. {\sl Phys. Rev. D} {\bf 1999}, {\it 59}, 123510.


\bibitem{Liddle:1998jc}
Liddle, A.R.; Mazumdar, A; Schunck, F.E.
  Assisted inflation.
  {\it Phys.\ Rev.\ D} {\bf 1998}, {\em 58}, 061301.
  
 \bibitem{Malik:1998gy}
  Malik, K.A.; Wands, D. 
  Dynamics of assisted inflation.
  {\it Phys.\ Rev.\ D} {\bf 1999}, {\em 59}, 123501.
 
\bibitem{Copeland:1999cs}
  Copeland, E.J.; Mazumdar, A.; Nunes, N.J.~
  Generalized assisted inflation.
  {\it Phys.\ Rev.\ D} {\bf 1999}, {\em 60}, 083506.

  \bibitem{Kim:2005ne}
  Kim, S.A.; Liddle, A.R.; Tsujikawa, S. 
  Dynamics of assisted quintessence.
  {\it Phys.\ Rev.\ D} {\bf 2005}, {\em 72}, 043506.

\bibitem{Tsujikawa:2006mw}
Tsujikawa, S.~
  General analytic formulae for attractor solutions of scalar-field dark energy models and their multi-field generalizations.
  {\it Phys.\ Rev.\ D} {\bf 2006}, {\em 73}, 103504.

\bibitem{Ohashi:2009xw}
  Ohashi, J.; Tsujikawa, S.~
 Assisted dark energy.
  {\it Phys.\ Rev.\ D} {\bf 2009}, {\em 80}, 103513.
 
\bibitem{Karwan:2010xw}
 Karwan, K.~
  Dynamics of entropy perturbations in assisted dark energy with mixed kinetic terms.
  {\it J. Cosmol. Astropart. Phys.} {\bf 2011}, {\em 1102}, 007.
  
  \bibitem{Zlatev:1998tr}
Zlatev, I.; Wang, L.M.; Steinhardt, P.J.
  Quintessence, cosmic coincidence, and the cosmological constant.
  {\it Phys.\ Rev.\ Lett.} {\bf 1999}, {\em 82}, 896.

\bibitem{Amendola:1999dr}
  Amendola, L. 
  Perturbations in a coupled scalar field cosmology.
  {\it Mon.\ Not.\ Roy.\ Astron.\ Soc.}  {\bf 2000}, {\em 312}, 521.
  
\bibitem{Holden:1999hm}
  Holden, D.J.; Wands, D. 
  Selfsimilar cosmological solutions with a nonminimally coupled scalar field.
  {\it Phys.\ Rev.\ D} {\bf 2000}, {\em 61}, 043506.
  
\bibitem{Amendola:1999er}
  Amendola, L. 
  Coupled quintessence.
  {\it Phys.\ Rev.\ D} {\bf 2000}, {\em 62}, 043511.

\bibitem{Brookfield:2007au}
  Brookfield, A.W.; van de Bruck, C.; Hall, L.M.H.
  New interactions in the dark sector mediated by dark energy.
  {\it Phys.\ Rev.\ D} {\bf 2008}, {\em 77}, 043006.
  
\bibitem{Baldi:2012kt}
  Baldi, M. 
  Multiple Dark Matter as a self-regulating mechanism for dark sector interactions.
  {\it Ann. Phys.} {\bf 2012}, {\em 524}, 602.

\bibitem{Amendola:2014kwa}
  Amendola, L.; Barreiro, T.; Nunes, N.J.~
  Multifield coupled quintessence.
  {\it Phys.\ Rev.\ D} {\bf 2014}, {\em 90} , 8,  083508.

  \bibitem{Piloyan:2013mla}
  Piloyan, A.; Marra, V.; Baldi, M.; Amendola, L. 
  Supernova constraints on Multi-coupled Dark Energy.
  {\it J. Cosmol. Astropart. Phys.} {\bf 2013}, {\em 1307}, 042.
  
\bibitem{Piloyan:2014gta}
Piloyan, A.; Marra, V.; Baldi, M.; Amendola, L. 
  Linear Perturbation constraints on Multi-coupled Dark Energy.
  {\it J. Cosmol. Astropart. Phys.} {\bf 2014}, {\em 1402}, 045.

\bibitem{Baldi:2014tja}
Baldi, M. 
  Cold dark matter halos in Multi-coupled Dark Energy cosmologies: Structural and statistical properties.
  {\it Phys.\ Dark Univ.} {\bf 2014}, {\em 3}, 4.

\bibitem{ArmendarizPicon:2000ah}
  Armendariz-Picon, C.; Mukhanov, V.F.; and Steinhardt, P.J.
  Essentials of k essence.
  {\it Phys.\ Rev.\ D} {\bf 2001}, {\it 63}, 103510.


\bibitem{Deffayet:2010qz}
  Deffayet, C.; Pujolas, O.; Sawicki, I.; Vikman, A.
  Imperfect Dark Energy from Kinetic Gravity Braiding.
  {\it J. Cosmol. Astropart. Phys.} {\bf 2010}, {\it 1010}, 026.

\bibitem{Woodard:2006nt}
  Woodard, R.P. 
  Avoiding dark energy with $1/r$ modifications of gravity.
  {\it Lect.\ Notes Phys.} {\bf 2007} {\it 720}, 403.

\bibitem{Deffayet:2011gz}
  Deffayet, C.; Gao, X.; Steer, D.A.; Zahariade, G.
  From k-essence to generalised Galileons.
  {\it Phys.\ Rev.\ D} {\bf 2011}, {\it 84}, 064039.

\bibitem{Deffayet:2009wt}
  Deffayet, C.; Esposito-Farese, G.; Vikman, A.
  Covariant Galileon.
  {\it Phys.\ Rev.\ D} {\bf 2009}, {\it 79}, 084003.

\bibitem{Nicolis:2008in}
  Nicolis, A.; Rattazzi, R.; Trincherini, E.
  The Galileon as a local modification of gravity.
  {\it Phys.\ Rev.\ D} {\bf 2009}, {\it 79}, 064036.

\bibitem{Zumalacarregui:2013pma}
  Zumalac\'arregui, M.; Garc\'{\i}a-Bellido, J.
  Transforming gravity: from derivative couplings to matter to second-order scalar-tensor theories beyond the Horndeski Lagrangian.
  {\it Phys.\ Rev.\ D} {\bf 2014}, {\it 89}, 064046.

\bibitem{Gleyzes:2014dya}
  Gleyzes, J.; Langlois, D.; Piazza, F.; Vernizzi, F.
  Healthy theories beyond Horndeski.
  {\it Phys.\ Rev.\ Lett.} {\bf 2015}, {\it 114}, 211101.

\bibitem{Gleyzes:2014qga}
  Gleyzes, J.; Langlois, D.; Piazza, F.; Vernizzi, F.
  Exploring gravitational theories beyond Horndeski.
  {\it J. Cosmol. Astropart. Phys.} {\bf 2015}, {\it 1502}, 018.

\bibitem{Weinberg:1988cp}
  Weinberg, S.
  The Cosmological Constant Problem.
  {\it Rev.\ Mod.\ Phys.} {\bf 1989}, {\bf 61}, 1.

\bibitem{Carroll:2000fy}
  Carroll, S.M.
  The Cosmological constant.
  {\it Living Rev.\ Rel.} {\bf 2001}, {\it 4}, 1.


\bibitem{Kaloper:2014dqa}
  Kaloper, M.; Padilla, A.
  Vacuum Energy Sequestering: The Framework and Its Cosmological Consequences.
  {\it Phys.\ Rev.\ D} {\bf 2014}, {\it 90},  084023.

\bibitem{Charmousis:2011bf}
  Charmousis, C.; Copeland, E.J.; Padilla, A.; Saffin, P.M.
  General second order scalar-tensor theory, self tuning, and the Fab Four.
  {\it Phys.\ Rev.\ Lett.} {\bf 2012}, {\it 108}, 051101.

\bibitem{attracted1}
  Mart\'{\i}n-Moruno, P.; Nunes, N.J.;  Lobo, F.S.N.
  Attracted to de Sitter: Cosmology of the linear Horndeski models.
 {\it J. Cosmol. Astropart. Phys.} {\bf 2015}, {\it 1505}, 05, 033.

\bibitem{attracted2}
Mart\'{\i}n-Moruno, P.; Nunes, N.J.
Attracted to de Sitter II: cosmology of the shift-symmetric Horndeski models.
  {\it J. Cosmol. Astropart. Phys.} {\bf 2015}, {\it 1509}, 056.


\bibitem{Theorems}
Penrose, R. Gravitational collapse and space-time singularities. 
{\it Phys. Rev. Lett} {\bf 1965}, {\it 14}, 57;

\bibitem{Theorems2}
Penrose, R. 
Gravitational collapse: The role of general relativity. 
{\it Riv. Nuovo Cim. Numero Speciale} {\bf 1969}, {\it 1}, 252.

\bibitem{Theorems3}
Hawking, S.W. 
Singularities in the universe. 
{\it Phys. Rev. Lett} {\bf 1966}, {\it 17}, 444.

\bibitem{SUSY}
Jungman, G.; Kamionkowski, M.; Griest, K. 
Supersymmetric dark matter.
  {\it Phys.\ Rept.} {\bf 1996}, {\it 267}, 195.

\bibitem{ST} 
Green, M.; Schwarz, J.; Witten, E. 
\textit{Superstring Theory}; Cambridge University Press: Cambridge, UK, 1987.

\bibitem{LQG}
Thiemann, T. 
Lectures on loop quantum gravity.  
{\it Lect.\ Notes Phys.} {\bf 2003}, {\it 631}, 41.


\bibitem{QFT2}
Birrell, N.D.; Davies, P.C.W. 
{\it Quantum Fields in Curved Space}; Cambridge University Press: Cambridge, UK, 1982.


\bibitem{first}  
Buchdahl, H.A.
  Non-linear Lagrangians and cosmological theory.
  {\it Mon.\ Not.\ Roy.\ Astron.\ Soc.} {\bf 1970}, {\it 150}, 1.

\bibitem{first2}
Barrow, J.D.; Ottewill, A.C.
  The Stability of General Relativistic Cosmological Theory.
  {\it J.\ Phys.\ A} {\bf 1983}, {\it 16}, 2757.


\bibitem{Palatini}
Ferraris, M.; Francaviglia, M.; Volovich, I. 
The Universality of vacuum Einstein equations with cosmological constant.
  {\it Class.\ Quant.\ Grav.} {\bf 1994}, {\it 11}, 1505.


\bibitem{Palatini2}
Vollick, D.N. 
  $1/R$ Curvature corrections as the source of the cosmological acceleration.
  {\it Phys.\ Rev.\ D} {\bf 2003}, {\it 68}, 063510.

\bibitem{Palatini3}
Meng, X.H.; Wang, P. 
Palatini formation of modified gravity with ln R terms.
  {\it Phys.\ Lett.\ B} {\bf 2004}, {\it 584}, 1.

\bibitem{Palatini4}
Poplawski, N.J. 
Interacting dark energy in f(R) gravity.
  {\it Phys.\ Rev.\ D} {\bf 2006}, {\it 74}, 084032.

\bibitem{Palatini5}
Li, B.; Chan, K.C.; Chu, M.C. Constraints on f(R) Cosmology in the Palatini Formalism.
  {\it Phys.\ Rev.\ D} {\bf 2007}, {\it 76}, 024002.

\bibitem{Palatini6}
Li, B.; Barrow, J.D.; Mota, D.F. 
The Cosmology of Ricci-Tensor-Squared gravity in the Palatini variational approach.
  {\it Phys.\ Rev.\ D} {\bf 2007}, {\it 76}, 104047.

\bibitem{Palatini7}
Iglesias, A.; Kaloper, N.; Padilla, A.; Park, M. 
How (Not) to Palatini.
  {\it Phys.\ Rev.\ D} {\bf 2007}, {\it 76}, 104001.


\bibitem{Amendola:2012ys} 
  Amendola, L.; Appleby, S.; Bacon, D.; Baker, T.; Baldi, M.; Bartolo, N.; Blanchard, A.; Bonvin, C.; Borgani, S.; Branchini, E.; et al. 
  Cosmology and fundamental physics with the Euclid satellite.
  {\it Living Rev.\ Rel.} {\bf 2013},  {\it 16}, 6.

\bibitem{Capozziello:2003tk}
  Capozziello, S.; Carloni, S.; Troisi, A. 
Quintessence without scalar fields.
  {\it Recent Res.\ Dev.\ Astron.\ Astrophys.} {\bf 2003}, {\it 1}, 625.

\bibitem{Sotiriou:2007yd}
  Sotiriou, T.P. 
  { Modified Actions for Gravity: Theory and Phenomenology.}  {\bf 2007},  
  arXiv:gr-qc/0710.4438.



\bibitem{Vikman:2004dc} 
  Vikman, A.
  Can dark energy evolve to the phantom?
  {\it Phys.\ Rev.\ D} {\bf 2005}, {\it 71}, 023515.

\bibitem{Cai:2009zp} 
  Cai, Y.F.; Saridakis, E.N.; Setare, M.R.; Xia, J.Q.
  Quintom Cosmology: Theoretical implications and observations.
  {\it Phys.\ Rept.} {\bf 2010},  {\it 493}, 1.


\bibitem{Bertolami:2007gv}
  Bertolami, O.; Boehmer, C.G.; Harko, T.; Lobo, F.S.N. 
Extra force in f(R) modified theories of gravity.
  {\it Phys.\ Rev.\ D} {\bf 2007}, {\it 75}, 104016.


\bibitem{Bertolami:2007vu}
Bertolami, O.; P\'aramos, J. 
Do f(R) theories matter?
  {\it Phys.\ Rev.\ D} {\bf 2008}, {\it 77}, 084018.

\bibitem{Bertolami:2008im}
Bertolami, O.; P\'{a}ramos, J. 
On the non-trivial gravitational coupling to matter.
  {\it Class.\ Quant.\ Grav.} {\bf 2008}, {\it 25}, 245017.


\bibitem{Faraoni:2007sn}
Faraoni, V.
A Viability criterion for modified gravity with an extra force.
  {\it Phys.\ Rev.\ D} {\bf 2007}, {\it 76}, 127501.

\bibitem{Odintsov}
Nojiri, S.; Odintsov, S.D. 
Gravity assisted dark energy dominance and cosmic acceleration.
  {\it Phys.\ Lett.\ B} {\bf 2004}, {\it 599}, 137.

\bibitem{Odintsov2}
Nojiri, S.; Odintsov, S.D. 
Dark energy and cosmic speed-up from consistent modified gravity.
  {\it PoS WC} {\bf 2004}, {\it 2004}, 024.

\bibitem{Odintsov3}
Allemandi, G.; Borowiec, A.; Francaviglia, M.; Odintsov, S.D.
Dark energy dominance and cosmic acceleration in first order formalism.
  {\it Phys.\ Rev.\ D} {\bf 2005}, {\it 72}, 063505.

\bibitem{Lambda}
Mukohyama, S.; Randall, L. 
A Dynamical approach to the cosmological constant.
  {\it Phys.\ Rev.\ Lett.} {\bf 2004}, {\it 92}, 211302.

\bibitem{Harko:2010hw} 
  Harko, T.; Koivisto, T.S.; Lobo, F.S.N.
  Palatini formulation of modified gravity with a nonminimal curvature-matter coupling.
  {\it Mod.\ Phys.\ Lett.\ A} {\bf 2011}, {\it 26}, 1467.

\bibitem{Koivisto}
Koivisto, T. 
Covariant conservation of energy momentum in modified gravities.
  {\it Class.\ Quant.\ Grav.} {\bf 2006}, {\it 23}, 4289.

\bibitem{Bertolami:2008zh} 
  Bertolami, O.; Paramos, J.; Harko, T.; Lobo, F.S.N. 
  Non-minimal curvature-matter couplings in modified gravity. { 2008},
  arXiv:0811.2876 [gr-qc].

\bibitem{Sotiriou:2008it}
  Sotiriou, T.P.; Faraoni, V. 
Modified gravity with R-matter couplings and (non-)geodesic motion.
  {\it Class.\ Quant.\ Grav.} {\bf 2008}, {\it 25}, 205002.

\bibitem{Schutz:1970my}
Schutz, B.F. 
Perfect Fluids in General Relativity: Velocity Potentials and a Variational Principle.
  {\it Phys.\ Rev.\ D} {\bf 1970}, {\it 2}, 2762.

\bibitem{Brown:1992kc}
Brown, J.D. 
Action functionals for relativistic perfect fluids.
  {\it Class.\ Quant.\ Grav.} {\bf 1993}, {\it 10}, 1579.

\bibitem{BLP}
  Bertolami, O.; Lobo, F.S.N.; P\'{a}ramos, J. 
  Non-minimum coupling of perfect fluids to curvature.
  {\it Phys.\ Rev.\ D} {\bf 2008}, {\it 78}, 064036.

\bibitem{HawkingEllis}
Hawking, S.W.; Ellis, G.F.R. \textit{The Large Scale Structure
of Spacetime}; Cambridge University Press: Cambridge, UK, 1973, 




\bibitem{Kroner}
Falk, J. Theory of elasticity of coherent inclusions by means of nonmetric geometry. {\it J. Elast.} {\bf 1981}, {\it 11}, 359.

\bibitem{Kroner2}
Kr\"oner, E.
The continuized crystal---A bridge between micro and macromechanics? {\it Z. Angew. Math. Mech.} {\bf 1986}, {\it 66}, T284--T292.

\bibitem{Kroner3}
Clayton, J.D. {\it Nonlinear Mechanics of Crystals}; Springer: Maryland, USA, 2011.

\bibitem{lor15}
Lobo, F.S.N.; Olmo, G.J.; Rubiera-Garcia, D. 
Crystal clear lessons on the microstructure of spacetime and modified gravity. {\it Phys. Rev. D} {\bf 2015}, {\it 91}, 124001.

\bibitem{lor152}
Olmo, G.J.; Rubiera-Garcia, D. 
The quantum, the geon, and the crystal. 
{\it Int. J. Mod. Phys. D} {\bf 2015} {\it 24}, 1542013.

\bibitem{Torsion} 
Olmo, G.J.; Rubiera-Garcia, D. 
Importance of torsion and invariant volumes in Palatini theories of gravity. {\it Phys. Rev. D} {\bf 2013}, {\it 88}, 084030.

\bibitem{or11}
Olmo, G.J.; Rubiera-Garcia, D. 
Palatini $f(R)$ black holes in nonlinear electrodynamics. {\it Phys. Rev. D} {\bf 2011}, {\it 84}, 124059.

\bibitem{Borunda}
Exirifard, Q.; Sheikh-Jabbari, M.M.
Lovelock gravity at the crossroads of Palatini and metric formulations. {\it Phys.\ Lett.\ B} {\bf 2008}, {\it 661}, 158.

\bibitem{Borunda2}
Borunda, M.; Janssen B.; Bastero-Gil, M. 
Palatini versus metric formulation in higher curvature gravity. {\it J. Cosmol. Astropart. Phys.} {\bf 2008}, {\it 2008}, 0811.

\bibitem{orbw}
Olmo, G.J.; Rubiera-Garcia, D. 
Brane-world and loop cosmology from a gravity-matter coupling perspective. 
{\it Phys. Lett. B} {\bf 2015}, {\it 740}, 73.

\bibitem{LQC}
Bojowald, M. 
Loop quantum cosmology. {\it Living Rev. Rel.} {\bf 2008}, {\it 11}, 4.

\bibitem{LQC2}
Ashtekar, A.; Singh, P. 
Loop Quantum Cosmology: A Status Report. 2011, {\it Class.\ Quant.\ Grav} {\bf 2011}, {\it 28}, 213001.

\bibitem{bws}
Maartens, R.; Koyama, K. 
Brane-World Gravity. {\it Living Rev.\ Rel} {\bf 2010}, {\it 13}, 5.

\bibitem{lqcs}
Ashtekar, A.; Pawlowski, T.; Singh, P. 
Quantum nature of the big bang. {\it Phys. Rev. Lett.} {\bf 2006}, {\it 96}, 141301.

\bibitem{bos}
Barragan, C.; Olmo, G.J.; Sanchis-Alepuz, H. 
Bouncing Cosmologies in Palatini $f(R)$ Gravity.
  {\it Phys.\ Rev.\ D} {\bf 2009}, {\it 80}, 024016.

\bibitem{orf}
Olmo, G.J.; Rubiera-Garcia, D. 
Nonsingular black holes in $f(R)$ theories. 
{\it Universe} {\bf 2015}, {\it 1}, 173.

\bibitem{orf2}
Bambi, C.; Cardenas-Avendano, A.; Olmo, G.J.; Rubiera-Garcia, D. 
Wormholes and nonsingular space-times in Palatini $f(R)$ gravity. 
{\it Phys. Rev. D} {\bf 2016}, {\it 93}, 064016.

\bibitem{bo}
Barragan, C.; Olmo, G.J.
  Isotropic and Anisotropic Bouncing Cosmologies in Palatini Gravity.
  {\it Phys.\ Rev.\ D} {\bf 2010}, {\it 82}, 084015.

\bibitem{DG}
Deser, S.; Gibbons, G.W. 
Born-Infeld-Einstein actions?. 
{\it Class. Quant. Grav} {\bf 1998}, {\it 15}, L35.

\bibitem{DG2}
Ba\~nados, M.; Ferreira, P.G. 
Eddington's theory of gravity and its progeny. 
{\it Phys. Rev. Lett} {\bf 2010}, {\it 105}, 011101.

\bibitem{jho}
Jim\'enez, J.B.; Heisenberg, L.; Olmo, G.J. 
Infrared lessons for ultraviolet gravity: the case of massive gravity and Born-Infeld.   {\it J. Cosmol. Astropart. Phys.} {\bf 2014}, {\it 1411}, 004.

\bibitem{or-sing}
Olmo, G.J.; Rubiera-Garcia, D.; Sanchez-Puente, A.  
Geodesic completeness in a wormhole spacetime with horizons. 
{\it Phys. Rev. D.} {\bf 2015}, {\it 92}, 044047.

\bibitem{or-sing2}
Bazeia, D.; Losano, L.; Olmo, G.J.; Rubiera-Garcia, D.; Sanchez-Puente, A. Classical resolution of black hole singularities in arbitrary dimension. 
{\it Phys. Rev. D} {\bf 2015}, {\it 92}, 044018.

\bibitem{or-sing3}
Olmo, G.J.; Rubiera-Garcia, D.; Sanchez-Puente, A. 
Classical resolution of black hole singularities via wormholes. 
{\it Eur. Phys. J. C} {\bf 2016}, {\it 76}, 143.

\bibitem{oor}
Odintsov, O.D.; Olmo, G.J.; Rubiera-Garcia, D. 
Born-Infeld gravity and its functional extensions. {\bf 2014}, \textit{90}, 044003.

\bibitem{Hawking}
Hawking, S. 
Gravitationally collapsed objects of very low mass. 
{\it Mon. Not. Roy. Astron. Soc.} {\bf 1971}, {\it 152}, 75.

\bibitem{lor13}
Lobo, F.S.N.; Olmo, O.J.; Rubiera-Garcia, D. 
Semiclassical geons as solitonic black hole remnants. 
{\it J. Cosmol. Astropart. Phys.} {\bf 2013}, \textit{1307}, 011.

\bibitem{Chen}
Chen, P.; Ong, Y.C.; Yeom, D.H.
  Black Hole Remnants and the Information Loss Paradox. {\bf 2014},
 arXiv:1412.8366 [gr-qc].


\bibitem{Capozziello:2012ny}
  Capozziello, S.; Harko, T.; Koivisto, T.S.; Lobo, F.S.N.; Olmo, G.J.
  Cosmology of hybrid metric-Palatini f(X)-gravity.
  {\it J. Cosmol. Astropart. Phys.} {\bf 2013}, {\it 1304}, 011.

\bibitem{Capozziello:2012hr}
Capozziello, S.; Harko, T.; Koivisto, T.S.; Lobo, F.S.N.; Olmo, G.J.
  Wormholes supported by hybrid metric-Palatini gravity.
  {\it Phys.\ Rev.\ D} {\bf 2012}, {\it 86}, 127504.

\bibitem{Capozziello:2012qt}
  Capozziello, S.; Harko, T.; Koivisto, T.S.; Lobo, F.S.N.; Olmo, G.J.
  The virial theorem and the dark matter problem in hybrid metric-Palatini gravity.
  {\it J. Cosmol. Astropart. Phys.} {\bf 2013}, {\it 1307}, 024.

\bibitem{Capozziello:2015lza} 
  Capozziello, S.; Harko, T.; Koivisto, T.S.; Lobo, F.S.N.; Olmo, G.J.
  Hybrid metric-Palatini gravity.
  {\it Universe} {\bf 2015}, {\it 1}, 199.



\bibitem{Mukhanov} 
Mukhanov, V. {\it Physical Foundations of Cosmology}; Cambridge University Press: Cambridge, UK, 2005.

\bibitem{Mukhanov2}
Liddle A.R.; Lyth D.H.  
{\it Cosmological Inflation and Large-Scale Structure}; Cambridge University Press: Cambridge, UK, 2000.

\bibitem{Mukhanov3}
Dodelson, S. 
{\it Modern Cosmology}; Academic Press: San Diego, California, USA, 1999.

\bibitem{Ade:2015lrj}
 Ade, P.A.R.; Aghanim, A.N.; Arnaud, M.; Arroja, F.; Ashdown, M.; Aumont, J.; Baccigalupi, C.; Ballardini, M.; Banday, A.J.; Barreiro, R.B.; et al. 
  Planck 2015 results. XX. Constraints on inflation. {\bf 2015},
 arXiv:1502.02114 [astro-ph.CO].


\bibitem{Lidsey:1995np} 
  Lidsey, J.E.; Liddle, A.R.; Kolb, E.W.; Copeland, E.J.; Barreiro, T.;  
  Abney, M.
  Reconstructing the inflation potential : An overview.
  {\it Rev.\ Mod.\ Phys.} {\bf 1997}, {\it 69}, 373.



  \bibitem{Elizalde:2008yf}
  Elizalde, E.; Nojiri, S.; Odintsov, S.D.; Saez-Gomez, D.; Faraoni, V.
  Reconstructing the universe history, from inflation to acceleration, with phantom and canonical scalar fields.
  {\it Phys.\ Rev.\ D} {\bf 2008}, {\it 77}, 106005.


\bibitem{Bamba:2014daa}
  Bamba, K.; Nojiri, S.; Odintsov, S.D.
Reconstruction of scalar field theories realizing inflation
consistent with the Planck and BICEP2 results.
 {\it Phys.\ Lett.\ B} {\bf 2014}, {\it 737} 374.


\bibitem{Bamba:2014daab}
Capozziello, S.; Faraoni, V.
\textit{Beyond Einstein Gravity};
Springer: Dordrecht, The Netherlands, 2010.

\bibitem{Bamba:2014daac}
de la Cruz-Dombriz, A.; Saez-Gomez, D.
  Black holes, cosmological solutions, future singularities,
  and their thermodynamical properties in modified gravity theories.
  {\it Entropy} {\bf 2012}, {\it 14}, 1717.



\bibitem{SaezGomez:2008uj}
  Saez-Gomez, D.
  Modified $f(R)$ gravity from scalar-tensor theory and inhomogeneous EoS dark energy.
  {\it Gen.\ Rel.\ Grav.} {\bf 2009}, {\bf 41}, 1527.

\bibitem{Nojiri:2009kx}
  Nojiri, S.; Odintsov, S.D.; Saez-Gomez, D.
  Cosmological reconstruction of realistic modified F(R) gravities.
  {\it Phys.\ Lett.\ B} {\bf 2009}, {\it 681}, 74.


  \bibitem{Bamba:2014wda}
  Bamba, K.; Nojiri, S.; Odintsov, S.D.; S\'aez-G\'omez, D.
  Inflationary universe from perfect fluid and $F(R)$ gravity and its comparison with observational data.
  {\it Phys.\ Rev.\ D} {\bf 2014}, {\it 90}, 124061.
  
  \bibitem{DeFelice:2011jm}
  De Felice, A.; Tsujikawa, S.; Elliston, G.; Tavakol, R.
  Chaotic inflation in modified gravitational theories.
  {\it J. Cosmol. Astropart. Phys.} {\bf 2011}, {\it 1108}, 021.
   
\bibitem{staro}
	Starobinsky, A.A.
  A New Type of Isotropic Cosmological Models Without Singularity.
  {\it Phys.\ Lett.\ B} {\bf 1980}, {\it 91}, 99.

\bibitem{Copeland:2013vva} 
  Copeland, E.J.; Rahmede, C.; Saltas, I.D. 
  Asymptotically Safe Starobinsky Inflation,
  {\it Phys.\ Rev.\ D} {\bf 2015}, {\it 91}, 103530.


\bibitem{delaCruz-Dombriz:2016bjj} 
  de la Cruz-Dombriz, A.; Elizalde, E.; Odintsov, S.D.;  Saez-Gomez, D.
  Spotting deviations from $R^2$ inflation. \textit{J. Cosmol. Astropart. Phys.} \textbf{2016}, 
  arXiv:1603.05537 [gr-qc].

\bibitem{Sasaki:1995aw} 
  Sasaki, M.; Stewart, E.D.
  A General analytic formula for the spectral index of the density perturbations produced during inflation.
  {\it Prog.\ Theor.\ Phys.} {\bf 1996}, {\it 95}, 71.


\bibitem{Avelino:2008mv}
  Avelino, P.P.; Menezes, R.; Sousa, L.
  p-brane dynamics in N+1-dimensional FRW universes.
  {\it Phys.\ Rev.\ D} {\bf 2009}, {\it 79}, 043519.
  
\bibitem{Sousa:2011ew}
  Sousa, L.; Avelino, P.P.
  p-brane dynamics in (N+1)-dimensional FRW universes: A unified framework.
  {\it Phys.\ Rev.\ D} {\bf 2011}, {\it 83}, 103507.
  
\bibitem{Sousa:2011iu}
  Sousa, L.; Avelino, P.P.
  The cosmological evolution of p-brane networks.
  {\it Phys.\ Rev.\ D} {\bf 2011}, {\it 84}, 063502.
  
\bibitem{Martins:1996jp}
  Martins, C.J.A.P.; Shellard, E.P.S.
  Quantitative string evolution.
  {\it Phys.\ Rev.\ D} {\bf 1996}, {\it 54}, 2535.
  
\bibitem{Martins:2000cs}
  Martins, C.J.A.P.; Shellard, E.P.S.
  Extending the velocity dependent one scale string evolution model.
  {\it Phys.\ Rev.\ D} {\bf 2002}, {\it 65}, 043514.
  
\bibitem{Leite:2012vn}
  Leite, A.M.M.; Martins, C.J.A.P.; Shellard, E.P.S.
  Accurate Calibration of the Velocity-dependent One-scale Model for Domain Walls.
  {\it Phys.\ Lett.\ B} {\bf 2013}, {\it 718}, 740.

\bibitem{Oliveira:2012nj}
  Oliveira, M.F.; Avgoustidis, A.;  Martins, C.J.A.P.
  Cosmic string evolution with a conserved charge.
  {\it Phys.\ Rev.\ D} {\bf 2012}, {\it 85}, 083515.
  
\bibitem{Avgoustidis:2007aa}
  Avgoustidis, A.; Shellard, E.P.S.
  Velocity-Dependent Models for Non-Abelian/Entangled String Networks.
  {\it Phys.\ Rev.\ D} {\bf 2008}, {\it 78}, 103510.
  
\bibitem{Nunes:2011sf}
  Nunes, A.S.; Avgoustidis, A.; Martins, C.J.A.P.; Urrestilla, J.
  Analytic Models for the Evolution of Semilocal String Networks.
  {\it Phys.\ Rev.\ D} {\bf 2011}, {\it 84}, 063504.
  
\bibitem{Martins:2009hj}
  Martins, C.J.A.P.
  Evolution of Hybrid Defect Networks.
  {\it Phys.\ Rev.\ D} {\bf 2009}, {\it 80}, 083527.

\bibitem{Avelino:2000iy}
  Avelino; P.P.; Martins, C.J.A.P.
  Primordial adiabatic fluctuations from cosmic defects.
  {\it Phys.\ Rev.\ Lett.} {\bf 2000}, {\it 85}, 1370.

\bibitem{Wu:1998mr}
  Wu, J.H.P.; Avelino, P.P.; Shellard, E.P.S.; Allen, B.
  Cosmic strings, loops, and linear growth of matter perturbations.
  {\it Int.\ J.\ Mod.\ Phys.\ D} {\bf 2002}, {\it 11}, 61.
  
\bibitem{Ade:2013xla}
Ade, P. A. R.; Aghanim, N.; Armitage-Caplan, C.; Arnaud, M.; Ashdown, M.; Atrio-Barandela, F.; Aumont, J.; Baccigalupi, C.; Banday, A. J.; Barreiro, R. B.; {et al.} 
  [Planck Collaboration],
  Planck 2013 results. XXV. Searches for cosmic strings and other topological defects.
  {\it Astron.\ Astrophys.} {\bf 2014}, {\it 571}, A25.
  
\bibitem{Bardeen:1980kt}
  Bardeen, J.M.
  Gauge Invariant Cosmological Perturbations.
  {\it Phys.\ Rev.\ D} {\bf 1980}, {\it 22}, 1882.
  
\bibitem{Martin:2014rqa}
  Martin, J.; Ringeval, C.; Vennin, V.
  How Well Can Future CMB Missions Constrain Cosmic Inflation?.
  {\it J. Cosmol. Astropart. Phys.} {\bf 2014}, {\it 1410}, 038.
  
\bibitem{Sousa:2015cqa}
  Sousa, L.; Avelino, P.P.
  Cosmic Microwave Background anisotropies generated by domain wall networks.
  {\it Phys.\ Rev.\ D} {\bf 2015} {\it 92}, 083520.

\bibitem{Pogosian:1999np}
  Pogosian, P.; Vachaspati, T.
  Cosmic microwave background anisotropy from wiggly strings.
  {\it Phys.\ Rev.\ D} {\bf 1999}, {\it 60}, 083504.
  
\bibitem{Pogosian:2006hg}
  Pogosian, L.; Wasserman, I.; Wyman, M.
  On vector mode contribution to CMB temperature and polarization from local strings. {\bf 2006},
  {astro-ph/0604141.}
  
\bibitem{Vilenkin:1981bx}
  Vilenkin, A.
  Gravitational radiation from cosmic strings.
  {\it Phys.\ Lett.\ B} {\bf 1981}, {\it 107}, 47.
  
\bibitem{Brandenberger:1986xn}
  Brandenberger, R.H.; Albrecht, A.; Turok, N.
  Gravitational Radiation From Cosmic Strings and the Microwave Background.
  {\it Nucl.\ Phys.\ B} {\bf 1986}, {\it 277}, 605.
  
\bibitem{Sanidas:2012ee}
  Sanidas, S.A.; Battye, R.A.; Stappers, B.W.
  Constraints on cosmic string tension imposed by the limit on the stochastic gravitational wave background from the European Pulsar Timing Array.
  {\it Phys.\ Rev.\ D} {\bf 2012}, {\it 85}, 122003.
  
\bibitem{Sousa:2013aaa}
  Sousa, L.; Avelino, P.P.
  Stochastic Gravitational Wave Background generated by Cosmic String Networks: Velocity-Dependent One-Scale model versus Scale-Invariant Evolution.
  {\it Phys.\ Rev.\ D} {\bf 2013}, {\it 88}, 023516.
  
\bibitem{TheLIGOScientific:2014jea}
   Aasi, J.; Abbott, B. P.; Abbott, R.; Abbott, T.; Abernathy, M. R.; Ackley, K.; Adams, C.; Adams, T.; Addesso, P.; Adhikari, R. X.; {et al.}
  {Advanced LIGO.}
  {\it Class.\ Quant.\ Grav.} {\bf 2015}, {\it 32}, 074001.
  
\bibitem{Accadia:2015pda}
  Accadia, T.; Agathos, M.; Allocca, A.; Astone, P.; Ballardin, G.; Barone, F.; Barsuglia, M.; Basti, A.; Bauer, T. S.; M. Bejger, M.; {et al.}
  Advanced Virgo Interferometer: A Second Generation Detector for Gravitational Waves Observation.
  
\bibitem{Dufaux:2012rs}
  Dufaux, J.F.
  Cosmological Backgrounds of Gravitational Waves and eLISA.
  {\it ASP Conf.\ Ser.} {\bf 2013}, {\it 467}, 91.
  
\bibitem{Manchester:2007mx}
  Manchester, R.N.
  The Parkes Pulsar Timing Array.
  {\it AIP Conf.\ Proc.} {\bf 2008}, {\it 983}, 584.
  
\bibitem{Ferdman:2010xq}
  Ferdman, R.D.; van Haasteren, R.; Bassa, C.G.; Burgay, M.; Cognard, I.; Corongiu, A.; D'Amico, N.; Desvignes, G.; Hessels, J.W.T.; Janssen, G.H.; et al.
  The European Pulsar Timing Array: current efforts and a LEAP toward the future.
  {\it Class.\ Quant.\ Grav.} {\bf 2010}, {\it 27}, 084014.
  
\bibitem{Hiramatsu:2013qaa}
  Hiramatsu, T.; Kawasaki, M.; Saikawa, K.
  On the estimation of gravitational wave spectrum from cosmic domain walls.
  {\it J. Cosmol. Astropart. Phys.} {\bf 2014}, {\it 1402}, 031.
  
\bibitem{Hiramatsu:2010yz}
  Hiramatsu, T.; Kawasaki, M.; Saikawa, K.
  Gravitational Waves from Collapsing Domain Walls.
  {\it J. Cosmol. Astropart. Phys.} {\bf 2010}, {\it 1005}, 032.







\bibitem[Bleem et al.(2015)]{2015ApJS..216...27B} 
Bleem, L.E.; Stalder, B.; de Haan, T.; Aird, K.A.; Allen, S.W.; Applegate, D.E.; Ashby, L.N.; Bautz, M.; Bayliss, M.; Benson, B.A.; et al. 
Galaxy Clusters discovered via the Sunyaev-Zel'dovic effect in the 
2500-square-degree SPT-SZ Survey. 
{\it Astrophys. J. Suppl.} {\bf 2015}, {\it 216}, 27.

\bibitem[Planck Collaboration et al.(2015)]{2015arXiv150201597P} 
Ade, P.A.R.; Aghanim, N.; Arnaud, M.; Ashdown, M.; Aumont, J.; Baccigalupi, C.; Banday, A.J.; Barreiro, R.B.; Bartlett, J.G.; Bartolo, N.; et al. Planck Collaboration, 
Planck 2015 results. XXIV. Cosmology from Sunyaev-Zeldovich cluster counts. { 2015}, arXiv:1502.01597

\bibitem[Andr{\'e} et al.(2014)]{2014JCAP...02..006A} 
Andr{\'e}, P; Baccigalupi, C; Banday, A; Barbosa, D; Barreiro, B; Bartlett, J; Bartolo, N; Battistelli, E; Battye, R; Bendo, G; { et al. }
PRISM (Polarized Radiation Imaging and Spectroscopy Mission): An extended white paper. 2014, {\it J. Cosmol. Astropart. Phys.} {\bf 2014}, {\it 2}, 006.


\bibitem[Jenkins et al.(2001)]{2001MNRAS.321..372J} 
Jenkins, A.; Frenk, C. S.; White, S. D. M.; Colberg, J. M.; Cole, S.; Evrard, A. E.; Couchman, H. M. P.; Yoshida, N.
The mass function of dark matter haloes.
{\it Mon. Not. Roy. Astron. Soc.} {\bf 2001}, {\it 321}, 372.

\bibitem[Ramos et al.(2012)]{2012ApJ...757...44R} 
Ramos, E.P.R.G.; da Silva, A.J.C.; Liu, G.-C. 
Cosmic Microwave Background Induced Polarization from Single Scattering by Clusters of Galaxies and Filaments. 
{\it Astrophys.  J.} {\bf 2012}, {\it 757}, 44.

\bibitem[Press \& Schechter(1974)]{1974ApJ...187..425P} 
Press, W.H.; Schechter, P. 
Formation of Galaxies and Clusters of Galaxies by Self-Similar Gravitational Condensation. 
{\it Astrophys.  J.} {\bf 1974}, {\it 187}, 425.

\bibitem[Kravtsov \& Borgani(2012)]{2012ARA&A..50..353K} 
Kravtsov, A.V.; Borgani, S. 
Formation of Galaxy Clusters. 
{\it Ann. Rev. Astron. Astrophys.} {\bf 2012}, {\it 50}, 353.

\bibitem[Wagner \& Verde(2012)]{2012JCAP...03..002W} 
Wagner, C.; Verde, L. 
N-body simulations with generic non-Gaussian initial conditions II: halo bias. 
{\it J. Cosmol. Astropart. Phys.} {\bf 2012}, {\it 3}, 002.

\bibitem[Achitouv et al.(2015)]{2015arXiv151101494A} 
Achitouv, I.; Baldi, M.; Puchwein, E.; Weller, J. 
The Imprint of $f(R)$ Gravity on Non-Linear Structure Formation. 
{2015}, arXiv:1511.01494

\bibitem[Nunes et al.(2006)]{2006A&A...450..899N} 
Nunes, N.J.; da Silva, A.C.; Aghanim, N. 
Number counts in homogeneous and inhomogeneous dark energy models. 
{\it Astron. Astrophys.} {\bf 2006}, \textit{450}, 899.

\bibitem[Kaiser(1986)]{1986MNRAS.222..323K} 
Kaiser, N. 
Evolution and clustering of rich clusters. 
{\it Mon. Not. Roy. Astron. Soc.} {\bf 1986}, {\it 222}, 323.

\bibitem[Kay et al.(2007)]{2007MNRAS.377..317K} 
Kay, S.~T.; da Silva, A.~C.; Aghanim, N.; Blanchard, A.; Liddle, A.~R.; Puget, J.-L.; Sadat, R.; Thomas, P.~A.
The evolution of clusters in the CLEF cosmological simulation: X-ray structural and scaling properties. {\it Mon. Not. Roy. Astron. Soc.} {\bf 2007}, {\it 377}, 317. 

\bibitem[da Silva et al.(2009)]{2009MNRAS.396..849D} 
da Silva, A.~C.; Catalano, A.; Montier, L.; Pointecouteau, E.; Lanoux, J.; Giard, M.
The impact of dust on the scaling properties of galaxy clusters.
{\it Mon. Not. Roy. Astron. Soc.} {\bf 2009}, {\it 396}, 849.

\bibitem[Aghanim et al.(2009)]{2009A&A...496..637A} 
Aghanim, N.; da Silva, A.C.; Nunes, N.J. 
Cluster scaling relations from cosmological hydrodynamic simulations in a dark-energy dominated universe.
{\it Astron. Astrophys.} {\bf 2009}, {\it 496}, 637.

\bibitem[Trindade \& da Silva(2016)]{2016arXiv160309270T} 
Trindade, A.M.M.; da Silva, A. Effect of Priomordial non-Gaussianities on Galaxy Clusters Scaling Relations. { 2016}, arXiv:1603.09270

\bibitem[Navarro et al.(1997)]{1997ApJ...490..493N} 
Navarro, J.F.; Frenk, C.S.; White, S.D.M. 
A Universal Density Profile from Hierarchical Clustering. 
{\it Astrophys.  J.} {\bf 1997}, {\it 490}, 493.

\bibitem[Nagai et al.(2007)]{2007ApJ...668....1N} 
Nagai, D.; Kravtsov, A.V.; Vikhlinin, A. 
Effects of Galaxy Formation on Thermodynamics of the Intracluster Medium. 
{\it Astrophys.  J.} {\bf 2007}, {\it 668}, 1.

\bibitem[da Silva et al.(2001)]{2001MNRAS.326..155D} 
da Silva, A.C.; Barbosa, D.; Liddle, A.R.; Thomas, P.A. 
Hydrodynamical simulations of the Sunyaev-Zel'dovich effect: the kinetic effect. 
{\it Mon. Not. Roy. Astron. Soc.} {\bf 2001}, {\it 326}, 155.



\bibitem[Planck Collaboration et al.(2015)]{2015arXiv150201598P} 
 Ade, P. A. R.; Aghanim, N.; Arnaud, M.; Ashdown, M.; Aumont, J.; Baccigalupi, C.; Banday, A. J.; Barreiro, R. B.; Barrena, R.; Bartlett, J. G.; {et al.} Planck Collaboration, Planck 2015 results. XXVII. The Second Planck Catalogue of Sunyaev-Zeldovich Sources. {\bf 2015}, arXiv:1502.01598.

\bibitem[Planck Collaboration et al.(2014)]{2014A&A...571A..20P}  
Ade, P. A. R.; Aghanim, N.; Armitage-Caplan, C.; Arnaud, M.; Ashdown, M.; Atrio-Barandela, F.; Aumont, J.; Baccigalupi, C.; Banday, A. J.; Barreiro, R. B.; {et al.} Planck Collaboration,
Planck 2013 results. XX. Cosmology from Sunyaev-Zeldovich cluster counts. 
{\it Astron. Astrophys.} {\bf 2014}, {\it 571}, A20.

\bibitem[Planck Collaboration et al.(2014)]{2014A&A...571A..21P} 
 Ade, P. A. R.; Aghanim, N.; Armitage-Caplan, C.; Arnaud, M.; Ashdown, M.; Atrio-Barandela, F.; Aumont, J.; Baccigalupi, C.; Banday, A. J.; Barreiro, R. B.; {et al.} Planck Collaboration, 
Planck 2013 results. XXI. Power spectrum and high-order statistics of the Planck all-sky Compton parameter map.  {\bf 2014}, {\em Astron. Astrophys.}, {\bf 571}, A21.

\bibitem[Planck Collaboration et al.(2015)]{2015arXiv150201596P}  
Aghanim, N.; Arnaud, M.; Ashdown, M.; Aumont, J.; Baccigalupi, C.; Banday, A. J.; Barreiro, R. B.; Bartlett, J. G.; Bartolo, N.; Battaner, E.; {et al.} Planck Collaboration,  Planck 2015 results. XXII. A map of the thermal Sunyaev-Zeldovich effect. {\bf 2015}, arXiv:1502.01596.


\bibitem[de Martino et al.(2012)]{2012ApJ...757..144D} 
de Martino, I.; Atrio-Barandela, F.; da Silva, A.; Ebeling, H.; Kashlinsky, A.; Kocevski, D.; Martins, C. J. A. P. Measuring the Redshift Dependence of the Cosmic Microwave Background Monopole Temperature with Planck Data. {\em Astrophys.  J.} {\bf 2012},  {\em 757}, 144.

\bibitem[de Martino et al.(2015)]{2015ApJ...808..128D} 
de Martino, I.; G{\'e}nova-Santos, R.; Atrio-Barandela, F.; Ebeling, H.; Kashlinsky, A.; Kocevski, D.; Martins, C. J. A. P.
Constraining the Redshift Evolution of the Cosmic Microwave Background Blackbody Temperature with PLANCK Data. {\bf 2015}, {\em Astrophys.  J.}, {\em 808}, 128.



%

\bibitem[Rudjord et al.(2009)]{2009ApJ...701..369R} 
Rudjord, {\O}.; Hansen, F.~K.; Lan, X.; Liguori, M.; Marinucci, D.; {Matarrese}, S.
An Estimate of the Primordial Non-Gaussianity Parameter $f_NL$ Using the Needlet Bispectrum from WMAP. {\em Astrophys.  J.} {\bf 2009}, {\em 701}, 369.

\bibitem[Komatsu et al.(2009)]{2009ApJS..180..330K} 
Komatsu, E.; Dunkley, J.; Nolta, M. R.; Bennett, C. L.; Gold, B.; Hinshaw, G.; Jarosik, N.; Larson, D.; Limon, M.; Page, L.; {et al.}
Five-Year Wilkinson Microwave Anisotropy Probe Observations: Cosmological Interpretation, {\em Astrophys. J. S.} {\bf 2009}, {\em 180}, 330.

\bibitem[Curto et al.(2009)]{2009ApJ...706..399C} 
Curto, A.; Mart{\'{\i}}nez-Gonz{\'a}lez, E.; Barreiro, R.B. 
Improved Constraints on Primordial Non-Gaussianity for the Wilkinson Microwave Anisotropy Probe 5-Year Data, {\em Astrophys.  J.} {\bf 2009}, {\em 706}, 399.


\bibitem[Planck Collaboration et al.(2014)]{2014A&A...571A..24P} 
Ade, P. A. R.; Aghanim, N.; Armitage-Caplan, C.; Arnaud, M.; Ashdown, M.; Atrio-Barandela, F.; Aumont, J.; Baccigalupi, C.; Banday, A. J.; Barreiro, R. B.; {et al.} Planck Collaboration, Planck 2013 results. XXIV. Constraints on primordial non-Gaussianity. {\em Astron. Astrophys.} {\bf 2014}, {\em 571}, A24.


\bibitem[Planck Collaboration et al.(2015)]{2015arXiv150201592P}  
Ade, P. A. R.; Aghanim, N.; Arnaud, M.; Arroja, F.; Ashdown, M.; Aumont, J.; Baccigalupi, C.; Ballardini, M.; Banday, A. J.; Barreiro, R. B.; {et al.} {Planck Collaboration, Planck 2015 results. XVII. Constraints on primordial non-Gaussianity.} {\bf 2015}, arXiv:1502.01592 [astro-ph.CO].

\bibitem[Sefusatti \& Komatsu(2007)]{2007PhRvD..76h3004S} 
Sefusatti, E.; Komatsu, E. 
Bispectrum of galaxies from high-redshift galaxy surveys: Primordial non-Gaussianity and nonlinear galaxy bias, {\em Phys. Rev. D} {\bf 2007}, {\em 76}, 083004.

\bibitem[Matarrese \& Verde(2008)]{2008ApJ...677L..77M} 
Matarrese, S.; Verde, L. 
The Effect of Primordial Non-Gaussianity on Halo Bias. {\em Astrophys.  J.. Lett.} {\bf 2008}, {\em 677}, L77.

\bibitem[Giannantonio et al.(2012)]{2012MNRAS.422.2854G} 
Giannantonio, T.; Porciani, C.; Carron, J.; Amara, A.; Pillepich, A. 	
Constraining primordial non-Gaussianity with future galaxy surveys. {\em Mon. Not. Roy. Astron. Soc.} {\bf 2012}, {\em 422}, 2854.

\bibitem[Sch{\"a}fer et al.(2012)]{2012MNRAS.421..797S} 
Sch{\"a}fer, B.M.; Grassi, A.; Gerstenlauer, M.; Byrnes, C.T. 
A weak lensing view on primordial non-Gaussianities. {\em Mon. Not. Roy. Astron. Soc.} {\bf 2012}, {\em 421}, 797.

\bibitem[Hilbert et al.(2012)]{2012MNRAS.426.2870H} 
Hilbert, S.; Marian, L.; Smith, R.E.; Desjacques, V. Measuring primordial non-Gaussianity with weak lensing surveys. {\em Mon. Not. Roy. Astron. Soc.} {\bf 2012}, {\em 426}, 2870.

\bibitem[Tashiro \& Ho(2013)]{2013MNRAS.431.2017T} 
Tashiro, H.; Ho, S. 
Constraining primordial non-Gaussianity with CMB-21 cm cross-correlations? {\em Mon. Not. Roy. Astron. Soc.} {\bf 2013}, {\em 431}, 2017.

\bibitem[Takeuchi et al.(2012)]{2012PhRvD..85d3518T} 
Takeuchi, Y.; Ichiki, K.; Matsubara, T. 
Application of cross correlations between CMB and large-scale structure to constraints on the primordial non-Gaussianity. {\em Phys. Rev. D} {\bf 2012}, {\em 85}, 043518.

\bibitem[Robinson \& Baker(2000)]{2000MNRAS.311..781R} 
Robinson, J.; Baker, J.E. 
Evolution of the cluster abundance in non-Gaussian models. {\em Mon. Not. Roy. Astron. Soc.} {\bf 2000}, {\em 311}, 781.

\bibitem[Matarrese et al.(2000)]{2000ApJ...541...10M} 
Matarrese, S.; Verde, L.; Jimenez, R. 
The Abundance of High-Redshift Objects as a Probe of Non-Gaussian Initial Conditions. {\em Astrophys.  J.} {\bf 2000}, {\em 541}, 10.

\bibitem[Kamionkowski et al.(2009)]{2009JCAP...01..010K} 
Kamionkowski, M.; Verde, L.; Jimenez, R. 
The void abundance with non-gaussian primordial perturbations. {\em J. Cosmol. Astropart. Phys.} {\bf 2009}, {\em 1}, 010.

\bibitem[Lam et al.(2009)]{2009MNRAS.399.1482L} 
Lam, T.Y.; Sheth, R.K.; Desjacques, V. 
The initial shear field in models with primordial local non-Gaussianity and implications for halo and void abundances. {\em Mon. Not. Roy. Astron. Soc.} {\bf 2009}, {\em 399}, 1482.

\bibitem[D'Amico et al.(2011)]{2011PhRvD..83b3521D} 
D'Amico, G.; Musso, M.; Nore{\~n}a, J.; Paranjape, A. 	
Excursion sets and non-Gaussian void statistics. {\em Phys. Rev. D} {\bf 2011}, {\em 83}, 023521.

\bibitem[Sekiguchi \& Yokoyama(2012)]{2012arXiv1204.2726S} 
Sekiguchi, T.; Yokoyama, S. 
Void bias from primordial non-Gaussianities. {2012}, arXiv:1204.2726 [astro-ph.CO].

\bibitem[Fergusson \& Shellard(2009)]{2009PhRvD..80d3510F} 
Fergusson, J.R.; Shellard, E.P.S.
Shape of primordial non-Gaussianity and the CMB bispectrum. {\em Phys. Rev. D} {\bf 2009}, {\em 80}, 043510.

\bibitem[LoVerde et al.(2008)]{2008JCAP...04..014L} 
LoVerde, M.; Miller, A.; Shandera, S.; Verde, L. 	
Effects of scale-dependent non-Gaussianity on cosmological structures. {\em J. Cosmol. Astropart. Phys.} {\bf 2008}, {\em 4}, 014.

\bibitem[Giannantonio \& Porciani(2010)]{2010PhRvD..81f3530G} 
Giannantonio, T.; Porciani, C. 
Structure formation from non-Gaussian initial conditions: Multivariate biasing, statistics, and comparison with N-body simulations. {\em Phys. Rev. D} {\bf 2010}, {\em 81}, 063530.

\bibitem[Maggiore \& Riotto(2010)]{2010ApJ...717..526M} 
Maggiore, M.; Riotto, A. 
The Halo Mass Function from Excursion Set Theory. III. Non-Gaussian Fluctuations. {\em Astrophys.  J.} {\bf 2010}, {\em 717}, 526.

\bibitem[D'Amico et al.(2011)]{2011JCAP...02..001D} 
D'Amico, G.; Musso, M.; Nore{\~n}a, J.; Paranjape, A. 
An improved calculation of the non-Gaussian halo mass function. {\em J. Cosmol. Astropart. Phys.} {\bf 2011}, {\em 2}, 001.

\bibitem[LoVerde \& Smith(2011)]{2011JCAP...08..003L} 
LoVerde, M.; Smith, K.M.
The non-Gaussian halo mass function with $f_NL$, $g_NL$ and $\tau_{NL}$. {\em J. Cosmol. Astropart. Phys.} {\bf 2011}, {\em 8}, 003.

\bibitem[Achitouv \& Corasaniti(2012)]{2012JCAP...02..002A} 
Achitouv, I.E.; Corasaniti, P.S. 	
Non-Gaussian halo mass function and non-spherical halo collapse: Theory vs. simulations. {\em J. Cosmol. Astropart. Phys.} {\bf 2012}, {\em 2}, 002.

\bibitem[D'Aloisio et al.(2013)]{2013MNRAS.428.2765D} 
D'Aloisio, A.; Zhang, J.; Jeong, D.; Shapiro, P.R. 
Halo statistics in non-Gaussian cosmologies: the collapsed fraction, conditional mass function and halo bias from the path-integral excursion set method. {\em Mon. Not. Roy. Astron. Soc.} {\bf 2013}, {\em 428}, 2765.


\bibitem[Zentner(2007)]{2007IJMPD..16..763Z} 
Zentner, A.R. 
The Excursion Set Theory of Halo Mass Functions, Halo Clustering, and Halo Growth. {\em Intern. J. Mod. Phys. D} {\bf 2007}, {\em 16}, 763.

\bibitem[Maggiore \& Riotto(2010)]{2010ApJ...711..907M} 
Maggiore, M.; Riotto, A. 
The Halo Mass Function from Excursion Set Theory. I. Gaussian Fluctuations with Non-Markovian Dependence on the Smoothing Scale. {\em Astrophys.  J.}, {\bf 2010}, {\em 711}, 907. 

\bibitem[Trindade et al.(2012)]{2012MNRAS.424.1442T} 
Trindade, A.M.M.; Avelino, P.P.; Viana, P.T.P. 
A new signature of primordial non-Gaussianities from the abundance of galaxy clusters. {\em Mon. Not. Roy. Astron. Soc.} {\bf 2012}, {\em 424}, 1442.

\bibitem[Trindade et al.(2013)]{2013MNRAS.435..782T} 
Trindade, A.M.M.; Avelino, P.P.; Viana, P.T.P. 
Biased cosmological parameter estimation with galaxy cluster counts in the presence of primordial non-Gaussianities. {\em Mon. Not. Roy. Astron. Soc.} {\bf 2013},  {\em 435}, 782.

\bibitem[da Silva et al.(2004)]{2004MNRAS.348.1401D} 
da Silva, A.C.; Kay, S.T.; Liddle, A.R.; Thomas, P.A. 
Hydrodynamical simulations of the Sunyaev-Zel'dovich effect: cluster scaling relations and X-ray properties. {\em Mon. Not. Roy. Astron. Soc.} {\bf 2004}, {\em 348}, 1401.





\bibitem{glbook} Schneider, P.; Ehlers, J.; Falco, E.E. {\em Gravitational Lenses}; Springer: Berlin, Germany, {1992}.

\bibitem{saasfee} Schneider, P.; Kochanek, C.S.; Wambsganss, J. 
{\em Gravitational
Lensing : Strong, Weak and Micro}; Springer: Berlin, Germany, {2006}.

\bibitem{narayan} Narayan, R.; Bartelmann, M. 
Gravitational Lensing. In {\em Formation of Structure in the Universe}; Dekel, A., Ostriker, J.P., Eds.; Cambridge University Press: Cambridge, UK, {1999}; {p. 360.}

\bibitem{bartelmann} Bartelmann, M. 
Gravitational Lensing. {\em Class. Quantum Gravity} {\bf 2010}, {\em 27}, 233001.

\bibitem{figtypesgl} Fort, B.; Mellier, Y.
Arc(let)s in clusters of galaxies. {\em Astron. Astrophys. Rev.} {\bf 1994}, {\em 5}, 239--292

\bibitem{cosmicshear80} Valdes, F.; Jarvis, J.F.; Tyson, J.A. 
Alignment of faint galaxy images---Cosmological distortion and rotation. {\em Astrophys.  J.} {\bf 1983}, {\em 271}, 431--441.

\bibitem{cosmicshear90} Mould, J.; Blandford, R.; Villumsen, J.; Brainerd, T.; Smail, I.; Small, T.; Kells, W. 
A search for weak distortion of distant galaxy images by large-scale structure. {\em Mon. Not. Roy. Astron. Soc.} {\bf 1994}, {\em 271}, 31--38.

\bibitem{schneiderobs} Schneider, P.; van Waerbeke, L.;  Jain, B. Kruse, G. A new measure for cosmic shear. {\em Mon. Not. Roy. Astron. Soc.} {\bf 1998}, {\em 296}, 873--892.

\bibitem{2000vw} 
Van Waerbeke, L.; Mellier, Y.; Erben, T.; Cuillandre, J. C.; Bernardeau, F.; Maoli, R.; Bertin, E.; McCracken, H. J.; Le F\`{e}vre, O.; Fort, B.; {et al. }
Detection of correlated galaxy ellipticities from CFHT data: first evidence for gravitational lensing by large-scale structures.  {\em Astron. Astrophys.} {\bf 2000}, {\em 358}, 30--44.

\bibitem{2000bacon} 
Bacon, D.J.;  Refregier, A.R.;  and Ellis, R.S. 
Detection of weak gravitational lensing by large-scale structure.  {\em Mon. Not. Roy. Astron. Soc.} {\bf 2000}, {\em 318}, 625--640.

\bibitem{2000wittman} 
Wittman, D.M.; Tyson, J.A.; Kirkman, D.; Dell'Antonio, I.; Bernstein, G.
Detection of weak gravitational lensing distortions of distant galaxies by cosmic dark matter at large scales.  {\em Nature} {\bf 2000}, {\em 405}, 143--148.

\bibitem{2000kaiser} 
Kaiser, N.; Wilson, G.; Luppino, G. 
Large-Scale Cosmic Shear Measurements. {2000}, arXiv:astro-ph/0003338.

\bibitem{cscompilation} 
Hetterscheidt, M.; Simon, P.; Schirmer, M.; Hildebrandt, H.; Schrabback, T.; Erben, T.; Schneider, P.
GaBoDS: The Garching-Bonn deep survey. VII. Cosmic shear analysis. {\em Astron. Astrophys.} {\bf 2007}, {\em 468}, 859--876.


\bibitem{fucfhtls} 
Fu, L.; Semboloni, E.; Hoekstra, H.; Kilbinger, M.; van Waerbeke, L.; Tereno, I.; Mellier, Y.; Heymans, C.; Coupon, J.; Benabed, K.; {et al.} 
Very weak lensing in the CFHTLS wide: Cosmology from cosmic shear in the linear regime. {\em Astron. Astrophys.} {\bf 2008}, {\em 479}, 9--25.


\bibitem{cfhtlenserben} 
Erben, T.; Hildebrandt, H.; Miller, L.; van Waerbeke, L.; Heymans, C.; Hoekstra, H.; Kitching, T. D.; Mellier, Y.; Benjamin, J.; Blake, C.; {et al.} 
CFHTLenS: The Canada-France-Hawaii Telescope Lensing Survey---Imaging data and catalogue products. {\em Mon. Not. Roy. Astron. Soc.} {\bf 2013}, {\em 433}, 2545--2563.

\bibitem{cfhtlenslensfit} 
Miller, L.; Heymans, C.; Kitching, T. D.; van Waerbeke, L.; Erben, T.; Hildebrandt, H.; Hoekstra, H.; Mellier, Y.; Rowe, B. T. P.; Coupon, J.; {et al.} 
Bayesian galaxy shape measurement for weak lensing surveys---III. Application to the Canada-France-Hawaii Telescope Lensing Survey. {\em Mon. Not. Roy. Astron. Soc.} {\bf 2013}, {\em 429}, 2858--2880.

\bibitem{cfhtlensphotoz} 
Benjamin, J.; Van Waerbeke, L.; Heymans, C.; Kilbinger, M.; Erben, T.; Hildebrandt, H.; Hoekstra, H.; Kitching, T.~D.; Mellier, Y.; Miller, L.; {et al.} CFHTLenS tomographic weak lensing: quantifying accurate redshift distributions.  {\em Mon. Not. Roy. Astron. Soc.} {\bf 2013}, {\em 431}, 1547--1564.

\bibitem{cfhtlensia} 
Heymans, C.; Van Waerbeke, L.; Miller, L.; Erben, T.; Hildebrandt, H.; Hoekstra, H.; Kitching, T. D.; Mellier, Y.; Simon, P.; Bonnett, C.;
{et al.} CFHTLenS: the Canada-France-Hawaii Telescope Lensing Survey. {\em Mon. Not. Roy. Astron. Soc.} {\bf 2012}, {\em 427}, 146--166.

\bibitem{cfhtlensmodgravsimpson} 
Simpson, F.; Heymans, C.; Parkinson, D.; Blake, C.; Kilbinger, M.; Benjamin, J.; Erben, T.; Hildebrandt, H.; Hoekstra, H.; Kitching, T.D.
{et al. }
CFHTLenS: Testing the laws of gravity with tomographic weak lensing and redshift-space distortions. {\em Mon. Not. Roy. Astron. Soc.} {\bf 2012}, {\em 429}, 2249--2263.

\bibitem{cfhtlensdossett} 
Dossett, J.N.; Ishak, M.; Parkinson, D.; Davis, T.M. Constraints and tensions in testing general relativity from Planck and CFHTLenS data including intrinsic alignment systematics. {\em Phys. Rev. D} {\bf 2015}, {\em 92}, 023003.

\bibitem{cfhtlensmodgravvw}
Harnois-D\'{e}raps, J.; Munshi, D.; Valageas, P.; van Waerbeke, L.; Brax, P.; Coles, P.; Rizzo, L.
Testing modified gravity with cosmic shear. {\em Mon. Not. Roy. Astron. Soc.} {\bf 2015}, {\em 454}, 2722--2735.


\bibitem{munshi} Munshi, D.; Valageas, P.; Van Waerbeke, L.; Heavens, A. Cosmology with Weak Lensing Surveys. {\em Phys. Rep.} {\bf 2008}, {\em 462}, 67--121.


\bibitem{kilbinger} Kilbinger, M.  
Cosmology with cosmic shear observations: A review. {\em Rep. Prog. Phys.} {\bf 2015}, {\em 78}, 086901.


\bibitem{mabertschinger} Ma, C.P.; Bertschinger, E.C
osmological Perturbation Theory in the Synchronous and Conformal Newtonian Gauges. {\em Astrophys.  J.} {\bf 1995}, {\em 455}, 7.

\bibitem{consistentPPF} Baker, T.; Ferreira, P.G.; Skordis, C.; Zuntz, J. 
Towards a fully consistent parameterization of modified gravity. {\em J. Cosmol. Astropart. Phys.} {\bf 2012}, {\em 6}, 32.

\bibitem{bertschingerPPF} Bertschinger, E. 
On the Growth of Perturbations as a Test of Dark Energy and Gravity. {\em Astrophys.  J.} {\bf 2006}, {\em 648}, 797--806.

\bibitem{caldwellPPF} Caldwell, R.; Cooray, A.; Melchiorri, A. 
Constraints on a new post-general relativity cosmological parameter. {\em Phys. Rev. D} {\bf 2007}, {\em 76}, 023507.

\bibitem{husawicki} Hu, W.; Sawicki, I. 
Parametrized post-Friedmann framework for modified gravity. {\em Phys. Rev. D} {\bf 2007}, {\em 76}, 104043.

\bibitem{amendolaPPF} Amendola, L.; Kunz, M.; Sapone, D. 
Measuring the dark side (with weak lensing). {\em J. Cosmol. Astropart. Phys.} {\bf 2008}, {\em 4}, 13.

\bibitem{thiswork} Tereno, I.; Semboloni, E.; Schrabback, T. 
COSMOS weak-lensing constraints on modified gravity. {\em Astron. Astrophys.} {\bf 2011}, {\em 530}, 68.

\bibitem{schrabbackCOSMOS} Schrabback, T.; Hartlap, J.; Joachimi, B.; Kilbinger, M.;  Simon, P. 
Evidence of the accelerated expansion of the Universe from weak lensing tomography with COSMOS. {\em Astron. Astrophys.} {\bf 2010}, {\em 516}, 63.

\bibitem{hutomo} Hu, W. 
Power Spectrum Tomography with Weak Lensing. {\em Astrophys.  J. Lett.} {\bf 1999}, {\em 522}, L21--L24.

\bibitem{daniel2010} 
Daniel, S.F.; Linder, E.V.; Smith, T.L.; Caldwell, R.~R.; Cooray, A.; Leauthaud, A.; Lombriser, L.
Testing general relativity with current cosmological data. {\em Phys. Rev. D} {\bf 2010}, {\em 81}, 123508.

\bibitem{lindergrowth} Linder, E.V. 
Cosmic growth history and expansion history. {\em Phys. Rev. D} {\bf 2005}, {\em 72}, 043529.

\bibitem{lindercahn} Linder, E.V.; Cahn, R.N. Parameterized beyond-Einstein growth. {\em Astropart. Phys.} {\bf 2007}, {\em 28}, 481--488.

\bibitem{camb} Lewis, A.; Bridle, S. Cosmological parameters from CMB and other data: A Monte Carlo approach. {\em Phys. Rev. D} {\bf 2002}, {\em 66}, 103511.

\bibitem{wmap7} 
Komatsu, E.; Smith, K. M.; Dunkley, J.; Bennett, C. L.; Gold, B.; Hinshaw, G.; Jarosik, N.; Larson, D.; Nolta, M. R.; Page, L.; {et al.} 
Seven-year Wilkinson Microwave Anisotropy Probe (WMAP) Observations: Cosmological Interpretation. {\em Astrophys. J. Suppl.} {\bf 2011}, {\em 192}, 18.

\bibitem{transfer} 
Eisenstein, D.; Hu, W. 
{Baryonic Features in the Matter Transfer Function}. {\em Astrophys. J.} {\bf 1998}, {\em 496}, 605--614.

\bibitem{halofit} 
Smith, R. E.; Peacock, J. A.; Jenkins, A.; White, S. D. M.; Frenk, C. S.; Pearce, F. R.; Thomas, P. A.; Efstathiou, G.; Couchman, H. M. P.
Stable clustering, the halo model and non-linear cosmological power spectra. {\em Mon. Not. Roy. Astron. Soc.} {\bf 2003}, {\em 341}, 1311--1332.

\bibitem{valageas} 
Valageas, P.; Nishimichi, T. Taruya, Atsushi. 
Matter power spectrum from a Lagrangian-space regularization of perturbation theory. {\em Phys. Rev. D} {\bf 2013}, {\em 87}, 083522.

\bibitem{tessacosmicshear} 
Leonard, C.D.; Baker, T.; Ferreira, P.G. 
Exploring degeneracies in modified gravity with weak lensing. {\em Phys. Rev. D} {\bf 2015}, {\em 91}, 083504.

\bibitem{halofitnew} 
Takahashi, R.; Sato, M.; Nishimichi, T.; Taruya, A.; Oguri, M. Revising the Halofit Model for the Nonlinear Matter Power Spectrum. 
{\em Astrophys. J.} {\bf 2012}, {\em 761}, 152--162.

\bibitem{coyote} 
Heitmann, K.; Lawrence, E.; Kwan, J.; Habib, S.; Higdon, D.
The Coyote Universe Extended: Precision Emulation of the Matter Power Spectrum. {\em Astrophys. J.} {\bf 2014}, {\em 780}, 111--129.

\bibitem{zhaopca} 
Zhao, G.B.; Pogosian, L.; Silvestri, A.; Zylberberg, J. Cosmological Tests of General Relativity with Future Tomographic Surveys. {\em Phys. Rev. Lett.} {\bf 2009}, {\em 103}, 241301.

\bibitem{schmidt} 
Schmidt, F. 
Weak lensing probes of modified gravity. {\em Phys. Rev. D} {\bf 2008}, {\em 78}, 043002.

\bibitem{egproposal} 
Zhang, P.; Liguori, M.; Bean, R.; Dodelson, S. Probing gravity at cosmological scales by measurements which test the relationship between gravitational lensing and matter overdensity. {\em Phys. Rev. Lett.} {\bf 2007}, {\em 99}, 141302.

\bibitem{egnew} 
Leonard, C.D.; Ferreira, P.G.; Heymans, C. 
Testing gravity with EG: Mapping theory onto observations. {\em arXiv preprint} {\bf 2015}, arXiv:1510.04287.

\bibitem{egdetection} 
Reyes, R.;  Mandelbaum, R.;  Seljak, U.; Baldauf, T.; Gunn, J.~E.; Lombriser, L.; Smith, R.~E.
Confirmation of general relativity on large scales from weak lensing and galaxy velocities. {\em Nature} {\bf 2010}, {\em 464}, 256--258.


\bibitem{betoule_2014}
Betoule, M.; Kessler, R.; Guy, J.; Mosher, J.; Hardin, D.; Biswas, R.; Astier, P.; El-Hage, P.; Konig, M.; Kuhlmann, S.; {et al.} 
Improved Cosmological Constraints from a Joint Analysis of the SDSS--II and SNLS Supernova Samples. 
{\it Astron. Astrophys.} {\bf 2014}, {\it 568}, 32.

\bibitem{kalus_2013}
Kalus, B.; Schwarz, D. J.; Seikel, M.; Wiegand, A.
Constraints on anisotropic cosmic expansion from supernova. 
{\it Astron. Astrophys.} {\bf 2013}, {\it 553}, A56.

\bibitem{wojtak_2014}
Wotjak, R.; Knebe, A.; Watson, W.~A.; Iliev, I.~T.; He{\ss}, S.; Rapetti, D.; Yepes, G.; Gottl{\"o}ber, S.
Cosmic Variance of the Local Hubble Flow in Large--scale Cosmological Simulations.  
{\it Mon. Not. Roy. Astron. Soc.} {\bf 2014}, {\it 438}, 1805. 

\bibitem{bengaly_2015}
Bengaly, C.A.P.; Bernui, A.; Alcaniz, J.S.
Probing cosmological isotropy with Type Ia Supernova. 
{\it Astrophys.  J.} {\bf 2015}, {\it 808}, 39.

\bibitem{cooray_2010}
Cooray, A.; Holz, D.E.; Caldwell, R. 
Measuring Dark Energy Spatial Inhomogeneity with Supernova Data. 
{\it J. Cosmol. Astropart. Phys.} {\bf 2010}, {\it 11}, 15. 

\bibitem{marra_2014}
Valkenburg, W.; Marra, V.; Clarkson, C. 
Testing the Copernican principle by constraining spatial homogeneity. 
{\it Mon. Not. Roy. Astron. Soc.} {\bf 2014}, {\it 438}, 6.

\bibitem{gorski_2005}
G\'orski, K.M.; Hivon, E.; Banday, A. J.; Wandelt, B. D.; Hansen, F. K.; Reinecke, M.; Bartelmann, M.
HEALPix: A Framework for High-Resolution Discretization and Fast Analysis of Data Distributed on the Sphere.
 {\it Astrophys.  J.} {\bf 2005}, {\it 622}, 759. 

\bibitem{rasanen_2006}
R\"as\"anen, S.
Accelerated Expansion from Structure Formation. 
 {\it J. Cosmol. Astropart. Phys.} {\bf 2006}, {\it 11}, 003.

\bibitem{carvalho_2016}
Carvalho, C.S.; Basilakos, S.
Angular distribution of cosmological parameters as a probe of inhomogeneities: A kinematic parametrisation.
{\bf 2016}, arXiv:1603.07519 [astro-ph.CO].







\end{thebibliography}


%


%

\end{document}